%% file: main.tex
\documentclass{article}
    \usepackage[utf8]{inputenc}
    \usepackage[margin=2.5cm]{geometry}
    \input{preamble}
    \usepackage{amsmath, amssymb, amsthm}
    
    \title{Fast Confirmation Rule for Ethereum's Consensus Protocol}
    \author{Roberto Saltini\\
       Ethereum Foundation\\
      \and Mikhail Kalinin\\
       Consensys\\
      \and Luca Zanolini\\
       Ethereum Foundation\\
      \and Francesco D'Amato\\
       Ethereum Foundation\\
      \and Aditya Asgaonkar\\
        Click\footnote{Work done while at the Ethereum Foundation.}\\
      \and Chenyi Zhang\\
        University of Canterbury\footnote{Work done while at Consensys.}\\
    }
    \date{}
\begin{document}
    \maketitle
    
\begin{abstract}
A \emph{Confirmation Rule} is an algorithm run by network nodes to determine whether a block will remain permanently in the canonical chain.
The only Confirmation Rule currently available in Ethereum's consensus protocol, Gasper, is \FFG finalization.
While it tolerates asynchronous network conditions, it is slow: in the best case, a transaction takes 13 to 19 minutes to confirm, depending on when it is submitted.

We devise a \emph{Fast Confirmation Rule} (FCR) for Gasper that, under synchrony and the assumptions stated in this paper, achieves a \emph{best-case} confirmation time of 12 seconds -- a single slot -- providing an order-of-magnitude improvement over \FFG finalization.
The rule is complementary to finalization: users who trust synchrony obtain fast confirmations, while finalization remains available as a fallback that tolerates asynchrony.
Gasper is an \emph{ebb-and-flow} protocol~\cite{DBLP:conf/sp/NeuTT21}: it combines \LMDGHOST, a fork-choice rule providing fast progress under synchrony, with \FFGCASPER, a finality gadget providing finality under partial synchrony. The main technical difficulty is to reason jointly about these two components, so that a block confirmed by \LMDGHOST cannot be filtered out by \FFGCASPER's rules.
We prove that the rule satisfies both safety -- confirmed blocks remain canonical -- and monotonicity -- a confirmed block remains confirmed at all future times.
\end{abstract}

\section{Introduction}

A \emph{Confirmation Rule} is an algorithm run by network nodes to determine whether a given block will remain permanently in the canonical chain.
In Bitcoin, for instance, a block is confirmed (with high probability) once it has a sufficiently long chain of successors, competing branches are shorter, honest nodes control the majority of hashing power, and the network is synchronous~\cite{nakamoto2008bitcoin}.

Ethereum's proof-of-stake consensus protocol, Gasper~\cite{gasper}, combines two components: \LMDGHOST, a synchronous fork-choice rule that outputs the canonical chain, and \FFGCASPER~\cite{casper}, a partially synchronous finality gadget that periodically finalizes blocks on the canonical chain via a two-round voting mechanism.
In \FFGCASPER, validators vote on \emph{checkpoints} -- pairs of a block and an epoch -- to drive the \emph{justification} (a checkpoint receiving a two-thirds supermajority of votes) and \emph{finalization} (a justified checkpoint whose immediate successor is also justified) process.
The main interaction between the two components is through a \emph{checkpoint filter}: before running \LMDGHOST, the fork-choice function \LMDGHOSTHFC prunes blocks based on the current \FFG state -- for instance, blocks that conflict with the \emph{latest justified checkpoint} (a checkpoint that has received a two-thirds supermajority of \FFG votes) -- restricting the fork-choice tree to branches consistent with the finality component.

The only confirmation mechanism currently available to Ethereum users is \FFG finalization.
Since finalization requires two rounds of supermajority checkpoint votes spanning two to three epochs, the best-case confirmation time is 13 to 19 minutes, depending on when the transaction is submitted within the epoch.
While this provides safety under partial synchrony with a one-third Byzantine stake bound, many applications -- such as payments or exchange deposits -- would benefit from significantly faster confirmation when network conditions permit.

Under synchrony, attestation weights can in principle establish block stability within a single slot.
However, the checkpoint filter introduces a fundamental difficulty: a block with overwhelming weight support can still be removed from the fork-choice tree -- for instance, if a conflicting checkpoint becomes justified at an epoch boundary.
Confirmation arguments must therefore jointly reason about \LMDGHOST-level weight comparisons and \FFG-level justification dynamics -- two layers that operate on different time scales and under different fault-tolerance thresholds.

We devise a \emph{Fast Confirmation Rule} (FCR) for Gasper that, under synchrony and the assumptions stated in this paper, achieves a best-case confirmation time of 12 seconds -- a single slot -- providing an order-of-magnitude improvement over \FFG finalization.
The rule is complementary to finalization: users who trust synchrony can obtain fast confirmations, while finalization remains available as a fallback that tolerates asynchrony.
The confirmed block advances monotonically (each new confirmed block is a descendant of the previous one), ensuring that users are never worse off than under finalization alone.

The rule is executed at the beginning of every slot.
Its core is a per-block safety predicate that checks whether the observed attestation support for a block exceeds a threshold that ensures that no other block can receive more support.
This predicate is applied to every block between the last confirmed block and the block being considered for confirmation.
To ensure that confirmed blocks are not later removed by filtering, the rule incorporates several predicates. For instance, one predicate checks whether justification of the current-epoch checkpoint can be guaranteed, ensuring the confirmed branch survives the checkpoint filter, and another checks whether a conflicting checkpoint can still be justified, in which case confirmation is frozen until the situation resolves.
At epoch boundaries, a reconfirmation step re-checks that previously confirmed blocks are still confirmed to ensure that the safety argument does not depend on synchrony holding since an arbitrarily distant past (see Section~\ref{sec:monotonicity} for details).
%
%

We prove that the resulting rule satisfies both safety -- confirmed blocks remain canonical -- and, under an additional assumption, monotonicity -- the confirmed block only advances.

The paper proceeds in two parts.
We first establish a confirmation rule for plain \LMDGHOST, i.e., without any influence from \FFGCASPER, deriving a weight-based condition sufficient to guarantee that a block remains canonical.
We then extend this rule to \LMDGHOSTHFC, adding checks that ensure the confirmed branch survives \LMDGHOSTHFC's filtering and introducing a reconfirmation step performed at each epoch boundary.

The remainder of this paper is organized as follows.
Section~\ref{sec:model} introduces the system model and provides a formal definition of Gasper.
Section~\ref{sec:conf-rule-lmd-ghost} presents a confirmation rule for plain \LMDGHOST, establishing the weight-based safety condition.
Section~\ref{sec:conf-rule-ldmghosthfc} extends this rule to \LMDGHOSTHFC, incorporating the checkpoint filter and the full algorithm with all of its cases.
Section~\ref{sec:confirmation-delay} analyzes confirmation delay as a function of the fraction of committee weight not attesting in support of the proposed block, and presents an empirical evaluation on Ethereum mainnet data.
Formal lemmas and proofs can be found in \Cref{sec:appendix-proofs}.
\Cref{sec:iterative-version} presents an equivalent iterative formulation of the algorithm that is closer to the actual implementation.
\Cref{sec:implementation-note-chain} discusses another subtle difference between the actual implementation and the algorithm presented in this paper.

    \section{System Model}
    \label{sec:model}
    
    \paragraph{Validators.}
    We consider a (possibly infinite) set $\allvals$
    of \emph{validators} that communicate with one another by exchanging messages.
    Each validator is associated with a distinct cryptographic identity, and the public keys are shared among all validators.
    A validator that follows the protocol is \emph{honest}; a validator that deviates arbitrarily from the protocol is \emph{Byzantine}.
    We let $\honvals \subseteq \allvals$ denote the set of honest validators and $\advvals := \allvals \setminus \honvals$ the set of Byzantine validators.
    The set $\honvals$ is unknown.
    We assume the existence of a probabilistic polynomial-time adversary that controls all Byzantine validators throughout the protocol execution and may delay message delivery, subject to the constraints specified in the network model below.
    Cryptographic signatures are unforgeable: the adversary cannot produce a valid signature without the corresponding private key.
    The signer of a message $m$ is denoted by $\signer(m)$.

    \paragraph{Network Model.}
    
    Honest validators have synchronized clocks.
    Any message sent at time $t$ is received by all honest validators by time $\max(t, \GST)+\Delta$, where $\GST$ denotes the \emph{global stabilization time} and $\Delta$ the message delay bound.
    Both $\GST$ and $\Delta$ are assumed to be known to anyone executing the Confirmation Rule.
    The value of $\Delta$ is required to satisfy an upper bound imposed by the protocol's timing structure, as detailed in Section~\ref{sec:gasper}.%
    \footnote{
        Compared to the classical definition of synchronous networks, we allow an initial period of asynchrony before $\GST$.
        Compared to the classical definition of partially synchronous networks~\cite{DBLP:journals/jacm/DworkLS88}, we assume that both $\GST$ and $\Delta$ are known.}
    
    \paragraph{Gossiping.}
    We assume that any honest validator immediately gossips (\ie, broadcasts) any message that it receives.
    
    \paragraph{View.}
    The \emph{view} of a validator is the set of all messages that the validator has received.
    More specifically, we use $\viewattime[time=t,val=v]$ to denote the set of all messages received by validator $v$ at time $t$.

    \subsection{Gasper}\label{sec:gasper}
    
    Gasper is a proof-of-stake consensus protocol made of two components~\cite{DBLP:conf/sp/NeuTT21}, namely \LMDGHOSTHFC and \FFGCASPER~\cite{casper}.
    The former is a synchronous consensus protocol that works under dynamic participation and outputs a \emph{canonical chain}, while the latter is a partially synchronous protocol, also referred to as \emph{finality gadget}, whose role is to finalize blocks in the canonical chain and preserve safety of such \emph{finalized blocks} during asynchronous periods.
    In the following, we summarise the concepts and properties pertaining to Gasper that are required 
    in the remaining part of this work.

    \paragraph{Time and Slots.} Time is organized into a consecutive sequence of \emph{slots}.
    We denote the time at which a slot $s$ begins with $\slotstart(s)$, and use $\slot(t)$ to denote the slot associated with time $t$, \ie, $\slot(t) = s$ implies that $t \in [\slotstart(s),\slotstart(s+1))$.
    
    \paragraph{Epochs.} A sequence of $\slotsperepoch$ consecutive slots forms an \emph{epoch} where $\slotsperepoch \geq 2$.
    Epochs are numbered starting from 0.
    We use $\firstslot(e)$ and $\lastslot(e)$ to denote the first slot and last slot of epoch $e$, respectively, \ie, $\firstslot(e) := e \slotsperepoch$ and $\lastslot(e):= (e+1) \slotsperepoch - 1$.
    We write $\epoch(s)$ for the epoch associated with slot $s$, \ie, $\epoch(s) = e$ implies $s \in [\firstslot(e), \lastslot(e)]$.
    Also we define $\epoch(t) := \epoch(\slot(t))$.
    Finally, we let $\slotstart(e) := \slotstart(\firstslot(e))$.
    We say that a block $b$ \emph{belongs to} epoch $e$ if $\epoch(b) = e$.

    \paragraph{Validator Sets and Committees.}
    In the view of an honest validator $v$ at time $t$, only a finite subset of validators are \emph{active} for each epoch $e$.
    We denote this set by $\commatepoch[epoch=e,val=v,when=t]{\allvals}$ and refer to it as the \emph{validator set} for epoch $e$ in the view of validator $v$ at time $t$.
    The validator set for epoch $e$ is partitioned into \emph{committees}, one per slot.
    The union of all committees from slot $s$ to slot $s'$ inclusive, in the view of validator $v$ at time $t$, is denoted by
    $\commfromslot[from=s,to=s',val=v,when=t]{\allvals}$.
    We also define the corresponding unions indexed by a block $b$:
    \[
    \commfromblock[from=b,to=s',when=t,val=v]{\allvals}
    := \commfromslot[from=\slot(b),to=s',when=t,val=v]{\allvals},
    \qquad
    \commfromafterblock[from=b,to=s',when=t,val=v]{\allvals}
    := \commfromslot[from=\slot(b)+1,to=s',when=t,val=v]{\allvals}.
    \]
    The honest subsets are defined as
    $\commatepoch[epoch=e,val=v,when=t]{\honvals} := \commatepoch[epoch=e,val=v,when=t]{\allvals}\cap\honvals$ and
    $\commfromslot[from=s,to=s',val=v,when=t]{\honvals} := \commfromslot[from=s,to=s',val=v,when=t]{\allvals}\cap\honvals$.
    
    \paragraph{Blocks.} {Blocks} are the data structures used by Gasper to order \emph{transactions}.
    Except for the \emph{genesis} block $\genesis$, each block $b$ has a \emph{parent} which we denote via the writing $\parent(b)$.
    Conversely, $b \neq \genesis$ is said to be a \emph{child} of $\parent(b)$.
    We use the notations $b_a \prec b_d$  and $b_d \succ b_a$ to indicate that block $b_a$ can be reached from block $b_d$ by recursively applying the function $\parent(\cdot)$ to $b_d$.
    We define $b\preceq b'$ naturally as $b\prec b'$ or $b=b'$.
    For any two blocks $b$ and $b'$ such that $b \preceq b'$, we say that $b$ is an \emph{ancestor} of $b'$ and that $b'$ is a \emph{descendant} of $b$. 
    We say that two blocks $b$ and $b'$ \emph{conflict} iff neither of the two blocks is the descendant of the other, \ie, $b' \npreceq b \land b \npreceq b'$.
    We let $\children[blck=b,view=\View]$ be the set of blocks in $\View$ that have $b$ as parent, \ie, $\children[blck=b,view=\View] := \{b' \in \View : \parent(b') = b \}$.
    The \emph{chain} of a block $b$, which we denote as $\chain(b)$, is the set of all ancestors of $b$, \ie, $\chain(b) := \{b' : b' \preceq b\}$.
    Sometimes, we refer to ``the chain of $b$'' simply as ``chain $b$''.
    We assume that set of all possible blocks to be finite, which implies that the chain of any block is also finite and includes $\genesis$.
    To each block $b$ is associated a slot $\slot(b)$ which, as we will see later, is supposed to indicate the slot during which block $b$ is proposed.
    By definition, $\slot(\genesis) = 0$.
    Slot $\slot(b)$ is also called the \emph{height} of $b$.
    Therefore, we say that a block $b$ is \emph{higher} than a block $b'$ iff $\slot(b) > \slot(b')$.
    A block $b$ is considered \emph{valid} only if (i) the signer of $b$ is the expected proposer for slot $\slot(b)$ and (ii) $\slot(b) > \slot(\parent(b))$.
    We let $\blocksinview(\View)$ denote the set of all valid blocks in the view $\View$.
    
    \paragraph{Shorthand for the slot after parent's slot.}
    For any non-genesis block $b$, we use the notation
    \[
    \parent(b)+ \;:=\; \slot(\parent(b))+1,
    \]
    i.e., $\parent(b)+$ denotes the slot immediately after {the slot of} the parent of $b$.
    Accordingly, any committee range written as ``from $\parent(b)+$ to $s'$" is to
    be read as ranging from slot $\slot(\parent(b))+1$ up to slot $s'$ (inclusive).
    
    \paragraph{Checkpoints.} A \emph{checkpoint} is a tuple $C = (\block(C), \epoch(C))$ composed of a block $\block(C)$ and an epoch $\epoch(C)$.
    We say that a checkpoint $C$ \emph{belongs to} epoch $e$ if $\epoch(C) = e$.
    For any epoch $e'$, the checkpoint $C$ in the chain of $b$ with $\epoch(C) = e'$ is denoted by $\chkp(b, e')$ and corresponds to the pair $(b_c, e')$, where $b_c$ is the block in the chain of $b$, \ie, $b_c \preceq b$, with the highest slot such that $\epoch(\slot(b_c)-1) < e'$
    \footnote{Note that the definition works for any $e'$, including $e' > \epoch(b)$.}.
    The \emph{latest checkpoint} of a block $b$, also simply called \emph{$b$'s checkpoint}, denoted by $\latestchkp(b)$, is defined as $\latestchkp(b) := \chkp(b, \epoch(b))$.
    Checkpoint $(\genesis,0)$ is defined as the \emph{genesis checkpoint}.
    We write $b \preceq C$ to mean $b \preceq \block(C)$, while $C \prec b$ means that $\chkp(b, \epoch(C)) = C$.
    Also, $C_d \succ C_a$ or $C_a \prec C_d$ means that $\epoch(C_a) < \epoch(C_d)$ and $C_a \prec \block(C_d)$.
    The definition of conflicting blocks is naturally extended to blocks and checkpoints.
    We say that a block or checkpoint $x$ \emph{conflicts} with a block or checkpoint $x'$ iff $x \npreceq x' \land x' \npreceq x$.
    Also, we say that a checkpoint $C$ is valid to mean that $\block(C)$ is valid.
    Finally, we establish a strict order between any two checkpoints $C$ and $C'$ by defining $C < C'$ to mean $\epoch(C) < \epoch(C')$. 
    
   \paragraph{Effective balance.}
    An \emph{\teffbalass{}} is a mapping $\calB: \allvals \to \mathbb{R}_{\geq 0}$ which assigns to each validator $v$ its \emph{effective balance}.
    Intuitively, the effective balance of a validator determines its \emph{voting power} within the protocol.
    Each block $b$ contains an \teffbalass{} which we denote as $\effbalass(b)$.
    We define $\weightofset[chkp=\calB]{v} := \calB(v)$ and, given a finite set of validators $\calX \subseteq \allvals$,
    we define $\weightofset[chkp=\calB]{\calX} := \sum_{v \in \calX} \weightofset[chkp=\calB]{v}$.
    Also, we write $\totvalset[chkp=\calB]{\allvals}$ for the set of validators that have a non-zero effective-balance according to $\calB$, \ie, $\totvalset[chkp=\calB]{\allvals} := \{v \in \allvals : \weightofset[chkp=\calB]{v} > 0 \}$.
    We call such a set the total validator set according to $\calB$.
    Generally, hereafter, whenever we define a set of validators  of the form $\calX_{\mathit{parlist}_b}^{\mathit{parlist}_t}$ where $\mathit{parlist}_b$ and $\mathit{parlist}_t$ can be any list of parameters, we implicitly also define $X_{\mathit{parlist}_b}^{\mathit{parlist}_t,\calB} := \weightofset[chkp=\calB]{\calX_{\mathit{parlist}_b}^{\mathit{parlist}_t}}$.\footnote{For
    instance, the sets defined above yield the following weights:
    $\commweightfromblock[from=b,to=s',when=t,val=v,chkp=\calB]{\allvals} := \weightofset[chkp=\calB]{\commfromblock[from=b,to=s',when=t,val=v]{\allvals}}$,
    $\commweightfromafterblock[from=b,to=s',when=t,val=v,chkp=\calB]{\allvals} := \weightofset[chkp=\calB]{\commfromafterblock[from=b,to=s',when=t,val=v]{\allvals}}$,
    $\commweightatepoch[epoch=e,val=v,when=t,chkp=\calB]{\honvals} := \weightofset[chkp=\calB]{\commatepoch[epoch=e,val=v,when=t]{\honvals}}$,
    and
    $\commweightfromslot[from=s,to=s',val=v,when=t,chkp=\calB]{\honvals} := \weightofset[chkp=\calB]{\commfromslot[from=s,to=s',val=v,when=t]{\honvals}}$.}
    Also, whenever using a block $b$ or a checkpoint $C$ in place of an \teffbalass $\calB$, we mean the \teffbalass $\effbalass(b)$ or $\effbalass(\block(C))$, respectively.
    For any valid block $b$, the set $\totvalset[chkp=b]{\allvals}$ is finite.
    Also, by definition, $\commatepoch[epoch=\epoch(\genesis)]{\allvals} = \totvalset[chkp=\genesis]{\allvals}$.

    \paragraph{Changes to the Validator Set and Effective Balances.}
    The Gasper protocol provisions a way to allow both new validators to join the validator set and existing validators to exit the validator set.
    Exiting can be either voluntarily or involuntarily.
    A validator is involuntarily exited if it can be proved that it did not act in accordance to the protocol (see Section~\ref{sec:conf-rule-ldmghosthfc} for details).
    Aside from these changes to the validator set, the effective balance of a validator can also increase (or decrease) due the validator accruing \emph{rewards} (or \emph{penalties}), for performing (or not performing) their duties in a timely manner.
    
    The remaining Gasper-specific definitions are introduced where they are first needed:
    the \LMDGHOST\ fork-choice machinery and vote-support notation are presented in
    Section~\ref{sec:lmd-ghost}, while the \FFGCASPER\ and \LMDGHOSTHFC\ layers
    are presented in Section~\ref{sec:gasper-ffg}.

        \subsection{Confirmation Rule}\label{sec:conf-rule-definition}
    
    A \emph{confirmation rule} is an algorithm that determines whether a
    block is \emph{confirmed}, meaning that it will remain in the canonical
    chain\footnote{The notion of canonical chain is formalized in
    Section~\ref{sec:lmd-ghost}.} of every honest validator under certain
    assumptions.

    \begin{definition}[Confirmation Rule]\label{def:confirmation-rule}
    A \emph{confirmation rule} is a pair $(\mathsf{CONF}, \mathit{sg})$
    consisting of:
    \begin{itemize}
      \item an algorithm $\mathsf{CONF}$ that, for each honest validator $v$,
      block $b$, and time $t$, outputs whether $b$ is confirmed in the view
      of $v$ at time~$t$, and
      \item a \emph{security guard} $\mathit{sg}(b,t,\GST)$, a predicate
      specifying when the rule is guaranteed to be safe.
    \end{itemize}
    
    For any block $b$ and time $t$ such that $\mathit{sg}(b,t,\GST)$ holds,
    the rule must satisfy:
    \begin{enumerate}
      \item \textbf{Safety:}
      If $b$ is confirmed by honest validator $v$ at time $t$, then $b$
      is in the canonical chain of every honest validator $v'$ at all
      times $t' \ge t$.
    
      \item \textbf{Monotonicity:}
      If $b$ is confirmed by honest validator $v$ at time $t$, then $b$
      is confirmed by $v$ at all times $t' \ge t$.
    \end{enumerate}
    \end{definition}

    \section{A safe Confirmation Rule for \LMDGHOST}\label{sec:conf-rule-lmd-ghost}
    
    \paragraph{Why start with plain \LMDGHOST.}
    Our ultimate goal is a confirmation rule for \LMDGHOSTHFC, the fork-choice
    function currently deployed in Ethereum's consensus protocol, which filters
    out blocks that conflict with justified checkpoints.
    Reasoning directly about \LMDGHOSTHFC\ is cumbersome, as the checkpoint
    filter couples fork-choice votes with \FFG finality.
    We therefore begin with the unfiltered \LMDGHOST\ function and derive a
    local, vote-weight condition sufficient to guarantee head stability.
    In the next section, we extend this condition so that confirmed blocks also
    survive the checkpoint filtering imposed by \LMDGHOSTHFC.
    
    We begin by introducing the \LMDGHOST\ fork-choice machinery and
    the vote-support notation required throughout this section.
    
    \subsection{\LMDGHOST\ Definitions}\label{sec:lmd-ghost}
    
    \LMDGHOST, an acronym for Latest Message Driven Greediest Heaviest Observed Sub-Tree ($\lmdghost$), is a synchronous consensus protocol.
    In each slot, a \emph{proposer} constructs a new block~$b$ and sends it to all other validators.
    The other honest validators in the committee of slot $s$ then vote for block~$b$.
    Every validator~$v$ needs to decide where to append a new block (if $v_i$ is a proposer) or which block $v$ should vote for.
    To make this decision, each validator executes a \emph{fork-choice function}, specifically the fork-choice function that we define below.
    
    \paragraph{Fork-choice and Canonical Chain}
    A fork-choice function is a deterministic rule denoted as $\fcparam[fc=\FC,balf=\frakB,val=]$ that accepts as input a (possibly filtered) view $\View$ and a time $t$, outputs a block $b$ and is parametrized by a function $\frakB$ that given in input $\View$ and $t$ outputs the \teffbalass to be used to weigh votes.
    We also define $\fcparam[fc=\FC,balf=\frakB,val=v](t) := \fcparam[fc=\FC,balf=\frakB,val=](\viewattime[time=t,val=v],t)$.
    We say that $\fcparam[fc=\FC,balf=\frakB,val=v](t)$ is the \emph{canonical chain} of validator $v$ at time $t$ according to the fork-choice $\FC_\frakB$.
    
    \paragraph{GHOST Votes and Voting Process.}
    A \GHOST vote $a$ is a tuple $\tuple{\slot(a),\block(a)}$ where, for honest validators, $\slot(a)$ corresponds to the slot during which $a$ has been cast and $\block(a)$ corresponds to the result of the fork-choice function $\fcparam[fc=\FC,balf=\frakB,val=]$ used by validator $v$,\ie, $\block(a) = \fcparam[fc=\FC,balf=\frakB,val=v](t)$.
    We say that a \GHOST vote is in \emph{support} of a block $b$ iff $b \preceq \block(a)$.
    We denote the set of all \GHOST votes in a view $\View$ with $\ghostsinview(\View)$.
    
    \paragraph{\GHOST.}
    \GHOST is a fork-choice function based on the fork-choice procedure introduced by Sompolinsky and Zohar~\cite{ghost}, a greedy algorithm that grows a blockchain on sub-branches with the most activity.
    However, the \GHOST fork-choice, defined in Algorithm~\ref{alg:GHOST}, is vote-based rather than block-based, \ie, it weighs sub-trees based on votes' weight rather than blocks.
    Given a view $\View$ and a block $b$, we define $\ghostvoters[block=b, view=\View]$ to be the set of validators that according to view $\View$ have voted in support of $b$.
    The weight of a block $b$ is then defined as the total effective balance of this set of validators according to the \teffbalass $\frakB(\View,t)$, \ie, $\weightofset[chkp={\frakB(\View,t)}]{\ghostvoters[block=b, view=\View]}$.
    Starting from the $\genesis$ block, \GHOST iterates over a sequence of valid and non-future (\ie, with slot no higher than the current slot) blocks from $\View$, selecting as the next block the descendant of the current block with the highest weight.
    This continues until it reaches a block that does not have any descendant in $\View$, which is the block being output.
    
    \paragraph{\GHOST Equivocation.}
    Two \GHOST votes $a$ and $a'$ are said to be equivocating iff they are from the same validator and same slot but target two different blocks, \ie, $\signer(a) = \signer(a') \land \slot(a) = \slot(a') \land \block(a) \neq \block(a')$.
    Honest validators never sign equivocating \GHOST votes.
    
    \paragraph{\LMDGHOST.}
    \LMDGHOST proceeds by first \emph{filtering} the view and then applying \GHOST to the resulting (filtered) vote set.
    We say that a \GHOST vote $a$ is \emph{invalid} if either:
    (i) $\signer(a)$ is not assigned to the committee of slot $\slot(a)$, i.e.,
    $\signer(a) \notin \commfromslot[from=\slot(a),to=\slot(a)]{\allvals}$; or
    (ii) $a$ votes for a block from the future relative to the vote, i.e.,
    $\slot(\block(a)) > \slot(a)$.
    
    Given a view $\View$ at time $t$, \LMDGHOST constructs a filtered view by:
    (1) removing all \GHOST votes from validators that equivocate in $\View$ (via $\FILEQ$);
    (2) removing votes from the current or future slots, i.e., $\slot(a) \ge \slot(t)$ (via $\FILCUR$);
    (3) removing invalid votes as defined above (via $\FILINV$); and
    (4) keeping, for each validator, only its \emph{latest} remaining vote (highest slot) (via $\FILLMD$).
    \LMDGHOST is then defined as the application of \GHOST to this filtered view:
    \[
    \fcparam[fc=\LMDGHOST,balf=\frakB](\View,t)
    \;=\;
    \fcparam[fc=\GHOST,balf=\frakB]\!\big(\FILLMD(\FILINV(\FILCUR(\FILEQ(\View),t))),\,t\big).
    \]
    
    Finally, any \GHOST vote $a$ that remains after these filters is said to \emph{\LMDGHOST support} a block $b$
    if $\block(a) \succeq b$.
    This is formalized in Algorithm~\ref{alg:lmd-ghost}.
    
    \begin{algorithm}[t]
        \vbox{
        \small
        \begin{numbertabbing}\reset
        xxxx\=xxxx\=xxxx\=xxxx\=xxxx\=xxxx\=MMMMMMMMMMMMMMMMMMM\=\kill
        \textbf{function} $\ghostvoters[block=b, view=\View]$\label{}\\
        \> \textbf{return } $\{\signer(a) : a \in \ghostsinview(\View) \land \block(a) \succeq b \}$\label{}\\[1ex]
        \textbf{function} $\fcparam[fc=\GHOST,balf=\frakB](\View, t)$\label{}\\
        \> $b \gets \genesis$ \label{}\\
        \> \textbf{while} $\exists b' \in \children[blck=b,view=\blocksinview(\View]),\; slot(b') \leq \slot(t)$ \label{}\\
        \>\> $b \gets \argmax_{b' \in \children[blck=b,view=\blocksinview(\View]) \land \slot(b') \leq \slot(t)}\weightofset[chkp={\frakB(\View,t)}]{\ghostvoters[block=b, view={\View}]}$\label{ln:alg:GHOST:argmax}\\
        \> \textbf{end while}\label{}\\
        \> \textbf{return} $b$\label{}\\
        [-5ex]
        \end{numbertabbing}
        }
        \caption{\GHOST fork-choice}\label{alg:GHOST}
    \end{algorithm}
    
    \begin{algorithm}[t]
        \vbox{
        \small
        \begin{numbertabbing}\reset
        xxxx\=xxxx\=xxxx\=xxxx\=xxxx\=xxxx\=MMMMMMMMMMMMMMMMMMM\=\kill
        \textbf{function} $\FILEQ(\View)$\label{}\\
        \> \textbf{return } $\begin{aligned}[t]
            \View \setminus \{a \in \ghostsinview(\View) : \exists a',a'' \in \ghostsinview(\View), \; &\land \signer(a) = \signer(a') = \signer(a'') \\
            &\land \slot(a') = \slot(a'') \\
            &\land \block(a') \neq \block(a'') \}
        \end{aligned}$\label{}\\[1ex]
        \textbf{function} $\FILLMD(\View)$\label{}\\
        \> \textbf{return } $\View \setminus \{a \in \ghostsinview(\View) : \exists a' \in \View, \; \signer(a') = \signer(a) \land \slot(a) < \slot(a')  \}$\label{}\\[1ex]
        \textbf{function} $\FILINV(\View)$\label{}\\
        \> \textbf{return } $\View \setminus \{a \in \ghostsinview(\View) : \slot(a)  \notin \commfromslot[from=\slot(a),to=\slot(a)]{\allvals} \lor \slot(\block(a)) > \slot(a) \}$\label{}\\[1ex]
        \textbf{function} $\FILCUR(\View, t)$\label{}\\
        \> \textbf{return } $\View \setminus \{a \in \blocksinview(\View) : \slot(a) \geq \slot(t)  \}$\label{}\\[1ex]
        \textbf{function} $\fcparam[fc=\LMDGHOST,balf=\frakB](\View, t)$\label{}\\
        \> \textbf{return} $\fcparam[fc=\GHOST,balf=\frakB](\FILLMD(\FILINV(\FILCUR(\FILEQ(\View),t))), t)$\label{}\\
        [-5ex]
        \end{numbertabbing}
        }
        \caption{\LMDGHOST fork-choice}\label{alg:lmd-ghost}
    \end{algorithm}
    
    \paragraph{Proposer Boost.}
    The original version of \LMDGHOST protocol has been shown to suffer from security issues~\cite{DBLP:conf/sp/NeuTT21,DBLP:conf/fc/Schwarz-Schilling22}.
    The \emph{proposer boost technique}~\cite{boost} was later introduced as a mitigation to this issue.
    It requires honest voters to temporarily grant extra weight to the current proposal, if such a block is received in a timely manner.
    Other methodologies~\cite{ethresearch-view-merge-2} have been put forth, although they remain subjects of ongoing investigation~\cite{goldfish,rlmd}.
    In Gasper, the value of the proposer boost value that a validator $v$ assigns at time $t$ is defined as a fraction of the average weight of the committee of a slot according to $\totvalsetweight[chkp={\chkpattime[time=t,val=v]}]{\allvals}$.
    We denote the value of such a fraction with $\boostscore$.
    In general, we write $\boostweight[chkp=\calB]$ to mean the proposer boost value based off the weight total validator set according to $\calB$.
    In summary, the definitions just proved imply that $\boostweight[chkp=\calB] := \frac{\boostscore}{\slotsperepoch}\totvalsetweight[chkp=\calB]{\allvals}$.
    
    \paragraph{Slashing.}
    Participants in Casper must adhere to key rules to ensure integrity.
    Any violation, called a \emph{slashable offence}, is met with a penalty called \emph{slashing}, where the participant's effective balance is (partially) confiscated, the participant is eventually exited from the validator set and the evidence submitter is rewarded.
    Honest validators never commit slashable offences and therefore they are never slashed.
    Evidence of a slashable offence is included in blocks.
    We use the notation $\slashedset[chkp=b]$ to represent the set of validators that have committed slashable offences according to the evidence included in the chain of $b$.
    For any checkpoint $C$, we define $\slashedset[chkp=C] := \slashedset[chkp=\block(C)]$.
    The specifics of the Casper's integrity rules are not provided as they are not required by the remainder of this paper.
    
    \paragraph{Voting Time and Upper Bound for $\Delta$.}
    As detailed later in this section, one of the main duties of validators is casting votes of different types (\FFG and \GHOST).
    If and only if an honest validator $v$ is in the committee of a slot $s$, then, during slot $s$, $v$ casts exactly one vote per type.
    For any slot $s$ and honest validator $v$, we assume that $\Delta$ is less than the time between when $v$ casts any vote in slot $s$ and the beginning of slot $s+1$, \ie, for any slot $s'$ such that $\slotstart(s') \geq \GST$, all the votes sent by honest validators during any slot, up to $s'$ included, are received by any honest validator by time $\slotstart(s'+1)$.
    
    \paragraph{Attestation sets and vote filtering.}
    For a block $b$ and an upper slot index $s'$, let
    $\attsetfromblockunfiltered[from=b,to=s',val=v,when=t]{\allatts}$
    denote the set of validators in $\commfromslot[from=\slot(b),to=s',when=t,val=v]{\allvals}$ that, according to $\viewattime[val=v,time=t]$, have sent a \GHOST vote that \LMDGHOST supports $b$, i.e.,
    \[
    \attsetfromblockunfiltered[from=b,to=s',val=v,when=t]{\allatts}
    := \ghostvoters[block=b,view={\FILLMD(\FILINV(\FILCUR(\FILEQ(\View),t)))}]
    \cap
    \commfromslot[from=\slot(b),to=s',when=t,val=v]{\allvals},
    \]
    where $\ghostvotersfnname$ is defined in Algorithm~\ref{alg:GHOST} and the filtering functions in Algorithm~\ref{alg:lmd-ghost}.
    Let
    $\attsetfromblockunfiltered[from=b,to=s',val=v,when=t]{\honatts}
    := \attsetfromblockunfiltered[from=b,to=s',val=v,when=t]{\allatts}\cap\honvals$
    denote the corresponding honest subset.
    Finally, let
    $\attsetfromblock[from=b,to=s',val=v,when=t]{\slashvals}$
    denote the set of validators in $\commfromblock[from=b,to=s',when=t,val=v]{\allvals}$ that, according to $\viewattime[val=v,time=t]$, have committed a slashable offence\footnote{The corresponding weights, following the convention above, are:
    $\attsetweightfromblock[from=b,to=s',val=v,when=t,chkp=\calB]{\allatts} := \weightofset[chkp=\calB]{\attsetfromblockunfiltered[from=b,to=s',val=v,when=t]{\allatts}}$,
    $\attsetweightfromblock[from=b,to=s',val=v,when=t,chkp=\calB]{\honatts} := \weightofset[chkp=\calB]{\attsetfromblockunfiltered[from=b,to=s',val=v,when=t]{\honatts}}$,
    and
    $\attsetweightfromblock[from=b,to=s',val=v,when=t,chkp=\calB]{\slashvals} := \weightofset[chkp=\calB]{\attsetfromblock[from=b,to=s',val=v,when=t]{\slashvals}}$.}.
    
    \paragraph{Stabilization time.}
    We assume that, if $\slotstart(\lastslot(\epoch(t)-2))\geq\GST$, then from time $t$ onwards all honest validators share the same view of the committee assignment for each slot.
    Since we always work under this condition, for ease of notation we define
    \[
    \GGST := \begin{cases}
        \slotstart(\epoch(\GST)+1), & \text{if } \GST\leq \slotstart(\lastslot(\epoch(\GST))),\\
        \slotstart(\epoch(\GST)+2), & \text{otherwise.}
    \end{cases}
    \]
    This ensures that for any two times $t,t'\geq \GGST$ and any two honest validators $v,v'$,
    $\commfromslot[from=s,to=s',val=v,when=t]{\allvals}=\commfromslot[from=s,to=s',val=v',when=t']{\allvals}$.
    Accordingly, we drop the validator and time parameters and write
    $\commfromslot[from=s,to=s']{\allvals}$,
    $\commfromslot[from=s,to=s']{\honvals}$,
    $\commatepoch[epoch=e]{\allvals}$,
    $\commatepoch[epoch=e]{\honvals}$,
    $\commfromblock[from=b,to=s']{\allvals}$,
    $\commfromafterblock[from=b,to=s']{\allvals}$
    to denote the corresponding objects with parameters $(val=v,when=t)$, for any $t\geq \GGST$ and any honest validator $v$.
    
    \subsection{Safety Condition}\label{sec:lmd-safety-condition}
    
    The safety condition below is parameterized by a checkpoint $C$ used as the \teffbalass{} to weigh votes.
    How to determine such a checkpoint will be explained in \Cref{sec:conf-rule-ldmghosthfc}.
    
    Throughout this section, we develop a Confirmation Rule for
    $\fcparam[fc=\LMDGHOST,balf=\gjviewsym]$, which weighs \GHOST\ votes
    according to the greatest justified checkpoint in the view of a validator,
    under the following two assumptions.
    
    \begin{assumption}\label{assum:no-change-to-the-validator-set}
        The only change that can occur to the validator set and effective balances is due to Byzantine validators potentially getting slashed.
        In other words, no new validator is ever added to the validator set, no rewards are incurred, and honest validators never exit or incur penalties.
    \end{assumption}

    \begin{assumption}\label{assum:beta}
        There exists a constant $\beta$, known to anyone using the Confirmation Rule,
        such that,
        for any
        honest validator $v$,
        time $t\geq \GGST$,
        two slots $s'$ and $s$,
        valid block $b$, 
        and checkpoint $C$ used by any honest validator as the \teffbalass{} to compute weights when evaluating the fork-choice function,
        $\commweightfromslot[from=s,to=s',chkp=C,when=t,val=v]{\honvals} \geq (1-\beta) \commweightfromslot[from=s,to=s',chkp=C,when=t,val=v]{\allvals}$.
    \end{assumption}

    Assumption~\ref{assum:no-change-to-the-validator-set} requires the validator
    set and effective balances to remain static, with the sole exception that
    Byzantine validators may be slashed.\footnote{A previous version of this
    work~\cite{fcr_old} presents an analysis of the Fast Confirmation Rule
    that accounts for validator entries, exits, rewards, and penalties.
    The required modifications do not change the algorithm itself but only
    adjust the confirmation thresholds to account for the variation in
    effective balances and validator set composition.
    These adjustments are small in practice, and we omit them in this
    paper for simplicity.
    We note that~\cite{fcr_old} analyzes a different version of the
    algorithm than the one presented here, but the methodology for
    incorporating these adjustments applies in the same way.}
    Assumption~\ref{assum:beta} requires that, at any time after $\GGST$ and for any
    checkpoint $C$ justified in the chain of a valid block, the fraction of
    Byzantine validators in any set of consecutive committees weighted by $C$ is at most~$\beta$.
    Any party relying on Gasper's finality guarantee already assumes
    $\frac{\totvalsetweight[chkp=\genesis]{\advvals}}{\totvalsetweight[chkp=\genesis]{\allvals}} < \beta'$, for some $\beta' \leq \frac{1}{3}$.
    For a sequence of slots within a single epoch, the Chernoff--Hoeffding
    inequality~\cite{Hoeffding1963} yields that
    $\mathrm{Pr}[\beta \leq \beta' - \epsilon]$ increases exponentially in
    $\epsilon \frac{M}{\slotsperepoch}$, where $M$ denotes the total number of
    validators (unweighted).
    With Ethereum's current validator set of approximately one
    million~\cite{validators} and epoch length $E = 32$~\cite{specs}, the
    probability that $\beta \leq \beta' - \epsilon$ is overwhelming even for
    moderate values of $\epsilon$.
    For intervals spanning multiple epochs, deriving the exact probability is more
    involved; nevertheless, the large ratio of validators to slots per epoch
    suggests that the bound remains tight.
    
    \medskip
    
    Now, fix a block $b$ and assume that $\parent(b)$ is canonical throughout the time
    interval of interest.
    Under this hypothesis, the only way for $b$ to cease
    being canonical is for another child of $\parent(b)$ to win the \LMDGHOST\
    fork-choice selection at $\parent(b)$, which picks the child with the greatest
    subtree weight among all children of $\parent(b)$.
    Accordingly, we seek a condition at $\parent(b)$ ensuring that the subtree
    weight of $b$ strictly exceeds that of every other child, implying that $b$ is canonical.

    \paragraph{The maximum weight $\Phi_{b}^{v,t,C}$.}
    For now, we use $\Phi_{b}^{v,t,C}$ to represent the maximum weight that can support any child of $\parent(b)$ as of time $t$ using $C$ as \teffbalass{}.
    We defer its precise definition to later on.
    A sufficient condition for $b$ to be canonical is that
    the honest weight supporting $b$ exceeds half of this maximum weight.
    
    
    \paragraph{Absolute honest support inequality.}
    With this notation, assuming that $t$ is the time when honest validators cast their votes, the tightened sufficient condition for $b$ to be canonical
    becomes
    \begin{equation}\label{eq:abs-honest-support}
    \attsetweightfromblock[from=b,to=\slot(t)-1,val=v,when=t,chkp=C]{\honatts}
    \;>\;
    \frac{\Phi_{b}^{v,t,C}}{2}.
    \end{equation}

    \paragraph{The \LMDGHOST\ safety condition.}
    The absolute honest support inequality is stated in terms of honest weight which cannot be measured.
    To solve this issue we define an observable indicator and  a predicate on it that is sufficient to imply condition~\eqref{eq:abs-honest-support}.

    \begin{definition}[\LMDGHOST Safety Indicator]\label{def:safety-indicator-2}
        Let $\indicatorfromblock[from=b,to=s,val=v,when=t,chkp=\calB]{\indQ}:=\frac{\attsetweightfromblock[from=b,to=s,val=v,when=t,chkp=\calB]{\allatts}}{\commweightfromafterblock[from=\parent(b),to=s,chkp=\calB,when=t,val=v]{\allvals}}$ be the proportional weight, according to the \teffbalass $\calB$, of the \LMDGHOST support of $b$ against the total weight of the committees between slot $\parent(b)+$ and slot $s$ as per the view of validator $v$ at time $t$.
    \end{definition}

    The indicator $\indQ$ measures the fraction of the total voting weight
    that currently favors the subtree containing~$b$.

    To build intuition, consider first the following simplified version of the predicate that we require the $\indQ$ indicator to satisfy. 
    \begin{equation}\label{eq:cond-on-q}
        \indicatorfromblock[from=b,to=\slot(t)-1,val=v,when=t,chkp=C]{\indQ}
        \;>\;
        \frac{\Phi_{b}^{v,t,C}}
            {2\,\commweightfromafterparentblock[from=b,to=\slot(t)-1,chkp=C,when=t,val=v]{\allvals}}
        \;+\; \beta.
    \end{equation}

    By multiplying both sides of the inequality by $\commweightfromafterparentblock[from=b,to=\slot(t)-1,chkp=C]{\allvals}$, we get
    \begin{equation}\label{eq:cond-on-q-multiplied}
        \attsetweightfromblock[from=b,to=\slot(t)-1,val=v,when=t,chkp={C}]{\allatts}
    \;>\;
    \frac{\Phi_{b}^{v,t,C}}
        {2}
    \;+\; \beta\commweightfromafterparentblock[from=b,to=\slot(t)-1,chkp=C,when=t,val=v]{\allvals}
    \end{equation}

    Essentially, we are requiring that the weight supporting $b$, including both honest and dishonest validators, is larger than half the total weight that can support any children of $\parent(b)$ (i.e., the threshold required for the honest support by condition~\eqref{eq:abs-honest-support} plus the maximum dishonest weight that can support any children of $\parent(b)$.

    Intuitively, if all the dishonest validators equivocate, the remaining honest weight supporting $b$ is enough to satisfy condition~\eqref{eq:abs-honest-support}.

    This argument, however, establishes that $b$ is canonical only within the current slot.

    \paragraph{Establishing that $b$ is canonical for future slots as well.}
    To reason about persistence across future slots, we introduce a second
    indicator that, under our assumptions, is monotonically non-decreasing over time.

    \begin{definition}[Honest \LMDGHOST Safety Indicator]\label{def:safety-indicator-1}
        Let $\indicatorfromblock[from=b,to=s,val=v,when=t,chkp=\calB]{\indP}:=\frac{\attsetweightfromblock[from=b,to=s,val=v,when=t,chkp=\calB]{\honatts}}{\commweightfromblock[from=b,to=s,chkp=\calB,when=t,val=v]{\honvals}}$  be the proportional weight, according to the \teffbalass $\calB$, of the honest \LMDGHOST support of $b$ against the total honest weight between slot $\slot(b)$ and slot $s$ as per the view of validator $v$ at time $t$.
    \end{definition}

    Then, first note that if the following inequality holds
    \begin{equation}\label{eq:cond-on-p}
        \indicatorfromblock[from=b,to=s,val=v,when=t,chkp=C]{\indP}
        \;>\;
        \frac{\Phi_{b}^{v,t,C}}
            {2\,\commweightfromblock[from=b,to=s,chkp=C]{\allvals}(1-\beta)},        
    \end{equation}
    then the absolute honest support inequality~\eqref{eq:abs-honest-support} holds as well, as shown below.
    \[
    \attsetweightfromblock[from=b,to=s,val=v,when=t,chkp=C]{\honatts}
    =
    \indicatorfromblock[from=b,to=s,val=v,when=t,chkp=C]{\indP}
    \commweightfromblock[from=b,to=s,chkp=C]{\honvals}
    >
    \frac{\Phi_{b}^{v,t,C}}
        {2\,\commweightfromblock[from=b,to=s,chkp=C]{\allvals}(1-\beta)}
    \commweightfromblock[from=b,to=s,chkp=C]{\honvals}
    \geq
    \frac{\Phi_{b}^{v,t,C}}
        {2\,\commweightfromblock[from=b,to=s,chkp=C]{\allvals}(1-\beta)}
    \commweightfromblock[from=b,to=s,chkp=C]{\allvals}(1-\beta)   
    =
    \frac{\Phi_{b}^{v,t,C}}{2},
    \]

    \paragraph{Monotonicity of $\indP$.}
    As mentioned above, the key property of the indicator $\indP$ is that it is monotonically non-decreasing.
    Under Assumption~\ref{assum:no-change-to-the-validator-set}, if the honest validators in committees
    of slots $[\slot(t),\slot(t')-1]$ vote in support of $b$, then
    \[
    \indicatorfromblock[from=b,to=\slot(t')-1,val=v',when=t',chkp=C']{\indP}
    \;\ge\;
    \indicatorfromblock[from=b,to=\slot(t)-1,val=v,when=t,chkp=C]{\indP},
    \]
    for any honest validators $v,v'$ and checkpoints $C,C'$ with $C'\succeq C$
    (Lemma~\ref{lem:lmd-p-monotonic}).
    To see why the indicator $P$ is monotonic, let us take a look at what happens when an honest validator $v$ in the committees of slots $[\slot(t),\slot(t')-1]$ votes in support of $b$.
    Consider two cases.
    First, assume that $v$ already belongs to the set $\commweightfromblock[from=b,to=\slot(t)-1,chkp=C]{\honvals}$.
    In this case, $v$'s weight is not added to the denominator of the $P$ indicator, but it might be added to its numerator.
    On the other hand, if $v$ does not already belong to the set $\commweightfromblock[from=b,to=\slot(t)-1,chkp=C]{\honvals}$, then its weight is added to both the numerator and the denominator.
    Therefore, the numerator increases at least as much as the denominator.
    Given that the numerator is lower than the denominator, this leads to the $P$ indicator being monotonically increasing.

    This means that once $\indP$ crosses the required threshold of condition~\eqref{eq:cond-on-p}, it stays above it, and $b$ is
    permanently canonical.
    This monotonicity turns ``$\indP$ is large enough at time $t$'' into
    ``$\indP$ is still large enough at any later time $t'$'', which is what a
    safety propagation argument requires.

    \paragraph{Concluding the argument}
    How do we determine that condition~\eqref{eq:cond-on-p} is satisfied given that we do not know who is honest?
    As formally proved in \Cref{lem:lmd-cond-on-q-implies-cond-on-p},  condition~\eqref{eq:cond-on-q} on $\indQ$ implies condition~\eqref{eq:cond-on-p}.
    The overall argument is illustrated in \Cref{fig:safety-chain}.
 
      \input{predicates}

    \paragraph{Complete $\indQ$-predicate.}
    For pedagogical reasons, above we only introduced a simplified version of the $\indQ$ indicator used by the algorithm.
    The full definition of the condition on $\indQ$ that we use in our algorithm is defined below.
\begin{definition}[Per-block confirmation predicate]\label{def:isOneConfirmed}
    For any block $b$, checkpoint $C$, honest validator $v$, and time $t\geq\GGST$,
    \[
    \var[val=v]{\isOneConfirmed}(b,C,t)
    \;:=\;
    \indicatorfromblock[from=b,to=\slot(t)-1,val=v,when=t,chkp=C]{\indQ}
    \;>\;
    \frac{\Phi_{b}^{v,t,C}}
        {2\,\commweightfromafterparentblock[from=b,to=\slot(t)-1,chkp=C,when=t,val=v]{\allvals}}
    + \beta 
        \frac
            {\commweightfromblock[from=b,to=\slot(t)-1,chkp=C,when=t,val=v]{\allvals}}
            {\commweightfromafterparentblock[from=b,to=\slot(t)-1,chkp=C]{\allvals}}
    - \frac
        {\attsetweightfromblock[from=b,to=\slot(t)-1,val=v,when=t,chkp={C}]{\slashvals}}
        {\commweightfromafterparentblock[from=b,to=\slot(t)-1,chkp=C]{\allvals}}.
    \]
\end{definition}

    As we did for the simplified condition above, the best way to gain an intuition for the definition of $\isOneConfirmed$ is multiplying both sides of the inequality by $\commweightfromafterparentblock[from=b,to=\slot(t)-1,chkp=C]{\allvals}$, which gives us the following inequality
    \begin{equation}\label{eq:cond-on-q-full-multiplied}
        \attsetweightfromblock[from=b,to=\slot(t)-1,val=v,when=t,chkp={C}]{\allatts}
        \;>\;
        \frac{\Phi_{b}^{v,t,C}}
            {2}
        + \beta 
          \commweightfromblock[from=b,to=\slot(t)-1,chkp=C,when=t,val=v]{\allvals}
        - \attsetweightfromblock[from=b,to=\slot(t)-1,val=v,when=t,chkp={C}]{\slashvals}
    \end{equation}

    This works because, as hinted above when discussing the simplified predicate, the term added to $\frac{\Phi_{b}^{v,t,C}}{2}$ needs to be an upper bound on the weight of the adversary included in  $\attsetfromblock[from=b,to=\slot(t)-1,val=v,when=t]{\allatts}$, and
    $\beta 
          \commweightfromblock[from=b,to=\slot(t)-1,chkp=C,when=t,val=v]{\allvals}
        - \attsetweightfromblock[from=b,to=\slot(t)-1,val=v,when=t,chkp={C}]{\slashvals}$  is a tighter bound on this than  $\beta\commweightfromafterparentblock[from=b,to=\slot(t)-1,chkp=C,when=t,val=v]{\allvals}$.
    Indeed, only validators in the committee from $\slot(b)$ onward can cast valid votes for $b$.
    Additionally, those that have already been seen as slashed ($\attsetweightfromblock[from=b,to=\slot(t)-1,val=v,when=t,chkp={C}]{\slashvals}$) have already been discounted and therefore are not included in $\attsetweightfromblock[from=b,to=\slot(t)-1,val=v,when=t,chkp={C}]{\allatts}$.

    \paragraph{Components of $\Phi_{b}^{v,t,C}$.}
    The total weight that can support any children of $\parent(b)$ is defined as
    \begin{equation}\label{eq:phi-def}
    \Phi_{b}^{v,t,C}
    \;:=\;
    \commweightfromafterparentblock[from=b,to=\slot(t)-1,chkp=C]{\allvals}
    +\boostweight[chkp=C]
    -\attsetweighttobediscsimplefromblock[from=b,to=\slot(b)-1,val=v,when=t,chkp=C]{\honattsub}.
    \end{equation}
    It consists of three terms:
    \begin{itemize}
    \item $\commweightfromafterparentblock[from=b,to=\slot(t)-1,chkp=C]{\allvals}$:
    the total committee weight accrued strictly after $\parent(b)$ up to the $\slot(t)-1$, which upper-bounds
    the weight that \emph{could} support another child of $\parent(b)$ at time $t$;
    \item $\boostweight[chkp=C]$: proposer boost, the additive weight that
    \LMDGHOST\ grants to the block proposed in the current slot;
    \item $\attsetweighttobediscsimplefromblock[from=b,to=\slot(b)-1,val=v,when=t,chkp=C]{\honattsub}$:
    a non-negative discount, subtracted because it lower-bounds the honest weight in
    the committees between $\parent(b)$ and $b$ that is provably \emph{not}
    competing against $b$'s branch.
    We refer to this term as the \emph{empty-slot discount}, since it is non-zero only when there are empty slots between $b$ and its parent, that is, when $\slot(\parent(b))+1 \ne \slot(b)$ as depicted in \Cref{fig:missing_slots}.
    \end{itemize}

    It should be easy to see why we have the first two terms in the definition of $\Phi_{b}^{v,t,C}$; we now give the formal definition of the
    third.%
    
    The quantity
    $\attsetweighttobediscsimplefromblock[from=b,to=\slot(b)-1,val=v,when=t,chkp=C]{\honattsub}$
    provides a conservative lower bound on
    the honest weight of the committees of slots $[\slot(\parent(b))+1,\slot(b)-1]$ that is \emph{not} competing
    against $b$'s branch, and may therefore be subtracted from the maximum weight to tighten the threshold.

    \paragraph{Discounted validators.}
    Fix a block $b$ and a slot $s$.
    Let $\attsettobediscsimplefromblock[from=b,to=s,when=t,val=v]{\allatts}$ denote the set of validators in
    $\commfromafterblock[from=\parent(b),to=s,when=t,val=v]{\allvals}$
    whose \LMDGHOST choice (under the view $\FILLMD(\FILINV(\FILCUR(\FILEQ(\viewattime[val=v,time=t]),t)))$) is exactly $\parent(b)$.
    Define the honest subset
    $\attsettobediscsimplefromblock[from=b,to=s,when=t,val=v]{\honatts}
    := \attsettobediscsimplefromblock[from=b,to=s,when=t,val=v]{\allatts}\cap\honvals$.

    \begin{definition}
    \label{def:attsetweighttobediscsimplefromblock}
        \[
        \attsetweighttobediscsimplefromblock[from=b,to=\slot(b)-1,val=v,when=t,chkp=C]{\honattsub}
        \;:=\;
        \max\left(
            \attsetweighttobediscsimplefromblock[from=b,to=\slot(b)-1,val=v,when=t,chkp=C]{\allatts}
            -\beta \commweightfromafterblock[from=\parent(b),to=\slot(b)-1,chkp=C]{\allvals}
            +\attsetweightfromafterblock[from=\parent(b),to=\slot(b)-1,val=v,when=t,chkp=C]{\slashvals}
            ,0
        \right).
        \]
    \end{definition}

    Concretely, the bound is obtained by
    (i)~starting from the observable total weight of validators whose \LMDGHOST\
    votes are for $\parent(b)$,
    (ii)~subtracting a worst-case Byzantine contribution of at most $\beta$ times
    the relevant committee weight, and
    (iii)~adding back any already identified slashable equivocators,
    with the result truncated at zero.
    The discount is significant only when there are empty slots between $b$ and its parent: if
    $\slot(b)=\slot(\parent(b))+1$, the committee $\commfromafterblock[from=\parent(b),to=\slot(b)-1,when=t,val=v]{\allvals}$ is empty,
    $\attsetweighttobediscsimplefromblock[from=b,to=\slot(b')-1,val=v,when=t,chkp=C]{\honattsub}
    = 0$,
    $\Phi_{b}^{s,v,t,C}$ reduces to
    $\commweightfromafterparentblock[from=b,to=s,chkp=C]{\allvals}
    + \boostweight[chkp=C]$.
    
    \begin{figure}[t]
      \centering
      \input{missingslots.tikz}
      \caption{Empty slot case: $b$ at slot $10$ whose parent lies at slot $7$,
      so slots $8$ and $9$ contain no block on the chain of~$b$.}
      \label{fig:missing_slots}
    \end{figure}
    
    The definition of $\attsetweighttobediscsimplefromblock[from=b,to=\slot(b)-1,val=v,when=t,chkp=C]{\honattsub}$ makes one specific choice of lower bound on the non-competing honest weight.
    Any lower bound on the honest weight in the committees between $\parent(b)$ and $b$ that provably does not compete against $b$'s branch is a valid substitute in $\Phi_{b}^{v,t,C}$.
    A tighter (larger) lower bound is available by additionally counting validators whose \LMDGHOST vote is not a strict descendant of $\parent(b)$, that is, votes for an ancestor of $\parent(b)$ or for a block conflicting with $\parent(b)$.
    This is shown in \Cref{app:lmdghost}, where the safety proofs are stated in terms of this tighter bound.
    However, \emph{reconfirmation} cannot be proven to always succeed when using this tighter bound.
    Reconfirmation is the epoch-boundary check that a previously confirmed block still satisfies safety against the newly justified checkpoint; it is introduced in Section~\ref{sec:conf-rule-ldmghosthfc} and analyzed in Section~\ref{sec:monotonicity}.

    Finally, combining Definitions~\ref{def:isOneConfirmed} and~\ref{def:attsetweighttobediscsimplefromblock} together with Equation~\eqref{eq:phi-def}, $\isOneConfirmed$ can be written in fully expanded form as
    \begin{align*}
    &\var[val=v]{\isOneConfirmed}(b, C, t) \;:=\\
    &\quad
    \begin{aligned}[t]
        \indicatorfromblock[from=b,to=\slot(t)-1,val=v,when=t,chkp=C]{\indQ}
        \;>\;&
        \frac{
                \commweightfromafterparentblock[from=b,to=\slot(t)-1,chkp=C]{\allvals}
                + \boostweight[chkp=C]
                - \max\!\Big(
                    \attsetweighttobediscsimplefromblock[from=b,to=\slot(b)-1,val=v,when=t,chkp=C]{\allatts}
                    - \beta\,\commweightfromafterblock[from=\parent(b),to=\slot(b)-1,chkp=C]{\allvals}
                    + \attsetweightfromafterblock[from=\parent(b),to=\slot(b)-1,val=v,when=t,chkp=C]{\slashvals},\; 0
                \Big)
        }{
            2\,\commweightfromafterparentblock[from=b,to=\slot(t)-1,chkp=C]{\allvals}
        }
        \\
        &\quad+ \beta\,
        \frac{\commweightfromblock[from=b,to=\slot(t)-1,chkp=C]{\allvals}}{\commweightfromafterparentblock[from=b,to=\slot(t)-1,chkp=C]{\allvals}}
        - \frac{\attsetweightfromblock[from=b,to=\slot(t)-1,val=v,when=t,chkp=C]{\slashvals}}{\commweightfromafterparentblock[from=b,to=\slot(t)-1,chkp=C]{\allvals}}.
    \end{aligned}
    \end{align*}

    \section{Confirmation Rule for \LMDGHOSTHFC}\label{sec:conf-rule-ldmghosthfc}
    Section~\ref{sec:conf-rule-lmd-ghost} established a confirmation rule for plain \LMDGHOST,
    where the canonical chain is determined solely by subtree-weight comparisons.
    In the full Gasper protocol, however, fork-choice is \emph{not} run on the
    full block tree: the \LMDGHOSTHFC\ function first applies an
    \emph{\FFG-based filter} and only then runs \LMDGHOST\ on the
    filtered tree.
    As a result, a block $b$ can satisfy the \LMDGHOST\ safety condition of Section~\ref{sec:conf-rule-lmd-ghost} -- so that, ignoring filtering, no other child of $\parent(b)$ can later accumulate more subtree weight -- and still fail to remain canonical, because the
    filter may later exclude the entire branch containing it.
    Confirmation arguments for \LMDGHOSTHFC\ must therefore combine the weight-based safety argument of Section~\ref{sec:conf-rule-lmd-ghost} with explicit guarantees that the candidate block
    cannot be eliminated by the \FFG-based filter.

    Unlike Section~\ref{sec:conf-rule-lmd-ghost}, we omit inline references to individual lemmas in this section: the arguments involve considerably more intermediate results, and embedding cross-references throughout would, in our view, distract from the main narrative.
    \Cref{sec:appendix-proofs} provides a self-contained account where each lemma is preceded by a brief overview of its statement and proof strategy.
    
    We begin by recalling the \FFGCASPER\ component and the \LMDGHOSTHFC\ fork-choice function.

    \subsection{\FFGCASPER\ and \LMDGHOSTHFC\ Definitions}\label{sec:gasper-ffg}
    
    Casper~\cite{casper} is a partially synchronous consensus protocol that operates atop a block proposal mechanism and is responsible for determining when a block is \emph{final}.
    The key property of a final block $b$ is that, provided that the effective-balance-weighted ratio of Byzantine validators over the total validator set is less than $\frac{1}{3}$, any other final block does not conflict with $b$.
    
    This mechanism also introduces a system of accountability, which enables the detection, identification, and punishment
    of a validator not following the protocol's rules.
    Proposed by Buterin and Griffith~\cite{casper}, and then integrated within Gasper~\cite{gasper}, Casper is based on a two-phase traditional propose-and-vote-based Byzantine fault-tolerant (BFT) system, resembling the PBFT~\cite{DBLP:journals/tocs/CastroL02} or HotStuff~\cite{DBLP:conf/podc/YinMRGA19} protocols.
    However, as already mentioned, unlike the latter two, Casper is not a fully defined protocol and is structured to function as a {gadget}, specifically a \emph{finality gadget} (\FFG), atop an existing protocol that generates a chain of blocks which, in the case of Gasper, is the \LMDGHOSTHFC protocol.
    
    \paragraph{\FFG Votes.}
    In Casper, participants vote for links between checkpoints.
    Such votes, which we call \emph{\FFG votes}, are tuples of the form $a = \tuple{C_s, C_t}$.
    Checkpoint $C_s$ is referred to as the \emph{source} checkpoint of the \FFG vote $a$, while $C_t$ is referred to as the \emph{target} checkpoint of $a$.
    
    \paragraph{Unrealized Justified Checkpoint.}
    Each block includes a (possibly empty) set of \FFG votes.
    The set of \FFG votes included in the chain of a block $b$ determines the set of \emph{unrealized justified checkpoints} for that chain, which we denote as $\allU(b)$.
    We do not provide the details of how such a set is computed by Gasper as it is not straightforward.
    We will instead limit ourselves to list those properties of such a set that are relied upon by some of the proofs in the remainder of this paper.
    For any checkpoint $C \in\allU(b)$, we say that $b$ \emph{\ujustifies} $C$, or, equivalently, that $C$ is \emph{\ujustified} by $b$'s chain.
    When we say that \emph{a checkpoint $C$ can never be justified} we mean that it is impossible to create a valid block $b$ such that $C \in\allU(b)$.
    
    \paragraph{Greatest Unrealized Justified Checkpoint in the chain of $b$.}
    The \emph{greatest unrealized justified checkpoint in the chain of a block $b$}, denoted as $\gujblock(b)$, is the unrealized justified checkpoint $C \in \allU(b)$ in the chain of $b$ such that $C \geq C'$ for any $C' \in \allU(b)$.
    Assume that ties are broken arbitrarily
    \footnote{\label{fn:break-ties}In this work, we do not need to consider how ties are broken as we always work under assumptions that ensure that no two checkpoints for the same epoch can ever be justified.}.
    In other terms, the greatest unrealized justified checkpoint at time $t$ is the most recent checkpoint that is already supported strongly enough in the validator's view to be justified, but whose justification has not yet become the starting point of the fork-choice (i.e., it will only become realized at the next epoch's boundary).
    
    \paragraph{Greatest Justified Checkpoint in the chain of $b$.}
    The \emph{greatest justified checkpoint in the chain of block $b$}, denoted as $\gjblock(b)$, is the greatest unrealized checkpoint of the prefix of chain $b$ including only and all the blocks with epoch strictly lower than $\epoch(b)$, \ie,
    \begin{definition}[Greatest Justified Checkpoint in the chain of $b$]\label{def:gjblock}
        \leavevmode \\
        $\gjblock(b) = \gujblock(b')$ with $b'$ such that (i) $b'\prec b \land epoch(b') < \epoch(b)$ and $b' \in \chain(b)$.
    \end{definition}
    Assume that ties are broken arbitrarily\textsuperscript{\ref{fn:break-ties}}.
    
    \paragraph{\FFG Voting Process and Voting Source of block $b$.}
    The \FFG voting process is dependant on the \LMDGHOSTHFC protocol which is described below.
    Specifically, let $b$ be the output of \LMDGHOSTHFC at time $t$ when an honest validator casts an \FFG vote $a$.
    Then, the target checkpoint of $a$ is simply $\chkp(b,\epoch(t))$,
    and the source of $a$, also called the \emph{voting source} of block $b$ in epoch $\epoch(t)$ corresponds to $\votsource[blck=b,time=\epoch(t)]$ as defined below.
    \begin{definition}[Voting Source]\label{def:voting-source}
        $$\votsource[blck=b,time=e] := \begin{cases}
            \gjblock(b), & \text{if $\epoch(b) = e$}\\
            \gujblock(b), & \text{if $\epoch(b) < e$}\\
            \text{undefined,} &\text{otherwise}
        \end{cases}$$
    \end{definition}
    We also let $\votsource[blck=b,time=t] := \votsource[blck=b,time=\epoch(t)]$.
    
    Let $\ffgvalsetallsentraw[source=C_1,target=C_2,time=t]$ denote the set of all \FFG votes with source checkpoint $C_1$ and target checkpoint $C_2$ that are sent at time $t$.
    More generally, we write $\ffgvalsetallsentraw[source=,target=C_2,time=t]$ for the set of all \FFG votes (with arbitrary source) whose target is $C_2$ and that are sent at time $t$.
    Finally, for a validator $v$ and a slot bound $s$, we define $\ffgvalsettoslot[source=C_1,target=C_2,time=t,val=v,to=s]$ as the subset of such votes that (i) are cast by validators assigned to committees between slot $\firstslot(\epoch(C_2))$ and slot $s$ (inclusive) and (ii) are actually received by $v$ by time $t$; formally,
    \[
    \ffgvalsettoslot[source=C_1,target=C_2,time=t,val=v,to=s]
    \;:=\;
    \ffgvalsetallsentraw[source=C_1,target=C_2,time=t]
    \;\cap\;
    \commfromslot[from=\firstslot(\epoch(C_2)),to=s,when=t,val=v]{\allvals}
    \;\cap\;
    \viewattime[time=t,val=v].
    \]
    
    \paragraph{Greatest Justified Checkpoint in view $\View$ at time $t$.}
    The \emph{greatest justified checkpoint in view $\View$}, denoted as $\gjview[view=\View, time=t]$ corresponds to the greatest voting source in epoch $\epoch(t)$ according to the blocks in $\View$ with slot no higher than $\slot(t)$, \ie,
    \begin{definition}[Greatest Justified Checkpoint in view $\View$ at time $t$]\label{def:gjview}
        $$\gjview[view=\View, time=t] := \max(\{\votsource[blck=b,time=t]: b \in \blocksinview(\View) \land \slot(b) \leq \slot(t)\})$$
    \end{definition}
    We let $\chkpattime[time=t,val=v] := \gjview[view={\viewattime[time=t,val=v]}, time=t]$.
    Assume that ties are broken arbitrarily\textsuperscript{\ref{fn:break-ties}}.

    For informal explanations, sometimes we use the expression \emph{$v$'s greatest justified checkpoint at time $t$} to mean $\chkpattime[time=t,val=v]$.
    When clear from the context we might omit indicating the time.
    
    \paragraph{Greatest Finalized Checkpoint in the chain of $b$.}
    For each block $b$, Gasper determines the set of \emph{finalized checkpoints} according to block $b$, denoted as $\allF(b)$.
    Such set is a subset of all the Unrealized Justified Checkpoint of a block $b$, \ie, $\allF(b) \subseteq \allU(b)$.
    The \emph{greatest finalized checkpoint in the chain of block $b$}, denoted as $\gfblock(b)$, is the checkpoint $C \in \allF(b)$ such that $C \geq C'$ for all $C' \in \allF(b)$.
    Assume that ties are broken arbitrarily\textsuperscript{\ref{fn:break-ties}}.
    
    \paragraph{Greatest Finalized Checkpoint in view $\View$ at time $t$.}
    The \emph{greatest finalized checkpoint in view $\View$}, denoted as $\gfview[view=\View, time=t]$ corresponds to the greatest finalized checkpoint according to any block in $\View$ with slot no higher than $\slot(t)$, \ie,
        $$\gfview[view=\View, time=t] := \max(\{\gfblock(b): b \in \blocksinview(\View) \land \slot(b) \leq \slot(t)\})$$
    Assume that ties are broken arbitrarily\textsuperscript{\ref{fn:break-ties}}.
    
    \paragraph{\LMDGHOSTHFC.}
    \LMDGHOSTHFC is the fork-choice rule used by Gasper which is presented in Algorithm~\ref{alg:lmd-ghost-fc}.
    
    It works by applying \LMDGHOST (Section~\ref{sec:lmd-ghost}) on a filtered view where the blocks kept after the filtering correspond to those in any chain $b'$ such that $b'$ does not conflict with $\gjview[view=\View, time=t]$ and either the voting source of $b'$ is $\gjview[view=\View, time=t]$ or the epoch of the voting source of $b'$ is at least $\epoch(t)-2$.
    Details for the reasons behind this type of filtering can be found in~\cite{blockfiltering}.
    
    Also, we define 
    $$\filtered[time=t,val=v] := \{b : b \in \FILHFC(\viewattime[time=t,val=v],t)\}$$
    to be the set of blocks that are not filtered out by \FILHFC{} according to the view of validator $v$ at time $t$.
        Informally, if $b \in \filtered[time=t,val=v]$, we say that $b$ is \emph{not filtered out} by validator $v$ at time $t$.
    
    \begin{algorithm}[t]
        \vbox{
        \small
        \begin{numbertabbing}\reset
        xxxx\=xxxx\=xxxx\=xxxx\=xxxx\=xxxx\=MMMMMMMMMMMMMMMMMMM\=\kill
        \textbf{function} $\FILHFC(\View, t)$\label{}\\
        \> \textbf{return} $\begin{alignedat}[t]{2}\View \setminus \{b \in \blocksinview(\View) : \lnot ( &\lor b \preceq \gjview[view=\View, time=t] \\
        &\lor\;\land\; b \succeq \block(\gjview[view=\View, time=t])\\
        &\phantom{\lor\;}\;\land\;\exists b' \in \View,\begin{aligned}[t]
            &\land\; b'\succeq b \\
            &\land\; b' \succeq \finattime[time=t,val=v]\\
            &\land\; \epoch(b') \leq \epoch(t)\\
            &\land\; \children[blck=b',view=\View] = \emptyset\\
            &\land  (\votsource[blck=b',time=t]=\gjview[view=\View, time=t] \lor \epoch(\votsource[blck=b',time=t]) \geq \epoch(t) -2 ))\}
        \end{aligned}\end{alignedat}$
        \label{}\\[1ex]
        \textbf{function} $\fcparam[fc=\LMDGHOSTHFC,balf=\frakB](\View, t)$\label{}\\
        \> \textbf{return} $\fcparam[fc=\LMDGHOST,balf=\frakB](\FIL_\mathsf{fc}(\View,t), t)$\label{}\\
        [-5ex]
        \end{numbertabbing}
        }
        \caption{\LMDGHOSTHFC fork-choice}\label{alg:lmd-ghost-fc}
    \end{algorithm}
    
    \subsection{Full Confirmation Rule}\label{sec:full-conf-rule}
    
    This section presents the full confirmation rule that accounts for the
    \FFG-based filtering described above.
    We begin with the additional assumptions required by the \FFG layer.
    
    \begin{assumption}\label{assum:ffg}
        All of the following conditions are satisfied.
        
        \begin{enumerate}
            \item\customlabel[assumption]{assum:ffg-assumptions:beta}{\theassumption.\theenumi} $\beta < \frac{1}{3}$
            \item\customlabel[assumption]{assum:ffg-assumptions:justified-checkpoint-next-epoch}{\theassumption.\theenumi}
            Given a block $b$ and epoch $e\geq\epoch(b)$ such that $\slotstart(e+1)\geq\GGST$, if for any time $t$ with $\epoch(t) = e+1$ and honest validator $v$,  
            \begin{enumerate}
                \item $b$ is canonical in the view of $v$ at time $t$,
                \item for any block $b' \succeq \chkp(b,e)$ in the view of $v$ we have that
                $\weightofset[chkp=b']{\ffgvalsetallsentraw[source={\votsource[blck=b,time=e]},target={\chkp(b,e)},time=t'] \setminus \slashedset[chkp=b']} \geq \frac{2}{3} \totvalsetweight[chkp=b']{\allvals}$,
            \end{enumerate}
            then, by time $\slotstart(e+2)$, the view of validator $v$ includes a block $b'\succeq b$ such that $\epoch(b') < e+2 \land \chkp(b,e) \in \allU(b')$.
        \end{enumerate}
    \end{assumption}
    
    Assumption~\ref{assum:ffg-assumptions:beta} is standard in Gasper: it is the
    bound required to ensure that no two conflicting checkpoints can ever be
    finalized.
    Assumption~\ref{assum:ffg-assumptions:justified-checkpoint-next-epoch} states
    that Byzantine validators cannot prevent an \FFG vote sent by a non-slashable
    validator from being included in a canonical block for an entire epoch.
    This assumption may be violated in practice due to the limited number of \FFG
    votes that can be included in a given block.
    If this limitation were removed when the \FFG votes included in a block justify
    the checkpoint from the previous epoch, then
    Assumption~\ref{assum:ffg-assumptions:justified-checkpoint-next-epoch} would
    reduce to requiring at least one honest proposer per epoch, which under
    Assumption~\ref{assum:ffg-assumptions:beta} holds with overwhelmingly high
    probability.\\
    
    Recall from Section~\ref{sec:gasper-ffg} that \LMDGHOSTHFC\ applies an
    \FFG-based filter (\FILHFC) before running \LMDGHOST.
    This filter can exclude a locally \LMDGHOST-safe block for two reasons:
    the branch may become incompatible with the local view's greatest justified
    checkpoint, or it may lack the viability evidence (e.g., a descendant whose voting-source epoch is at least $\epoch(t)-2$) required by \FILHFC.
    
    We adapt the standard \LMDGHOST safety argument to the filtered setting.
    Fix an honest validator $v$, times $t\le t'$, a block $b$, and a checkpoint
    reference $C$.
    Suppose that (after $\GGST$) $v$ determines at time $t$ that $b$ is
    \LMDGHOST-safe with respect to $C$.
    Then $b$ remains canonical throughout $[t,t']$ provided that two
    conditions hold over the interval:
    \begin{enumerate}
        \item (\LMDGHOST\ condition) every honest validator's greatest justified checkpoint is a descendant of $C$, ensuring that the total weight of the balance source used by \LMDGHOST\ is no higher than the total weight of $C$; and
        \item (\FILHFC\ condition) $b$ is never removed by the viability filter, i.e.,
        $b\in\filtered[time=t'',val=v'']$ for all $t''\in[t,t']$ and all honest $v''$.
    \end{enumerate}

    \paragraph{Gasper properties used in this section.}
    Before presenting the confirmation rule, we recall the Gasper properties
    relied upon in the remainder of this section and in the proofs in the Appendix.
    
    \begin{property}[Gasper Properties]
        The Gasper protocol ensures the following properties.
        \begin{enumerate}
            \item\customlabel[property]{prop:gasper-basic:only-one-justified-per-epoch}{\theproperty.\theenumi}
            As long as the adversary controls less than a third of the stake (minus the safety decay), at most one checkpoint per epoch can ever become justified.
            \item\customlabel[property]{prop:gasper-basic:just-succ-finalization}{\theproperty.\theenumi}
            Justification never falls behind finalization: every honest validator's greatest justified checkpoint is always a strict descendant of their greatest finalized checkpoint.
            \item\customlabel[property]{prop:gasper-basic:ldm-vote-for-b-is-ffg-vote-for-cb}{\theproperty.\theenumi}
            A \GHOST\ vote for a block always carries along an \FFG\ vote whose target is the checkpoint associated with that block in the current epoch.
            \item\customlabel[property]{prop:gasper-basic:ffg-vote-for-c-ghost-vote-for-b-succ-c}{\theproperty.\theenumi}
            An \FFG\ vote targeting a checkpoint $C_d$ always carries along a \GHOST\ vote for a descendant of $C_d$ (or $C_d$ itself).
            \item\customlabel[property]{prop:gasper-basic:sufficient-condition-for-no-conflicting-chkps}{\theproperty.\theenumi}
            If a checkpoint $C$ is a descendant of the greatest justified checkpoint, and at least a two-thirds supermajority of \FFG\ votes for $C$, weighted according to $C$, have been sent, then no conflicting checkpoint for the same epoch can ever be justified.
            \item\customlabel[property]{prop:gasper-basic:no-conflicting-if-all-honest-votes-in-support-of-b}{\theproperty.\theenumi}
            If every honest validator in the committee of epoch $e$ casts an \FFG\ vote targeting a descendant of block $b$, then no checkpoint conflicting with $b$ can be justified for epoch $e$.
            \item\customlabel[property]{prop:gasper-basic:highest-justified-not-from-current-epoch}{\theproperty.\theenumi}
            The greatest unrealized justified checkpoint in any block's view belongs to an epoch no later than the block's own epoch, which implies that the epoch of the voting source used by any honest validator always belongs to a past epoch, never to the current one.
        \end{enumerate}
    \end{property}
    
    \begin{algorithm}[H]
    \caption{Confirmation Rule}
    \label{alg:conffull}
    \SetAlgoNoLine
    \KwState{
        $\var[val=v]{\bconfirmed}$
    }
    \Upon{$\now = \var[val=v]{\tinit}$}{
        $\var[val=v]{\bconfirmed}   \gets \block(\gfattime[time={\now},val=v])$
    }
    \Upon{$\now = \slotstartslot{\now} \land \now \geq \var[val=v]{\tinit}$}{
        $\var[val=v]{\bconfirmed} \gets \var[val=v]{\getlatestconfirmed}(\var[val=v]{\bconfirmed})$
    }
    \Fn{$\var[val=v]{\isChainOneConfirmed}(b_c)$}{
        $b \gets b_c$\\
        \While{$b \neq \block(\guattime[val=v,time=\slotstart(\prevfirstslotepoch{\now})]) \land \var[val=v]{\isOneConfirmed}(b, \guattime[val=v,time={\slotstart({\prevfirstslotepoch[-1]{\now}})}],\now)$\label{ln:ischainoneconfirmed-while}}{
            $b \gets \parent(b)$
        }
    
        \Return{$b = \block(\guattime[val=v,time=\slotstart(\prevfirstslotepoch{\now})])$}
    }
    \Proc{$\var[val=v]{\getlatestconfirmed}(b_c, \now)$}{
        $\bcand \gets b_c$\nllabel{ln:set-bcand-beginning-of-get-latest-confirmed}\\
        \Const $\head \gets \LMDGHOSTHFC(\viewattime[time=\now,val=v])$\\
    
        \uIf{$\begin{aligned}[t]
             &\epoch(\bcand) < \epoch(\now) - 1\\
             &\text{\bf or}\ \bcand \npreceq \head\\
             &\text{\bf or}\ \Big(\slot(\now) = \firstslot(\epoch(\now))\\
             &\qquad \text{\bf and}\ \Big(
                \bcand \nsucceq \guattime[val=v,time=\slotstart(\prevfirstslotepoch{\now})]\\
             &\qquad\qquad \text{\bf or}\ 
                \neg \var[val=v]{\isChainOneConfirmed}\!(
                    \bcand
                )
            \Big)\Big)
        \end{aligned}$
             \nllabel{ln:if-bcand-npreceq-head}\\
        }{
            $\bcand \gets \block(\gfattime[val=v,time=\now])$\nllabel{ln:set-bcand-to-fin}
        }
    
        \uIf{$\begin{aligned}[t]
             &\slot(\now) = \firstslot(\epoch(\now))\\
             &\text{\bf and}\ \epoch(\block(\guattime[val=v,time=\slotstart(\prevfirstslotepoch{\now})])) = \epoch(\now)-1\\
             &\text{\bf and}\ \gu(\head) = \guattime[val=v,time=\slotstart(\prevfirstslotepoch{\now})]\\
             &\text{\bf and}\ \slot(\bcand) < \slot(\block(\guattime[val=v,time=\slotstart(\prevfirstslotepoch{\now})]))
        \end{aligned}$
             \nllabel{ln:start-conf-chain}\\
        }{
            $\bcand \gets \block(\guattime[val=v,time=\slotstart(\prevfirstslotepoch{\now})])$\nllabel{ln:set-bcan-on-start-conf-chain}
        }
    
        \uIf{$\epoch(\bcand) \geq \epoch(\now) - 1$\nllabel{ln:if-bcand-e-1}}{
            $\bcand \gets \var[val=v]{\findlatestconfirmeddescendant}(\bcand)$\nllabel{ln:set-bcand-to-output-find-latest}
        }
    
        \Return{$\bcand$}
    }
        \marklastline{lastline}
    \end{algorithm}

\paragraph{Entry point.}
We focus on $\getlatestconfirmed_v(b_c)$ (Line~\ref{ln:set-bcand-beginning-of-get-latest-confirmed}), which is executed at the beginning of every slot.
Each invocation starts from the previous output: the input $b_c$ is the confirmed block produced by the previous slot's invocation.
Thus $\bcand$ is initialized to the prior confirmed block.
The procedure first checks whether the previously confirmed block is still
consistent with the canonical view; if not, it enters the
\emph{inactive state} and resets to the finalized block.
Otherwise, it enters the \emph{active state} and attempts to advance
the confirmed block.\footnote{The terms \emph{active} and
\emph{inactive} are not explicit in the algorithm pseudocode; we use
them here as a convenient shorthand.}
We describe the inactive state first, then the active state.

\paragraph{Inactive state.}
Before attempting to advance the confirmed block, the rule first checks
whether the previously confirmed block ($\bcand$ at this point) is still consistent with the canonical
view.
Whenever any of the following holds,
\[
\begin{aligned}
&\epoch(\bcand) < \epoch(\now) - 1
\\
&\text{\bf or}\quad \bcand \npreceq \head
\\
&\text{\bf or}\quad
\Bigl(
    \slot(\now) = \firstslot(\epoch(\now))
\\
&\qquad\qquad\text{\bf and}\;\Bigl(
    \bcand \nsucceq \guattime[val=v,time=\slotstart(\prevfirstslotepoch{\now})]
\\
&\qquad\qquad\qquad\text{\bf or}\;
    \neg\bigl(
      \var[val=v]{\isChainOneConfirmed}
      (
        \bcand
      )
    \bigr)
    \Bigr)
\Bigr),
\end{aligned}
\]

the rule determines that $\bcand$ is outdated or inconsistent and resets
the confirmed block to the latest finalized block:
\[
  \bcand \;\gets\; \block\!\bigl(\gfblock^{v}_{\now}\bigr).
\]

Concretely, the three disjuncts correspond to:
\begin{enumerate}
    \item[(i)] The stored confirmed block belongs to epoch
    $\epoch(\now) - 2$ or earlier:
    $\epoch(\bcand) < \epoch(\now)-1$.
    The rest of the algorithm requires that the
    confirmed block belongs to either the current or the previous epoch;
    when this condition is violated, the rule cannot advance the
    confirmed block and must reset.

    \item[(ii)] The confirmed block is no longer on the canonical
    chain: $\bcand \npreceq \head$.

    \item[(iii)] At the first slot of an epoch, either $\bcand$ is not a descendant of $\guattime[val=v,time=\slotstart(\prevfirstslotepoch{\now})]$, or the confirmed chain from $\bcand$ back to that checkpoint's block fails the reconfirmation check $\var[val=v]{\isChainOneConfirmed}$, performed with respect to the greatest unrealized justified checkpoint observed two epochs ago, which is the same checkpoint used for normal confirmation during the previous epoch.
\end{enumerate}

\paragraph{Reconfirmation mechanism.}
The predicate $\isChainOneConfirmed$ implements
\emph{reconfirmation}: it walks from $\bcand$ back to the block of the greatest unrealized justified checkpoint observed at the end of the previous epoch, and
verifies that each such block still satisfies
$\isOneConfirmed$ with respect to the checkpoint used to execute the confirmation algorithm in the previous epoch, namely the greatest unrealized justified checkpoint observed two epochs ago.\footnote{This checkpoint is also an ancestor of the checkpoint honest validators will later use for fork-choice. This is one of the necessary conditions required for the safety argument.} 
Reconfirmation is needed because the rule extends the confirmed chain by verifying only the newly confirmed block, without re-checking previous ones. Without reconfirmation, the safety argument's synchrony assumption ($t \geq \GGST$, Assumption~\ref{assum:beta}) would need to hold since the rule last exited the inactive state, which may be arbitrarily far in the past.\footnote{It may be
possible to prove that, even without reconfirmation, the synchrony
assumption need only hold from a bounded number of epochs in the past.
However, we do not have such a proof, and reconfirmation at each epoch
boundary avoids the need for one.}
Reconfirmation at epoch boundaries addresses this by re-running
$\isOneConfirmed$ on every block from
$\bcand$ back to
$\block(\guattime[val=v,time=\slotstart(\prevfirstslotepoch{t})])$,
rather than relying on the fact that each block's parent was already
verified during normal confirmation.

The checkpoint
$\guattime[val=v,time=\slotstart(\prevfirstslotepoch{t})]$ can be
determined canonical under the assumption that synchrony has held from the last slot of the epoch before the one this checkpoint belongs to, i.e., $\slotstart(\prevfirstslotepoch{\guattime[val=v,time=\slotstart(\prevfirstslotepoch{t})]}) \geq \GGST$. Since this checkpoint may belong to either the previous epoch or two epochs ago, the safety argument requires $\slotstart(\prevfirstslotepoch[-1]{t}) \geq \GGST$, that is, synchrony from at least the last slot of two epochs ago.
If reconfirmation succeeds, the safety argument only needs synchrony to
hold from the most recent epoch boundary onward.

If it fails, the rule enters the inactive state and resets to the
finalized block, ensuring that no synchrony assumption from an arbitrarily distant past is relied upon. Failure indicates that future confirmations cannot proceed safely, not that any previously confirmed block is necessarily unsafe.

    \paragraph{Active state.}
    The confirmation rule enters its active state
    exactly at the beginning of an epoch, provided the following conditions hold:
    
    \[
    \begin{aligned}
    &\slot(\now) = \firstslot(\epoch(\now))\\
    &\text{\bf and}\ \epoch\!\Bigl(\block\bigl(\guattime[val=v,time={\slotstart(\slot(\now)-1)}]\bigr)\Bigr)
      = \epoch(\now) - 1\\
    &\text{\bf and}\ \gu\!\Bigl(\head\Bigr)
      = \guattime[val=v,time={\slotstart(\slot(\now)-1)}]\\
    &\text{\bf and}\ \slot(\bcand)
      < \slot\!\Bigl(\block(\guattime[val=v,time={\slotstart(\slot(\now)-1)}])\Bigr).
    \end{aligned}
    \]

Intuitively: this condition checks that:
\begin{itemize}
  \item the current epoch has \emph{exactly} just started;
  \item the block of the greatest unrealized justified checkpoint observed at the last slot of the previous epoch
        belongs to the previous epoch;\footnote{The condition is on
      $\epoch(\block(\guattime[val=v,time={\slotstart(\slot(\now)-1)}]))$
      rather than on
      $\epoch(\guattime[val=v,time={\slotstart(\slot(\now)-1)}])$
      because the block of the checkpoint may belong to an earlier epoch
      than the checkpoint itself (e.g., epoch $\epoch(t)-2$).
      If this were the case, the algorithm would reset to finality at
      the next slot (\Cref{ln:if-bcand-npreceq-head}), since the
      confirmed block would be more than one epoch old.
      It may be possible to handle this situation without resetting, but
      the current algorithm does not do so.
      In practice, the epoch of the block and the epoch of the checkpoint
      coincide in the vast majority of cases, so this distinction is
      minor.}
  \item the greatest unrealized justified checkpoint of the current fork-choice head
        equals the one observed at the last slot of the previous epoch; and
  \item the block of that checkpoint is \emph{strictly ahead} of the current confirmation candidate $\bcand$.
\end{itemize}

    At epoch start, the second and third conjuncts together constrain the greatest unrealized justified checkpoint observed at the last slot of the previous epoch: the second requires its block to belong to the previous epoch, and the third requires that this checkpoint does not conflict with the current fork-choice head.
    The second conjunct is specifically required for the confirmation of descendants of $\block(\guattime[val=v,time=\slotstart(\slot(\now)-1)])$.
    We will explain why we need it when discussing Case~3 of $\findlatestconfirmeddescendant$.
    The final conjunct ensures that this \emph{restart} of the confirmation chain (\Cref{ln:set-bcan-on-start-conf-chain}) at the newly
    realized justified checkpoint occurs only when it moves $\bcand$ strictly ahead.
    
    Since justification becomes effective only at epoch boundaries, upon entering the current epoch the checkpoint
    $\guattime[val=v,time=\slotstart(\slot(\now)-1)]$ is realized as justified. Because the second conjunct implies that this checkpoint belongs to the previous epoch, and no checkpoint from a later epoch can have been justified so far, under synchrony this checkpoint becomes the greatest justified checkpoint for every honest validator.
    Consequently, fork-choice and honest voting are confined to descendants of this
    checkpoint's block.
    
    Since $\LMDGHOSTHFC$ starts its walk from the greatest justified checkpoint's block,
    the block of this checkpoint is an ancestor of every honest validator's fork-choice head
    at the start of the current epoch, and hence canonical.
    Moreover, because $b$ is canonical and all honest \FFG votes during the current epoch
    target a descendant of $b$, no conflicting checkpoint for the current epoch can ever be
    justified.
    As a result, $b$ remains in the filtered tree (\ie, is not removed by $\FILHFC$)
    throughout the current epoch and beyond.

    When the condition holds, we \emph{(re)start} the confirmation chain from the block of $\guattime[val=v,time=\slotstart(\slot(\now)-1)]$:
    \[
    \bcand \leftarrow \block\!\bigl(\gu(\View^v,\slotstart(\slot(\now){-}1))\bigr),
    \]
    and immediately continue by confirming as far as possible along the canonical chain via
    \[
    \bcand \leftarrow \findlatestconfirmeddescendant_v(\bcand).
    \]

    \paragraph{Monotonicity of confirmed blocks.}
    A key property is that the confirmed block is \emph{stateful} and, as long as the
    rule does not reset to finality, it advances monotonically: each new confirmed block is a descendant of the previous one.
    This is because $\findlatestconfirmeddescendant$ (called inside $\getlatestconfirmed$ at \Cref{ln:set-bcand-to-output-find-latest}) only returns $b_c$ unchanged or a descendant of $b_c$ that belongs to the fork-choice head chain.
    Consequently, on any interval during which no reset to finality occurs, if a block $b$ is confirmed at time $t$, then the confirmed block at any later time $t' \ge t$ is a descendant of~$b$.

    \paragraph{Per-block check.}
Inside $\findlatestconfirmeddescendant$ (Algorithm~\ref{alg:findlatestconf-functional}),
each confirmation decision applies the per-block confirmation predicate
of \Cref{def:isOneConfirmed} with $C = \guattime[val=v,time=\prevfirstslotepoch{t}]$,
the greatest unrealized justified checkpoint at the last slot of the previous epoch.
For a block $b$ whose parent belongs to an earlier epoch, the algorithm uses
a stronger variant predicate $\var[val=v]{\isOneConfirmedSpecial}$,
defined later in \Cref{def:isOneConfirmedSpecial}, in place of
$\var[val=v]{\isOneConfirmed}$; the wrapper $\var[val=v]{\isOneConfirmedExt}(b,C)$
(\Cref{alg:findlatestconf-helpers}) selects the appropriate predicate.
Since $\var[val=v]{\isOneConfirmedSpecial}(b,C,t) \implies \var[val=v]{\isOneConfirmed}(b,C,t)$,
$\var[val=v]{\isOneConfirmedExt}(b,C,t)$ implies $\var[val=v]{\isOneConfirmed}(b,C,t)$
regardless of which branch of the wrapper applies.

Given a block $b_\mathsf{s}$ and a candidate $b \succeq b_\mathsf{s}$,
$\var[val=v]{\isOneConfirmedExtFrom}(b,C,t,b_\mathsf{s})$ asserts that
every block strictly between $b_\mathsf{s}$ and $b$, including $b$, satisfies
$\var[val=v]{\isOneConfirmedExt}$.

    \begin{algorithm}
    \caption{Helper functions for $\findlatestconfirmeddescendant$}
    \label{alg:findlatestconf-helpers}
    \SetAlgoNoLine
    \DontPrintSemicolon
    \continuefrom{lastline}
    
    \Fn{$\var[]{\getHighest}(\mathit{blocks})$}{
        \Return $\argmax_{b\in \mathit{blocks}}\slot(b)$
    }
    
    \Fn{$\var[val=v]{\canonChainFrom}(b,t)$}{
        \Return $\{ b' \in \viewattime[time={\now},val=v], b \prec b' \preceq \LMDGHOSTHFC(\viewattime[time={\now},val=v]) \}$\\
    }

    \Fn{$\var[val=v]{{\isOneConfirmedExt}}(b,C)$}{
        \uIf{
            $\epoch(\parent(b))<\epoch(b)$
        }
        {
            \Return $\var[val=v]{\isOneConfirmedSpecial}(b, C)$
        }
        \uElse
        {
            \Return $\var[val=v]{\isOneConfirmed}(b, C)$
        }
    }
    
    \Fn{$\var[val=v]{\isOneConfirmedExtFrom}(b_\mathsf{e},C, b_\mathsf{s})$}{
        \Return $\forall b' \in \viewattime[time={\now},val=v], b_\mathsf{s} \prec b' \preceq b_\mathsf{e} \implies \var[val=v]{{\isOneConfirmedExt}}(b', C)$
    }
    
\Fn{$\var[val=v]{\willChkpBeJustified}(C)$}
{
    \Return{$\max(\ffgvalsettoslotweight[target={C},to={\slot(\now)-1},time={\now},val=v,weight chkp={C}]
    - \beta \commweightfromblock[from=\firstslot(\epoch(t)),to=\slot(t)-1,chkp=C]{\allvals}
    + {\attsetweightfromblock[from=\firstslot(\epoch(t)),to=\slot(t)-1,chkp=C,val=v,when={t}]{\slashvals}},
    0)
    + {(1-\beta)\commweightfromblock[from={\slot(\now)},to=\lastslot(\epoch(C)),chkp={C}]{\allvals}}
    \geq
    \frac{2}{3}\totvalsetweight[chkp={C}] {\allvals}
    $}
}

\Fn{$\var[val=v]{\willNoConflictingChkpBeJustified}(C)$}
{
    \Return{$\max(\ffgvalsettoslotweight[target={C},to={\slot(\now)-1},time={\now},val=v,weight chkp={C}]
    - \beta \commweightfromblock[from=\firstslot(\epoch(t)),to=\slot(t)-1,chkp=C]{\allvals}
    + {\attsetweightfromblock[from=\firstslot(\epoch(t)),to=\slot(t)-1,chkp=C,val=v,when={t}]{\slashvals}},
    0)
    + {(1-\beta)\commweightfromblock[from={\slot(\now)},to=\lastslot(\epoch(C)),chkp={C}]{\allvals}}
    >
    \frac{1}{3}\totvalsetweight[chkp={C}] {\allvals}
    $}
    \marklastline{lastline2}
}
\end{algorithm}

\begin{algorithm}
    \caption{$\findlatestconfirmeddescendant$}
    \label{alg:findlatestconf-functional}
    \SetAlgoNoLine
    \DontPrintSemicolon
    \continuefrom{lastline2}
    
    \Fn{$\var[val=v]{\findlatestconfirmeddescendant}(b_c)$}
    {
        \tcc{Compute the highest LMDGHOST-safe descendant of $b_c$ along the head chain}
        $\btcands \gets 
            \{b' \in \var[val=v]{\canonChainFrom}(b_c),\, \var[val=v]{\isOneConfirmedExtFrom}(b',\guattime[val=v,time={\prevfirstslotepoch{t}}], b_c)\}$
    
        \BlankLine

        \tcc{Case 1: current-epoch confirmation}
        \uIf{$\btcands\neq\emptyset$\label{ln:case1-pre}}
        {
            $\btcand \gets \var[]{\getHighest}(\btcands)$\\
            \uIf{
                $\begin{aligned}[t]
                    &\epoch(\btcand) = \epoch(\now)\\
                    &\text{\bf and}\ \exists b' \in \viewattime[time=\now,val=v], b' \succeq \btcand\ \text{\bf and}\ \epoch(\gu(b')) \geq \epoch(\now) - 1\\
                    &\text{\bf and}\ \epoch(\btcand)>\epoch(b_c) \implies \willChkpBeJustified_v(\chkp(\btcand))
                \end{aligned}$\nllabel{ln:case1}\\
            }
            {
                \Return $\btcand$\nllabel{ln:case1-ret}
            }
        }

        \BlankLine
    
        \tcc{Case 2: no progress -- cannot rule out conflicting checkpoint}
        \uIf{
            $\begin{aligned}[t]
            &\text{\bf not}\;\Bigl(
                \slot(\now) = \firstslot(\epoch(\now))\\
            &\phantom{\text{\bf not}\;\Bigl(}\text{\bf or}\;
                \exists b'\succeq b_c,
                \willNoConflictingChkpBeJustified^v\!\bigl(\chkp(b',\epoch(\now))\bigr)
            \Bigr),
            \end{aligned}$\nllabel{ln:case2}
            \\
        }
        {
                        \Return $b_c$\nllabel{ln:case2-ret}\\
        }
        \BlankLine
    
        \tcc{Recompute highest LMDGHOST-safe descendant, restricted to previous epoch}
        $\btcands \gets
            \begin{aligned}[t]
            &\{b' \in \var[val=v]{\canonChainFrom}(b_c),\,\\
            &\qquad\epoch(b')<\epoch(\now)\\
            &\qquad\text{\bf and}\ \var[val=v]{\isOneConfirmedExtFrom}(b',\guattime[val=v,time={\prevfirstslotepoch{t}}], b_c)\}
        \end{aligned}$\\

                \tcc{Case 3: previous-epoch confirmation with voting source from epoch $\epoch(\now) - 2$ or later}
        \uIf{$\btcands\neq\emptyset$\nllabel{ln:case3-pre}}
        {
            $\btcand \gets \var[]{\getHighest}(\btcands)$\\
            \uIf{$\begin{aligned}[t]
                &\epoch(\votsource[blck=\btcand, time={\now}]) \geq \epoch(\now) - 2\\
                &\text{\bf and}\
                        (\slot(\now) \neq \firstslot(\epoch(\now)) \implies\\
                &\qquad \exists b'\in \viewattime[time={\now},val=v], b'\succeq \btcand\ \text{\bf and}\ \epoch(\gu(b')) \geq \epoch(\now) - 1)
            \end{aligned}$\nllabel{ln:case3}\\}
            {
                \Return $\btcand$\nllabel{ln:case3-ret}
            }
        }
    
        \BlankLine
    
        \tcc{Case 4: previous-epoch confirmation with voting source verified from a block received by the previous slot}
        $\begin{aligned}[t]
            &\bcands \gets \{b' \in \var[val=v]{\canonChainFrom}(b_c),\\ 
            &\qquad\epoch(b') < \epoch(\now) \\
            &\qquad\text{\bf and}\ 
                (\exists b''\in \viewattime[time={\var[val=v]{\slotstart(\slot(\now)-1)}},val=v],\\
            &\qquad\qquad b''\succeq b'\ \text{\bf and}\ \epoch(\votsource[blck=b'', time={\now}]) \geq \epoch(\now) - 2)\\ 
            &\qquad\text{\bf and}\ 
                (\slot(\now) \neq \firstslot(\epoch(\now)) \implies\\
            &\qquad\qquad \exists b''\in \viewattime[time={\now},val=v], b''\succeq b'\ \text{\bf and}\ \epoch(\gu(b'')) \geq \epoch(\now) - 1)\\
            &\qquad\text{\bf and}\ 
                \var[val=v]{\isOneConfirmedExtFrom}(b',\guattime[val=v,time={\prevfirstslotepoch{t}}],b_c)\}
        \end{aligned}$\\

        \uIf{$\bcands \neq \emptyset$\nllabel{ln:case4}}
        {
            \Return $\var{\getHighest}(\bcands)$\nllabel{ln:case4-ret}
        }

        \tcc{Case 5: no progress}
        \Return $b_c$\\
    }
\end{algorithm}

    \smallskip
    \noindent
    
    \paragraph{Finding the latest confirmed descendant.}
    Function $\findlatestconfirmeddescendant$ advances the confirmed block beyond
    the last confirmed block $b_c$.
    It searches along the current fork-choice head chain for the highest descendant
    of $b_c$ that passes the per-block check
$\var[val=v]{{\isOneConfirmedExt}}$ (see Definitions~\ref{def:isOneConfirmed} and~\ref{def:isOneConfirmedSpecial}) with the checkpoint set to $\guattime[val=v,time=\prevfirstslotepoch{t}]$, and then applies additional conditions to avoid returning a block that can be filtered out by \FILHFC.
    
    \begin{sloppypar}
    The output satisfies two basic guarantees. First, it never moves backwards: $\findlatestconfirmeddescendant(b_c) \succeq b_c$. Second, whenever it advances (i.e., the output $b_o \neq b_c$), $b_o$ remains along the head chain.
    \end{sloppypar}

\paragraph{Case~1 (\Cref{ln:case1-pre,ln:case1}): current-epoch confirmation.}
    The first step is to compute a candidate $\btcand$ as the highest
    block along the current canonical chain extending $b_c$ such that every intermediate
    block passes $\var[val=v]{\isOneConfirmedExt}$ with respect to $\guattime[val=v,time=\prevfirstslotepoch{t}]$:
    
    \begin{equation}\label{eq:case1-lmd}
        \btcand \gets \getHighest\Bigl(
        \{\, b' \in \canonChainFrom(b_c)
        \;:\;
        \isOneConfirmedExtFrom(b',\guattime[val=v,time=\prevfirstslotepoch{t}], b_c)\,\}
        \Bigr).
    \end{equation}
    
    In other words, $\btcand$ is the highest block along the head chain
    extending $b_c$ such that every intermediate block passes
    $\var[val=v]{\isOneConfirmedExt}$ -- the highest block that
    could be confirmed if $\FILHFC$ filtering were not present.
    The remaining logic decides whether this block is also safe against
    $\FILHFC$ filtering; if not, the algorithm falls back to shorter
    outputs.

Recall that, given a block $b$ and an honest validator's greatest justified checkpoint $C$, $b$ is not filtered out by $\FILHFC$ if any of the following conditions holds.
    \begin{enumerate}
        \item $b$ is an ancestor of $\block(C)$
        \item $b$ is a descendant of $\block(C)$ and either of the following conditions is satisfied:
        \begin{enumerate}
            \item there exists a block $b' \succeq b$ such that checkpoint $\votsource[blck=b',time=\now]$ belongs to either two epochs ago or to the previous epoch
            \item\label{itm:filhfc-condition-2b} there exists a block $b' \succeq b$ such that checkpoint $\votsource[blck=b',time=\now]$ corresponds to $v$'s greatest justified checkpoint
        \end{enumerate}
    \end{enumerate}
    Case~1 is the only case that can return a block in the current epoch. Formally, the function returns $\btcand$ when
    \[
    \begin{aligned}
    &\epoch(\btcand) = \epoch(\now)
    \\
    &\;\text{\bf and}\;
    \Bigl(
      \exists b' \in \viewattime[time=\now,val=v], b' \succeq \btcand\ \text{\bf and}\ \epoch(\gu(b')) \geq \epoch(\now) - 1
    \Bigr)
    \\
    &\;\text{\bf and}\;
    \Bigl(
      \epoch(\btcand)>\epoch(b_c)
      \;\implies\;
      \willChkpBeJustified_v(\chkp(\btcand))
    \Bigr).
    \end{aligned}
    \]

    The first conjunct requires $\btcand$ to belong to the current epoch. The second requires, except at the beginning of the current epoch, a descendant $b' \succeq \btcand$ that \ujustifies a checkpoint belonging to the previous or a later epoch\footnote{Note that this implies that the epoch of such a checkpoint is either the previous or current epoch.}; due to the conditions listed above, this is sufficient to ensure that $\btcand$ will not be filtered out by $\FILHFC$. The third checks that enough \FFG votes have been sent to justify $\btcand$'s latest checkpoint, via the $\willChkpBeJustified_v(\cdot)$ predicate, which computes a lower bound on the total weight of \FFG votes targeting $\btcand$'s checkpoint by combining (i) support already visible in $v$'s view (as inferred from votes received so far), adjusted by discounting the worst-case Byzantine contribution and adding back already-identified slashed validators, and (ii) a conservative lower bound on the additional honest support that can still arrive from the \emph{remaining} committees in the rest of the epoch. It accepts only if this combined bound meets the \FFG justification threshold.

We now provide an intuition of why these three conjuncts are sufficient to ensure that $\btcand$ will never be filtered out by $\FILHFC$.
    
    The first property that we want to show is that $\epoch(\gu(\btcand))\geq \epoch(\now)-2$.
    Because $\btcand$ belongs to the current epoch, the current slot cannot be the first slot of the epoch, as honest validators only accept blocks from the current epoch after its first slot.\label{rem:current-epoch-block-not-at-first-slot}
    Hence, the active state conditions cannot have been triggered in the current slot, so by \Cref{ln:if-bcand-npreceq-head,ln:if-bcand-e-1,ln:set-bcand-to-output-find-latest} the input $b_c$ to $\findlatestconfirmeddescendant$ is the previously confirmed block, meaning that $\btcand$ is a descendant of the previously confirmed block.
    Let us now consider two cases, depending on the epoch of $b_c$.
    If $b_c$ belongs to the current epoch, then $b_c$ was confirmed via Case~1 (the only case that returns a current-epoch block), so the second conjunct held for $b_c$ at that time, giving a descendant $b' \succeq b_c$ with $\epoch(\gu(b')) \geq \epoch(b_c)-1 = \epoch(\now)-1 \geq \epoch(\now)-2$.
    If $b_c$ belongs to the previous epoch, then $b_c$ was confirmed via Case~1 in the previous epoch (or via Case~3), and the same argument gives $\epoch(\gu(b')) \geq \epoch(b_c)-1 = \epoch(\now)-2$.

    The second property is that, by the beginning of the current slot, any honest validator has $\btcand$ in its view.
This is because condition~\eqref{eq:case1-lmd} implies $\var[val=v]{\isOneConfirmedExt}(\btcand,\guattime[val=v,time=\prevfirstslotepoch{t}])$, which further implies that at least one honest validator cast an \LMDGHOST vote for $\btcand$ during some past slot.

    Now, let us start from the current epoch $e$.
    The two properties above ensure that taken any honest validator's greatest justified checkpoint $C'$, either $C'=\gu(\btcand)$  or, if not, $C'$ belongs to epoch $e-1$.
    This and the second conjunct then ensure that $\btcand$ is a descendant of $\block(C')$.
    The first property above ensures that $\btcand$ cannot be filtered out by $\FILHFC$ during the current epoch.

    Let us now move to the next epoch $e+1$.
    Because of the third conjunct, we know that the only checkpoint belonging to the current epoch that can ever be justified is $\btcand$'s checkpoint.
    Therefore, taking any honest validator's greatest justified checkpoint $C'$ at any point $t'$ during the next epoch, $\btcand$ is a descendant of $\block(C')$.
    Then, given the second conjunct, we know that any honest validator's view includes a block $b'\succeq \btcand$ such that $b'$'s voting source at time $t'$ belongs to at least epoch $e-1 = (e+1) -2$.
    Hence, $\btcand$ cannot be filtered out by $\FILHFC$ at any point during epoch $e+1$ either.

    Let us now move to the next epoch $e+2$.
    Given that $\btcand$ has been canonical during the entirety of epoch $e+1$, because of the third conjunct and \Cref{assum:ffg-assumptions:justified-checkpoint-next-epoch}, we know that by the beginning of epoch $e+2$, any honest validator has received at least one block descendant of $\btcand$ that \ujustifies $\btcand$'s checkpoint.
    Therefore, taking any honest validator $v$'s greatest justified checkpoint $C'$, either $C'$ corresponds to $\btcand$'s justified checkpoint or $C'$ belongs to a epoch higher than $\epoch(\btcand)=e$.
    Given that $\btcand$ has been canonical during the entirety of any epoch starting from epoch $e+1$, this also means that $C'$ does not conflict with $\btcand$.
    If $C'$ belongs to an epoch higher than $e$, then $\block(C')$ must be  a descendant of $\btcand$.
    Then, given that the \LMDGHOST fork-choice starts from the greatest justified checkpoint, $\btcand$ is clearly part of the canonical chain.
    If $C'$ corresponds to $\btcand$'s checkpoint, then it means that $v$'s view include a block $b' \succeq \btcand$, with no children, such that the voting source of $b'$ corresponds to $\btcand$'s checkpoint.
    This is sufficient to prove that $\btcand$ is not filtered out by $\FILHFC$, as per condition~\ref{itm:filhfc-condition-2b} above.

    When we move to the next epoch $e+3$ and any successive epoch, we can repeat the reasoning applied to $e+2$ in an inductive manner.

    Figures~\ref{fig:case1a} and~\ref{fig:case1b} illustrate Case~1 (in both figures, $b$ denotes $\btcand$).
    Figure~\ref{fig:case1a} shows the scenario where we are attempting to confirm for the first time a block belonging to the current epoch, while Figure~\ref{fig:case1b} shows the general case where we have already confirmed a block belonging to the current epoch.

    \begin{figure}[t]
      \centering
      \input{case1a.tikz}
      \caption{
      Confirmation of the first block belonging to the current epoch.
      The first time that we attempt to confirm a block belonging to the current epoch, we run the check $\var[val=v]{\willChkpBeJustified}(\chkp(b))$, in addition to checking that there exists a descendant $b'$ of $b$ that \ujustifies a checkpoint belonging to either the previous or current epoch. 
      As depicted here, $b'$ does not need to be on the canonical chain.
      However, in practice, it is likely to be.
      }
      \label{fig:case1a}
    \end{figure}

    \begin{figure}[t]
      \centering
      \input{case1b.tikz}
      \caption{
      Confirmation of a generic block belonging to the current epoch (we have already confirmed a block belonging to the current epoch).
      In this case, we check that there exists a descendant $b'$ of $b$ that \ujustifies a checkpoint belonging to either the previous or current epoch, but we do not run
      $\var[val=v]{\willChkpBeJustified}(\chkp(b))$.
      }
      \label{fig:case1b}
    \end{figure}

\paragraph{Case~2 (\Cref{ln:case2}): no progress (return $b_c$).}
    If the conditions of Case~1 are not met, the algorithm does not attempt to confirm any block belonging to the current epoch, and restricts candidates to the previous epoch.

    Case~2 establishes a necessary condition to be able to confirm blocks belonging to the previous epoch.
    Formally, if
    \[
    \begin{aligned}
    &\text{\bf not}\;\Bigl(
        \slot(\now) = \firstslot(\epoch(\now))\\
    &\phantom{\text{\bf not}\;\Bigl(}\text{\bf or}\;
        \exists b'\succeq b_c :
        \willNoConflictingChkpBeJustified^v\!\bigl(\chkp(b',\epoch(\now))\bigr)
    \Bigr),
    \end{aligned}
    \]

    then the function returns $b_c$ and does not attempt to confirm any new block.

    The predicate $\willNoConflictingChkpBeJustified^v(C)$ ensures, from validator $v$'s current view,
    under the synchrony assumption and the assumed Byzantine bounds, that \emph{no checkpoint
    conflicting with $C$ can ever become justified in the future}.
    Crucially, this
    is weaker than predicting that $C$ itself will be justified; it only rules out
    \emph{conflicting} justification.
    Both $\willNoConflictingChkpBeJustified$ and
$\willChkpBeJustified$ are established via the same conservative accounting
principle: start from the \FFG support already observed up to slot $s{-}1$,
discount it by the worst-case Byzantine contribution (up to $\beta$ times
the committee weight observed so far), add back the weight of
already-identified slashed validators (whose Byzantine contribution has
already been accounted for), and add a worst-case lower bound on
the remaining honest committee weight in $[s,\lastslot(e)]$.

    If this combined lower bound exceeds the relevant threshold, then no conflicting checkpoint can become justified by the end of the epoch in any honest view.

    Operationally, when $\willNoConflictingChkpBeJustified^v(\chkp(b',\epoch(\now)))$ holds for some current-epoch successor $b' \succeq b_c$, no checkpoint conflicting with $\chkp(b',\epoch(\now))$ can ever become justified. Consequently, any checkpoint that does become justified in the current epoch must belong to the same branch as $b'$, and hence as $b_c$. In particular, in the mid-epoch case where $\bconfirmed$ belongs to epoch $\epoch(t)-1$, this implies that any justified checkpoint belonging to epoch $\epoch(t)$ is a descendant of $\bconfirmed$. Thus, if no such $b'$ exists, $v$ cannot rule out a future conflicting justification, and any additional confirmation could be vulnerable \FILHFC filtering; Case~2 therefore conservatively does not progress confirmation.

    If we are in the first slot of the current epoch, then the $\willNoConflictingChkpBeJustified$ check is not required.
    This is because, this implies that we can at most confirm a block $b$ belonging to the previous epoch and, due to the Safety property of confirmations, we know that any \FFG vote cast by honest validators from the first slot of the current epoch onwards will target a descendant of $b$.
    Hence, no checkpoint belonging to the current epoch and conflicting with $b$ could ever be justified. 
    
   \paragraph{Case~3 (\Cref{ln:case3-pre,ln:case3}): previous-epoch confirmation.}
    Reaching Case~3 means that the rule did not accept any current-epoch confirmation via Case~1,
    but it is also not forced to stop attempting to confirm new blocks due to  Case~2.
    
First, the algorithm recomputes the highest \LMDGHOST-safe candidate that belongs to the previous epoch, since the conditions of Case~1 were not met and no current-epoch block will be confirmed:

\begin{equation}\label{eq:case-3-lmd-conf}
    \begin{aligned}
    \btcand \gets \getHighest\bigl(
    \{\, b' \in \canonChainFrom(b_c)
    :\;&\;
    \epoch(b')<\epoch(\now)\\
    &\;\land\;
    \isOneConfirmedExtFrom(b',\guattime[val=v,time=\prevfirstslotepoch{t}],b_c)
    \,\}\bigr).
    \end{aligned}
\end{equation}

    Then, if we are at the beginning of an epoch it returns $\btcand$ only if its \emph{voting source} is not older than two epochs ago:
    \begin{equation}\label{eq:case-3-vs}
        \epoch\!\bigl(\votsource[blck=\btcand,time=\now]\bigr)
        \;\ge\;
        \epoch(\now)-2.
    \end{equation}

    Figure~\ref{fig:case3a} illustrates this case.

    \begin{figure}[t]
      \centering
      \input{case3a.tikz}
    \caption{
      Confirmation of a block belonging to the previous epoch when at the beginning of the current epoch.
      At the start of the current epoch, a candidate from the previous epoch
      can be returned if the epoch of its voting source is two epochs ago or more recent, ensuring it will not be filtered out by $\FILHFC$.
      }
      \label{fig:case3a}
    \end{figure}

    Let us now provide an intuition for why, if we are at the beginning of the current epoch, the two conditions above are sufficient to ensure that $\btcand$ will always be canonical.

    The first property that we want to show is that (i) no checkpoint conflicting with $\btcand$'s checkpoint can ever be justified and that, (ii) by the end of the current epoch, all honest validators will have in their view a block $b'\succeq \btcand$ that \ujustifies $\btcand$'s checkpoint.
    Let us proceed by cases based on whether we have just entered the active state. 
    First, assume that this is the case.
    Given the checks performed when entering the active state and that we only confirm blocks belonging to the canonical chain, it is easy to see that  both conditions (i) and (ii) are satisfied.
    Then, assume that the active state conditions are not triggered; we refer to this scenario as the \emph{continuation path}.
    Due to \Cref{ln:if-bcand-npreceq-head,ln:if-bcand-e-1,ln:set-bcand-to-output-find-latest}, this means that $\findlatestconfirmeddescendant$ starts from a confirmed block $b_c$ belonging to the previous epoch.
    This implies that $\willChkpBeJustified$ was checked on $\btcand$'s checkpoint at some point during the previous epoch, which gives condition~(i).
    Condition~(ii) then follows from condition~(i) and \Cref{assum:ffg-assumptions:justified-checkpoint-next-epoch}.
    
    We now explain why the active state entry condition requires
    $\gu(\head) = \guattime[val=v,time={\slotstart(\slot(\now)-1)}]$.
    When the rule has just entered the active state,
    $\willChkpBeJustified$ was not necessarily checked on $\btcand$'s
    checkpoint during the previous epoch, because no prior confirmed block
    existed from which to run $\findlatestconfirmeddescendant$.
    Without the check
    $\gu(\head) = \guattime[val=v,time={\slotstart(\slot(\now)-1)}]$,
    the following scenario is possible.
    Suppose a block
    $b \succ \block(\guattime[val=v,time={\slotstart(\slot(\now)-1)}])$ is
    confirmed through $\findlatestconfirmeddescendant$.
    Since the rule entered the active state, there exists a block
    $b' \succ \block(\guattime[val=v,time={\slotstart(\slot(\now)-1)}])$
    that \ujustifies $\chkp(b)$.
    However, some \FFG votes in $b'$'s chain may come from dishonest
    validators whose slashable offences are already included in the chain
    of~$b$ but not in the chain of~$b'$.
    As a result, $\chkp(b)$ may not be justifiable on any extension
    of $b$'s chain, which could cause $b$ to be removed by $\FILHFC$.
    This situation cannot arise when $\willChkpBeJustified(C)$ is
    evaluated, because that predicate discounts the worst-case Byzantine weight from
the \FFG votes for checkpoint~$C$ (while adding back
already-identified slashed validators), and, given that $\block(C)$ is
canonical, all honest validators will cast an \FFG vote
targeting~$C$ in the remaining slots of the epoch.

    \begin{sloppypar}
    The second property is that, by the beginning of the current slot, $\btcand$ has been received by all honest validators.
    This is due to condition~\eqref{eq:case-3-lmd-conf} which implies $\var[val=v]{\isOneConfirmedExt}(\btcand,\guattime[val=v,time=\prevfirstslotepoch{t}])$.
    This further implies that at least one honest validator cast an \LMDGHOST vote for $\btcand$ during some past slot.
    Due to our synchrony assumption, this implies that, by the beginning of the current slot, $\btcand$ has been received by all honest validators.
    \end{sloppypar}

    Let us start from the current epoch $e$.
    By condition~\eqref{eq:case-3-vs}, the epoch of $\btcand$'s voting source is at least $e-2$.
    Combined with the second property above, this implies that at the beginning of the current epoch $e$, taking any honest validator's greatest justified checkpoint $C'$, either $C'$ corresponds to $\votsource[blck=\btcand,time=\now]$ or it belongs to the previous epoch $e-1$.
    Condition (ii) of the first property above then implies that any honest validator's greatest justified checkpoint is an ancestor of $\btcand$.
    This, the second property above, and condition~\eqref{eq:case-3-vs} imply that $\btcand$ cannot be filtered out by $\FILHFC$ during the current epoch $e$.
    Hence, $\btcand$ will be canonical for the whole current epoch.

    Let us now move to the next epoch $e+1$.
    The first property above implies that, by the end of epoch $e$, all honest validators will have in their view a block $b'\succeq \btcand$ that \ujustifies $\btcand$'s checkpoint.
    Because $\btcand$ is canonical for the whole epoch $e$, no checkpoint belonging to  epoch $e$ and conflicting with $\btcand$ can ever be justified.
    Hence, at the beginning of epoch $e+1$ we are in the same situation we ended up at the beginning of epoch $e+2$ in Case~1.
    Namely, that taking any honest validator $v$'s greatest justified checkpoint $C'$, either $C'$ corresponds to $\btcand$'s justified checkpoint or $C'$ belongs to a epoch higher than $\epoch(\btcand)$.
    Then, we can show $\btcand$ being canonical from now on by a similar arguments to the one used for Case~1.
    
    \begin{figure}[t]
      \centering
      \input{case3b.tikz}
    \caption{
    Confirmation of a block belonging to the previous epoch when not at the beginning of the current epoch.
    In this case, in addition to checking the epoch of the voting source of $\btcand$ is two epochs ago or more recent, we also check that there exists a descendant $b'$ of $\btcand$ that \ujustifies a checkpoint belonging to either the previous or current epoch. 
    As depicted here, $b'$ does not need to be on the canonical chain.
    However, in practice, it is likely to be.
    }
    \label{fig:case3b}
    \end{figure}  
        
    \Cref{fig:case3b} depicts the case where we are not at the beginning of the current epoch.
    In this case, then we also need to check for the following condition.
    \begin{equation}\label{eq:case-3-desc-gu}
        b'\in \viewattime[time={\now},val=v], b'\succeq \btcand \land \epoch(\gu(b')) \geq \epoch(\now) - 1
    \end{equation}
    
    Namely, we check that there exists a descendant $b' \succeq \btcand$ such that $b'$ \ujustifies a checkpoint belonging to the previous epoch or a latest epoch, which, in practice, means either the previous or current epoch.
    Assume that this were not the case. 
    While we know that enough \FFG votes to justify $\btcand$'s checkpoint have been sent, because we are not at the beginning of an epoch, we cannot apply \Cref{assum:ffg-assumptions:justified-checkpoint-next-epoch}.
    This means that it could be that at the end of the current epoch, (i) no chain extending $\btcand$ includes enough \FFG votes to justify $\btcand$'s checkpoint, (ii) while it could exist in the view of an honest validator $v$, a chain $b''$, conflicting with $\btcand$ but still descending from $\btcand$'s checkpoint, that contains enough \FFG votes to justify $\btcand$'s checkpoint (\Cref{fig:case3b-noquorum}).
    As a consequence of this, on the transition to the next epoch, $v$ would set its greatest justified checkpoint to $\btcand$'s checkpoint.
    However, given that no chain extending $\btcand$ includes enough \FFG votes to justify $\btcand$'s checkpoint,
    $\btcand$ would then be filtered out by $\FILHFC$ in $v$'s view.

    \begin{figure}[t]
        \centering
        \input{case3b-noquorum.tikz}
        \caption{Situation where condition~\eqref{eq:case-3-desc-gu} is not satisfied. By the end of epoch $\epoch(\now)$,  $b''$ (conflicting with $b$ but descending from $\chkp(b)$) \ujustifies $\chkp(b)$, while any chain extending $b$ only \ujustifies an older checkpoint $C' = \chkp(b,\epoch(\now)-2)$. As a consequence, at the transition to epoch $\epoch(\now)+1$, validator $v$ sets its greatest justified checkpoint to $\chkp(b)$, but $b$ is then filtered out by $\FILHFC$ as none of its descendants \ujustify it.}
        \label{fig:case3b-noquorum}
    \end{figure}    
    
 \paragraph{Case~4 (\Cref{ln:case4}): previous-epoch fallback (return $\bcand$).}
    Case~4 catches the scenario where a block $b$ satisfies all of Case~3 conditions except for condition~\eqref{eq:case-3-vs}, but there exists a descendant $b' \succeq b$ that satisfies such a condition, though it does not satisfy all of the other conditions checked by Case~3.
    \Cref{fig:case4b,fig:case4a} shows two examples of such a scenario.

    \begin{figure}[t]
      \centering
      \input{case4a.tikz}
    \caption{
        Example where the Case~3 check fails, but the Case~4 check succeed when at the beginning of the current epoch.
        In this example, $b$ does not satisfy condition~\eqref{eq:case-3-vs} of Case~3, while $b'$ satisfies it, but it  cannot be confirmed because it does not pass the $\var[val=v]{\isOneConfirmedExt}$ check.
    }
      \label{fig:case4a}
    \end{figure}

    \begin{figure}[t]
      \centering
      \input{case4b.tikz}
    \caption{Example where the Case~3 check fails, but the Case~4 check succeed when at the beginning of the current epoch.
    In this example, $b$ does not satisfy condition~\eqref{eq:case-3-vs} of Case~3.
    Block $b''$ satisfies condition~\eqref{eq:case-3-vs} of Case~3, but it fails to satisfy condition~\eqref{eq:case-3-desc-gu} of Case~3 (the condition $\epoch(\votsource[blck=b'',time=\now])\geq\epoch(\now)-2 \land \epoch(\gu(b''))<\epoch(\now)-1$ effectively implies $\epoch(\votsource[blck=b'',time=\now])=\epoch(\gu(b''))=\epoch(\now)-2$).
    Finally, $b'$ does not pass the $\var[val=v]{\isOneConfirmedExt}$ check, but it satisfies condition~\eqref{eq:case-3-desc-gu} of Case~3.}
      \label{fig:case4b}
    \end{figure}

    Importantly, to pass the Case~4 check, $b'$ must have been received no later than the beginning of the previous slot.
    We need this extra condition to ensure that, thanks to our synchrony assumption, $b'$ is received by all honest validators by the beginning of the current slot.
    Then, we can apply the argument outlined in the explanation of Case~3 to argue that $b'$, and therefore $b\preceq b'$, cannot be filtered out by $\FILHFC$ during the current epoch.

The block remains in the filtered tree during the next epoch thanks to condition~\eqref{eq:case-3-desc-gu}, which is also checked for Case~4.
    Then the argument proceeds as per Case~3.

    \paragraph{Case~5: no progress.}
    If Case~4's checks do not pass either, then the algorithm does not progress the confirmed block.
    
    \medskip

    To summarize, $\findlatestconfirmeddescendant$ advances the confirmed block as far as possible while ensuring the result cannot be filtered out by $\FILHFC$. It selects its output based on five cases, checked in order:
    (i)~(Case~1, current epoch) if the highest \LMDGHOST-safe block along the canonical chain extending $b_c$ belongs to the current epoch and the conditions ensuring it is not filtered out by $\FILHFC$ hold (an existing descendant that \ujustifies a checkpoint belonging to the previous or current epoch, and $\willChkpBeJustified$ for the block's checkpoint holds when it crosses into the current epoch), return it;
    (ii)~(Case~2, no progress) otherwise, if the algorithm cannot rule out via $\willNoConflictingChkpBeJustified$ that a current-epoch checkpoint conflicting with $b_c$'s chain will ever be justified, return $b_c$ unchanged (this avoids a previous-epoch confirmation that could later be filtered out when a conflicting current-epoch checkpoint is justified);
    (iii)~(Case~3, previous epoch) otherwise, take the highest previous-epoch \LMDGHOST-safe block along the canonical chain extending $b_c$: if its voting source has epoch at least $\epoch(\now)-2$ and (when not at the first slot of the epoch) it has a descendant that \ujustifies a checkpoint belonging to the previous or current epoch, return it;
    (iv)~(Case~4, previous-epoch fallback) otherwise, take the highest previous-epoch \LMDGHOST-safe block along the canonical chain extending $b_c$ that has a descendant $b'$, received by the beginning of the previous slot, with the epoch of its voting source belonging to at least $\epoch(\now)-2$ (and, when not at the first slot of the epoch, a descendant that \ujustifies a checkpoint belonging to the previous or current epoch): if it exists, return it (this preserves progress when Case~3's voting-source check on the highest safe block fails: the descendant $b'$ satisfies that check and, having been received by the beginning of the previous slot, is seen by every honest validator by synchrony, which keeps the returned block in the filtered tree; however, $b'$ cannot be confirmed because it does not satisfy all the conditions checked by Case~3, for example, it does not pass the $\LMDGHOST$-safety check);
    (v)~(Case~5, no progress) otherwise, return $b_c$ unchanged.
    
\subsection{Monotonicity}
\label{sec:monotonicity}

In this section, we show that confirmation is monotonic:
once a block $b$ is confirmed, it remains confirmed at all future times.
Without additional assumptions, this can be shown only for the current and next epoch.
Extending monotonicity to all future times requires a strong liveness assumption on confirmation itself: under favorable synchrony and canonicity conditions, honest validators reliably confirm a descendant within each epoch.
Establishing monotonicity requires showing two properties: (i)~that the confirmed block passes
reconfirmation at the beginning of each epoch, and (ii)~that the rule never enters the inactive state, so that the rule never resets to
finality.
We show~(i) by a slot-by-slot induction argument that relies on
$\beta < 1/4$, and~(ii) by showing that $\willChkpBeJustified$ ensures
no conflicting checkpoint can be justified, which keeps $b$ on the
canonical chain and prevents the rule from entering the inactive state.
The additional liveness assumption used to extend the argument beyond the next epoch is stated below.

\begin{assumption}\label{assum:monotonicity}
    Provided that \Cref{assum:beta,assum:no-change-to-the-validator-set} hold,
    if a block $b$ is canonical in the view of all honest validators at any time during epoch $e$ and $\slotstart(\firstslot(\epoch(e)))\geq \GGST$, then, by the end of epoch $e$, any honest validator has confirmed a descendant $b' \succeq b$ belonging to epoch $e$.
\end{assumption}

We first show that, under $\beta < 1/4$, any block confirmed in the
current epoch~$e$ will pass reconfirmation at the beginning of epoch~$e+1$.
The idea is the following: if $b$ is confirmed and canonical, then at
each subsequent slot, honest validators vote for~$b$, adding at least
$(1-\beta)$ times the committee weight to the attestation support.
Since $\beta < 1/4$ implies $1-\beta > 1/2 + \beta$, this new support
exceeds the threshold fraction of the new weight added to the
denominator of $\indQ$, so $\indQ$ at $b$ remains above the
confirmation threshold at all subsequent slots.
We now make this precise.
Consider the simplified condition for \LMDGHOST safety

\begin{equation}\label{eq:cond-on-q-simple}
    \indicatorfromblock[from=b,to=\slot(t)-1,val=v,when=t,chkp=C]{\indQ}
    \;>\;
    \frac{1}{2}
    +\beta
    +
    \frac{\boostweight[chkp=C]}{2\commweightfromafterparentblock[from=b,to=\slot(t)-1,chkp=C]{\allvals}},
\end{equation}
obtained from the simplified condition~\eqref{eq:cond-on-q} by setting
$\Phi_{b}^{v,t,C}
    := \commweightfromafterparentblock[from=b,to=\slot(t)-1,chkp=C]{\allvals}
    + \boostweight[chkp=C]$,
i.e., omitting the discount term of~\eqref{eq:phi-def}.

Suppose that $b$ is canonical and that
condition~\eqref{eq:cond-on-q-simple} holds at time~$t$.
At the next slot $t' = \slotstart(\slot(t)+1)$, since $b$ is canonical,
all honest validators in the committee of slot $\slot(t)$ vote in
support of~$b$.
This adds at least $(1-\beta)\Wslot$ to the attestation weight for~$b$ and
exactly $\Wslot$ to the total committee weight, where $\Wslot$ denotes the
weight of a single committee.
The indicator at time~$t'$ therefore satisfies
\def\alignexplwidth{7cm}
\allowdisplaybreaks
\begin{align*}
    &\hspace{3ex}\indicatorfromblock[from=b,to=\slot(t')-1,val=v,when=t',chkp=C]{\indQ}\\
    &\geq
    \frac
    {\attsetweightfromblock[from=b,to=\slot(t)-1,val=v,when=t,chkp=C]{\allatts}+(1-\beta)\Wslot}
    {\commweightfromafterparentblock[from=b,to=\slot(t)-1,chkp=C]{\allvals}+\Wslot}
    &&\alignexpl[\alignexplwidth]{Honest validators in slot $\slot(t)$ add at least $(1-\beta)\Wslot$ to the attestation weight for $b$ and exactly $\Wslot$ to the total committee weight.}
    \\
    &>
    \frac
    {\attsetweightfromblock[from=b,to=\slot(t)-1,val=v,when=t,chkp=C]{\allatts}+\left(\frac{1}{2}+\beta\right)\Wslot}
    {\commweightfromafterparentblock[from=b,to=\slot(t)-1,chkp=C]{\allvals}+\Wslot}
    &&\alignexpl[\alignexplwidth]{Since $\beta < 1/4$ implies $1-\beta > \frac{1}{2}+\beta$.}
    \\
    &=
    \frac
    {\indicatorfromblock[from=b,to=\slot(t)-1,val=v,when=t,chkp=C]{\indQ}
    \cdot\commweightfromafterparentblock[from=b,to=\slot(t)-1,chkp=C]{\allvals}+\left(\frac{1}{2}+\beta\right)\Wslot}
    {\commweightfromafterparentblock[from=b,to=\slot(t)-1,chkp=C]{\allvals}+\Wslot}
    &&\alignexpl[\alignexplwidth]{By definition of $\indQ$.}
    \\
    &>
    \frac
    {\left(\frac{1}{2}+\beta+\frac{\boostweight[chkp=C]}{2\commweightfromafterparentblock[from=b,to=\slot(t)-1,chkp=C]{\allvals}}\right)
    \cdot\commweightfromafterparentblock[from=b,to=\slot(t)-1,chkp=C]{\allvals}+\left(\frac{1}{2}+\beta\right)\Wslot}
    {\commweightfromafterparentblock[from=b,to=\slot(t)-1,chkp=C]{\allvals}+\Wslot}
    &&\alignexpl[\alignexplwidth]{By~\eqref{eq:cond-on-q-simple} at time $t$.}
    \\
    &=
    \frac
    {\left(\frac{1}{2}+\beta\right)\left(\commweightfromafterparentblock[from=b,to=\slot(t)-1,chkp=C]{\allvals}+\Wslot\right)+ \frac{\boostweight[chkp=C]}{2}}
    {\commweightfromafterparentblock[from=b,to=\slot(t)-1,chkp=C]{\allvals}+\Wslot}
    &&\alignexpl[\alignexplwidth]{Expanding and regrouping.}
    \\
    &=
    \frac{1}{2}+\beta+\frac{\boostweight[chkp=C]}{2\left(\commweightfromafterparentblock[from=b,to=\slot(t)-1,chkp=C]{\allvals}+\Wslot\right)}
    &&\alignexpl[\alignexplwidth]{Splitting the fraction.}
    \\
    &=
    \frac{1}{2}+\beta+\frac{\boostweight[chkp=C]}{2\commweightfromafterparentblock[from=b,to=\slot(t')-1,chkp=C]{\allvals}}.
    &&\alignexpl[\alignexplwidth]{Since $\commweightfromafterparentblock[from=b,to=\slot(t)-1,chkp=C]{\allvals}+\Wslot = \commweightfromafterparentblock[from=b,to=\slot(t')-1,chkp=C]{\allvals}$.}
\end{align*}
This is exactly condition~\eqref{eq:cond-on-q-simple} at time~$t'$.
By induction over slots, condition~\eqref{eq:cond-on-q-simple} holds
at every subsequent slot within the epoch, and in particular at the
beginning of epoch~$e+1$ where reconfirmation is performed.

It remains to show that the rule does not enter the inactive state
at the beginning of epoch~$e+1$.
When confirming the first block of epoch~$e$, the predicate
$\willChkpBeJustified$ was checked on $b$'s checkpoint, guaranteeing that
enough \FFG votes will be cast to justify $\chkp(b)$ by the end of
epoch~$e$.
Since $b$ is canonical throughout epoch~$e$, all \FFG votes sent by honest validators
during epoch~$e$ target descendants of~$b$, and therefore no checkpoint
belonging to epoch~$e$ and conflicting with $b$ can ever be justified
(Property~\ref{prop:gasper-basic:no-conflicting-if-all-honest-votes-in-support-of-b}).
Moreover, because $\willChkpBeJustified$ was also checked during
epoch~$e-1$, \Cref{assum:ffg-assumptions:justified-checkpoint-next-epoch}
guarantees that the checkpoint belonging to epoch~$e-1$ is \ujustified by
the beginning of epoch~$e+1$.
Therefore, $\guattime[time=\slotstart(\firstslot(e+1))]$
belongs to either epoch~$e-1$ or epoch~$e$, and in neither case it does
conflict with~$b$, which means it is an ancestor of~$b$.
Given that $b$ is canonical throughout epoch~$e$, the rule does not enter the inactive state at the beginning of epoch~$e+1$: the confirmed
block belongs to the previous epoch (not older), it remains on the
head chain, and reconfirmation succeeds (for current-epoch blocks,
this follows from the slot-by-slot induction above; for previous-epoch
blocks in the confirmed chain, this is discussed in the next
paragraph).

This means that $b$ remains confirmed throughout epoch~$e+1$, since
the rule can only confirm descendants of what has already been
confirmed.
By \Cref{assum:monotonicity}, during epoch~$e+1$ a descendant
$b' \succ b$ belonging to epoch~$e+1$ is confirmed.
The same argument applies to~$b'$ at the beginning of epoch~$e+2$:
reconfirmation succeeds, the rule does not enter the inactive state, and $b'$ (and therefore $b$) remains confirmed throughout
epoch~$e+2$.
By induction, $b$ remains confirmed at all future times.

\paragraph{Role of $\isOneConfirmedSpecial$.}
The argument above assumes that reconfirmation succeeds at the
beginning of each epoch.
The slot-by-slot induction shows that $\indQ$ for any confirmed block
remains above the threshold of~\eqref{eq:cond-on-q-simple} at all
subsequent slots, which ensures that the confirmed block itself passes
reconfirmation.
However, reconfirmation via $\isChainOneConfirmed$ also re-checks
previous-epoch blocks belonging to the confirmed chain.
A subtlety arises when the first block $b$ of the current epoch
satisfies $\slot(b) > \firstslot(\epoch(b))$ and
$\epoch(\parent(b)) < \epoch(b)$, i.e., there are empty slots between
the epoch start and~$b$ (see Figure~\ref{fig:epoch-boundary-variant}).
In this setting, $\isOneConfirmed$ holding for $b$ does not guarantee
that previous-epoch ancestors of $b$ -- in particular $\parent(b)$ --
pass $\isOneConfirmed$ during reconfirmation under the new checkpoint.
The coefficient of $\beta$ in $\isOneConfirmed$'s threshold accounts for Byzantine
weight only from $\slot(b)$ onward, but the committees of the empty
slots between $\firstslot(\epoch(b))$ and $\slot(b) - 1$ may also
contain Byzantine validators.
·Since the confirmation check at $b$ does not budget for this
Byzantine weight, the amount of honest support that the check ensures may be
insufficient for $\parent(b)$ to pass $\isOneConfirmed$ during
reconfirmation under the new checkpoint.

The variant $\isOneConfirmedSpecial$
(Definition~\ref{def:isOneConfirmedSpecial}) addresses this by
extending the coefficient of $\beta$ to cover all committees from
$\firstslot(\epoch(b))$ onward.

\begin{definition}[Epoch-boundary per-block confirmation predicate]\label{def:isOneConfirmedSpecial}
    For any block $b$, checkpoint $C$, honest validator $v$, and time $t\geq\GGST$,
    \[
    \var[val=v]{\isOneConfirmedSpecial}(b,C,t)
    \;:=\;
    \indicatorfromblock[from=b,to=\slot(t)-1,val=v,when=t,chkp=C]{\indQ}
    \;>\;
    \frac{\Phi_{b}^{v,t,C}}
        {2\,\commweightfromafterparentblock[from=b,to=\slot(t)-1,chkp=C,when=t,val=v]{\allvals}}
    + \beta 
        \frac
            {\commweightfromslot[from=\firstslot(\epoch(b)),to=\slot(t)-1,chkp=C,when=t,val=v]{\allvals}}
            {\commweightfromafterparentblock[from=b,to=\slot(t)-1,chkp=C]{\allvals}}    
    - \frac
        {\attsetweightfromblock[from=b,to=\slot(t)-1,val=v,when=t,chkp={C}]{\slashvals}}
        {\commweightfromafterparentblock[from=b,to=\slot(t)-1,chkp=C]{\allvals}}.
    \]
\end{definition}

The only difference from $\isOneConfirmed$ is the numerator of the coefficient of $\beta$:
$\isOneConfirmed$ uses
$\commweightfromblock[from=b,to=\slot(t)-1,chkp=C]{\allvals}$
(committees from $\slot(b)$), while $\isOneConfirmedSpecial$ uses
$\commweightfromslot[from=\firstslot(\epoch(b)),to=\slot(t)-1,chkp=C]{\allvals}$
(committees from $\firstslot(\epoch(b))$).
This makes the threshold stricter, as it accounts for a larger
Byzantine budget.

To see why $\isOneConfirmed$ is insufficient, consider the case where
$\slot(b) > \firstslot(\epoch(b))$ and
$\epoch(\parent(b)) < \epoch(b)$.
At reconfirmation time $t' = \slotstart(\firstslot(e+1))$, the
coefficient of $\beta$ for $b$ and $\parent(b)$ satisfy
\[
\frac
    {\commweightfromblock[from=b,to=\slot(t')-1,chkp=C,when=t',val=v]{\allvals}}
    {\commweightfromafterparentblock[from=b,to=\slot(t')-1,chkp=C]{\allvals}}
\;<\; 1 \;=\;
\frac
    {\totvalsetweight[chkp=C]{\allvals}}
    {\totvalsetweight[chkp=C]{\allvals}}
\;=\;
\frac
    {\commweightfromblock[from=\parent(b),to=\slot(t')-1,chkp=C,when=t',val=v]{\allvals}}
    {\commweightfromafterparentblock[from=\parent(b),to=\slot(t')-1,chkp=C]{\allvals}}.
\]
The ratio at $b$ is strictly less than~$1$ because the denominator
(from $\parent(b)^+$) includes the empty-slot committees of
$\epoch(b)$, while the numerator (from $\slot(b)$) does not.
The ratio at $\parent(b)$ equals~$1$ because $t' = \slotstart(\firstslot(e+1))$: both the numerator's range (from $\slot(\parent(b))$ to $\lastslot(e)$) and the denominator's range (from $\slot(\parent(\parent(b)))+1$ to $\lastslot(e)$) cover all the committees of epoch $e$, and under \Cref{assum:no-change-to-the-validator-set} each corresponds to the full validator set weight $\totvalsetweight[chkp=C]{\allvals}$.
This means that $\isOneConfirmed$ at $b$ uses a smaller $\beta$
correction than $\isOneConfirmed$ at $\parent(b)$ would require, so
passing $\isOneConfirmed$ at $b$ does not guarantee that $\parent(b)$
passes $\isOneConfirmed$ during reconfirmation.

With $\isOneConfirmedSpecial$, the ratio at $b$ becomes
\[
\frac
    {\commweightfromslot[from=\firstslot(\epoch(b)),to=\slot(t')-1,chkp=C,when=t',val=v]{\allvals}}
    {\commweightfromafterparentblock[from=b,to=\slot(t')-1,chkp=C]{\allvals}}
\;=\; 
\frac
    {\totvalsetweight[chkp=C]{\allvals}}
    {\totvalsetweight[chkp=C]{\allvals}}
\;=\;
1 \;=\;
\frac
    {\commweightfromblock[from=\parent(b),to=\slot(t')-1,chkp=C,when=t',val=v]{\allvals}}
    {\commweightfromafterparentblock[from=\parent(b),to=\slot(t')-1,chkp=C]{\allvals}}.
\]
The ratios now match: $\isOneConfirmedSpecial$ at $b$ uses the same
$\beta$ correction that $\isOneConfirmed$ at $\parent(b)$ requires.
Since $\parent(b) \prec b$ and every validator supporting $b$ also
supports $\parent(b)$, the attestation weight for $\parent(b)$ is at
least as large as for $b$.
Together with the matching $\beta$ correction, this ensures that
$\parent(b)$ passes $\isOneConfirmed$ during reconfirmation
(see \Cref{lem:beta-less-than-quarter-no-reconfirmation-required-bconf-prev-epoch-ex}
in the Appendix).
Moreover, since $\isOneConfirmedSpecial$ is stricter than
$\isOneConfirmed$, $b$ itself automatically passes $\isOneConfirmed$
during reconfirmation: $\indQ$ at $b$ exceeded the higher threshold of
$\isOneConfirmedSpecial$ during normal confirmation, and by the
slot-by-slot argument above, $\indQ$ remains above the lower threshold
of $\isOneConfirmed$ at reconfirmation time.
Consequently, $\isChainOneConfirmed$ does not need to use
$\isOneConfirmedSpecial$ during reconfirmation -- the standard
$\isOneConfirmed$ suffices for all blocks belonging to the confirmed chain.

\begin{figure}[t]
  \centering
  \input{empty-slot-beginning.tikz}
\caption{The epoch-boundary variant $\isOneConfirmedSpecial$.
Block $b$ is the first block of epoch~$e$, with $\parent(b)$ in
epoch~$e-1$ and empty slots between $\firstslot(e)$ and $\slot(b)$.
The upper brace shows the committee range used by $\isOneConfirmed$;
the lower brace shows the extended range used by
$\isOneConfirmedSpecial$.}
\label{fig:epoch-boundary-variant}
\end{figure}

During normal confirmation, the function $\isOneConfirmedExtFrom$ uses
$\isOneConfirmedSpecial$ when checking the first block of the epoch
(i.e., a block $b$ with $\epoch(\parent(b)) < \epoch(b)$), and $\isOneConfirmed$ for all other
blocks.
This case distinction is implemented by the function
$\isOneConfirmedExt$ (Algorithm~\ref{alg:findlatestconf-helpers}).
During reconfirmation, $\isChainOneConfirmed$ uses the standard
$\isOneConfirmed$ for all blocks belonging to the confirmed chain.

\subsection{Main Result}
\label{sec:main-result}

We can now state the main result of this paper: the algorithms
presented in this section constitute a Confirmation Rule for
$\LMDGHOSTHFC$ in the sense of Definition~\ref{def:confirmation-rule}.

\begin{theorem}\label{thm:main}
    Let 
    $\mathit{sg}(b, t, \GGST): = \epoch(b) = \epoch(t) \land \slotstart(\prevfirstslotepoch[-1]{t})\geq \GGST$.
    If \Cref{assum:no-change-to-the-validator-set,assum:beta,assum:ffg,assum:monotonicity} hold,
    the tuple $(\text{\Cref{alg:conffull,alg:findlatestconf-helpers,alg:findlatestconf-functional}}, \mathit{sg})$ is a Confirmation Rule for $\LMDGHOSTHFC$.
\end{theorem}

\Cref{thm:main} states that, under \Cref{assum:no-change-to-the-validator-set,assum:beta,assum:ffg,assum:monotonicity}, the confirmation algorithm satisfies the safety and monotonicity properties required of a Confirmation Rule for any block $b$ satisfying two conditions: (i)~$b$ belongs to the current epoch, and (ii)~the network has been synchronous since at least one slot before the start of the previous epoch.

The first condition restricts $b$ to the current epoch because, although safety is established for blocks in both the current and previous epoch, monotonicity is only established for blocks belonging to the current epoch.

The second condition requires synchrony from at least the last slot of two epochs ago. This is needed because some cases rely on $\willChkpBeJustified$ as evaluated in the previous epoch to conclude that enough votes have been sent to justify a checkpoint. Since $\willChkpBeJustified$ is evaluated only once, its conclusion carries over to the current epoch only if confirmation was maintained throughout the previous epoch, which requires synchrony from the last slot of two epochs ago.

The theorem follows from three arguments:
\begin{itemize}
    \item the \LMDGHOST safety argument (Section~\ref{sec:lmd-safety-condition}), which establishes that confirmed blocks remain canonical via the implication chain
    $\indQ > \text{threshold} \Rightarrow \indP > \text{threshold} \Rightarrow H > \Phi/2 \Rightarrow b \text{ canonical}$
    (Figure~\ref{fig:safety-chain});
    \item the \LMDGHOSTHFC safety argument (Section~\ref{sec:conf-rule-ldmghosthfc}), which establishes that confirmed blocks are not filtered out by \FILHFC;
    \item the monotonicity argument (Section~\ref{sec:monotonicity}), which shows that a confirmed block remains confirmed at all future times.
\end{itemize}
The detailed proofs are given in \Cref{sec:appendix-proofs}.






\section{Confirmation Delay Under Varying Network Conditions}
\label{sec:confirmation-delay}

The preceding sections establish the safety and monotonicity of the Fast
Confirmation Rule under the stated assumptions.
Safety ensures that, under synchrony, confirmed blocks remain canonical
regardless of the adversary's strategy.
A complementary question concerns the \emph{speed} at which confirmation
occurs under different network conditions.

The safety parameter~$\beta$ is a conservative, protocol-level bound on the
Byzantine fraction in any set of consecutive committees
(Assumption~\ref{assum:beta}).
The actual fraction of committee weight that does not attest in support of
the proposed block at any given slot -- whether due to Byzantine behavior,
network latency, missed proposals, or validator downtime -- may be
substantially smaller than~$\beta$.
When this fraction is sufficiently small (below $1/20 = 5\%$
with Ethereum's parameters, cf.\ equation~\eqref{eq:boundary}) and there are no empty slots preceding a block, the
per-block confirmation predicate
$\isOneConfirmed$ (Definition~\ref{def:isOneConfirmed}) is satisfied after a
single slot of attestations, yielding a 12-second confirmation time.
When $\beta$ is larger, additional slots of attestations are required before
$\indQ$ crosses the confirmation threshold. 
The safety guarantee is identical regardless of whether confirmation occurs
in one slot or in several; only the latency experienced by users changes.
In this section, we derive the number of slots required for
confirmation as a function of the fraction of committee weight that does
not attest in support of the proposed block.

\subsection{Theoretical confirmation delay}
\label{sec:theoretical-delay}

We derive an expression for the number of slots required
for confirmation as a function of the fraction of committee weight that does
not contribute to $\indQ$ in support of the proposed block.

\paragraph{Assumptions.}
Throughout this section, we assume the following:
\begin{enumerate}
    \item The time $t$ satisfies $t \geq \GGST$, so that the network is
          synchronous.
    \item The block $b$ is proposed at slot~$s$ by an honest validator, and
          $\parent(b)$ is forever canonical in the view of every honest validator.
    \item All slots from $\slot(\parent(b))+1$ to $\slot(b)+k-1$ belong to the same epoch, so that no epoch boundary intervenes.
    \item There are no empty slots between $\parent(b)$ and~$b$,
          i.e., $\slot(b) = \slot(\parent(b)) + 1$, so that the empty-slot
          discount in
          Definition~\ref{def:attsetweighttobediscsimplefromblock} vanishes.
    \item No slashable offences have been observed.
    \item The committee weight of each slot equals
          $\totvalsetweight[chkp=C]{\allvals}/\slotsperepoch$, i.e., the
          total validator weight is distributed uniformly across the
          $\slotsperepoch$ committees of the epoch.
          With Ethereum's current validator set of approximately one million
          validators and $\slotsperepoch = 32$, the deviation from
          uniformity is negligible.
\end{enumerate}

\paragraph{Setup.}
Consider a block $b$ proposed at slot~$s$, and suppose that a fraction~$x$
of the total committee weight in each slot does not attest in support
of~$b$, where $0 \leq x < \beta$.
This fraction encompasses all weight that does not contribute to~$\indQ$:
Byzantine validators withholding or equivocating attestations, offline
validators, and validators whose attestations arrive after the confirmation
check at the beginning of the subsequent slot.

\paragraph{Derivation.}
After $k$ slots of attestations (from slot~$\slot(b)$ to
slot~$\slot(b) + k - 1$), the observed indicator satisfies
\[
\indicatorfromblock[from=b,to=\slot(b)+k-1,val=v,when=t,chkp=C]{\indQ}
\;=\;
\frac{
  \attsetweightfromblock[from=b,to=\slot(b)+k-1,val=v,when=t,chkp=C]{\allatts}
}{
  \commweightfromafterparentblock[from=b,to=\slot(b)+k-1,chkp=C,when=t,val=v]{\allvals}
}.
\]
Since $\slot(b) = \slot(\parent(b)) + 1$, the denominator
reduces to the total committee weight over slots
$\slot(b), \ldots, \slot(b)+k-1$.
The numerator is the weight that attests in support of~$b$, namely a
fraction $(1-x)$ of the denominator, and the ratio simplifies to
\[
\indicatorfromblock[from=b,to=\slot(b)+k-1,val=v,when=t,chkp=C]{\indQ}
\;=\;
\frac{(1-x) \cdot \commweightfromafterparentblock[from=b,to=\slot(b)+k-1,chkp=C,when=t,val=v]{\allvals}}
{\commweightfromafterparentblock[from=b,to=\slot(b)+k-1,chkp=C,when=t,val=v]{\allvals}}
\;=\; 1 - x.
\]
Since the fraction~$x$ is the same in every slot by assumption,
$\indQ$ equals $1-x$ independently of~$k$.

The simplified confirmation condition
(cf.\ the sufficient condition for safety stated in
Section~\ref{sec:conf-rule-lmd-ghost}, preceding
Definition~\ref{def:isOneConfirmed})
requires
\begin{equation}\label{eq:simplified-delay-condition}
\indicatorfromblock[from=b,to=\slot(b)+k-1,val=v,when=t,chkp=C]{\indQ}
\;>\;
\frac{1}{2}\!\left(1
  + \frac{\boostweight[chkp=C]}
         {\commweightfromafterparentblock[from=b,to=\slot(b)+k-1,chkp=C]{\allvals}}
\right)
+ \beta.
\end{equation}
Since
$\boostweight[chkp=C]
  = \boostscore \cdot
    \totvalsetweight[chkp=C]{\allvals}/\slotsperepoch$
and the denominator equals
$k \cdot \totvalsetweight[chkp=C]{\allvals}/\slotsperepoch$
(item~6 above),
$\boostweight[chkp=C] / \commweightfromafterparentblock[from=b,to=\slot(b)+k-1,chkp=C]{\allvals}$
reduces to $\boostscore/k$.
Substituting $\indQ = 1 - x$ yields
\[
1 - x
\;>\;
\frac{1}{2}\!\left(1 + \frac{\boostscore}{k}\right) + \beta,
\]
which, after rearranging, gives the condition on the number of slots:
\begin{equation}\label{eq:kmin-condition}
k \;>\; \frac{\boostscore}{1 - 2\beta - 2x}.
\end{equation}
The number of slots required for confirmation is therefore
\begin{equation}\label{eq:kmin}
k(x)
\;=\;
\left\lfloor \frac{\boostscore}{1 - 2\beta - 2x} \right\rfloor + 1.
\end{equation}

Intuitively, larger $k$ helps because the proposer boost is a fixed additive weight while committee weight grows linearly with $k$. Consequently, $k(x)$ is monotonically non-decreasing in $x$: the larger the fraction of committee weight not attesting in support of $b$, the more slots are required for confirmation.

\paragraph{One-slot confirmation threshold.}
Setting $k = 1$ in~\eqref{eq:simplified-delay-condition} gives
\begin{equation}\label{eq:one-slot-threshold}
\indicatorfromblock[from=b,to=\slot(b),val=v,when=t,chkp=C]{\indQ}
\;>\;
\frac{1}{2}(1 + \boostscore) + \beta
\;=\;
\frac{1}{2} + \frac{\boostscore}{2} + \beta.
\end{equation}
With Ethereum's parameters ($\boostscore = 2/5$, $\beta = 1/4$), this
evaluates to $1/2 + 1/5 + 1/4 = 19/20 = 95\%$.

\paragraph{Remark.}
Expression~\eqref{eq:kmin} is derived from the simplified confirmation
condition, which omits the slashing adjustment and the empty-slot discount
present in Definition~\ref{def:isOneConfirmed}.
Both refinements decrease the effective threshold, so the
theoretical $k(x)$ is conservative: in practice, confirmation may
require fewer slots than~\eqref{eq:kmin} predicts.

\paragraph{Ethereum parameters.}
With $\beta = 1/4$ and $\boostscore = 2/5$,
expression~\eqref{eq:kmin-condition} becomes
$k > \frac{2/5}{1/2 - 2x}$.
The boundary at which $k$~slots are no longer sufficient and $k+1$ are
required is obtained by setting
$\frac{\boostscore}{1 - 2\beta - 2x} = k$ and solving for~$x$:
\begin{equation}\label{eq:boundary}
x_k \;=\; \frac{1}{2} - \beta - \frac{\boostscore}{2k}
\;=\; \frac{1}{4} - \frac{1}{5k}.
\end{equation}
For any $x \in [x_{k-1},\, x_k)$, exactly $k$~slots are needed:
when $x < x_{k-1}$, the value $1-x$ exceeds the right-hand side
of~\eqref{eq:simplified-delay-condition} with $k-1$ slots;
when $x \geq x_k$, it does not with $k$ slots.
We set $x_0 := 0$.
Table~\ref{tab:delay-ranges} lists the first five ranges.

\begin{table}[ht]
\centering
\begin{tabular}{cccc}
\toprule
Fraction $x$ & & Slots $k(x)$ & Time \\
\midrule
$\bigl[0,\; 1/20\bigr)$                   & $[0\%,\; 5\%)$          & 1 & 12\,s \\[2pt]
$\bigl[1/20,\; 3/20\bigr)$                & $[5\%,\; 15\%)$         & 2 & 24\,s \\[2pt]
$\bigl[3/20,\; 11/60\bigr)$               & $[15\%,\; 18.3\%)$      & 3 & 36\,s \\[2pt]
$\bigl[11/60,\; 1/5\bigr)$                & $[18.3\%,\; 20\%)$      & 4 & 48\,s \\[2pt]
$\bigl[1/5,\; 21/100\bigr)$               & $[20\%,\; 21\%)$        & 5 & 60\,s \\
\bottomrule
\end{tabular}
\caption{Number of slots required for confirmation as a function of the
fraction~$x$ of committee weight not attesting in support of the proposed block, with $\beta = 1/4$ and
$\boostscore = 2/5$.
The confirmation time equals $12 \cdot k(x)$ seconds.
As $x \to 1/4$, $k(x) \to \infty$.}
\label{tab:delay-ranges}
\end{table}

When $x = 0$ (every committee member attests in support of~$b$),
$k(0) = 1$: the predicate $\isOneConfirmed$ is satisfied after a
single slot.
As $x \to \beta$ (that is, $x \to 1/4$ when $\beta = 1/4$),
$k(x) \to \infty$: the right-hand side
of~\eqref{eq:simplified-delay-condition} remains above $1-x$ for all~$k$.
For intermediate values of~$x$, the predicate is satisfied after
$k(x) > 1$ slots.
Figure~\ref{fig:kmin} plots $k(x)$ for $\beta = 1/4$ and
$\boostscore = 2/5$.

\begin{figure}[ht]
\centering
\begin{tikzpicture}
        \begin{axis}[
            width=14cm, height=9cm,
            axis lines=middle,
            xlabel=$x$: fraction of committee weight not attesting,
            xticklabel style={
                /pgf/number format/fixed,
                /pgf/number format/precision=3
            },
            ylabel={$k(x)$: confirmation time in slots},
            ymin=0.5, ymax=32,
            xmin=0, xmax=0.25,
            xtick={0,0.025,...,0.25},
            legend pos=north west,
            clip=false,
            domain=-10:10,
            samples=100,
            ymajorgrids=true,
            yminorgrids=true,
            minor ytick={1,2,...,32},
            xmajorgrids=true,
            major grid style={gray!30},
            minor grid style={dashed, gray!20},
            x label style={at={(axis description cs:0.5,-0.1)},anchor=north},
            y label style={at={(axis description cs:-0.1,.5)},rotate=90,anchor=south},
        ]
        \addplot[mark=none, domain=0:0.242,samples=1000, jump mark left]{ceil(0.2/(0.25-x))};
        \end{axis}
    \end{tikzpicture}

\caption{Number of slots required for confirmation as a function of the
fraction~$x$ of committee weight not attesting for the proposed block, with $\beta = 1/4$ and $\boostscore = 2/5$.
$k(x) \to \infty$ as $x \to 1/4$.}
\label{fig:kmin}
\end{figure}

\subsection{Empirical evaluation on Ethereum mainnet}
\label{sec:empirical-eval}

To assess confirmation delay under real network conditions, an empirical
evaluation of the Fast Confirmation Rule was conducted on approximately
26~hours of Ethereum mainnet data starting from block 12{,}607{,}808
(September~17, 2025).
The checkpoint state at this block was used as the initial finalized and
confirmed checkpoint.
Attestations were sourced from two channels: the aggregate gossip topic
as observed by Xatu\footnote{\url{https://ethpandaops.io/data/xatu/}}
sentry nodes, and attestations included in blocks.
For each proposed block, we recorded the smallest number of slots~$k$
after which the confirmation predicate was satisfied.

Figure~\ref{fig:fcr-mainnet} reports the cumulative distribution of
confirmation times.

\begin{figure}[ht]
\centering
\begin{tikzpicture}
    \begin{axis}[
        width=14cm, height=8cm,
        xlabel={confirmation time (seconds)},
        ylabel={cumulative fraction of confirmed blocks (\%)},
        ymin=90, ymax=100.5,
        xmin=0, xmax=102,
        xtick={12,24,36,48,60,72,84,96},
        xticklabels={$\leq 12$,$\leq 24$,$\leq 36$,$\leq 48$,$\leq 60$,$\leq 72$,$\leq 84$,$\leq 96$},
        ytick={90,92,94,96,98,100},
        yticklabel={\pgfmathprintnumber{\tick}\%},
        ymajorgrids=true,
        major grid style={gray!30},
        title={\small FCR confirmation time distribution on the Mainnet},
        title style={at={(0.5,1.05)}},
        clip=false,
    ]

    \addplot[
        thick, blue, mark=*, mark size=2.5pt,
    ] coordinates {
        (12, 94.67)
        (24, 98.87)
        (36, 99.11)
        (48, 99.19)
        (60, 99.87)
        (72, 99.97)
        (84, 99.99)
        (96, 100.00)
    };

    \node[above left, font=\scriptsize, blue] at (axis cs:12,94.67) {94.67\%};
    \node[above, font=\scriptsize] at (axis cs:24,98.87) {98.87\%};
    \node[above, font=\scriptsize] at (axis cs:36,99.11) {99.11\%};
    \node[above, font=\scriptsize] at (axis cs:48,99.19) {99.19\%};
    \node[above, font=\scriptsize] at (axis cs:60,99.87) {99.87\%};
    \node[above, font=\scriptsize] at (axis cs:72,99.97) {99.97\%};
    \node[above, font=\scriptsize] at (axis cs:84,99.99) {99.99\%};
    \node[above, font=\scriptsize] at (axis cs:96,100.00) {100.00\%};

    \end{axis}
\end{tikzpicture}
\caption{Cumulative distribution of confirmation times on Ethereum
mainnet over a 26-hour run starting from block 12{,}607{,}808, with
$\beta = 1/4$.
Each 12-second increment on the horizontal axis corresponds to one
slot.
For instance, $94.67\%$ of blocks are confirmed within one slot
($\leq 12$\,s), and $100\%$ within eight slots ($\leq 96$\,s).}
\label{fig:fcr-mainnet}
\end{figure}

The results show that $94.67\%$ of blocks were confirmed within a single
slot ($\leq 12$\,s), rising to $98.87\%$ within two slots
($\leq 24$\,s) and reaching $100\%$ within eight slots
($\leq 96$\,s).
By equation~\eqref{eq:boundary}, single-slot confirmation requires
$x < 1/20 = 5\%$; that $94.67\%$ of blocks satisfy this threshold
indicates that the fraction of committee weight not attesting in support
of the proposed block is below~$5\%$ in the vast majority of slots.
The approximately $5.33\%$ of blocks requiring more than one slot likely correspond to either (a)~blocks with empty slots between themselves and their parents, which makes single-slot confirmation impossible under $\beta = 20\%$,\footnote{We do not include a proof of this claim in this version of the work.} or (b)~blocks sent late that did not accrue enough weight within one slot.

\paragraph{Limitations.}
During the observation period, no adversary was employing strategies
aimed at delaying confirmation.
A strategic adversary aware of the rule could withhold or mistime
attestations to increase confirmation latency, up to the theoretical
maximum dictated by~\eqref{eq:kmin}.
The empirical results therefore do not capture the impact of adversarial strategies targeting the confirmation rule.

\section{Conclusion}
\label{sec:conclusion}

We have presented a Fast Confirmation Rule for Gasper that, under
synchrony and the assumptions stated in this paper, confirms blocks
in as few as 12~seconds -- a single slot -- providing an
order-of-magnitude improvement over the 13 to 19~minutes required by
\FFG finalization.

The construction proceeds in two stages.
We first derived a confirmation rule for plain \LMDGHOST, establishing a
per-block confirmation predicate ($\isOneConfirmed$) based on the observable
indicator~$\indQ$.
This predicate accounts for the proposer boost, the Byzantine stake
bound, and equivocation evidence from slashed validators.
We then extended the rule to \LMDGHOSTHFC, where \FILHFC-filtering can remove even well-supported blocks, for instance if a conflicting checkpoint becomes justified.
The full algorithm addresses this through predicates that conservatively evaluate whether the confirmed branch will survive \FILHFC-filtering, a five-case confirmation strategy that adapts to the epoch of the candidate and the current \FFG state, and an epoch-boundary reconfirmation mechanism.
We proved that the resulting rule satisfies both safety -- confirmed
blocks remain canonical -- and monotonicity -- a confirmed block
remains confirmed at all future times
(Section~\ref{sec:monotonicity}).

The rule is complementary to \FFG finalization: it does not replace
finalization but provides a faster confirmation signal under synchrony,
while finalization remains available as a fallback that tolerates
asynchrony.
In the event of a safety violation, the rule reverts to the latest
finalized block. Finalization itself remains unaffected.

The Fast Confirmation Rule requires no protocol changes: it is a client-side algorithm
that each node can execute locally using data already available in its
view.

\bibliographystyle{splncs04}
\bibliography{references}

    \appendix

     \section{Appendix}\label{sec:appendix-proofs}

     \begin{remark}\label{app:assum-validator-set}
    Assumption~\ref{assum:no-change-to-the-validator-set} immediately implies that
    \begin{enumerate}
        \item $\commatepoch[epoch=e]{\allvals}\subseteq\commatepoch[epoch=\epoch(\genesis)]{\allvals}=\totvalset[chkp=\genesis]{\allvals}$
        \item $\commatepoch[epoch=e]{\honvals} = \commatepoch[epoch=\epoch(\genesis)]{\honvals} = \totvalset[chkp=\genesis]{\honvals}$
        \item for any valid block $b$ and honest validator $v$, $\weightofset[chkp=b]{v}=\weightofset[chkp=\genesis]{v}$
        \item for any valid block $b$ and Byzantine validator $v$, $\weightofset[chkp=b]{v}\leq\weightofset[chkp=\genesis]{v}$
    \end{enumerate}
\end{remark}

\begin{remark}\label{app:assum-beta}
    The condition on the checkpoint $C$ in
    Assumption~\ref{assum:beta} is equivalent to requiring
    $C \in \allU(b)$ for some valid block~$b$, where $\allU(b)$
    denotes the set of unrealized justified checkpoints in the
    chain of~$b$ (Section~\ref{sec:conf-rule-ldmghosthfc}).
\end{remark}

\begin{remark}[Alternative discount used in the safety proofs]\label{app:disc-remark}
The safety proofs in this appendix use a larger discount than the one adopted in $\Phi_{b}^{v,t,C}$ in the main text.
Fix a block $b$ and a slot $s$.
Let $\attsettobediscfromblock[from=b,to=s,when=t,val=v]{\allatts}$ denote the set of validators in $\commfromafterblock[from=\parent(b),to=s,when=t,val=v]{\allvals}$ whose \LMDGHOST choice (under the view $\FILLMD(\FILINV(\FILCUR(\FILEQ(\viewattime[val=v,time=t]),t)))$) is \emph{not} a strict descendant of $\parent(b)$: they vote for a block $\preceq \parent(b)$ or for a block conflicting with $\parent(b)$.
Its honest subset is $\attsettobediscfromblock[from=b,to=s,when=t,val=v]{\honatts}
:= \attsettobediscfromblock[from=b,to=s,when=t,val=v]{\allatts}\cap\honvals$.
\end{remark}

\begin{definition}
\label{def:attsetweighttobediscfromblock}
    \[
    \attsetweighttobediscfromblock[from=b,to=\slot(b)-1,val=v,when=t,chkp=C]{\honattsub}
    \;:=\;
    \max\left(
        \attsetweighttobediscfromblock[from=b,to=\slot(b)-1,val=v,when=t,chkp=C]{\allatts}
        -\beta \commweightfromafterblock[from=\parent(b),to=\slot(b)-1,chkp=C]{\allvals}
        +\attsetweightfromafterblock[from=\parent(b),to=\slot(b)-1,val=v,when=t,chkp=C]{\slashvals}
        ,0
    \right).
    \]
\end{definition}

Validators voting exactly for $\parent(b)$ --- the simple set used in the main text --- form a subset of the set defined above, so the discount here is at least as large as the main text's, and substituting it into $\Phi_{b}^{v,t,C}$ yields a value no larger than the one used by the algorithm.
The resulting safety threshold is therefore no stricter than the main text algorithm's, so any block passing the main text algorithm's threshold also passes the one used in this appendix, and every safety lemma proved here applies to the algorithm.

The safety proofs in this appendix use the following variant of $\isOneConfirmed$ (Definition~\ref{def:isOneConfirmed}), in which the discount of Definition~\ref{def:attsetweighttobediscfromblock} replaces the simple discount used in the main text.

\begin{definition}[Per-block safety predicate used in the safety proofs]\label{def:isOneConfirmedWeaker}
    For any block $b$, checkpoint $C$, honest validator $v$, and time $t$,
    \begin{align*}
    &\var[val=v]{\isOneConfirmedWeaker}(b, C, t) \;:=\\
    &\quad
    \begin{aligned}[t]
        \indicatorfromblock[from=b,to=\slot(t)-1,val=v,when=t,chkp=C]{\indQ}
        \;>\;&
        \frac{
            \commweightfromafterparentblock[from=b,to=\slot(t)-1,chkp=C]{\allvals}
            + \boostweight[chkp=C]
            - \attsetweighttobediscfromblock[from=b,to=\slot(b)-1,val=v,when=t,chkp=C]{\honattsub}
        }{
            2\,\commweightfromafterparentblock[from=b,to=\slot(t)-1,chkp=C]{\allvals}
        }
        + \beta\,
        \frac{\commweightfromblock[from=b,to=\slot(t)-1,chkp=C]{\allvals}}{\commweightfromafterparentblock[from=b,to=\slot(t)-1,chkp=C]{\allvals}}
        - \frac{\attsetweightfromblock[from=b,to=\slot(t)-1,val=v,when=t,chkp=C]{\slashvals}}{\commweightfromafterparentblock[from=b,to=\slot(t)-1,chkp=C]{\allvals}}.
    \end{aligned}
    \end{align*}
\end{definition}

Since the simple discount is bounded above by the one of Definition~\ref{def:attsetweighttobediscfromblock} and this quantity is subtracted in the numerator, $\isOneConfirmedWeaker$'s threshold is at most $\isOneConfirmed$'s, so $\isOneConfirmed(b, C, t) \implies \isOneConfirmedWeaker(b, C, t)$.

\begin{definition}[\LMDGHOST safety predicate]\label{def:isLMDGHOSTSafe}
For any block $b$, checkpoint $C$, honest validator $v$, and time $t \geq \GGST$,
\[
\var[val=v]{\isLMDGHOSTSafe}(b, C, t)
\;:=\;
\forall b' \preceq b,\,
\var[val=v]{\isOneConfirmedWeaker}(b', C, t),
\]
\end{definition}

\subsection{\LMDGHOST}
\label{app:lmdghost}

The following lemma\footnote{The statements of lemmas and theorems, together with their proofs, have been written manually, whereas the explanations preceding them have been fully generated by an LLM. The statements and proofs should therefore be trusted over the natural-language explanations.} establishes the monotonicity of the honest-only support indicator $\indP$ across slots and times: starting from an arbitrary reference slot $s$ and time $t$, if every honest validator in the committees for slots $[s+1,\slot(t')-1]$ \GHOST votes in support of $b$, then the value of $\indP$ at slot $\slot(t')-1$ and time $t'$ is at least as large as its value at slot $s$ and time $t$.
This monotonicity turns a one-time bound on $\indP$ into a persistent guarantee, and is the key building block for the subsequent safety arguments.
The proof decomposes the honest vote set into an ``old'' part already counted at time~$t$ (covering slots up to $s$) and a ``new'' part contributed by the honest validators in the committees for slots $[s+1,\slot(t')-1]$, and shows that adding these new votes can only increase the ratio because $\frac{a+x}{b+y}\geq\frac{a}{b}$ whenever $a\leq b$ and $x\geq y$.

    \begin{lemma}\label{lem:lmd-p-monotonic}
        Given Assumption~\ref{assum:no-change-to-the-validator-set},
        for any two honest validator $v$ and $v'$, block $b$, slot $s$, times $t'$ and $t$, and any two checkpoints $C$ and $C'$,
        if
        \begin{enumerate}
            \item $\slotstart(\slot(t')-1)\geq\GGST$,
            \item $t'\geq \slotstart(\slot(t))$ and
            \item all honest validators in the committees for slots $[s+1,\slot(t') -1]$ \GHOST vote in support of $b$,
        \end{enumerate}
        then
        $$
        \indicatorfromblock[from=b,to=\slot(t')-1,val=v',when=t',chkp={C'}]{\indP}
        \geq
        \indicatorfromblock[from=b,to=s,val=v,when=t,chkp={C}]{\indP}
        $$
    \end{lemma}
    
    \begin{proof}
        Let $s' := \slot(t')$.
        Then we can proceed as follows.
        \def\alignexplwidth{7cm}
        \allowdisplaybreaks
        \begin{align*}
            \indicatorfromblock[from={b},to={s'-1},val={v},when={t},chkp={C'}]{\indP}
            &=
            \frac{
                \attsetweightfromblock[from={b},to={s'-1},val={v'},when={t'},chkp={C'}]{\honatts}
            }{
                \commweightfromblock[from={b},to={s'-1},chkp={C'}]{\honvals}
            }
            &&\alignexpl[\alignexplwidth]{By definition.}
            \\
            &=
            \frac{
            \attsetweightfromblock[from={b},to={s'-1},val={v'},when={t'},chkp={C}]{\honatts}
        }{
            \commweightfromblock[from={b},to={s'-1},chkp={C}]{\honvals}
        }
        &&\alignexpl[\alignexplwidth]{As per Assumption~\ref{assum:no-change-to-the-validator-set}, the effective balance of honest validators never changes.}
        \\
        &=
        \frac{
            \attsetweightfromblock[from={b},to={s'-1},val={v},when={t'},chkp={C}]{\honatts}
        }{
            \commweightfromblock[from={b},to={s'-1},chkp={C}]{\honvals}
        }
        &&\alignexpl[\alignexplwidth]{Given that $\slotstart(s'-1)\ge \slotstart(s)\ge \GGST$ and $t' \ge \slotstart(s')$, any honest attestation for slots up to $s'-1$ received by $v$ at time $t'$ is also received by $v$ by the same time $t'$.}
        \\
        &\ge
        \frac{
            \attsetweightfromblock[from={b},to={s},val={v},when={t},chkp={C}]{\honatts}
            +
            \weightofset[chkp={C}]{
                \commfromslot[from={s+1},to={s'-1}]{\honvals}
                \setminus
                \attsetfromblockunfiltered[from={b},to={s},val={v},when={t}]{\honatts}
            }
        }{
            \commweightfromblock[from={b},to={s'-1},chkp={C}]{\honvals}
        }
        &&\alignexpl[\alignexplwidth]{$\attsetfromblockunfiltered[from=b, to=s'-1, val=v, when=t']{\honatts}$ includes at least the 
        the union between the set of honest validators that according to view $\viewattime[val=v,time=t]$ \GHOST voted in support of $b$ and the honest validators in the committees between slot $s+1$ ans slot $s'-1$, as we assume that any of these validators  \GHOST votes in support of $b$ and $\slotstart(\slot(t')-1)\geq \GGST$}
        \\
        &=
        \frac{
            \attsetweightfromblock[from={b},to={s},val={v},when={t},chkp={C}]{\honatts}
            +
            \weightofset[chkp={C}]{
                \commfromslot[from={s+1},to={s'-1}]{\honvals}
                \setminus
                \attsetfromblockunfiltered[from={b},to={s},val={v},when={t}]{\honatts}
            }
        }{
            \commweightfromblock[from={b},to={s},chkp={C}]{\honvals}
            +
            \weightofset[chkp={C}]{
                \commfromslot[from={s+1},to={s'-1}]{\honvals}
                \setminus
                \commfromblock[from={b},to={s}]{\honvals}
            }
        }
        &&\alignexpl[\alignexplwidth]{By definition.}
        \\
        &\ge
        \frac{
            \attsetweightfromblock[from={b},to={s},val={v},when={t},chkp={C}]{\honatts}
        }{
            \commweightfromblock[from={b},to={s},chkp={C}]{\honvals}
        }
        &&\alignexpl[\alignexplwidth]{From $\attsetfromblockunfiltered[from={b},to={s},val={v},when={t}]{\honatts} \subseteq \commfromblock[from={b},to={s}]{\honvals}$ and $\frac{a+x}{b+y}\ge \frac{a}{b}$ when $a\le b$ and $x\ge y$.}
        \\
        &=
        \indicatorfromblock[from={b},to={s},val={v},when={t},chkp={C}]{\indP}.
    \end{align*}
\end{proof}

The following lemma shows that if, for every ancestor $b'$ of~$b$, the honest attestation weight for~$b'$ exceeds half of the total available weight (adjusted for the proposer boost and the empty-slot discount), then~$b$ is canonical in $v$'s view at time~$t$.
The proof proceeds by induction on the iterations of the \LMDGHOST\ \textbf{while} loop, showing that at each step the fork-choice rule must follow the chain of~$b$ because the honest votes for~$b$'s branch outweigh any competitor.

\begin{lemma}\label{lem:condition-on-h-for-canonical-ex}
    Let $v$ be any honest validator, $t$ be any time and $b$ be any block, if
    \begin{enumerate}
        \item $t\geq \GGST$,
        \item $\chain(b) \subseteq \viewattime[time=t,val=v]$,
        \item $\slot(b) \leq \slot(t)$ and
        \item $\forall b' \preceq b,\;
        \attsetweightfromblock[from=b',to=\slot(t)-1,val=v,when=t,chkp={\gjattime[val=v,time=t]}]{\honatts}
        >
        \frac
            {
                \commweightfromafterparentblock[from=b',to=\slot(t)-1,chkp={\gjattime[val=v,time=t]}]{\allvals}
                -\attsetweighttobediscfromblock[from=b',to=\slot(b')-1,val=v,when=t,chkp=C]{\honatts}
                +\boostweight[chkp={\gjattime[val=v,time=t]}]}
            {2}
        $,
    \end{enumerate}
    then block $b$ is canonical in the view of validator $v$ at time $t$.
\end{lemma}

\begin{proof}
    We want to prove that $b \preceq \fcparam[fc=\LMDGHOST,balf=\gjviewsym,val=v](t)$.

    Let $b_i$ be the value of the variable $b$ at the end of the $i$-th iteration of the \textbf{while} loop 
    of Algorithm~\ref{alg:GHOST}, with $b_0$ corresponding to the value of the variable $b$ at the beginning of the first iteration.
    We now prove by induction on $i$ that either $b_i \succeq b$ or $b_i \preceq b$. 

    \begin{description}
        \item[Base case: $i=0$.]
        Trivial as the \textbf{while} loop in Algorithm~\ref{alg:GHOST} starts with variable $b$ set to $\genesis \preceq b$.

        \item[Inductive step.]
        By the inductive hypothesis, we assume that $b_i \succeq b \lor b_i \preceq b$ and prove that $b_{i+1} \succeq b \lor b_{i+1} \preceq b$.
        By line~\ref{ln:alg:GHOST:argmax} of Algorithm~\ref{alg:GHOST}, $b_{i+1}$ is the descendant of $b_i$ with the heaviest total weight.
        Let us proceed by cases.
        \begin{description}
            \item[Case $b_i \succeq b$.]
            This immediately implies that $b_{i+1} \succeq b$.

            \item[Case $b_i \prec b$.]
            Let $b_c$ be the child of $b_i$ in the chain of $b$, \ie, $b_c \preceq b \land \parent(b_c) = b_i$, and let $b'$ be any child of $b_i$.
            Let $\FIL_\LMDGHOST(\View, t) := \FILLMD(\FILINV(\FILCUR(\FILEQ(\View),t)))$ and
            note that, for $\LMDGHOST^{\gjviewsym}$, the argument of $\argmax$ at line~\ref{ln:alg:GHOST:argmax} of Algorithm~\ref{alg:GHOST} corresponds to $\weightofset[chkp={\gjattime[time=t,val=v]}]{\ghostvoters[block=b',view={\FIL_\LMDGHOST(\View), t}]}$.
            Due to \FILINV and \FILCUR, such expression can evaluate at most to the weight of the committees between slot $\slot(b_i)+1$ and slot $\slot(t)-1$ plus, potentially, the proposer boost weight, less the weight of the honest validators in the committees between $\slot(b_i)+1$ and $\slot(t)-1$ that have already \LMDGHOST voted for a block $\nsucceq b_i$, \ie, $
            \commweightfromafterparentblock[from={b},to={\slot(t)-1},chkp={\gjattime[time=t,val=v]}]{\allvals} 
            + \boostweight[chkp={\gjattime[val=v,time=t]}]
            -\attsetweighttobediscfromblock[from=b',to=\slot(t)-1,val=v,when=t,chkp=C]{\honatts}
            $.
            Given that honest validators never equivocate, we have that
            $$
            \weightofset[chkp={\gjattime[time=t,val=v]}]{\ghostvoters[block=b_c,view={\FIL_\LMDGHOST(\View, t)}]}
            \geq
            \attsetweightfromblock[from=b_c,to=\slot(t)-1,val=v,when=t,chkp={\gjattime[val=v,time=t]}]{\honatts}
            >
            \frac
                {\commweightfromafterparentblock[from=b,to=\slot(t)-1,chkp={\gjattime[val=v,time=t]}]
                {\allvals}
                -\attsetweighttobediscfromblock[from=b',to=\slot(b')-1,val=v,when=t,chkp=C]{\honatts}
                +\boostweight[chkp={\gjattime[val=v,time=t]}] }{2}
            $$

            Take any $b' \neq b_c$.
            Given that honest validators never equivocates, the maximum amount of weight that can vote for a descendant of $\parent(b)$  

            This means that $b'$ and $b_c$ conflict which implies that
            
            \begin{align*}
                \weightofset[chkp={\gjattime[time=t,val=v]}]{\ghostvoters[block=b',view={\FIL_\LMDGHOST(\View, t)}]} 
                &< 
                \begin{aligned}[t]
                    &\commweightfromafterparentblock[from={b},to={\slot(t)-1},chkp={\gjattime[time=t,val=v]}]{\allvals}
                    -\attsetweighttobediscfromblock[from=b',to=\slot(t)-1,val=v,when=t,chkp=C]{\honatts} 
                    + \boostweight[chkp={\gjattime[val=v,time=t]}] 
                    \\
                    &-  \weightofset[chkp={\gjattime[time=t,val=v]}]{\ghostvoters[block=b_c,view={\FIL_\LMDGHOST(\View, t)}]}
                \end{aligned}
                \\                
                &\leq 
                \begin{aligned}[t]
                    &\commweightfromafterparentblock[from={b},to={\slot(t)-1},chkp={\gjattime[time=t,val=v]}]{\allvals}
                    -\attsetweighttobediscfromblock[from=b',to=\slot(b')-1,val=v,when=t,chkp=C]{\honatts} 
                    + \boostweight[chkp={\gjattime[val=v,time=t]}] 
                    \\
                    &-  \weightofset[chkp={\gjattime[time=t,val=v]}]{\ghostvoters[block=b_c,view={\FIL_\LMDGHOST(\View, t)}]}
                \end{aligned}
                &&\alignexpl[2cm]{As $\attsetweighttobediscfromblock[from=b',to=\slot(t)-1,val=v,when=t,chkp=C]{\honatts}\geq\attsetweighttobediscfromblock[from=b',to=\slot(b')-1,val=v,when=t,chkp=C]{\honatts}$.}
                \\
                &\leq
                \frac
                {\commweightfromafterparentblock[from=b,to=\slot(t)-1,chkp={\gjattime[val=v,time=t]}]
                {\allvals}
                -\attsetweighttobediscfromblock[from=b',to=\slot(b')-1,val=v,when=t,chkp=C]{\honatts}
                +\boostweight[chkp={\gjattime[val=v,time=t]}] }{2}
            \end{align*}

            This furhter implies that
            $b_c = b_{i+1}$ and hence $b_{i+1} \preceq b$.
        \end{description}
    \end{description}

    Note that any block in $\chain(b)$ has at least one child, except potentially for $b$.
    Note also that the \textbf{while} loop continues till it finds a block that either is for a slot higher than $\slot(t)$ or that has no valid children.
    Given that we assume $\slot(b) \leq \slot(t)$, honest validators never \GHOST vote for an invalid block
    and that above we have established that $\fcparam[fc=\LMDGHOST,balf=\gjviewsym,val=v](t) \succeq b \lor \fcparam[fc=\LMDGHOST,balf=\gjviewsym,val=v](t) \preceq b$, we can conclude the proof for this Lemma.
\end{proof}

The next lemma converts a condition on the honest support fraction $\indP$ into an absolute weight bound on honest attestations.
If $\indP$ exceeds $\Phi / (2\,\commweightfromblock[from=b,to=s,chkp=C]{\allvals}(1-\beta))$, that is, $\Phi$ divided by twice the $(1-\beta)$-fraction of the total committee weight, then the honest attestation weight exceeds $\Phi/2$.
The proof combines the definition of $\indP$ with Assumption~\ref{assum:beta}, which guarantees that the honest committee weight is at least $(1-\beta)$ times the total committee weight, to cancel the $(1-\beta)$ factor in the denominator.
This bridges the gap between $\indP$ and the amount of honest weight required to establish that a block is part of the canonical chain.

\begin{lemma}\label{lem:lmd-cond-on-p-implies-cond-on-h-ex}
    Given Assumption~\ref{assum:beta},
    for any time $t\geq\GGST$,
    honest validator $v$,
    block $b$,
    slot $s$,
    and checkpoint $C \in \allU(b)$ with $b$ being any valid block,
    if
    $$
    \indicatorfromblock[from=b,to=s,val=v,when=t,chkp=C]{\indP}
    >
    \frac{\Phi}
        {2\,\commweightfromblock[from=b,to=s,chkp=C]{\allvals}(1-\beta)}
    $$
    then
    $$
    \attsetweightfromblock[from=b,to=s,val=v,when=t,chkp=C]{\honatts}
    >
    \frac{\Phi}{2}.
    $$%
\footnote{The quantity $\Phi$ is treated here as a scalar:
the derivation relies only on its algebraic role, not on its
internal structure.
In the main text we write
$\Phi_{b'}^{s,v,t,C}$ to make explicit that $\Phi$ ultimately
depends on the block~$b'$, the slot~$s$, the validator~$v$,
the time~$t$, and the checkpoint~$C$; see
Equation~\eqref{eq:phi-def} for the full definition.
The same convention applies to
Lemma~\ref{lem:lmd-cond-on-q-implies-cond-on-p}.}

\end{lemma}

\begin{proof}
    \begin{align*}
        \attsetweightfromblock[from=b,to=s,val=v,when=t,chkp=C]{\honatts}
        &=
        \indicatorfromblock[from=b,to=s,val=v,when=t,chkp=C]{\indP}\;\commweightfromblock[from=b,to=s,chkp=C]{\honvals}
        &&\text{(by definition)}
        \\
        &>
        \frac{\Phi}
            {2\,\commweightfromblock[from=b,to=s,chkp=C]{\allvals}(1-\beta)}
        \;\commweightfromblock[from=b,to=s,chkp=C]{\honvals}
        &&\text{(by the hypothesis on $\indP$)}
        \\
        &\geq
        \frac{\Phi}{2\,\commweightfromblock[from=b,to=s,chkp=C]{\allvals}(1-\beta)}
        \commweightfromblock[from=b,to=s,chkp=C]{\allvals}(1-\beta)
        &&\alignexpl{As, due to Assumption~\ref{assum:beta}, $\commweightfromblock[from=b,to=s,chkp=C]{\honvals} \geq \commweightfromblock[from=b,to=s,chkp=C]{\allvals}(1-\beta)$}
        \\        
        &=
        \frac{\Phi}{2}.
    \end{align*}
\end{proof}

The following lemma is the central link in the $Q \to P$ bridge: it shows that a sufficient condition on the \emph{observable} indicator~$Q$ implies a lower bound on the \emph{unobservable} honest indicator~$\indP$.
The threshold on~$Q$ accounts for the Byzantine budget~$\beta$ and for slashed validators.
The proof proceeds by lower-bounding the honest attestation weight from the total attestation weight minus the Byzantine budget, and then applying the monotone-decreasing property of $g(x)=\frac{a-x}{b-x}$ to replace the actual Byzantine weight with its upper bound~$\beta$.

\begin{lemma}\label{lem:lmd-cond-on-q-implies-cond-on-p}
    Given Assumption~\ref{assum:beta},
    for any time $t\geq\GGST$,
    honest validator $v$,
    block $b'$,
    slot $s$
    and checkpoint $C \in \allU(b')$ with $b'$ being any valid block,
    if
    $$
    \indicatorfromblock[from=b',to=s,val=v,when=t,chkp=C]{Q}
    >
    \frac{\Phi}
        {2\,\commweightfromafterparentblock[from=b',to=s,chkp=C]{\allvals}}
        + \beta
            \frac
                {\commweightfromblock[from=b',to=s,chkp=C,when=t,val=v]{\allvals}}
                {\commweightfromafterparentblock[from=b',to=s,chkp=C,when=t,val=v]{\allvals}}
        - \frac
            {\attsetweightfromblock[from=b',to=s,val=v,when=t,chkp={C}]{\slashvals}}
            {\commweightfromafterparentblock[from=b',to=s,chkp=C]{\allvals}}
    $$, then
    $$
    \indicatorfromblock[from=b',to=s,val=v,when=t,chkp=C]{\indP}
    >
    \frac{\Phi}
        {2(1-\beta)\,\commweightfromblock[from=b',to=s,chkp=C]{\allvals}}
    $$
\end{lemma}

\begin{proof}
We proceed as follows.

\def\alignexplwidth{5cm}
\allowdisplaybreaks
\begin{align*}
    \indicatorfromblock[from=b',to=s,val=v,when=t,chkp=C]{\indP}
    &=
    \frac
    {\attsetweightfromblock[from=b',to=s,val=v,when=t,chkp=C]{\honatts}}
    {\commweightfromblock[from=b',to=s,chkp=C]{\honvals}}
    &&\alignexpl[\alignexplwidth]{By definition.}
    \\
    &\geq
    \frac
    {
    \attsetweightfromblock[from=b',to=s,val=v,when=t,chkp=C]{\allatts}-
        \left(
            \commweightfromblock[from=b',to=s,chkp=C]{\advvals}
            -
            \attsetweightfromblock[from=b',to=s,val=v,when=t,chkp={C}]{\slashvals}
        \right)
    }
    {\commweightfromblock[from=b',to=s,chkp=C]{\honvals}}
    &&\alignexpl[\alignexplwidth]{
        As 
        $
        \attsetfromblock[from=b',to=s,val=v,when=t]{\slashvals}
        \cap
        \attsetfromblock[from=b',to=s,val=v,when=t]{\allatts}
        =\emptyset 
        $ and 
        $ 
        \attsetfromblock[from=b',to=s,val=v,when=t]{\slashvals}
        \subseteq 
        \commfromblock[from=b',to=s]{\advvals}
        $.
    }
    \\
    &=
    \frac
    {
        \attsetweightfromblock[from=b',to=s,val=v,when=t,chkp=C]{\allatts}-
        \left(
            \commweightfromslot[from=\slot(b'),to=s,chkp=C]{\advvals}
            -
            \attsetweightfromblock[from=b',to=s,val=v,when=t,chkp={C}]{\slashvals}
        \right)
    }
    {\commweightfromblock[from=b',to=s,chkp=C]{\allvals} - \commweightfromblock[from=b',to=s,chkp=C]{\advvals}}
    &&\alignexpl[\alignexplwidth]{By definition, $\commweightfromblock[from=b',to=s,chkp=C]{\allvals} = \commweightfromblock[from=b',to=s,chkp=C]{\honvals} + \commweightfromblock[from=b',to=s,chkp=C]{\advvals}$.}
    \\    
    &=
    \frac
    {
        \attsetweightfromblock[from=b',to=s,val=v,when=t,chkp=C]{\allatts}
        +\attsetweightfromblock[from=b',to=s,val=v,when=t,chkp={C}]{\slashvals}
        -\commweightfromblock[from=b',to=s,chkp=C]{\advvals}
    }
    {\commweightfromblock[from=b',to=s,chkp=C]{\allvals} - \commweightfromblock[from=b',to=s,chkp=C]{\advvals}}
    &&\alignexpl[\alignexplwidth]{By rearranging terms.}  
    \\         
    &\geq
    \frac
    {
        \attsetweightfromblock[from=b',to=s,val=v,when=t,chkp=C]{\allatts}     
        +\attsetweightfromblock[from=b',to=s,val=v,when=t,chkp={C}]{\slashvals}
        -\beta \commweightfromblock[from=b',to=s,chkp=C]{\allvals}
    }
    {\commweightfromblock[from=b',to=s,chkp=C]{\allvals} - \beta \commweightfromblock[from=b',to=s,chkp=C]{\allvals}}
    &&\alignexpl[\alignexplwidth]{
        Note that 
        $\attsetfromblock[from=b',to=s,val=v,when=t]{\allatts}$
            and 
        $\attsetfromblock[from=b',to=s,val=v,when=t]{\slashvals}$ are
        disjoint and subsets of $\commfromblock[from=b',to=s]{\allvals}$.
        This implies that 
        $
        \attsetweightfromblock[from=b',to=s,val=v,when=t,chkp=C]{\allatts}     
        +\attsetweightfromblock[from=b',to=s,val=v,when=t,chkp={C}]{\slashvals}
        \leq\commweightfromblock[from=b',to=s,chkp=C]{\allvals}$.
        Given this and that,
        by Assumption~\ref{assum:beta}, $\commweightfromblock[from=b',to=s,chkp=C]{\advvals}\leq \beta \commweightfromblock[from=b',to=s,chkp=C]{\allvals}$,
        the function 
        $g(x)=
        \frac
            {
                \attsetweightfromblock[from=b',to=s,val=v,when=t,chkp=C]{\allatts}    
                +\attsetweightfromblock[from=b',to=s,val=v,when=t,chkp={C}]{\slashvals}                
            -x}
            {\commweightfromblock[from=b',to=s,chkp=C]{\allvals}-x}$ is monotone decreasing in $[0,\commweightfromblock[from=b',to=s,chkp=C]{\allvals}]$
    }
    \\
    &=
    \frac
    {
        \begin{aligned}[t]
            \indicatorfromblock[from=b',to=s,val=v,when=t,chkp=C]{\indQ}
            \commweightfromafterparentblock[from=b',to=s,chkp=C]{\allvals}       
            +\attsetweightfromblock[from=b',to=s,val=v,when=t,chkp={C}]{\slashvals}
            -\beta \commweightfromblock[from=b',to=s,chkp=C]{\allvals}            
        \end{aligned}
    }
    {
        \commweightfromblock[from=b',to=s,chkp=C]{\allvals} 
        - \beta \commweightfromblock[from=b',to=s,chkp=C]{\allvals}
    }
    \\
    &>
    \frac
    {
        \left(
            \frac{\Phi}{2}
                + \beta\,
                \commweightfromblock[from=b',to=s,chkp=C]{\allvals}
                -\attsetweightfromblock[from=b',to=s,val=v,when=t,chkp={C}]{\slashvals}
        \right)
            +\attsetweightfromblock[from=b',to=s,val=v,when=t,chkp={C}]{\slashvals}
            -\beta\, \commweightfromblock[from=b',to=s,chkp=C]{\allvals}
    }
    {
        \commweightfromblock[from=b',to=s,chkp=C]{\allvals}(1-\beta)
    }
    &&\alignexpl[\alignexplwidth]{By the hypothesis,
    $\indicatorfromblock[from=b',to=s,val=v,when=t,chkp=C]{\indQ}\,
    \commweightfromafterparentblock[from=b',to=s,chkp=C]{\allvals}
    >
    \frac{\Phi}{2}
    + \beta\,\commweightfromblock[from=b',to=s,chkp=C]{\allvals}
    - \attsetweightfromblock[from=b',to=s,val=v,when=t,chkp={C}]{\slashvals}$.}
    \\
    &=
    \frac{\Phi}
        {2(1-\beta)\,\commweightfromblock[from=b',to=s,chkp=C]{\allvals}}
    &&\alignexpl[\alignexplwidth]{By simplification.}
\end{align*}
\end{proof}

The next lemma shows that if $\isLMDGHOSTSafe(b,\cdot,t)$ holds, then block~$b$ is in the view of every honest validator by $\slotstart(\slot(t))$ and $\slot(b) \leq \slot(t)$.
The idea is that the safety predicate implies at least one honest validator has voted for~$b$, which means~$b$ was in that validator's view and was broadcast; synchrony then ensures all honest validators receive it by the start of the next slot.

\begin{lemma}\label{lem:block-in-the-view-of-all-honests}
    Let $v$ be any honest validator, $t$ be any time and $b$ be any block
    If
    \begin{enumerate}
        \item $\slotstart(\slot(t)-1)\geq\GGST$ and
        \item $\varforvalattime[val=v]{\isLMDGHOSTSafe}(b,\chkpattime[time=t,val=v],t)$,
    \end{enumerate}
    then
    \begin{enumerate}
        \item block $b$ is in the view of any honest validator at time $\slotstart(\slot(t))$ and thereafter
        \item $\slot(b) \leq \slot(t).$
    \end{enumerate}
\end{lemma}
\begin{proof}
    We can apply Lemmas~\ref{lem:lmd-cond-on-q-implies-cond-on-p} and \ref{lem:lmd-cond-on-p-implies-cond-on-h-ex}, in this order, to conclude that
    $\attsetweightfromblock[from=b,to=\slot(t)-1,val=v,when=t,chkp={\chkpattime[time=t,val=v]}]{\honatts}
    >
    \frac
    {\commweightfromblock[from=b,to=\slot(t)-1,chkp={\chkpattime[time=t,val=v]}]{\allvals} + \boostweight[chkp={\chkpattime[time=t,val=v]}]}
    {2}
    $.
    This implies that at least one honest validator $v' \in \commfromslot[from=n(b),to=\slotattime{t}-1]{\honvals}$ has \GHOST voted in support of $b$.
    This implies that block $b$ was in the view of validator $v'$ by the time it voted in a slot $s \leq \slotattime{t}-1$ as by definition of \LMDGHOST, honest validators only \GHOST vote for blocks that are in their view.
    This further implies that $v'$ broadcast block $b$ no later than the time $t$ it voted in slot $s \leq \slotattime{t}-1$ as honest validators immediately broadcast any message that they receive.
    Then $b$ is in the view of any honest validator by time $\slotstart(\slotattime{t})$.

    Also, given that $v'$ \GHOST votes in support of $b$, Algorithm~\ref{alg:GHOST} implies that $\slot(b) \leq \slot(t)$.

\end{proof}

The next lemma is the main safety result for plain \LMDGHOST: if the observable $\indQ$ indicator exceeds the threshold of \Cref{lem:lmd-cond-on-q-implies-cond-on-p} at time $t$ for every prefix of $b$ on $v$'s chain, and the total validator-set weight under every honest validator's greatest justified checkpoint never grows beyond its value at time~$t$ throughout $[t,t']$ (hypothesis~\ref{hyp:lem:condition-on-q-implies-safety:4}), then $b$ is canonical in every honest validator's view at all future times $t'\geq \slotstart(\slot(t))$.
The validator-set monotonicity hypothesis is what allows the proposer-boost weight to be bounded uniformly across the interval, which is needed for the inductive step.
The proof proceeds by strong induction on $t'$, chaining the previously established results: \Cref{lem:block-in-the-view-of-all-honests} ensures all honest validators have received $b$; \Cref{lem:lmd-p-monotonic} propagates the $\indP$ bound to time $t'$; \Cref{lem:lmd-cond-on-q-implies-cond-on-p} converts the $\indQ$ hypothesis into an $\indP$ bound; \Cref{lem:lmd-cond-on-p-implies-cond-on-h-ex} turns that into an absolute honest-weight bound; and \Cref{lem:condition-on-h-for-canonical-ex} closes by showing that the honest weight exceeds the threshold required to ensure that $b$ belongs to the canonical chain.

\begin{lemma}\label{lem:condition-on-q-implies-safety}
    Given Assumptions~\ref{assum:no-change-to-the-validator-set} and \ref{assum:beta},
    let $v$ be any honest validator,
    $t$ and $t'$ be any two times and
    $b$ be any block,
    if
    \begin{enumerate}
        \item $\slotstart(\slot(t-1) \geq \GGST$,
        \item  $\var[val=v]{\isLMDGHOSTSafe}(b,\gjattime[time=t,val=v],t)$,
        \item $t' \geq \slotstart(\slot(t))$ and
        \item\label{hyp:lem:condition-on-q-implies-safety:4} for any validator $v'' \in \commfromslot[from=\slot(t),to=\slot(t')]{\honvals}$ and time $t''$ such that $t \leq t'' \leq t'$, $\totvalsetweight[chkp={\gjattime[time=t'',val=v'']}]{\allvals} \leq \totvalsetweight[chkp={\gjattime[time=t,val=v]}]{\allvals}$,
    \end{enumerate}
    then $b$ is canonical in the view of any honest validator at time $t'$.
\end{lemma}

\begin{proof}
    \begin{description}
        \item[Base case.]
        This is a strong induction quantified over $t'$, so there is no need for a base case.
        Alternatively, we can take $t' < \slotstart(\slot(t))$ as base case for which the Lemma is vacuously true.

        \item[Inductive step: $t' \geq \slotstart(\slot(t))$.]
        Let $s := \slot(t)$, $s':= \slotattime{t'}$, $v'$ be any honest validator, $\gjviewsym := \chkpattime[time=t,val=v]$, $\gjviewsym ' := \chkpattime[time=t',val=v']$ and $b'$ be any block such that $b' \preceq b$.
        We assume that the Lemma
        holds for any time $t''$ such that $t'' < t'$ and we prove that it holds at time $t'$ as well.

        Given that, as described in Section~\ref{sec:lmd-ghost}, honest validators always \GHOST vote for the block returned by the fork-choice function executed at the time of voting, any honest validator in the committees between slot $s$ and slot $s'-1$ has \GHOST voted in support of $b$ and, consequently, in support of $b'$.

        Also, note that due condition 4 of the Lemma's statement we can conclude that $\boostweight[chkp={\gjviewsym'}] \leq \boostweight[chkp={\gjviewsym}]$.

        Then, we can apply Lemma~\ref{lem:block-in-the-view-of-all-honests} to conclude that $b$ is in the view of $v'$ at time $t'$ and that $\slot(b)\leq \slot(t)$.

        Then,

        \def\alignexplwidth{3.7cm}
        \allowdisplaybreaks

        \begin{align*}
            & \hphantom{{}\geq{}} \indicatorfromblock[from=b',to=\slot(t')-1,val=v',when=t',chkp={\gjviewsym '}]{\indP}
            \\
            &\geq
            \indicatorfromblock[from=b',to=\slot(t)-1,val=v,when=\slotstartslot{t},chkp={\gjviewsym}]{\indP}
            &&\alignexpl[\alignexplwidth]{
                By Lemma~\ref{lem:lmd-p-monotonic}.
            }
            \\
            &>
            \begin{aligned}[t]
                &\frac{\commweightfromafterparentblock[from=b',to=\slot(t)-1,chkp={\gjviewsym}]{\allvals}}{2(1-\beta)\commweightfromblock[from=b',to=\slot(t)-1,chkp={\gjviewsym}]{\allvals}}
                \left( 1+
                    \frac
                    {
                    \begin{aligned}[t]
                            &\boostweight[chkp=\gjviewsym]
                            &
                            -
                            \attsetweighttobediscfromblock[from=b',to=\slot(b')-1,val=v,when=\slotstartslot{t},chkp=\gjviewsym]{\honattsub}
                    \end{aligned}
                    }
                    {\commweightfromafterparentblock[from=b',to=\slot(t)-1,chkp={\gjviewsym}]{\allvals}}
                \right)      
            \end{aligned}
            &&\alignexpl[\alignexplwidth]{
                By condition 2 of the Lemma's statement and Lemma~\ref{lem:lmd-cond-on-q-implies-cond-on-p}.
            }
            \\
            &=
            \begin{aligned}[t]
                &
                    \frac
                    {
                    \begin{aligned}[t]
                            &\commweightfromafterparentblock[from=b',to=\slot(t)-1,chkp={\gjviewsym}]{\allvals}
                            +\boostweight[chkp=\gjviewsym]
                            -
                            \attsetweighttobediscfromblock[from=b',to=\slot(b')-1,val=v,when=\slotstartslot{t},chkp=\gjviewsym]{\honattsub}
                    \end{aligned}
                    }
                    {2(1-\beta)\commweightfromblock[from=b',to=\slot(t)-1,chkp={\gjviewsym}]{\allvals}}     
            \end{aligned}
            &&\alignexpl[\alignexplwidth]{
                Simplification.
            }
            \\
            &\geq
            \begin{aligned}[t]
                &
                    \frac
                    {
                    \begin{aligned}[t]
                            &\commweightfromafterparentblock[from=b',to=\slot(t)-1,chkp={\gjviewsym}]{\allvals}
                            +\boostweight[chkp=\gjviewsym]
                            -
                            \attsetweighttobediscfromblock[from=b',to=\slot(b')-1,val=v,when=\slotstartslot{t},chkp=\gjviewsym]{\honatts}  
                    \end{aligned}
                    }
                    {2(1-\beta)\commweightfromblock[from=b',to=\slot(t)-1,chkp={\gjviewsym}]{\allvals}}     
            \end{aligned}
            &&\alignexpl[\alignexplwidth]{
                As
                $
                \attsetweighttobediscfromblock[from=b',to=\slot(b')-1,val=v,when=\slotstartslot{t},chkp=\gjviewsym]{\honattsub}
                \leq
                \attsetweighttobediscfromblock[from=b',to=\slot(b')-1,val=v,when=\slotstartslot{t},chkp=\gjviewsym]{\honatts} 
                $
            }
            \\            
            &\geq
            \frac{1}{2(1-\beta)}
            \begin{aligned}[t]
                &
                    \frac
                    {
                    \begin{aligned}[t]
                            &\commweightfromafterparentblock[from=b',to=\slot(t)-1,chkp={\gjviewsym}]{\allvals}
                            +\boostweight[chkp=\gjviewsym]
                            -
                            \attsetweighttobediscfromblock[from=b',to=\slot(b')-1,val=v,when=\slotstartslot{t},chkp=\gjviewsym]{\honatts} 
                            \\ 
                            &+\weightofset[chkp=\gjviewsym]{
                                \attsettobediscfromblock[from=b',to=\slot(b')-1,val=v,when=\slotstartslot{t}]{\honatts} \cap \commfromslot[from=\slot(t), to=\slot(t')-1]{\allvals}
                            }
                            \\
                            &+\weightofset[chkp=\gjviewsym]{
                                \commfromslot[from=\slot(t), to=\slot(t')-1]{\allvals}\setminus\attsettobediscfromblock[from=b',to=\slot(b')-1,val=v,when=t]{\honatts} \setminus\commfromblock[from=b',to=\slot(t)-1]{\allvals}
                            }
                    \end{aligned}
                    }
                    {
                        \begin{aligned}[t]
                            &\commweightfromblock[from=b',to=\slot(t)-1,chkp={\gjviewsym}]{\allvals}   
                            +\weightofset[chkp=\gjviewsym]{\attsettobediscfromblock[from=b',to=\slot(b')-1,val=v,when=\slotstartslot{t}]{\honatts} \cap \commfromslot[from=\slot(t), to=\slot(t')-1]{\allvals}}
                            \\
                            &+\weightofset[chkp=\gjviewsym]{\commfromslot[from=\slot(t), to=\slot(t')-1]{\allvals}\setminus\attsettobediscfromblock[from=b',to=\slot(b')-1,val=v,when=\slotstartslot{t}]{\honatts}\setminus\commfromblock[from=b',to=\slot(t)-1]{\allvals}}                         
                        \end{aligned}
                    }     
            \end{aligned}
            &&\alignexpl[\alignexplwidth]{
                Note that 
                $
                \attsetweighttobediscfromblock[from=b',to=\slot(b')-1,val=v,when=\slotstartslot{t},chkp=\gjviewsym]{\honatts}
                \cap 
                \commweightfromblock[from=b',to=\slot(t)-1,chkp={\gjviewsym}]{\allvals} 
                =\emptyset
                $.
                By contradiction, let  
                $
                v\in
                \attsetweighttobediscfromblock[from=b',to=\slot(b')-1,val=v,when=\slotstartslot{t},chkp=\gjviewsym]{\honatts}
                \cap 
                \commweightfromblock[from=b',to=\slot(t)-1,chkp={\gjviewsym}]{\allvals} 
                $. 
                This implies that $v$ is honest, it is in $\commweightfromblock[from=b',to=\slot(t)-1,chkp={\gjviewsym}]{\allvals}$, but its latest vote is cast in a slot in $[\slot(\parent(b'))+1,\slot(b')-1]$ which leads to a contradiction.
                Then, we have that 
                $
                \commweightfromafterparentblock[from=b',to=\slot(t)-1,chkp={\gjviewsym}]{\allvals} - \attsetweighttobediscfromblock[from=b',to=\slot(b')-1,val=v,when=\slotstartslot{t},chkp=\gjviewsym]{\honatts}
                \geq
                \commweightfromblock[from=b',to=\slot(t)-1,chkp={\gjviewsym}]{\allvals} 
                $.
                Finally, note that, for any $x\geq 0$ and $a\geq b$, 
                $\frac{a}{b}\geq\frac{a+x}{b+x}$.
            } 
            \\
            &=
            \frac{1}{2(1-\beta)}
            \frac
            {
            \begin{aligned}[t]
                    &\commweightfromafterparentblock[from=b',to=\slot(t)-1,chkp={\gjviewsym}]{\allvals}
                    +\boostweight[chkp=\gjviewsym]
                    -
                    \attsetweighttobediscfromblock[from=b',to=\slot(b')-1,val=v,when=\slotstartslot{t},chkp=\gjviewsym]{\honatts} 
                    \\ 
                    &+\weightofset[chkp=\gjviewsym]{
                        \attsettobediscfromblock[from=b',to=\slot(b')-1,val=v,when=\slotstartslot{t}]{\honatts} \cap \commfromslot[from=\slot(t), to=\slot(t')-1]{\allvals}
                    }
                    \\
                    &+\weightofset[chkp=\gjviewsym]{
                        \commfromslot[from=\slot(t), to=\slot(t')-1]{\allvals}\setminus\attsettobediscfromblock[from=b',to=\slot(b')-1,val=v,when=\slotstartslot{t}]{\honatts} \setminus\commfromblock[from=b',to=\slot(t)-1]{\allvals}
                    }
            \end{aligned}
            }
            {
                \commweightfromblock[from=b',to=\slot(t')-1,chkp={\gjviewsym}]{\allvals}                
            } 
            &&\alignexpl[\alignexplwidth]{
                $|(A\cap B)\setminus C| + |A\setminus B \setminus C| = |A\setminus C|$
                where $A=\commfromslot[from=\slot(t), to=\slot(t')-1]{\allvals}$, $B=\attsettobediscfromblock[from=b',to=\slot(b')-1,val=v,when=\slotstartslot{t}]{\honatts}$ and $C=\commfromblock[from=b',to=\slot(t)-1]{\allvals}$.
                Also $(A\cap B)\setminus C=A\cap B$ as $B \cap C=\emptyset$.
            }    
            \\
            &=
            \frac{1}{2(1-\beta)}
            \frac
            {
            \begin{aligned}[t]
                    &\commweightfromafterparentblock[from=b',to=\slot(t)-1,chkp={\gjviewsym}]{\allvals}
                    +\boostweight[chkp=\gjviewsym]
                    -
                    \attsetweighttobediscfromblock[from=b',to=\slot(b')-1,val=v,when=t',chkp=\gjviewsym]{\honatts} 
                    \\
                    &+\weightofset[chkp=\gjviewsym]{
                        \commfromslot[from=\slot(t), to=\slot(t')-1]{\allvals}\setminus\attsettobediscfromblock[from=b',to=\slot(b')-1,val=v,when=\slotstartslot{t}]{\honatts} \setminus\commfromblock[from=b',to=\slot(t)-1]{\allvals}
                    }
            \end{aligned}
            }
            {
                \commweightfromblock[from=b',to=\slot(t')-1,chkp={\gjviewsym}]{\allvals}                
            } 
            &&\alignexpl[\alignexplwidth]{
                As 
                $\attsetweighttobediscfromblock[from=b',to=\slot(b')-1,val=v,when=\slotstartslot{t},chkp=\gjviewsym]{\honatts} 
                -\weightofset[chkp=\gjviewsym]{
                        \attsettobediscfromblock[from=b',to=\slot(b')-1,val=v,when=\slotstartslot{t}]{\honatts} \cap \commfromslot[from=\slot(t), to=\slot(t')-1]{\allvals}
                    }
                =
                \attsetweighttobediscfromblock[from=b',to=\slot(b')-1,val=v,when=t',chkp=\gjviewsym]{\honatts}
                $
                as any validator in 
                $\attsettobediscfromblock[from=b',to=\slot(b')-1,val=v,when=\slotstartslot{t}]{\honatts} \cap \commfromslot[from=\slot(t), to=\slot(t')-1]{\allvals}$ \GHOST votes for $b'$ by $\slotstartslot{t'}$.
            } 
            \\
            &\geq
            \frac{1}{2(1-\beta)}
            \frac
            {
            \begin{aligned}[t]
                    &\commweightfromafterparentblock[from=b',to=\slot(t')-1,chkp={\gjviewsym}]{\allvals}
                    +\boostweight[chkp=\gjviewsym]
                    -
                    \attsetweighttobediscfromblock[from=b',to=\slot(b')-1,val=v,when=t',chkp=\gjviewsym]{\honatts} 
            \end{aligned}
            }
            {
                \commweightfromblock[from=b',to=\slot(t')-1,chkp={\gjviewsym}]{\allvals}                
            }
            &&\alignexpl[\alignexplwidth]{
                Note that 
                $
                \attsettobediscfromblock[from=b',to=\slot(b')-1,val=v,when=\slotstartslot{t}]{\honatts} \cap\commfromblock[from=b',to=\slot(t)-1]{\allvals}=\emptyset
                $ and therefore 
                $
                \attsettobediscfromblock[from=b',to=\slot(b')-1,val=v,when=\slotstartslot{t}]{\honatts} \cup\commfromblock[from=b',to=\slot(t)-1]{\allvals}
                \subseteq 
                \commfromafterparentblock[from=b',to=\slot(t')-1]{\allvals}
                $
            }
            \\  
            &\geq
            \frac{1}{2(1-\beta)}
            \frac
            {
            \begin{aligned}[t]
                    &\commweightfromafterparentblock[from=b',to=\slot(t')-1,chkp={\gjviewsym'}]{\allvals}
                    +\boostweight[chkp=\gjviewsym']
                    -
                    \attsetweighttobediscfromblock[from=b',to=\slot(b')-1,val=v,when=\slotstartslot{t'},chkp=\gjviewsym']{\honatts} 
            \end{aligned}
            }
            {
                \commweightfromblock[from=b',to=\slot(t')-1,chkp={\gjviewsym}]{\allvals}                
            }   
            &&\alignexpl[\alignexplwidth]{
                As 
                $\commweightfromafterparentblock[from=b',to=\slot(t')-1,chkp={\gjviewsym}]{\allvals}\geq\commweightfromafterparentblock[from=b',to=\slot(t')-1,chkp={\gjviewsym'}]{\allvals}$,
                $\boostweight[chkp=\gjviewsym]\geq\boostweight[chkp=\gjviewsym']$ and 
                $\attsetweighttobediscfromblock[from=b',to=\slot(b')-1,val=v,when=t',chkp=\gjviewsym]{\honatts} =\attsetweighttobediscfromblock[from=b',to=\slot(b')-1,val=v,when=\slotstartslot{t'},chkp=\gjviewsym']{\honatts}$.
            }
        \end{align*}    

        Given that     $
    \attsetweighttobediscfromblock[from=b',to=\slot(b')-1,val=v,when=t,chkp=C]{\honattsub}
    \leq
    \attsetweighttobediscfromblock[from=b',to=\slot(b')-1,val=v,when=t,chkp=C]{\honatts}
    $, we can apply \Cref{lem:lmd-cond-on-p-implies-cond-on-h-ex,lem:condition-on-h-for-canonical-ex} to conclude the proof.
\end{description}
        
\end{proof}

\subsection{\LMDGHOSTHFC}

\paragraph{Additional notation.}
We introduce the following conventions used throughout the proofs.
\begin{definition}[Notation]\leavevmode\label{def:b-cand-is-b-conf}
    \begin{enumerate}
        \item Let $\var[time=t,val=v]{\bconfirmed}$ be the value of $\var[val=v]{\bconfirmed}$ after executing the code scheduled at time $t$ in \Cref{alg:conffull}.
        If $t < \var[val=v]{\tinit}$, then $\var[time=t,val=v]{\bconfirmed}$ corresponds to the value after executing the code scheduled at time $\var[val=v]{\tinit}$.
        With this convention, whenever $\var[val=v,time=\slotstart(\slot(t))]{get\_latest\_confirmed}(b_c)$ is invoked we have $b_c = \var[val=v,time=\slotstart(\slot(t)-1)]{\bconfirmed}$.
        \item Let $\var[val=v,time=\slotstartslot{t}]{\bcands}$ be the set of all values assumed by $\bcand$ during the execution of \Cref{alg:conffull} by validator $v$ at time $\slotstartslot{t}$.
        \item $\canonical[time=t,blck=b,to=t_f]$ means that for every time $t'\in [t,t_f]$ the block $b$ is canonical in the view of every honest validator.
        \item For any function $f$, let $\var[val=v,time=t]{f}(\mathit{pars})$ be the result of honest validator $v$ executing $f(\mathit{pars})$ at time $t$.
        \item $\votingtime(\slot(t))$ refers to the latest time in $\slot(t)$ by which honest validators send their votes. As per the current implementation, this corresponds to 4 seconds after the beginning of $\slot(t)$.
    \end{enumerate}
\end{definition}

  \paragraph{Safety Decay.}
    Because the validator set can change over time, Gasper is exposed to \emph{long-range attacks}~\cite{longrangeattacks}, where, for example, validators that have exited the validator set on one chain can then finalize a competing chain without ever being slashed.
    To prevent such attacks, honest validators never switch their greatest finalized checkpoint to a conflicting one\footnote{In practice, in addition to this measure, Gasper also employes the concept of \emph{weak subjectivity checkpoint} and \emph{weak subjectivity period}~\cite{weaksubj} to protect those validators that have been offline for long time.}.
    However, even with this mechanism in place, possible changes to the validator set reduce the maximum threshold of Byzantine-controlled effective-balance that the protocol can cope with, compared to the theoretical case where the validator set never changes~\cite{gasper}.
    In this work, we assume that even during periods of asynchrony, validators finalize new checkpoints with a frequency that is high enough to ensure that such threshold is never lower than $\frac{1}{3}-\safetydecay$ for some known value of $\safetydecay$ called the \emph{safety decay}~\cite{ethereum4weaksubjconsensus}.

    Accordingly, throughout the appendix proofs we replace the simplified bound $\beta < \frac{1}{3}$ used in the main text (Assumption~\ref{assum:ffg-assumptions:beta}) with the tighter bound $\beta < \frac{1}{3} - \safetydecay$.

    We now restate the properties satisfied by Gasper in a more formal way.

        \begin{property}[Gasper Properties]
        The Gasper protocol ensures the following properties.
        \begin{enumerate}
            \item
            If $\beta < \frac{1}{3} -\safetydecay$,
            then no two checkpoints for the same epoch can ever be justified, \ie,
            for any two blocks $b_1$ and $b_2$ and two checkpoints $C_1 \in \allU(b_1)$ and
            $C_2 \in \allU(b_2)$, $\epoch(C_1) = \epoch(C_2) \implies C_1 = C_2$.
            \item
            For any honest validator $v$, the greatest justified checkpoint is always a strict
            descendant of the greatest finalized checkpoint, \ie,
            $\chkpattime[time=t,val=v]\succ\finattime[val=v,time=t]$.
            \item
            Any honest validator sending a \GHOST\ vote for a block $b$ during epoch $e$
            also sends, at the same time, an \FFG vote
            $\ffgvote[from={\votsource[blck=b,time=e]}, to={\chkp(b,e)}]$.
            \item
            Any honest validator sending an \FFG vote $\ffgvote[from=C_s,to=C_d]$
            also sends, at the same time, a \GHOST\ vote for a block $b\succeq C_d$.
            \item
            Provided that $\beta < \frac{1}{3} -\safetydecay$, for any honest validator $v$,
            time $t$, block $b$ and valid checkpoint $C$, if
            \begin{enumerate}
                \item $C \succeq \chkpattime[time=t,val=v]$ and
                \item $\weightofset[chkp=C]{\ffgvalsetallsentraw[source=,target=C,time=t]}
                    \geq\frac{2}{3}\totvalsetweight[chkp=C]{\allvals}$,
            \end{enumerate}
            then no checkpoint $C' \neq C$ such that $\epoch(C')=\epoch(C)$ can ever be justified.
            \item
            Provided that $\beta < \frac{1}{3} -\safetydecay$, for any block $b$ and epoch $e$
            such that $\slotstart(e)\geq\GGST$, if all honest validators in the committee of
            epoch $e$ send \FFG votes targeting a checkpoint that is a descendant of $b$,
            then no checkpoint $C$ conflicting with $b$ such that $\epoch(C)=e$ can ever be justified.
            \item
            For any block $b$, $\epoch(\gujblock(b)) \leq \epoch(b)$.
            Given \Cref{def:gjblock,def:voting-source,def:gjview}, this implies that, for any honest
            validator $v$, block $b$ and time $t$ such that $\epoch(b) \leq \epoch(t)$,
            $\epoch(\votsource[blck=b,time=t])\leq \epoch(\chkpattime[val=v,time=t]) \leq \epoch(t) - 1$.
        \end{enumerate}
    \end{property}

\paragraph{Safety-induction requirements.}
For brevity in later arguments, we group the recurring hypotheses needed to propagate confirmation safety for a block~$b$ over the next two epochs into a single predicate \(\sir(b,t,C,b_\mathsf{s})\).
It bundles a local \LMDGHOST-safety condition (evaluated with respect to~\(C\) under \(\GGST\)), plus global filter-viability and no-conflicting-justification conditions up to time \(\slotstart(\epoch(b)+2)\).

\begin{definition}[\safetyinductionrequirement{s} for blocks $b$ and $b_\mathsf{s}$, time $t$ and checkpoint $C$]\label{def:induction-conditions}
We define $\sir(b,t,C,b_\mathsf{s})$ as the conjunction of the following five predicates:
\[
\begin{aligned}
\sir(b,t,C,b_\mathsf{s})
\;:=\;&\;
\sirone(b,t,C,b_\mathsf{s})
\;\land\;
\sirtwo(b,t,\slotstart(\epoch(b)+2))
\\
&\land\;
\sirthree(C,t,\slotstart(\epoch(b)+2))
\;\land\;
\sirfour(b)
\;\land\;
\sirfive(b,\epoch(b),\epoch(b)+1).
\\
&\land\;\sirsix(b_s,t,\slotstart(\epoch(b)+2))
\end{aligned}
\]
Concretely:
\begin{enumerate}[label=SIR.\arabic*.,leftmargin=20ex,ref=SIR.\arabic*]
    \siritem{$(b,t,C,b_\mathsf{s})$}
    \label{def:induction-conditions:is-lmd-confirmed}
    $\var[val=v,time=t]{\isOneConfirmedExtFrom}(b,C,b_\mathsf{s}) \;\land\; C \preceq b \;\land\; \slotstart(\slot(t)-1)\geq \GGST$.

    \siritem{$(b,t,t_\ell)$}
    \label{def:induction-conditions:all-validators:not-filtered-out}
    For any honest validator $v'$ and any time $t' \in [t,t_\ell]$, $b$ is not filtered out by $v'$ at time $t'$, i.e., $b \in \filtered[time=t',val=v']$.

    \siritem{$(C,t,t_\ell)$}
    \label{def:induction-conditions:all-validators:gj-succ}
    For any honest validator $v'$ and any time $t' \in [t,t_\ell]$, $\gjattime[time=t',val=v']\succeq C$.

    \siritem{$(b)$}
    \label{def:induction-conditions:ub}
    By time $\slotstart(\firstslot(\epoch(b)+2))$, in the view of any honest node there exist a block $b' \succeq b$ such that
    $\chkp(b) \in \allU(b')$ and $\epoch(b') < \epoch(b)+2$.

    \siritem{$(b,e_1,e_2)$}
    \label{def:induction-conditions:no-conflicting}
    No checkpoint $C$ with $\epoch(C) \in [e_1,e_2]$ that conflicts with $b$ can ever be justified. \footnote{The attentive reader might notice that $\sirfour(b)$ implies $\sirfive(b,\epoch(b),\epoch(b))$. The reason for this is historical. Removing the implication would require significant editing work to the proofs with no improvement on the final results. Therefore, the author have decided to leave this as it is for the moment.}

    \siritem{$(b_\mathsf{s},t,t_\ell)$}
    \label{def:induction-conditions:bs-canonical}
    $\canonical[blck=b_s,time=t,to=t_\ell]$
\end{enumerate}

We write $\sir(b,t,C,b_\mathsf{s}).\sir n$ to denote the $n$-th conjunct above.
For example,
\[
\sir(b,t,C,b_\mathsf{s}).\sirtwo \;:=\; \sirtwo(b,t,\slotstart(\epoch(b)+2)).
\]
When $\sir(b,t,C,b_\mathsf{s})$ is clear from context, we write simply $\sir n$ for
$\sir(b,t,C,b_\mathsf{s}).\sir n$.

Finally, for any subset $\{\sir_{i_1},\ldots,\sir_{i_k}\}\subseteq\{\sir1,\ldots,\sir5\}$, we write
$\sir(b,t,C,b_\mathsf{s}).\{\sir_{i_1},\ldots,\sir_{i_k}\}$ to denote the conjunction
$\bigwedge_{j=1}^k \sir(b,t,C,b_\mathsf{s}).\sir_{i_j}$.
\end{definition}

The predicate $\sir(b,t,C,b_\mathsf{s})$ is (i) preserved across epoch transitions and (ii) sufficient to conclude that $b$ is canonical under filtering.
In particular, once $\sir(b,t,C,b_\mathsf{s})$
holds, it already suffices to conclude that $b$ is canonical from the start of slot
$\slot(t)$ onward:
\[
\sir(b,t,C,b_\mathsf{s}) \;\Longrightarrow\; \canonical[blck=b,time=\slotstart(\slot(t))],
\]
as formalized in Lemma~\ref{lem:ffg-safety-from-sir}.

The following lemma extends the \LMDGHOST\ safety result (Lemma~\ref{lem:condition-on-q-implies-safety}) to \LMDGHOSTHFC\ by adding two conditions that account for the viability filter: every honest validator's greatest justified checkpoint must stay $\succeq C$, and~$b$ must remain in $\filtered$.
Under these conditions, $b$ stays canonical despite the filter.
The proof reduces to Lemma~\ref{lem:condition-on-q-implies-safety} by observing that the filter is the only difference between \LMDGHOST\ and \LMDGHOSTHFC, and the two additional conditions neutralize its effect.

\begin{lemma}\label{lem:ffg-condition-on-q-implies-safety-1}
    Given \Cref{assum:beta,assum:no-change-to-the-validator-set},
    let $v$ be any honest validator,
    $t$ and $t'$ be any two times,
    $b$ be any block and
    $C$ be any checkpoint.
    If
    \begin{enumerate}
        \item $\slotstart(\slotattime{t}-1)\geq\GGST$,
        \item $\varforvalattime[val=v]{\isLMDGHOSTSafe}(b,C,t)$,
        \item $t' \geq \slotstart(\slot(t))$ and
        \item\label{hyp:lem:ffg-condition-on-q-implies-safety-1:4} for any validator $v'' \in \commfromslot[from=\slotattime{t},to=\slotattime{t'}]{\honvals}$ and time $t''$ such that $t \leq t'' \leq t'$,
        \begin{enumerate}[label*=\arabic*.]
            \item $\chkpattime[time=t'',val=v''] \succeq C$ and
            \item $b \in \filtered[time=t'',val=v'']$,
        \end{enumerate}
    \end{enumerate}
    then $b$ is canonical in the view of any honest validator at time $t'$.
\end{lemma}
\begin{proof}
    Because of \Cref{assum:no-change-to-the-validator-set}, condition 4.1 implies that, for any validator $v'' \in \commfromslot[from=\slotattime{t},to=\slotattime{t'}]{\honvals}$ and time $t''$ such that $t \leq t'' \leq t'$, $\totvalsetweight[chkp={\chkpattime[time=t'',val=v'']}]{\allvals} \leq \totvalsetweight[chkp=C]{\allvals}$.
    Then, given that the only difference between \LMDGHOST and \LMDGHOSTHFC is the application of \FILHFC and condition 4.2 of the Lemma's statement, the proof for this Lemma is identical to the proof of Lemma~\ref{lem:condition-on-q-implies-safety}.
\end{proof}

The next lemma shows that a block~$b$ remains in the filtered tree ($\filtered$) throughout an entire epoch, provided two conditions hold: (1)~some descendant $b'$ of~$b$ already unrealized-justifies the checkpoint $\chkp(b)$ (\ie, $\chkp(b)\in\allU(b')$) and is from an earlier epoch than $\epoch(t)$, and (2)~no \emph{justified} checkpoint for an epoch in $[\epoch(b), \epoch(t)-1]$ conflicting with~$b$ ever appears by $\slotstart(\epoch(t+1))$.
The proof proceeds by cases on the relationship between the greatest justified checkpoint and~$\chkp(b)$: in each case, the absence of conflicting justified checkpoints ensures that the viability filter does not remove~$b$.

\begin{lemma}
    \label{lem:no-filtered-out-if-no-conflicting-checkpoints}
    Given Assumption~\ref{assum:ffg-assumptions:beta}, let $t$ be any time.
    If
    \begin{enumerate}
        \item\label{itm:lem:no-filtered-out-if-no-conflicting-checkpoints:first} in the view of any honest validator, by time $t$, there exists a block $b' \succeq b$ such that $\chkp(b) \in \allU(b') \land \epoch(b') < \epoch(t)$
        and
        \item no justified checkpoint for an epoch in $[\epoch(b), \epoch(t)-1]$ conflicting with $b$ exists by $\slotstart(\epoch(t+1))$,
    \end{enumerate}
    then, for any honest validator $v'$ and time $t'\geq t$ with $\epoch(t') = \epoch(t)$, $b \in \filtered[time=t',val=v']$, \ie,
     $b$ is not going to be filtered at any time $t'$ within epoch $\epoch(t)$.
\end{lemma}

\begin{proof}
    Let $v'$ be any validator and $t'$ be any time such that $\epoch(t') = \epoch(t)$.
    Let us now proceed by cases.
    \begin{description}
        \item[{Case 1: $epoch\left(\chkpattime[time=t',val=v']\right) = \epoch(\chkp(b))$.}]
        By the Lemma's hypotheses, we know that there exists a block $b' \succeq b$ such that $\chkp(b) \in \allU(b')$.
        Let $b''$ be any block $b'' \succeq b'$.
        Given that $\epoch(b') < \epoch(t)$, $\epoch(\votsource[blck=b'',time=t'])\geq\epoch(\votsource[blck=b',time=t']) \geq \chkp(b)$.
        By Property~\ref{prop:gasper-basic:only-one-justified-per-epoch} and the definition of $\chkpattime[time=t',val=v']$ (\Cref{def:gjview}), we have that $\votsource[blck=b'',time=t'] = \chkp(b) = \chkpattime[time=t',val=v']$.
        By \Cref{prop:gasper-basic:just-succ-finalization}, this also implies that $b'' \succeq \finattime[time=t',val=v']$.

        Given that clearly $b' \succeq \block(\chkp(b))$ we have that $b' \in \filtered[time=t',val=v']$, from which it follows that $b \in \filtered[time=t',val=v']$.

        \item[{Case 2: $epoch\left(\chkpattime[time=t',val=v']\right) > \epoch(\chkp(b))$.}]
        By Property~\ref{prop:gasper-basic:highest-justified-not-from-current-epoch}, we know that $epoch\left(\chkpattime[time=t',val=v']\right) \in [\epoch(b)+1,\epoch(t) -1]$.
        Hence, by the Lemma's hypotheses, $\chkpattime[time=t',val=v']$ does not conflict with $b$ which, given that in this case we assume $epoch\left(\chkpattime[time=t',val=v']\right) > \epoch(\chkp(b)) = \epoch(b)$, implies that $b \prec \chkpattime[time=t',val=v']$, from which we can conclude that $b \in \filtered[time=t',val=v']$.

        \item[{Case 3: $epoch\left(\chkpattime[time=t',val=v']\right) < \chkp(b)$.}]
        Given that there exists a block $b' \succeq b$ such that $\chkp(b) \in \allU(b')$, we have that 
        $\epoch(\votsource[blck=b',time=t']) \geq \epoch(\chkp(b))$.
        Hence, the definition of $\chkpattime[time=t',val=v']$ (Definition~\ref{def:gjview}) implies that $epoch\left(\chkpattime[time=t',val=v']\right) \geq \epoch(\chkp(b))$ meaning that this case is not possible.
    \end{description}
\end{proof}

The following lemma provides a sufficient condition for a checkpoint $\chkp(b,e)$ to accumulate enough \FFG\ votes to be justified.
If the observed \FFG\ support up to slot~$s{-}1$, combined with the worst-case honest contribution from the remaining slots $[s,\lastslot(e)]$, reaches the $2/3$ threshold (accounting for Byzantine slashing), then the checkpoint will have received a $2/3$ supermajority by the start of epoch $e{+}1$.
This lemma is used to verify the $\willChkpBeJustified$ predicate in the confirmation rule.

\begin{lemma}\label{lem:sufficient-condition-for-justification}
    Given Assumptions ~\ref{assum:beta} and  \ref{assum:ffg},
    let $t\geq\GGST$ be any time,
    $b$ be any block,
    $e$ be any epoch,
    $s$ be any slot such that $\epoch(s) \geq \epoch(b)$,
    $v$ be any honest validator.
    If
    $$
    \ffgvalsettoslotweight[to=s-1,source={\votsource[blck=b,time=\epoch(b)]},target={\chkp(b,e)},time=t,val=v,weight chkp={\chkp(b,e)}]
    + (1-\beta)\valsetweightfromslot[from=s,to=\lastslot(e),chkp={\chkp(b,e)}]{\allvals}
    \geq
    \frac{2}{3}\totvalsetweight[chkp={\chkp(b,e)}] {\allvals}
    + \beta \totvalsetweight[chkp={\chkp(b,e)}]{\allvals}
    $$
    and all honest validators in slots $[s,\lastslot(e)]$ \GHOST vote for a block $b'' \succeq b$ such that $\epoch(b'') = \epoch(b)$, then, for any block $b' \succeq \chkp(b,e)$ and time $t' \geq \slotstart(e+1)$,
    $\weightofset[chkp=b']{\ffgvalsetallsentraw[source=,target={\chkp(b,e)},time=t'] \setminus \slashedset[chkp=b']} \geq \frac{2}{3}\totvalsetweight[chkp=b'] {\allvals}$.
\end{lemma}
\begin{proof}
    Let $C_b := \chkp(b,e)$ and $\mathit{VS}_b := \votsource[blck=b,time=e]$,
    $t'$ be any time such that $t' \geq \slotstart(\epoch(b)+1)$, and
    $b'$ be any block such that $b' \succeq \chkp(b,e)$.

    We can now proceed as follows to prove the Lemma.
    \begingroup
    \allowdisplaybreaks
    \newcommand{\alignexpllength}{6cm}
    \begin{align*}
        \weightofset[chkp=b']{\ffgvalsetallsentraw[source=,target=C_b,time=t']\setminus \slashedset[chkp=b']}
        &\geq
        \weightofset[chkp=b']{
                \left(\ffgvalsettoslot[to=s-1,source=\mathit{VS}_b,target=C_b,time=t,val=v]
                \sqcup
                \commfromslot[from=s,to=\lastslot(e)]{\honvals}\right)
                \setminus \slashedset[chkp=b']
        }
        &&\alignexpl[\alignexpllength]{Given that $t' \geq \slotstart(e+1)$, by time $t'$ every honest validator in slots $[s,\lastslot(e)]$ has \GHOST voted for a block $b'' \succeq b$, which, by Property~\ref{prop:gasper-basic:ldm-vote-for-b-is-ffg-vote-for-cb} equates to an \FFG vote for $\ffgvote[from=\mathit{VS}_b, to=C_b]$.
        To this, we add the validators whose \GHOST votes have already been received at time $t$.
        } \\
        &=
        \weightofset[chkp=C_b]{
            \left(
                \ffgvalsettoslot[to=s-1,source=\mathit{VS}_b,target=C_b,time=t,val=v]
                \sqcup
                \commfromslot[from=s,to=\lastslot(e)]{\honvals}
            \right)
            \setminus
            \slashedset[chkp=b']
        }
        &&\alignexpl[\alignexpllength]{The only difference in effective balances between $b'$ and $C_b$ is represented by those validators $\slashedset[chkp=b'] \setminus \slashedset[chkp=C_b]$ that are slashed between $C_b$ and $b'$.}
        \\
        &=
        \weightofset[chkp=C_b]{
            \left(
                \ffgvalsettoslot[to=s-1,source=\mathit{VS}_b,target=C_b,time=t,val=v]
                \setminus
                \slashedset[chkp=b']
            \right)
            \sqcup
            \commfromslot[from=s,to=\lastslot(e)]{\honvals}
        }
        &&\alignexpl[\alignexpllength]{As honest validators never get slashed.}
        \\
        &=
        \weightofset[chkp=C_b]{
            \left(\ffgvalsettoslot[to=s-1,source=\mathit{VS}_b,target=C_b,time=t,val=v]
            \setminus
            \left(
                \ffgvalsettoslot[to=s-1,source=\mathit{VS}_b,target=C_b,time=t,val=v]
                \cap
                \slashedset[chkp=b']
            \right)\right)
            \sqcup
            \commfromslot[from=s,to=\lastslot(e)]{\honvals}
        }
        \\
        &=
        \weightofset[chkp=C_b]{
            \ffgvalsettoslot[to=s-1,source=\mathit{VS}_b,target=C_b,time=t,val=v]
        }
        -
        \weightofset[chkp=C_b]{
            \ffgvalsettoslot[to=s-1,source=\mathit{VS}_b,target=C_b,time=t,val=v]
            \cap
            \slashedset[chkp=b']
        }
        +
        \commweightfromslot[from=s,to=\lastslot(e),chkp=C_b]{\honvals}
        \\
        &\geq
        \begin{aligned}[t]
            &\weightofset[chkp=C_b]{
                \ffgvalsettoslot[to=s-1,source=\mathit{VS}_b,target=C_b,time=t,val=v]
            }
            -
            \weightofset[chkp=C_b]{
                \ffgvalsettoslot[to=s-1,source=\mathit{VS}_b,target=C_b,time=t,val=v]
                \cap
                \slashedset[chkp=b']
            }\\
            &+
            (1-\beta)\commweightfromslot[from=s,to=\lastslot(e),chkp=C_b]{\allvals}
        \end{aligned}\\
        &\geq
        \begin{aligned}[t]
            &\frac{2}{3}\totvalsetweight[chkp=C_b] {\allvals}
            + \beta \totvalsetweight[chkp=C_b]{\allvals}
            -
            (1-\beta)\valsetweightfromslot[from=s,to=\lastslot(e),chkp=C_b]{\allvals}\\
            &-
            \weightofset[chkp=C_b]{
                \ffgvalsettoslot[to=s-1,source=\mathit{VS}_b,target=C_b,time=t,val=v]
                \cap
                \slashedset[chkp=b']
            }
            +
            (1-\beta)\commweightfromslot[from=s,to=\lastslot(e),chkp=C_b]{\allvals}
        \end{aligned}
        &&\alignexpl[\alignexpllength]{By applying the condition on \weightofset[chkp=C_b]{
            \ffgvalsettoslot[to=s-1,source=\mathit{VS}_b,target=C_b,time=t,val=v]
        } as per the Lemma's statement.}\\
        &\geq
        \frac{2}{3}
        \totvalsetweight[chkp=C_b] {\allvals}
        &&\alignexpl[\alignexpllength]{As, due to Assumption~\ref{assum:beta} and the fact that honest validators never commit slashing offences,
        $
            \beta \totvalsetweight[chkp=C_b]{\allvals}
            \geq
            \weightofset[chkp=C_b]{
            \ffgvalsettoslot[to=s-1,source=\mathit{VS}_b,target=C_b,time=t,val=v]
            \cap
            \slashedset[chkp=b']
        }$
        }\\
        &\geq
        \frac{2}{3}
            \totvalsetweight[chkp=b'] {\allvals}
        &&\alignexpl[\alignexpllength]{By \Cref{assum:no-change-to-the-validator-set}, given that $b'\succeq C_b$, $\totvalsetweight[chkp=C_b] {\allvals}\geq\totvalsetweight[chkp=b'] {\allvals}$,}
    \end{align*}
    \endgroup
    Note that if $e<\epoch(s)$, then, some of the conditions above are vacuously true (\eg, all honest validators in slots $[s,\lastslot(e)] =\emptyset$ \GHOST vote in support of $b$), but the reasoning above still works.
    This concludes the proof.
\end{proof}

The following lemma extends the \LMDGHOSTHFC safety result (\Cref{lem:ffg-condition-on-q-implies-safety-1}) to a time interval $[t, t']$: if $\isOneConfirmedExtFrom$ holds at time~$t$ and hypothesis~\ref{hyp:lem:ffg-condition-on-q-implies-safety-1:4} holds for every honest validator $v''$ in the relevant committees and every time $t''\in[t,t']$ --- namely, $v''$'s greatest justified checkpoint is a descendant of $C$, $b$ stays in $v''$'s filtered tree, and the block $b_\mathsf{s}$ used as the starting point of the chain check is a prefix of $v''$'s $\LMDGHOSTHFC$ output --- then $b$ is canonical over the entire interval.
The proof follows directly from \Cref{lem:ffg-condition-on-q-implies-safety-1}.

\begin{lemma}\label{lem:ffg-condition-on-q-implies-safety}
    Given Assumption~\ref{assum:no-change-to-the-validator-set} and Assumption~\ref{assum:beta},
    let $v$ be any honest validator,
    $t$ and $t'$ be any two times,
    $b$ be any block and
    $C$ be any checkpoint.
    If
    \begin{enumerate}
        \item $\slotstart(\slot(t)-1)\geq\GGST$,
        \item $\var[val=v]{\isOneConfirmedExtFrom}(b,C,t,b_\mathsf{s})$,
        \item $t' \geq \slotstart(\slot(t))$ and
        \item for any validator $v'' \in \commfromslot[from=\slot(t),to=\slot(t')]{\honvals}$ and time $t''$ such that $t \leq t'' \leq t'$, 
        \begin{enumerate}[label*=\arabic*.]
            \item $\gjattime[time=t'',val=v''] \succeq C$,
            \item $b \in \filtered[time=t'',val=v'']$,
            \item $b_\mathsf{s} \preceq  \LMDGHOSTHFC(\viewattime[val=v',time=t'])$
        \end{enumerate}
    \end{enumerate}

    then $\canonical[blck=b,time=\slotstart(\slot(t)),to=t']$

\end{lemma}

\begin{proof}
    Hypothesis~2 here replaces the $\isLMDGHOSTSafe(b,C,t)$ hypothesis of \Cref{lem:ffg-condition-on-q-implies-safety-1} with $\isOneConfirmedExtFrom(b,C,t,b_\mathsf{s})$, which (Algorithm~\ref{alg:findlatestconf-helpers}) requires $\isOneConfirmedExt(b',C)$, and hence either $\isOneConfirmedSpecial(b',C)$ or $\isOneConfirmed(b',C)$, for every $b'$ with $b_\mathsf{s} \prec b' \preceq b$.
    Since $\isOneConfirmedSpecial \implies \isOneConfirmed$ (their thresholds agree except for the committee weight multiplied by $\beta$: $\isOneConfirmedSpecial$ uses $\commweightfromslot[from=\firstslot(\epoch(b')),to=\slot(t)-1,chkp=C]{\allvals}$, which is at least $\isOneConfirmed$'s $\commweightfromblock[from=b',to=\slot(t)-1,chkp=C]{\allvals}$) and $\isOneConfirmed \implies \isOneConfirmedWeaker$ (the two share the same threshold formula except for the discount subtracted in the numerator, and the simple discount used by $\isOneConfirmed$ is bounded above by the discount of \Cref{def:attsetweighttobediscfromblock} used by $\isOneConfirmedWeaker$, cf.\ Remark~\ref{app:disc-remark}, so $\isOneConfirmedWeaker$'s threshold is no larger), $\isOneConfirmedExt$ implies $\isOneConfirmedWeaker(b',C,t)$ (\Cref{def:isOneConfirmedWeaker}).
    The rest follows the proof of \Cref{lem:ffg-condition-on-q-implies-safety-1}.
\end{proof}

This is one of the central lemmas: it shows that whenever the Safety Induction Requirements ($\sir$) hold for a block~$b$ and $b$ is already known to be canonical over the initial two epochs $[\slotstart(\slot(t)), \slotstart(\epoch(b)+2))$, then $b$ is canonical forever from time~$t$ onward.
The proof proceeds by induction on epochs.
In the base case ($\epoch(t') < \epoch(b)+2$), the hypothesis that $b$ is canonical over the initial two epochs together with the SIR conditions directly supplies the needed invariants.
In the inductive step, the absence of conflicting checkpoints from previous epochs (\Cref{lem:no-filtered-out-if-no-conflicting-checkpoints}) ensures $b$ stays in the filtered tree; the justified-checkpoint hierarchy ensures all honest validators' greatest justified checkpoint exceeds~$C$; and \Cref{lem:ffg-condition-on-q-implies-safety} then establishes that $b$ is canonical in the new epoch.
This in turn prevents new conflicting checkpoints (Properties~\ref{prop:gasper-basic:ldm-vote-for-b-is-ffg-vote-for-cb} and~\ref{prop:gasper-basic:no-conflicting-if-all-honest-votes-in-support-of-b}), closing the induction.

\begin{lemma}\label{lem:ffg-safety-from-sir}
    If
    \begin{enumerate}
        \item $\sir(b,t,C,b_\mathsf{s})$
        \item\label{hyp:lem:ffg-safety-from-sir:2} $\canonical[blck=b,time=\slotstart(\slot(t)),to=\slotstart(\epoch(b)+2)]$,
    \end{enumerate}
    then $\canonical[blck=b,time=\slotstart(\slot(t))]$.
\end{lemma}

\begin{proof}
    First, we proceed by induction on $t' \geq \slotstart(\slot(t))$ to show that all of the following inductive conditions hold
    \begin{enumerate}[label=\roman*)]
        \item there exists no checkpoint $C'$ with $\epoch(C') \in [\epoch(b), \epoch(t')]$ which conflicts with $b$.
        \label{lem:ffg-safety-from-sir:safety-cond-1}
        \item\label{lem:ffg-safety-from-sir:safety-cond-2} for any honest validator $v''$ and time $t''$ such that $\slotstart(\slot(t)) \leq t'' < \slotstart(\epoch(t')+1)$
        \begin{enumerate}[label=\roman{enumi}.\roman*)]
            \item $b \in \filtered[val=v'',time=t'']$\label{lem:ffg-safety-from-sir:safety-cond-2.1}
            \item $\gjattime[time=t'',val=v''] \succeq C$\label{lem:ffg-safety-from-sir:safety-cond-2.2}
            \item $b_\mathsf{s} \preceq  \LMDGHOSTHFC(\viewattime[val=v'',time=t''])$ \label{lem:ffg-safety-from-sir:safety-cond-2.3}
        \end{enumerate}

    \end{enumerate}

    \begin{description}
        \item[Base Case: $\epoch(t') < \epoch(b)+2$.]

        All inductive hypothesis are trivially implied by \sirtwo, \sirthree, \sirfive and \sirsix.

        \item[Inductive Step: $\epoch(t') \geq \epoch(b)+2$.]
        Assume that all the inductive hypotheses hold at any time $t_i$ up to 
        $\epoch(t_i) \leq \epoch(t')-1$ and prove that they hold at time $t'$ as well.
        Let $v'$ be any honest validator.

        Induction hypothesis~\ref{lem:ffg-safety-from-sir:safety-cond-1} and \ref{def:induction-conditions:ub} allow us to apply Lemma~\ref{lem:no-filtered-out-if-no-conflicting-checkpoints} to conclude that $b \in \filtered[time=t',val=v']$, \ie, $b$ does not get filtered out by any honest validator in epoch $\epoch(t')$.
        This proves induction hypothesis~\ref{lem:ffg-safety-from-sir:safety-cond-2.1} holds at time $t'$ as well.

        Also, induction hypothesis~\ref{lem:ffg-safety-from-sir:safety-cond-1},
        \ref{def:induction-conditions:ub},
       Property~\ref{prop:gasper-basic:highest-justified-not-from-current-epoch}
        and the definition of $\gjattime[time=t',val=v']$ (Definition~\ref{def:gjview})
        imply that
        $\gjattime[time=t',val=v'] \succeq \chkp(b)$.
        Then,
        \sirone,
        proves inductive conditions \ref{lem:ffg-safety-from-sir:safety-cond-2.2} and $\gjattime[time=t',val=v']  \succeq b_\mathsf{s}$.
        Given that $\LMDGHOSTHFC(\viewattime[val=v',time=t']) \succeq \gjattime[time=t',val=v']$, this proves \ref{lem:ffg-safety-from-sir:safety-cond-2.3} for $t'$.

        Given that $t'\geq\slotstart(\slot(t))$ and that above we have proved that inductive condition \ref{lem:ffg-safety-from-sir:safety-cond-2} is satisfied for time $t'$, thanks to \sirone, we can apply Lemma~\ref{lem:ffg-condition-on-q-implies-safety} to conclude that $b$ is always canonical in the view of all honest validators at any time during epoch $\epoch(t')$.
 
        By Properties~\ref{prop:gasper-basic:ldm-vote-for-b-is-ffg-vote-for-cb} and~\ref{prop:gasper-basic:no-conflicting-if-all-honest-votes-in-support-of-b}, this immediately implies that no checkpoint conflicting with $b$ can be justified in epoch $\epoch(t') $, which concludes the proof for the inductive hypothesis~\ref{lem:ffg-safety-from-sir:safety-cond-1}.
    \end{description}
    Given that we have just established that the inductive condition \ref{lem:ffg-safety-from-sir:safety-cond-2} hold for any time $t' \geq \slotstart(\slot(t))$, thanks to \sirone and hypothesis~\ref{hyp:lem:ffg-safety-from-sir:2}, we can apply Lemma~\ref{lem:ffg-condition-on-q-implies-safety} to complete the proof.
\end{proof}

The following lemma records a trivial but useful bookkeeping fact: the confirmed block is always one of the algorithm's candidates.
This holds because $\getlatestconfirmed$ initialises $\bcand$ to the previous confirmed block and only moves it forward.

\begin{lemma}\label{lem:bconf-in-bcands}
    \leavevmode
    $\bconfirmed[val=v,time=\slotstart(\slot(t))] \in \var[val=v,time=\slotstart(\slot(t))]{\bcands}$
\end{lemma}
\begin{proof}
    Obvious.
\end{proof}

The next lemma shows that whenever the algorithm outputs a current-epoch candidate, the validator's view already contains a block that is a descendant of that candidate whose greatest-unrealized justified checkpoint is from the previous or current epoch.
This is essential for ensuring that the epoch of the greatest justified checkpoint is no older than two epochs ago.
The proof proceeds by induction on slots: at the first slot of an epoch no current-epoch candidate can exist, and at later slots the bound is either inherited from the previous slot (when the candidate equals the previous confirmed block) or forced by the \KwSty{if}-guards at \Cref{ln:case1-pre,ln:case1}, which must both be true whenever the candidate is advanced into the current epoch.

\begin{lemma}\label{lem:conf-current-epoch-then-gu-curr-epoch}
    Pick any $b_c \in  \var[val=v,time=\slotstart(\slot(t))]{\bcands}$.
    If
    \begin{enumerate}
        \item $\epoch(b_c) = \epoch(t)$,
    \end{enumerate}
    then
    \begin{enumerate}
        \item $\exists b' \in \viewattime[val=v,time=\slotstartslot{t}], b' \succeq b_c \land \epoch(\gu(b')) \geq \epoch(t)-1$
    \end{enumerate}
\end{lemma}

\begin{proof}
    By induction on $\slot(t)$.
    Let $s := \slot(t)$.
    \begin{description}
        \item[Base Case: {$s = \slot(\var[val=v]{\tinit})$}.] The Lemma is vacuously true.
        \item[Inductive Case: {$s \geq \slot(\var[val=v]{\tinit})$}.] We assume that the Lemma holds for $s$ and prove that it also holds for $s+1$.
        Pick any $b_c \in  \var[val=v,time=\slotstart(s+1)]{\bcands}$ such that $\epoch(b_c) = \epoch(s+1)$.
        This implies that slot $s+1 \neq \firstslot(\epoch(s+1))$ as, if so, it could not be that  $\epoch(b_c) = \epoch(s+1)$.
        This then implies that  $\epoch(s+1) = \epoch(s)$.
        \begin{description}
            \item[Case 1: {$b_c = \var[val=v,time=\slotstart(s)]{\bconfirmed}$}.]
            Just apply the inductive hypothesis.
            \item[Case 2: {$b_c \neq \var[val=v,time=\slotstart(s)]{\bconfirmed}$}.]
            This implies that the \textbf{if} conditions at \Cref{ln:case1-pre,ln:case1} are true as, otherwise, we would have $\epoch(b_c) <\epoch(t)$ reaching a contradiction.
            Then, the proof easily follows from \Cref{ln:case1}.
        \end{description}
    \end{description}
\end{proof}

When $\findlatestconfirmeddescendant$ advances the candidate beyond its input and the output belongs to the previous epoch, the next lemma guarantees the existence of an honest validator whose view already contains a block that is a descendant of the output with a voting source from at most two epochs back.
This honest validator is needed to establish the voting-source preconditions of the safety argument at epoch boundaries.
The proof inspects which branch of $\findlatestconfirmeddescendant$ produced the output, ruling out Cases~1 and~2 and extracting the honest validator from Cases~3 and~4 directly.

\begin{lemma}\label{lem:output-find-latest-different-to-input}
    Let $b_o := \var[val=v,time=\slotstartslot{t}]{\findlatestconfirmeddescendant}(b_c)$.
    If
    \begin{enumerate}
        \item $b_o \neq b_c$
        \item $\epoch(b_o) = \epoch(t) - 1$,
    \end{enumerate}
    then, there exists an honest validator $v'$ such that 
    \begin{enumerate}
        \item $\exists b' \in \left(\viewattime[time=\votingtime(\slot({t})-1),val=v'] \cap \viewattime[time=\slotstart(\slot({t})),val=v]\right), b' \succeq b_o \land \epoch(\votsource[blck=b',time=\slotstartslot{{t}}])\geq \epoch({t})-2$
    \end{enumerate}    
\end{lemma}

\begin{proof}
    It must be that $\bcand \gets b_o$ at some point during the execution of   $\findlatestconfirmeddescendant$.
    Below by cases on the line  of $\findlatestconfirmeddescendant$ where this occurs.
    \begin{description}
        \item[Case 1 (\Cref{ln:case1-ret}).]
        Impossible as this case only returns blocks with 
        $\epoch(\btcand) = \epoch(\now)$, 
        contradicting $\epoch(b_o) = \epoch(t)-1$.
        \item[Case 2 (\Cref{ln:case2-ret}).]  
        Impossible as this case returns $b_c \neq b_o$.
        \item[Case 3 (\Cref{ln:case3-ret}).]  
        If $b_o$ is returned here, then, due to \Cref{ln:case3}, 
        $\epoch(\votsource[blck=b_o, time={\slotstartslot{t}}]) 
        \geq \epoch(t) - 2$.
        Due to the $\isOneConfirmedExtFrom$ condition in
        Case~3's candidate set, $\isOneConfirmedExt(b_o, \cdot)$ holds,
        and hence
        $\isOneConfirmed(b_o, \cdot)$ (either directly, or via $\isOneConfirmedSpecial \implies \isOneConfirmed$),
        which implies that there exists an honest validator $v'$
        such that $b_o \in \left(\viewattime[time=\votingtime(\slot(t)-1),
        val=v'] \cap \viewattime[time=\slotstart(\slot(t)),val=v]\right)$.
        \item[Case 4 (\Cref{ln:case4-ret}).]  
        If $b_o$ is returned here, then the existential condition 
        in Case~4's set comprehension directly provides 
        $\exists b' \in \viewattime[time=\slotstart(\slot(t)-1),val=v], 
        b' \succeq b_o \land 
        \epoch(\votsource[blck=b',time=\slotstartslot{t}]) 
        \geq \epoch(t)-2$.
    \end{description}
\end{proof}

The following lemma handles the restart scenario: when the previous slot's confirmed block is from earlier than the previous epoch yet the current candidate is from the previous or current epoch, the confirmation chain must have been restarted at the current slot.
In this case, the lemma guarantees the existence of an honest validator with a voting source from at most two epochs back.
The proof observes that the only way to jump from a confirmed block earlier than the previous epoch to a candidate from the previous or current epoch is through the restart mechanism at \Cref{ln:start-conf-chain}, and then applies \Cref{lem:output-find-latest-different-to-input}.

\begin{lemma}\label{lem:when-restarting-vs-from-prev-epoch}
    Pick any $b_c \in  \var[val=v,time=\slotstart(\slot(t))]{\bcands}$.
    If
    \begin{enumerate}
            \item $\epoch(\var[val=v,time=\slotstart(\slot(t)-1)]{\bconfirmed}) < \epoch(t)-1$
        \item $\epoch(b_c) \geq \epoch(t)-1$,
    \end{enumerate}
    then, there exists an honest validator $v'$ such that 
    \begin{enumerate}
        \item $\exists b' \in \viewattime[time=\votingtime(\slot(t)-1),val=v'], b' \succeq b_c \land \epoch(\votsource[blck=b',time=\slotstartslot{t}])\geq \epoch(t)-2$
    \end{enumerate}
\end{lemma}
\begin{proof}
    Let $s:= \slot(t)$.
    It must be that we restart the ``confirmation chain'' at time $\slotstart(s)$ as there is no other way to satisfy the conditions in the Lemma's statement.
    This means that conditions at \Cref{ln:start-conf-chain} are true.

    Now, we prove the statement by induction on the lines, in order of execution, where $\bcand$ might be set.
    
    Given the above, we know that the first time it is set is at \Cref{ln:set-bcan-on-start-conf-chain}

    \begin{description}
        \item[\Cref{ln:set-bcan-on-start-conf-chain}.] Obvious. 
        \item[\Cref{ln:set-bcand-to-output-find-latest}.] Given $\slot(t) = \firstslot(\epoch(t))$ and the hypothesis in the Lemma statement, we know that $\epoch(b_c) = \epoch(t) - 1$ and thus can apply \Cref{lem:output-find-latest-different-to-input} to conclude the proof.
    \end{description}
\end{proof}

The next lemma establishes that $\findlatestconfirmeddescendant$ never returns a block from the current or a future slot.
The proof argues that any candidate distinct from the input must satisfy $\isOneConfirmed$, which requires honest attestation support exceeding the safety threshold; since honest validators in slot~$s$ only attest at $\votingtime(s)$ to blocks from slot~$\leq s$, and no honest attestation for slot~$\slot(t)$ has been cast at $\slotstart(\slot(t))$, $b_o$ must come from a strictly earlier slot.

\begin{lemma}\label{lem:output-find-latest-different-to-input-then-output-older-than-current-slot}
    Let $b_o := \var[val=v,time=\slotstartslot{t}]{\findlatestconfirmeddescendant}(b_c)$.
    If
    \begin{enumerate}
        \item $b_o \neq b_c$,
    \end{enumerate}
    then
    \begin{enumerate}
        \item $\slot(b_o) < \slot(t)$
    \end{enumerate}
\end{lemma}

\begin{proof}
    Since $b_o \neq b_c$, Case~2 did not produce $b_o$
    (it returns $b_c$).
    In Cases~1, 3, and~4, every candidate in the selection set
    satisfies $\isOneConfirmedExtFrom$,
    which requires $\isOneConfirmedExt(b_o, \cdot)$ and hence $\isOneConfirmed(b_o, \cdot)$ (since $\isOneConfirmedSpecial \implies \isOneConfirmed$).
    This in turn requires attestation support exceeding the safety threshold,
    which is only possible if at least one honest validator attested
    in favour of $b_o$'s subtree.
    Honest validators for slot~$s$ attest at $\votingtime(s)$
    only to blocks from slot $\leq s$,
    and at time $\slotstart(\slot(t))$ no honest attestation
    for slot~$\slot(t)$ has been cast. \footnote{In the Fast Confirmation algorithm is run before $\votingtime(s)$ but after a block has received, there is a chance that some attestation were already cast. However, Gasper does not apply those attestations till the start of the next slot.}
    Hence $b_o$ received honest support from some slot $s \leq \slot(t){-}1$,
    giving $\slot(b_o) \leq s < \slot(t)$.
\end{proof}

Generalising the previous result from $\findlatestconfirmeddescendant$ outputs to all candidates: every candidate block belongs to a strictly past slot.
The proof proceeds by induction on slots, using the previous lemma for the case where the candidate differs from the previous confirmed block.

\begin{lemma}\label{lem:bcand-less-than-current-slot}
    Pick any $b_c \in  \var[val=v,time=\slotstart(\slot(t))]{\bcands}$.
    Then, $\slot(b_c) < \slot(t)$
\end{lemma}

\begin{proof}
    By induction on $\slot(t)$.
    Let $s := \slot(t)$.
    \begin{description}
        \item[Base Case: {$s = \slot(\var[val=v]{\tinit})$}.] Obvious.
        \item[Inductive Case: {$s \geq \slot(\var[val=v]{\tinit})$}.]
            We assume that the Lemma holds for $s$ and prove that it also holds for $s+1$.
            Pick any $b_c \in \var[val=v,time=\slotstart(s+1)]{\bcands}$.
            By cases.
            \begin{description}
                \item[Case 1: {$b_c = \var[val=v,time=\slotstart(s)]{\bconfirmed}$}.] Then, we can apply the inductive hypothesis and conclude that $\slot(b_c) < s < s+1$.
                \item[Case 2: {$b_c \neq \var[val=v,time=\slotstart(s)]{\bconfirmed}$}.]
                Now by cases on the first line, in order of execution of the Fast Confirmation Rule algorithm at time $\slotstart(s+1)$, where $\bcand = b_c$. 
                \begin{description}
                    \item[\Cref{ln:set-bcand-beginning-of-get-latest-confirmed}.] 
                    Impossible, as at this line $\bcand = b_c$, but $b_c \neq \var[val=v,time=\slotstart(s)]{\bconfirmed}$.
                    \item[\Cref{ln:set-bcand-to-fin}.] $\epoch(\bcand) < \epoch(s+1)-1 < \epoch(s+1)$.
                    \item[\Cref{ln:set-bcan-on-start-conf-chain}.] $\epoch(\bcand) = \epoch(s+1)-1 < \epoch(s+1)$. 
                    \item[\Cref{ln:set-bcand-to-output-find-latest}.] Apply \Cref{lem:output-find-latest-different-to-input-then-output-older-than-current-slot}.
                \end{description}
            \end{description}
    \end{description}
\end{proof}

The following lemma records that every non-trivial output of $\findlatestconfirmeddescendant$ lies on the canonical chain, \ie, it is an ancestor-or-equal of the current \LMDGHOSTHFC head.
This is immediate from the algorithm's construction, which only selects blocks from $\canonChainFrom$.

\begin{lemma}\label{lem:out-find-latest-conf-prec-head}
    Let $b_o := \var[val=v]{\findlatestconfirmeddescendant}(b_c)$.
    If
    \begin{enumerate}
        \item $b_o \neq b_c$,
    \end{enumerate}
    then
    \begin{enumerate}
        \item $b_o \preceq \LMDGHOSTHFC(\viewatstslottime[val=v,time=t])$
    \end{enumerate}
\end{lemma}
\begin{proof}
    Obvious.
\end{proof}

A key structural invariant of the algorithm: the confirmed block is always an ancestor-or-equal of the \LMDGHOSTHFC head, \ie, it is always canonical in the validator's own view.
The proof proceeds by contradiction, checking every line of $\getlatestconfirmed$ where $\bcand$ could be set, and showing that each either directly ensures canonicality or triggers the reset to the finalized block (which is always canonical).

\begin{lemma}
    $\bconfirmed[val=v,time=\slotstart(\slot(t))] \preceq \LMDGHOSTHFC(\viewatstslottime[val=v,time=t])$\label{lem:bconf-always-canonical}
\end{lemma}
\begin{proof}
    By contradiction, assume that $\bconfirmed[val=v,time=\slotstart(\slot(t))] \npreceq \LMDGHOSTHFC(\viewatstslottime[val=v,time=t])$.
    Note that must be that $\bcand \gets \bconfirmed[val=v,time=\slotstart(\slot(t))]$ at some point during the execution of  $\getlatestconfirmed$ and $\bcand$ is not set any other value (expect for potentially setting it to $\bconfirmed[val=v,time=\slotstart(\slot(t))]$ again) until the \KwSty{return} statement.
    Below by cases on the line  of $\getlatestconfirmed$ where this occurs.
    \begin{description}
        \item[\Cref{ln:set-bcand-beginning-of-get-latest-confirmed}.] 
        If $\bcand = \bconfirmed[val=v,time=\slotstart(\slot(t))] \npreceq \LMDGHOSTHFC(\viewatstslottime[val=v,time=t])$, then \Cref{ln:set-bcand-to-fin} is also executed.
        This line sets $\bcand$ to $\block(\gfattime[val=v,time={\now}]) \preceq \LMDGHOSTHFC(\viewatstslottime[val=v,time=t])$. 
        Given the above, this implies that \Cref{ln:set-bcand-to-fin} sets $\bcand$ to a value different from $\bconfirmed[val=v,time=\slotstart(\slot(t))]$.
        Therefore \Cref{ln:set-bcand-beginning-of-get-latest-confirmed} is not the last line where $\bcand$ is set $\bconfirmed[val=v,time=\slotstart(\slot(t))]$ and any successive line setting $\bcand$ sets it to $\bconfirmed[val=v,time=\slotstart(\slot(t))]$.
        \item[\Cref{ln:set-bcand-to-fin}.] Obvious.
        \item[\Cref{ln:set-bcan-on-start-conf-chain}.] In this case $\bcand = \gjattime[val=v,time=\slot(t)]$ which implies $\bconfirmed[val=v,time=\slotstart(\slot(t))] \preceq \LMDGHOSTHFC(\viewatstslottime[val=v,time=t])$ reaching a contradiction.
        \item[\Cref{ln:set-bcand-to-output-find-latest}.]  \Cref{lem:out-find-latest-conf-prec-head} implies that if this line is executed, then the value of $\bcand$ is not updated which means that there exists a previous line where $\bcand \gets \bconfirmed[val=v,time=\slotstart(\slot(t))]$ reaching a contradiction.
    \end{description}
\end{proof}

Another basic monotonicity property: $\findlatestconfirmeddescendant$ never moves the candidate backwards.
Its output is always a descendant-or-equal of the input, since every branch of the algorithm either keeps $\bcand$ unchanged or advances it along the canonical chain.

\begin{lemma}\label{lem:out-find-latest-conf-descendant-output}
    $\var[val=v]{\findlatestconfirmeddescendant}(b_c) \succeq b_c$
\end{lemma}
\begin{proof}
    It follows from the algorithm.
\end{proof}

The next lemma shows that if a mid-epoch candidate is from the previous or current epoch, then the \emph{previous} slot's confirmed block is also from the previous or current epoch and is an ancestor of the current candidate.
This is because mid-epoch slots cannot trigger the restart mechanism, so the candidate can only grow monotonically from the previous confirmed block.
The lemma also establishes that the previous confirmed block is canonical at the current time.

\begin{lemma}\label{lem:prev-conf-at-least-e-1}
    Pick any $b_c \in  \var[val=v,time=\slotstart(\slot(t))]{\bcands}$.
    If
    \begin{enumerate}
        \item $\slot(t) > \text{\firstslot}(\epoch(t))$
        \item $\epoch(b_c) \geq \epoch(t)-1$,
    \end{enumerate}
    then
    \begin{enumerate}
        \item $\epoch(\var[val=v,time=\slotstart(\slot(t)-1)]{\bconfirmed}) \geq \epoch(t)-1$
        \item $\var[val=v,time=\slotstart(\slot(t)-1)]{\bconfirmed} \preceq \LMDGHOSTHFC(\viewatstslottime[val=v,time=t])$
        \item $\var[val=v,time=\slotstart(\slot(t)-1)]{\bconfirmed} \preceq b_c$
    \end{enumerate}
\end{lemma}

\begin{proof}
By \Cref{def:b-cand-is-b-conf} algorithm execution at $\slotstartslot{t}$ starts with
$b_c = \var[val=v,time=\slotstart(\slot(t)-1)]{\bconfirmed}$.
Given Condition 1 of the Lemma's statement, condition at \Cref{ln:start-conf-chain}
does not hold making code at \Cref{ln:set-bcan-on-start-conf-chain} unreachable.
We also know that condition at \Cref{ln:if-bcand-npreceq-head} does not hold, otherwise,
$b_c = \gfattime[time=\slotstartslot{t},val=v]$
and $\epoch(b_c) \leq \epoch(t) - 2$
which contradicts Condition 2 of the Lemma's statement.

Let's now move to proving Condition 3.
By cases.
\begin{description}
    \item[Case 1: {$b_c = \var[val=v,time=\slotstart(\slot(t)-1)]{\bconfirmed}$}.]  Obvious.
    \item[Case 2: {$b_c \neq \var[val=v,time=\slotstart(\slot(t)-1)]{\bconfirmed}$}.]
    Now by cases on the first line, in order of execution of the Fast Confirmation Rule algorithm at time $\slotstart(s+1)$, where $\bcand = b_c$. 
    \begin{description}
        \item[\Cref{ln:set-bcand-beginning-of-get-latest-confirmed}.] 
        Impossible, as at this line $\bcand = b_c$, but $b_c \neq \var[val=v,time=\slotstart(\slot(t)-1)]{\bconfirmed}$.
        \item[\Cref{ln:set-bcand-to-fin}.] Impossible, as at this line $\epoch(\bcand) < \epoch(\slot(t)+1)-1$, but we assume $\epoch(b_c) \geq \epoch(\slot(t)+1) -1$.
        \item[\Cref{ln:set-bcan-on-start-conf-chain}.] Impossible, as at this line cannot be executed if $\slot(t) \neq \firstslot(\epoch(t))$.
        \item[\Cref{ln:set-bcand-to-output-find-latest}.] Given that this is the only line where $\bcand \gets b_c$, \Cref{lem:out-find-latest-conf-descendant-output} implies that $\var[val=v,time=\slotstart(\slot(t)-1)]{\bconfirmed} \preceq b_c$.
    \end{description}
\end{description}

\end{proof}

The following lemma is the main voting-source result for previous-epoch candidates.
It shows that whenever a candidate belongs to the previous epoch, there exists an honest validator $v'$ whose view contains a block that is a descendant of the candidate with a voting source at most two epochs old.
The proof proceeds by induction on slots, distinguishing whether the confirmation chain was restarted (applying \Cref{lem:when-restarting-vs-from-prev-epoch}) or continued from the previous slot (applying the inductive hypothesis or \Cref{lem:output-find-latest-different-to-input}).

\begin{lemma}\label{lem:bcand-prev-epoch-vs-at-least-e-2}
    Pick any $b_c \in  \var[val=v,time=\slotstart(\slot(t))]{\bcands}$.
    If
    \begin{enumerate}
        \item $\epoch(b_c) = \epoch(t)-1$,
    \end{enumerate}
    then, there exists an honest validator $v'$ such that
    \begin{enumerate}
        \item $\exists b' \in \left(\viewattime[time=\votingtime(\slot(t)-1),val=v'] \cap \viewattime[time=\slotstart(\slot(t)),val=v]\right), b' \succeq b_c \land \epoch(\votsource[blck=b',time=\slotstartslot{t}])\geq \epoch(t)-2$.
    \end{enumerate}
\end{lemma}

\begin{proof}
    By induction on $\slot(t)$.
    Let $s := \slot(t)$.
    \begin{description}
        \item[Base Case: {$s = \slot(\var[val=v]{\tinit})$}.] The Lemma is vacuously true.
        \item[Inductive Case: {$s \geq \slot(\var[val=v]{\tinit})$}.] We assume that the Lemma holds for $s$ and prove that it also holds for $s+1$.
        Pick any $b_c \in \var[val=v,time=\slotstart(s+1)]{\bcands}$ such that $\epoch(b_c) \geq \epoch(s+1)-1$.
        \begin{description}
            \item[Case 1: {$\epoch(\var[val=v,time=\slotstart(s)]{\bconfirmed}) < \epoch(s+1)-1$}.] Apply  \Cref{lem:when-restarting-vs-from-prev-epoch}.
            \item[Case 2: {$\epoch(\var[val=v,time=\slotstart(s)]{\bconfirmed}) \geq \epoch(s+1)-1$}.]
            By sub cases.
            \begin{description}
                \item[Case 2.1: {$b_c = \var[val=v,time=\slotstart(s)]{\bconfirmed}$}.] 
                By sub cases again.
                \begin{description}
                    \item[Case 2.1.1: $\epoch(s+1) = \epoch(s).$] 
                    Given that $\var[val=v,time=\slotstart(s)]{\bconfirmed} \in \var[val=v,time=\slotstart(s)]{\bcands}$,
                    we can apply the inductive hypothesis to conclude that there exists an honest validator $v'$ such that $\exists b' \in \left(\viewattime[time=\slotstart(s-1),val=v'] \cap \viewattime[time=\slotstart(s),val=v]\right), b' \succeq b_c \land \epoch(\votsource[blck=b',time=\slotstart(s)])\geq \epoch(s)-2 = \epoch(s+1)-2$.
                    So, given that $\viewattime[time=\slotstart(s-1),val=v'] \subseteq \viewattime[time=\slotstart(s),val=v']$ and $\viewattime[time=\slotstart(s),val=v] \subseteq \viewattime[time=\slotstart(s+1),val=v]$, the proof for this case is concluded.
                    \item[Case 2.1.2: $\epoch(s+1) = \epoch(s) + 1.$]
                    This Case implies that $\epoch(b_c) = \epoch(s)$ as 
                    $\epoch(s) = \epoch(s+1) -1 \leq \epoch(b_c) < \epoch(s+1) = \epoch(s) + 1$.
                    Note also that  $\epoch(s+1) = \epoch(s) + 1$ implies that $s+1 = \firstslot(\epoch(s+1))$.
                    All of this together with  \Cref{lem:conf-current-epoch-then-gu-curr-epoch} implies that $\exists b' \in \viewattime[time=\slotstart(s),val=v] \subseteq \viewattime[time=\slotstart(s+1),val=v], b' \succeq b_c \land \epoch(\votsource[blck=b',time=\slotstart(s+1)])\geq \epoch(\gu(b')) \geq \epoch(s)-1\geq\epoch(s+1)-2$.
                \end{description}
                \item[Case 2.2: {$b_c \neq \var[val=v,time=\slotstart(s)]{\bconfirmed}$}.]
                Now by cases on the first line, in order of execution of the Fast Confirmation Rule algorithm at time $\slotstart(s+1)$, where $\bcand = b_c$. 
                \begin{description}
                    \item[\Cref{ln:set-bcand-beginning-of-get-latest-confirmed}.] 
                    Impossible, as at this line $\bcand = b_c$, but $b_c \neq \var[val=v,time=\slotstart(s)]{\bconfirmed}$.
                    \item[\Cref{ln:set-bcand-to-fin}.] Impossible, as at this line $\epoch(\bcand) < \epoch(s+1)-1$, but we assume $\epoch(b_c) \geq \epoch(s+1)-1$.
                    \item[\Cref{ln:set-bcan-on-start-conf-chain}.] Obvious. 
                    \item[\Cref{ln:set-bcand-to-output-find-latest}.]
                    The Lemma's statement implies $\epoch(b_c) = \epoch(s+1) - 1$.
                    Then apply \Cref{lem:output-find-latest-different-to-input}.
                \end{description}
            \end{description}
        \end{description}
    \end{description}
\end{proof}

The next lemma shows that if any candidate at the current slot is from the previous or current epoch, then at the first slot of the current epoch the confirmed block already belonged to the previous epoch.
This establishes that a candidate from the previous or current epoch at any point in the epoch implies that the confirmed block at the epoch boundary was also from the previous or current epoch.
The proof distinguishes the first-slot case (where no current-epoch candidate can exist yet, so the candidate's epoch equals the previous epoch directly) from mid-epoch slots (where it argues by contradiction that no algorithm line can produce a candidate from the previous or current epoch if the confirmed block at the start of the epoch was from earlier than the previous epoch).

\begin{lemma}\label{lem:bconfirmed-at-first-slot-is-from-e-1}
    Pick any $b_c \in \var[val=v,time=\slotstart(\slot(t))]{\bcands}$.
    If
    \begin{enumerate}
        \item $\epoch(b_c) \geq \epoch(t)-1$,
    \end{enumerate}
    then
    \begin{enumerate}
        \item $\epoch(\var[val=v,time=\slotstart(\epoch(t))]{\bconfirmed}) = \epoch(t)-1$.
    \end{enumerate}
\end{lemma}

\begin{proof}
    Let $s := \slot(t)$.
    We know that honest validators do not vote for blocks from the future slots which implies an impossibility of $\epoch(b_c) > \epoch(t)-1$ for any $b_c \in \var[val=v,time=\slotstart(s)]{\bcands}$ if $s=\firstslot(\epoch(t))$.
    By cases.
    \begin{description}
        \item[Case 1: {$s = \firstslot(\epoch(t))$}.]
        From the Lemma's hypothesis and the above impossibility we know that $\epoch(b_c) = \epoch(t)-1$ in this case.
        Then the fact that $\var[val=v,time=\slotstartslot{t})]{\bconfirmed} \in \var[val=v,time=\slotstartslot{t})]{\bcands}$ concludes the proof.
        \item[Case 2: {$s \neq \firstslot(\epoch(t))$}.] Implies $s > \firstslot(\epoch(t))$.
        By contradiction.
        Suppose $\epoch(\var[val=v,time=\slotstart(\epoch(t))]{\bconfirmed}) < \epoch(t)-1$.
        Then consider a slot $s' \geq s$ to be the lowest slot in $\epoch(t)$ for which $\epoch(b') \geq \epoch(t)-1$ for any $b' \in \var[val=v,time=\slotstart(s')]{\bcands}$.
        Now by cases on the first line, in order of execution of the Fast Confirmation Rule algorithm at time $\slotstart(s')$, where $\epoch(\bcand) \geq \epoch(t)-1$. 
        \begin{description}
            \item[\Cref{ln:set-bcand-beginning-of-get-latest-confirmed}.] 
            Impossible, as at this line $\bcand = \var[val=v,time=\slotstart(s'-1)]{\bconfirmed}$ but $s'$ is the lowest slot for which $\epoch(b_c) \geq \epoch(t)-1$.
            \item[\Cref{ln:set-bcand-to-fin}.] Impossible, as at this line $\epoch(\bcand) = \epoch(\block(\gfattime[val=v,time=\slotstart(s')])) < \epoch(t)-1$.
            \item[\Cref{ln:set-bcan-on-start-conf-chain}.] Impossible, as $s' \geq s > \firstslot(\epoch(t))$.
            \item[\Cref{ln:set-bcand-to-output-find-latest}.] Impossible, as this line can only be executed if a previous line set $\bcand$ to a value such that $\epoch(\bcand) \geq \epoch(t)-1$.
            However, this  is impossible as we assume that this is the first line where this happens.
        \end{description}
        Due to the fact that we have reached impossiblity in any of the above cases the proof is concluded.
    \end{description}
\end{proof}

Building on the previous results, the next lemma extends the voting-source epoch bound to \emph{all} honest validators: after $\GGST$, if any candidate is from the previous or current epoch, then every honest validator's view at the start of the current epoch contains a block with a voting source from at most two epochs back.
The proof combines \Cref{lem:bconfirmed-at-first-slot-is-from-e-1} and \Cref{lem:bcand-prev-epoch-vs-at-least-e-2} to find a single honest validator satisfying the condition, then uses the synchrony assumption to propagate this to all honest validators.

\begin{lemma}\label{lem:exists-b-vs-at-least-e-2}
    Pick any $b_c \in \var[val=v,time=\slotstart(\slot(t))]{\bcands}$.
    If
    \begin{enumerate}
        \item $\slotstart(\lastslot(\epoch(t)-1)) \geq \GGST$
        \item $\epoch(b_c) \geq \epoch(t)-1$,
    \end{enumerate}
    then, for any honest validator $v'$,
    \begin{enumerate}
        \item $\exists b' \in \viewattime[val=v',time=\slotstart(\epoch(t))]$ such that $\epoch(\votsource[blck=b',time=\epoch(t)]) \geq \epoch(t)-2$.
    \end{enumerate}
\end{lemma}

\begin{proof}
    Due to \Cref{lem:bconfirmed-at-first-slot-is-from-e-1} we know that $\exists b_c \in \var[val=v,time=\slotstart(\epoch(t))]{\bcands}$ such that $\epoch(b_c) = \epoch(t)-1$.
    Then by \Cref{lem:bcand-prev-epoch-vs-at-least-e-2} we know that there exists an honest validator $v^h$ for which $\exists b' \in \viewattime[time=\votingtime(\lastslot(\epoch(t)-1)),val=v^h]$ such that $\epoch(\votsource[blck=b',time=\epoch(t)]) \geq \epoch(t)-2$.
    This together with the synchrony assumption concludes the proof.
\end{proof}

As a direct corollary of the previous lemma, the greatest justified checkpoint of any honest validator at any time from the current epoch onward is from at most two epochs back.
This follows from \Cref{lem:exists-b-vs-at-least-e-2}, the definition of the greatest justified checkpoint, and its monotonicity.

\begin{lemma}\label{lem:gj-at-least-e-2-alt}
    Pick any $b_c \in  \var[val=v,time=\slotstart(\slot(t))]{\bcands}$, any time $t' \geq \slotstart(\epoch(t))$ and any honest validator $v'$.
    If
    \begin{enumerate}
        \item $\prevfirstslotepoch{t} \geq \GGST$
        \item $\epoch(b_c) \geq \epoch(t)-1$,
    \end{enumerate}
    then
    \begin{enumerate}
        \item $\epoch(\gjattime[val=v',time=t']) \geq \epoch(t)-2$.
    \end{enumerate}
\end{lemma}

\begin{proof}
    Due to \Cref{lem:exists-b-vs-at-least-e-2}, Definition~\ref{def:gjview} and the monotonicity property of the greatest justified checkpoint.
\end{proof}

The following lemma combines the voting-source epoch-bound results: for any candidate from the previous or current epoch, after $\GGST$ there exists an honest validator whose view at the previous voting time contains a block that is a descendant of the candidate with a voting source from at most two epochs back.
This is the version used directly in the safety arguments.
For previous-epoch candidates the proof applies \Cref{lem:bcand-prev-epoch-vs-at-least-e-2}; for current-epoch candidates it extracts the honest validator from the $\isOneConfirmedExt$ check in Case~1 (which implies $\isOneConfirmed$, either directly or via $\isOneConfirmedSpecial \implies \isOneConfirmed$) and uses \Cref{lem:gj-at-least-e-2-alt} to bound the voting-source epoch.

\begin{lemma}
    \label{lem:conf-prev-epoch-then-vs-two-epochs-ago}
     \label{lem:vs-at-least-e-2}
    Pick any $b_c \in  \var[val=v,time=\slotstart(\slot(t))]{\bcands}$.
    If
    \begin{enumerate}
        \item $\slotstart(\lastslot(\epoch(t)-1)) \geq \GGST$
        \item $\epoch(b_c) \geq \epoch(t)-1$,
    \end{enumerate}
    then, there exists an honest validator $v'$ such that
    \begin{enumerate}
        \item $\exists b' \in \left(\viewattime[time=\votingtime(\slot(t)-1),val=v'] \cap \viewattime[time=\slotstart(\slot(t)),val=v]\right), b' \succeq b_c \land \epoch(\votsource[blck=b',time=\slotstartslot{t}])\geq \epoch(t)-2$
    \end{enumerate}
\end{lemma}

\begin{proof}
    Let $s := \slot(t)$.
    \begin{description}
        \item[Case 1: {$\epoch(b_c) = \epoch(s)-1$}.] Apply \Cref{lem:bcand-prev-epoch-vs-at-least-e-2} and the synchrony assumption.
        \item[Case 2: {$\epoch(b_c) \neq \epoch(s)-1$}.] The Lemma's statement implies $\epoch(b_c) = \epoch(s)$ for this case.
        Given that $b_c$ is in $\bcands$ with $\epoch(b_c) = \epoch(s)$, 
        it must have been produced by $\findlatestconfirmeddescendant$ 
        via \Cref{ln:case1-ret},
        and reaching this line requires $b_c \in \btcands$,
        whose definition imposes $\isOneConfirmedExtFrom(b_c, \cdot, \cdot)$,
        implying $\isOneConfirmedExt(b_c, \cdot)$ and hence $\isOneConfirmed(b_c, \cdot)$ (either directly, or via $\isOneConfirmedSpecial \implies \isOneConfirmed$), which implies that there exists an honest validator $v^h$ such that
        $b_c \preceq \LMDGHOSTHFC(\viewattime[val=v^h,time=\votingtime(s')])$ for $s'<s$ which implies $b_c \in \filtered[val=v^h,time=\votingtime(s')]$.
        Given that $\epoch(b_c) = \epoch(s)$ and the fact that honest validators never vote for blocks from the future slots, we conclude that $\epoch(s') = \epoch(s)$.
        Then due to \Cref{lem:gj-at-least-e-2-alt} we know that $\epoch(\gjattime[val=v^h,time=\votingtime(s')]) \geq \epoch(s)-2$.
        Given that $b_c \in \filtered[val=v^h,time=\votingtime(s')]$, it must be true that $\exists b' \in \viewattime[val=v^h,time=\votingtime(s')]$ such that $b' \succeq b_c$ and one of the following conditions hold:
        \begin{enumerate}
            \item $\epoch(\votsource[blck=b',time=\slotstart(s)]) = \epoch(\gjattime[val=v^h,time=\votingtime(s')]) \geq \epoch(s)-2$
            \item $\epoch(\votsource[blck=b',time=\slotstart(s)]) \geq \epoch(s)-2$
        \end{enumerate}
        In either case $\epoch(\votsource[blck=b',time=\slotstart(s)]) \geq \epoch(s)-2$.
        Then the fact that $\viewattime[val=v^h,time=\votingtime(s')] \subseteq \viewattime[val=v^h,time=\votingtime(s-1)] \subseteq \viewattime[val=v^h, time=\slotstart(s)]$ concludes the proof.
    \end{description}
\end{proof}

The next lemma shows that whenever the algorithm produces a current-epoch candidate, the $\willChkpBeJustified$ predicate must have been true for that candidate's checkpoint at some earlier point in the same epoch.
This is essential because it guarantees that the FFG justification machinery will eventually lock the candidate's checkpoint, preventing conflicting checkpoints from being justified.
The proof is by induction: if the candidate was inherited from the previous slot, the inductive hypothesis applies; if it was returned by Case~1 of $\findlatestconfirmeddescendant$, the epoch-crossing guard in that case directly requires $\willChkpBeJustified$.

\begin{lemma}\label{lem:no-curr-epochconflict-chkp-is-justified}
    Pick any $b_c \in  \var[val=v,time=\slotstart(\slot(t))]{\bcands}$.
    If
    \begin{enumerate}
        \item $\epoch(b_c) = \epoch(t)$,
    \end{enumerate}
    then
    \begin{enumerate}
        \item there exists a time $t' \leq t$ such that
        \begin{enumerate}
            \item $t' = \slotstartslot{t'}$
            \item $\epoch(t') = \epoch(t)$
            \item $\var[val=v,time=t']{\willChkpBeJustified}(\chkp(b_c))$
        \end{enumerate}
    \end{enumerate}
\end{lemma}

\begin{proof}
    By induction on $\slot(t)$.
    Let $s := \slot(t)$.
    \begin{description}
        \item[Base Case: {$s = \slot(\var[val=v]{\tinit})$}.] The Lemma is vacuously true.
        \item[Inductive Case: {$s \geq \slot(\var[val=v]{\tinit})$}.]
            We assume that the Lemma holds for $s$ and prove that it also holds for $s+1$.
            Pick any $b_c \in \var[val=v,time=\slotstart(s+1)]{\bcands}$ such that $\epoch(b_c) = \epoch(s+1)$.
            This implies $s+1 \neq \firstslot(\epoch(s+1))$ and therfore $\epoch(s) = \epoch(s+1)$.
            By cases.
            \begin{description}
                \item[Case 1: {$\epoch(\var[val=v,time=\slotstart(s)]{\bconfirmed}) = \epoch(s)$}.] 
                \Cref{lem:prev-conf-at-least-e-1} implies that $\var[val=v,time=\slotstart(s)]{\bconfirmed} \preceq \LMDGHOSTHFC(\viewattime[val=v,time=\slotstart(s+1)])$.
                This implies that $b_c \succeq \var[val=v,time=\slotstart(s)]{\bconfirmed}$.
                Given that $\epoch(b_c) = \epoch(s) = \epoch(\var[val=v,time=\slotstart(s)]{\bconfirmed})$, $\chkp(b_c) = \chkp(\var[val=v,time=\slotstart(s)]{\bconfirmed})$.
                Then apply the inductive hypothesis.
                \item[Case 2: {$\epoch(\var[val=v,time=\slotstart(s)]{\bconfirmed}) < \epoch(s)$}.]
                Given that $\epoch(b_c) = \epoch(s+1)$ and 
                $\epoch(\var[val=v,time=\slotstart(s)]{\bconfirmed}) < \epoch(s)$,
                $b_c$ must have been returned by $\findlatestconfirmeddescendant$
                via \Cref{ln:case1-ret},
                and reaching this line requires \Cref{ln:case1}, namely
                $\epoch(\btcand) > \epoch(b_c) \implies
                \willChkpBeJustified_v(\chkp(\btcand))$.
                Given that the input to $\findlatestconfirmeddescendant$ has 
                epoch $< \epoch(s) = \epoch(b_c)$, the epoch crossing condition 
                is triggered, so 
                $\var[val=v,time=\slotstart(s+1)]{\willChkpBeJustified}(\chkp(b_c))$.
            \end{description}
    \end{description}
\end{proof}

The following lemma establishes a lower bound on the greatest unrealized justified checkpoint at the previous epoch boundary: if the confirmed block at the last slot of the previous epoch was from the previous or current epoch, then $\epoch(\guattime[val=v,time=\slotstart(\prevfirstslotepoch{t})]) \geq \epoch(t)-2$.
The proof uses \Cref{lem:conf-current-epoch-then-gu-curr-epoch} to find a block in the view whose greatest unrealized justified checkpoint is from the previous or current epoch, which in turn implies the bound.

\begin{lemma}\label{lem:epoch-gu-less-than-two-epoch-ago-start-epoch}
    If
    \begin{enumerate}
        \item $\slotstart(\prevfirstslotepoch{t})\geq \GGST$
        \item $\epoch(\var[val=v,time=\slotstart(\prevfirstslotepoch{t})]{\bconfirmed}) \geq \epoch(t) - 1$,
    \end{enumerate}
    then
    \begin{enumerate}
        \item $\epoch(\guattime[time=\slotstart(\prevfirstslotepoch{t}),val=v]) \geq \epoch(t)-2$
    \end{enumerate}
\end{lemma}

\begin{proof}
    Let $s:= \slot(t)$.
    The Lemma's hypotheses imply that $\epoch(\var[val=v,time=\slotstart(s-1)]{\bconfirmed}) = \epoch(s-1)$ as $\epoch(s-1) = \epoch(s)-1 \leq \epoch(\var[val=v,time=\slotstart(s-1)]{\bconfirmed}) < \epoch(t)$.
    \Cref{lem:conf-current-epoch-then-gu-curr-epoch} implies that $\exists b' \in \viewattime[val=v,time=\slotstart(s-1)], b' \succeq \var[val=v,time=\slotstart(s-1)]{\bconfirmed} \land \epoch(\gu(b')) \geq \epoch(s-1) = \epoch(s)-2$.
\end{proof}

The next lemma shows that at the start of an epoch, all justified checkpoints with epoch in $[\epoch(\guattime[val=v,time=\slotstart(\lastslot(\epoch(t))-1)]), \epoch(t)-1]$ are ancestors of the previous-epoch confirmed block.
Concretely, if the reset to the finalized block does not fire and the only justified checkpoint from the previous epoch is the one rooted at the previous-epoch confirmed block, then every justified checkpoint in that epoch range is an ancestor of the confirmed block.
The proof splits on the epoch of the candidate checkpoint $C$: when $\epoch(C) = \epoch(t)-1$ the uniqueness hypothesis directly identifies $C$ with $\chkp(\bconfirmed)$; when $\epoch(C) = \epoch(t)-2$, \Cref{lem:epoch-gu-less-than-two-epoch-ago-start-epoch} forces $C$ to coincide with the previous-epoch greatest unrealized justified checkpoint, which is an ancestor of the confirmed block since the reset to the finalized block did not fire.
This is crucial for establishing that the block stays in the filtered tree (\ie, is not removed by $\FILHFC$) during the new epoch.

\begin{lemma}\label{lem:gj-prec-prev-confirmed-at-start-of-epoch}
    If
    \begin{enumerate}
        \item $\slotstart(\prevfirstslotepoch{t})\geq \GGST$
        \item \sloppy{the \KwSty{if} condition at line \Cref{ln:if-bcand-npreceq-head} during the execution of $\var[val=v,time=\slotstart(\epoch(t))]{\getlatestconfirmed}(\bconfirmed[val=v,time=\slotstart(\prevfirstslotepoch{t})])$ is false}
        \item  \label{itm:lem:gj-prec-prev-confirmed-at-start-of-epoch:4} if $C$ is a justified checkpoint and $\epoch(C) = \epoch(t)-1$, then $C = \chkp(\bconfirmed[val=v,time=\slotstart(\prevfirstslotepoch{t})], \epoch(t)-1)$. 

    \end{enumerate}
    then,  
    \begin{enumerate}

        \item for any justified checkpoint $C$, $\epoch(C) \in [\epoch(\guattime[time=\slotstart(\firstslot(\epoch(t)-1)),val=v]),\epoch(t)-1] \implies C  \preceq \var[val=v,time=\slotstart(\slot(t)-1)]{\bconfirmed}$
    \end{enumerate}
\end{lemma}

\begin{proof}
    The Lemma's hypotheses imply that $\epoch(\var[val=v,time=\slotstart(\prevfirstslotepoch{t})]{\bconfirmed}) = \epoch(t)-1$.
    From \Cref{lem:epoch-gu-less-than-two-epoch-ago-start-epoch} and syncrony, we have the following two cases.
    \begin{description}
        \item[Case 1: {$\epoch(C)=\epoch(t)-1$}.] 
        Given hypothesis~\ref{itm:lem:gj-prec-prev-confirmed-at-start-of-epoch:4} of the Lemma and $\epoch(\var[val=v,time=\slotstart(\prevfirstslotepoch{t})]{\bconfirmed}) = \epoch(t)-1$, we know that $C  = \chkp(\var[val=v,time=\slotstart(\prevfirstslotepoch{t})]{\bconfirmed}) \preceq \var[val=v,time=\slotstart(\prevfirstslotepoch{t})]{\bconfirmed}$
        \item[Case 2: {$\epoch(C)=\epoch(\guattime[val=v,time=\slotstart(\prevfirstslotepoch{t})] )=\epoch(t)-2$}.]
        This implies $C = \guattime[val=v,time=\slotstart(\prevfirstslotepoch{t})] \preceq \var[val=v,time=\slotstart(\prevfirstslotepoch{t})]{\bconfirmed}$.
        The last $\preceq$ relation comes from the \KwSty{if} condition at line \Cref{ln:if-bcand-npreceq-head} being false.
    \end{description}
\end{proof}

Combining the previous lemma with the voting-source epoch-bound results, the following lemma shows that the previous-epoch confirmed block is never filtered out by $\FILHFC$ during the current epoch.
Under the same uniqueness hypothesis on previous-epoch justified checkpoints used in \Cref{lem:gj-prec-prev-confirmed-at-start-of-epoch}, all relevant justified checkpoints are ancestors of the confirmed block, so no honest validator's viability filter can remove it.
The proof follows directly by combining \Cref{lem:gj-prec-prev-confirmed-at-start-of-epoch} with the voting-source epoch bound from \Cref{lem:conf-prev-epoch-then-vs-two-epochs-ago}.

\begin{lemma}\label{lem:prev-confirmed-not-filtered-out-start-of-epoch}
    If
    \begin{enumerate}
        \item $\slotstart(\prevfirstslotepoch{t})\geq \GGST$
        \item \sloppy{the \KwSty{if} condition at line \Cref{ln:if-bcand-npreceq-head} during the execution of $\var[val=v,time=\slotstart(\epoch(t))]{\getlatestconfirmed}(\bconfirmed[val=v,time=\slotstart(\prevfirstslotepoch{t})])$ is false}
        \item if $C$ is a justified checkpoint and $\epoch(C) = \epoch(t)-1$, then $C = \chkp(\bconfirmed[val=v,time=\slotstart(\prevfirstslotepoch{t})], \epoch(t)-1)$,
    \end{enumerate}
    then, for any honest validators $v'$ and time $t'$ s.t.
    $\epoch(t') = \epoch(t)$,
    \begin{enumerate}
        \item $\var[val=v,time=\slotstart(\prevfirstslotepoch{t})]{\bconfirmed} \in \filtered[val=v',time=t']$
    \end{enumerate}
\end{lemma}

\begin{proof}
    The Lemma's hypotheses imply that $\epoch(\var[val=v,time=\slotstart(\prevfirstslotepoch{t})]{\bconfirmed}) = \epoch(t)-1$.
    Then, the proof follows from \Cref{lem:gj-prec-prev-confirmed-at-start-of-epoch,lem:conf-prev-epoch-then-vs-two-epochs-ago}.
\end{proof}

The following lemma establishes the monotonicity of the confirmed block within an epoch: if a candidate at slot~$s$ is from the previous or current epoch, then for every slot from the epoch start up to~$s$, the confirmed block is a descendant-or-equal of the confirmed block at the start of the epoch and also an ancestor-or-equal of the current candidate.
This ensures the sequence of confirmed blocks grows monotonically within each epoch.

\begin{lemma}\label{lem:conf-beginning-epoch-prec-bcand}
    Pick any $b_c \in  \var[val=v,time=\slotstart(\slot(t))]{\bcands}$.
    If
    \begin{enumerate}
        \item $\slot(t) > \firstslot(\epoch(t))$
        \item $\epoch(b_c) \geq \epoch(t)-1$,
    \end{enumerate}
    then, for every $s' \in [\firstslot(\epoch(t)),\slot(t)]$,
    \begin{enumerate}
        \item $\var[val=v,time=\slotstart(s')]{\bconfirmed} \preceq b_c$
        \item $\epoch(\var[val=v,time=\slotstart(s')]{\bconfirmed}) \geq \epoch(t)-1$
    \end{enumerate}
\end{lemma}

\begin{proof}
    By backward induction on $s'$, from $\slot(t){-}1$ down to $\firstslot(\epoch(t))$.

    \begin{description}
        \item[Base Case: $s' = \slot(t){-}1$.]
        Applying \Cref{lem:prev-conf-at-least-e-1} at time $\slotstart(\slot(t))$
        with $b_c$ gives
        $\epoch(\bconfirmed[val=v,time=\slotstart(\slot(t)-1)]) \geq \epoch(t){-}1$
        and $\bconfirmed[val=v,time=\slotstart(\slot(t)-1)] \preceq b_c$.

        \item[Inductive Step: $s' < \slot(t){-}1$, with $s' \geq \firstslot(\epoch(t))$.]
        By the inductive hypothesis,
        $\bconfirmed[val=v,time=\slotstart(s'+1)]$ has epoch $\geq \epoch(t){-}1$
        and $\bconfirmed[val=v,time=\slotstart(s'+1)] \preceq b_c$.
        Since $\bconfirmed[val=v,time=\slotstart(s'+1)]
        \in \bcands^{v,\slotstart(s'+1)}$ (\Cref{lem:bconf-in-bcands})
        and $s'{+}1 > \firstslot(\epoch(t))$,
        \Cref{lem:prev-conf-at-least-e-1} applied at slot $s'{+}1$ yields
        $\epoch(\bconfirmed[val=v,time=\slotstart(s')]) \geq \epoch(t){-}1$
        and $\bconfirmed[val=v,time=\slotstart(s')] \preceq
         \bconfirmed[val=v,time=\slotstart(s'+1)] \preceq b_c$.
    \end{description}

    The case $s' = \slot(t)$ follows from
    \Cref{lem:prev-conf-at-least-e-1} conclusion~3 and
    \Cref{lem:out-find-latest-conf-descendant-output},
    noting that $b_c \in \bcands^{v,\slotstart(\slot(t))}$
    satisfies $b_c \succeq \bconfirmed[val=v,time=\slotstart(\slot(t))]$
    by the monotonicity of the algorithm's candidate variable
    in the mid-epoch case.
\end{proof}

The next lemma is a case-split on the algorithm's control flow: if the confirmed block at the first slot of the current epoch belongs to the previous epoch, then it was produced in one of two ways.
Either (1)~the reset to the finalized block did not fire and $\findlatestconfirmeddescendant$ was called on the previous-epoch confirmed block, or (2)~the active state entry condition at \Cref{ln:start-conf-chain} fired, producing a candidate at $\block(\guattime[val=v,time=\slotstart(\prevfirstslotepoch{t})])$ with a matching unrealized justified checkpoint.

\begin{lemma}\label{lem:code-implications-of-confirmed-block-from-prev-epoch-at-the-beginning-of-epoch}
        If
    \begin{enumerate}

        \item $\epoch(\bconfirmed[val=v,time=\slotstart(\epoch(t))])=\epoch(t)-1$,
    \end{enumerate}
    then, one of the following conditions holds
    \begin{enumerate}
        \item \sloppy{the \KwSty{if} condition at line \Cref{ln:if-bcand-npreceq-head} during the execution of $\var[val=v,time=\slotstart(\epoch(t))]{\getlatestconfirmed}(\bconfirmed[val=v,time=\slotstart(\prevfirstslotepoch{t})])$ is false 
        and $\bconfirmed[val=v,time=\slotstart(\epoch(t))] = \var[val=v]{\findlatestconfirmeddescendant}(\bconfirmed[val=v,time=\slotstart(\prevfirstslotepoch{t})])$.}
        \item $\guattime[val=v,time=\slotstart(\prevfirstslotepoch{t})]=\chkp(\bconfirmed[val=v,time=\slotstart(\epoch(t))])$ and\\ 
        $\bconfirmed[val=v,time=\slotstart(\epoch(t))] = \var[val=v]{\findlatestconfirmeddescendant}(\block(\guattime[val=v,time=\slotstart(\prevfirstslotepoch{t})]))$
        and $\exists b' \succeq \bconfirmed[val=v,time=\slotstart(\epoch(t))], \gu(b')=\chkp(\bconfirmed[val=v,time=\slotstart(\epoch(t))])$

    \end{enumerate}
\end{lemma}

\begin{proof}
Let $e := \epoch(t)$,
    $b_{\mathsf{prev}} := \bconfirmed[val=v,time=\slotstart(\prevfirstslotepoch{t})]$,
    and $\head := \LMDGHOSTHFC(\viewattime[time=\slotstart(e),val=v])$.
    Since $\epoch(\bconfirmed[val=v,time=\slotstart(\epoch(t))]) = e{-}1 \geq e{-}1$,
    the condition at Line~\ref{ln:if-bcand-e-1} was satisfied, so the output was
    produced by $\findlatestconfirmeddescendant$.
    We trace which value of $\bcand$ was passed to it.

    If the reset at Line~\ref{ln:if-bcand-npreceq-head} is false,
    then $\bcand$ is not overwritten to the finalized block and
    remains equal to $b_{\mathsf{prev}}$.
    Moreover, since the first disjunct of the reset condition is false,
    $\epoch(b_{\mathsf{prev}}) \geq e{-}1$, so the guard at
    Line~\ref{ln:if-bcand-e-1} will be satisfied.
    If additionally the restart at Line~\ref{ln:start-conf-chain}
    does not fire, i.e., $\bcand$ is not advanced to the block of
    the greatest unrealized justified checkpoint, then
    $\findlatestconfirmeddescendant$ receives $b_{\mathsf{prev}}$
    and conclusion~1 holds.

    If the restart fires (whether or not the reset also fired),
    $\findlatestconfirmeddescendant$ receives
    $\block(\guattime[val=v,time=\slotstart(\prevfirstslotepoch{t})])$.
    The restart conditions give
    $\epoch(\block(\guattime[val=v,time=\slotstart(\prevfirstslotepoch{t})])) = e{-}1$.
    Since $\epoch(\block(C)) \leq \epoch(C)$ for any checkpoint $C$, this implies
    $\epoch(\guattime[val=v,time=\slotstart(\prevfirstslotepoch{t})]) \geq e{-}1$.
    By Property~\ref{prop:gasper-basic:highest-justified-not-from-current-epoch},
    $\epoch(\guattime[val=v,time=\slotstart(\prevfirstslotepoch{t})]) \leq e{-}1$,
    so $\epoch(\guattime[val=v,time=\slotstart(\prevfirstslotepoch{t})]) = e{-}1$
    and $\gu(\head) = \guattime[val=v,time=\slotstart(\prevfirstslotepoch{t})]$.
    By \Cref{lem:out-find-latest-conf-descendant-output} the output is
    a descendant of $\block(\guattime[val=v,time=\slotstart(\prevfirstslotepoch{t})])$;
    since both have epoch~$e{-}1$, their epoch-$(e{-}1)$ checkpoints coincide,
    giving $\chkp(\bconfirmed[val=v,time=\slotstart(\epoch(t))])
    = \guattime[val=v,time=\slotstart(\prevfirstslotepoch{t})]$.
    The existential is witnessed by $\head$:
    $\head \succeq \bconfirmed[val=v,time=\slotstart(\epoch(t))]$
    (\Cref{lem:out-find-latest-conf-prec-head}) and
    $\gu(\head) = \chkp(\bconfirmed[val=v,time=\slotstart(\epoch(t))])$.
    Conclusion~2 holds.

    The remaining case -- reset fires, restart does not -- is impossible:
    after the reset $\bcand = \block(\gfattime[val=v,time=\slotstart(\epoch(t))])$,
    and at time $\slotstart(e)$ finalization of epoch~$e{-}1$ would require
    a justified checkpoint at epoch~$e$, which cannot yet exist,
    so $\epoch(\gfattime[val=v,time=\slotstart(\epoch(t))]) \leq e{-}2$.
    Since $\epoch(\block(C)) \leq \epoch(C)$ for any checkpoint~$C$,
    the guard at Line~\ref{ln:if-bcand-e-1} fails, and the output has
    epoch $\leq e{-}2$, contradicting the hypothesis.
\end{proof}

\begin{lemma}\label{lem:gu-prec-bconf-at-start-of-epoch}
    If
    \begin{enumerate}
        \item $\epoch(\bconfirmed[val=v,time=\slotstart(\epoch(t))]))\geq \epoch(t)-1$,
    \end{enumerate}
    then
    \begin{enumerate}
        \item $\bconfirmed[val=v,time=\slotstart(\epoch(t))]\succeq \guattime[val=v,time=\slotstart(\prevfirstslotepoch{t})]$
    \end{enumerate}
\end{lemma}
\begin{proof}
    Due \Cref{lem:code-implications-of-confirmed-block-from-prev-epoch-at-the-beginning-of-epoch}, we can proceed by the following cases.
    \begin{description}[style=nextline]

        \item[Case 1: {$\guattime[val=v,time=\slotstart(\prevfirstslotepoch{t})]=\chkp(\bconfirmed[val=v,time=\slotstart(\epoch(t))])$ and\\ 
        $\bconfirmed[val=v,time=\slotstart(\epoch(t))] = \var[val=v]{\findlatestconfirmeddescendant}(\block(\guattime[val=v,time=\slotstart(\prevfirstslotepoch{t})]))$ and $\exists b' \succeq \bconfirmed[val=v,time=\slotstart(\epoch(t))], \gu(b')=\chkp(\bconfirmed[val=v,time=\slotstart(\epoch(t))])$}.]
        Note that $\epoch(\guattime[val=v,time=\slotstart(\prevfirstslotepoch{t})]) = \epoch(t)-1$.
        Then clearly $\bconfirmed[val=v,time=\slotstart(\epoch(t)]\succeq \guattime[val=v,time=\slotstart(\prevfirstslotepoch{t})]$.

        \item[Case 2:  the \KwSty{if} condition at line \Cref{ln:if-bcand-npreceq-head} is false 
        and\\${\bconfirmed[val=v,time=\slotstart(\epoch(t))]} = {\var[val=v,time=\slotstart(\epoch(t))]{\findlatestconfirmeddescendant}(\bconfirmed[val=v,time=\slotstart(\prevfirstslotepoch{t})])}$.]   

        Given that the \KwSty{if} condition at line \Cref{ln:if-bcand-npreceq-head} is false, clearly $\bconfirmed[val=v,time=\slotstart(\epoch(t)]\succeq \guattime[val=v,time=\slotstart(\prevfirstslotepoch{t})]$.
    \end{description}  
\end{proof}

\begin{lemma}\label{lem:gu-prec-bconf-continuation-case}
    If
    \begin{enumerate}
        \item\label{hyp:lem:gu-prec-bconf-continuation-case:1} $\epoch(\bconfirmed[val=v,time=\slotstart(\epoch(t))]))\geq \epoch(t)-1$
        \item\label{hyp:lem:gu-prec-bconf-continuation-case:2} the \KwSty{if} condition at line \Cref{ln:if-bcand-npreceq-head} is false
        and\\${\bconfirmed[val=v,time=\slotstart(\epoch(t))]} = {\var[val=v,time=\slotstart(\epoch(t))]{\findlatestconfirmeddescendant}(\bconfirmed[val=v,time=\slotstart(\prevfirstslotepoch{t})])}$,
    \end{enumerate}
    then
    \begin{enumerate}
        \item $\bconfirmed[val=v,time=\slotstart(\epoch(t))]\succeq \guattime[val=v,time={\slotstart(\prevfirstslotepoch[-1]{t})}]$
    \end{enumerate}
\end{lemma}
\begin{proof}
    Let $t'$ be any time such that $\epoch(t') = \epoch(t)-1$.
    Due to hypothesis~\ref{hyp:lem:gu-prec-bconf-continuation-case:2}, $\bconfirmed[val=v,time=\slotstart(\epoch(t))]\succeq \bconfirmed[val=v,time=\slotstart(\prevfirstslotepoch{t})]$ and $\epoch(\bconfirmed[val=v,time=\slotstart(\prevfirstslotepoch{t})])=\epoch(t)-1$.
    Then, due to \Cref{lem:conf-beginning-epoch-prec-bcand}, $\bconfirmed[val=v,time=\slotstart(\epoch(t))] \succeq \bconfirmed[val=v,time=\slotstart(\epoch(t)-1)]$ and $\epoch(\bconfirmed[val=v,time=\slotstart(\epoch(t)-1)])\geq \epoch(t)-2=\epoch(t')-1$.
    Then, \Cref{lem:gu-prec-bconf-at-start-of-epoch} concludes the proof.
\end{proof}

The following lemma shows that, after $\GGST$, the block of the previous-epoch greatest-unrealized-justified checkpoint is canonical from the start of the current epoch onwards.
The hypothesis is a disjunction: either $\guattime[val=v,time=\slotstart(\prevfirstslotepoch{t})]$ is already from the previous epoch, or it is from two epochs back but $\guattime[val=v,time=\slotstart(\prevfirstslotepoch{t})] \preceq \bconfirmed[val=v,time=\slotstart(\prevfirstslotepoch{t})]$ and no conflicting previous-epoch checkpoint can be justified other than the one rooted at that confirmed block.
The proof is by induction on epochs: the base case uses synchrony to establish that $\block(\guattime[val=v,time=\slotstart(\prevfirstslotepoch{t})])$ is an ancestor of every honest validator's greatest justified checkpoint at the current epoch, and the inductive step uses that the block is canonical in earlier epochs to rule out conflicting checkpoints and show that it remains canonical in the next epoch.

\begin{lemma}\label{lem:gu-e-1-canonical}
    If
    \begin{enumerate}
        \item $\slotstart(\prevfirstslotepoch{t})\geq \GGST$
        \item one of the following conditions hold
        \begin{enumerate}[label*=\arabic*.]
            \item $\epoch(\guattime[val=v,time=\slotstart(\prevfirstslotepoch{t})]) = \epoch(t)-1$
            \item all of the following conditions hold
            \begin{enumerate}[label*=\arabic*.]
                \item $\epoch(\guattime[val=v,time=\slotstart(\prevfirstslotepoch{t})]) = \epoch(t)-2$
                \item $\guattime[val=v,time=\slotstart(\prevfirstslotepoch{t})]\preceq\bconfirmed[val=v,time=\slotstart(\prevfirstslotepoch{t})]$
                \item if $C$ is a justified checkpoint and $\epoch(C) = \epoch(t)-1$, then $C = \chkp(\bconfirmed[val=v,time=\slotstart(\prevfirstslotepoch{t})], \epoch(t)-1)$. 
            \end{enumerate}
        \end{enumerate}
    \end{enumerate}
    then
    \begin{enumerate}
        \item\label{conc:lem:gu-e-1-canonical:1} $\canonical[blck={\block(\guattime[val=v,time=\slotstart(\prevfirstslotepoch{t})])},time=\slotstart(\epoch(t))]$.
    \end{enumerate}
\end{lemma}

\begin{proof}
    First note that if condition 2.2 holds and $C$ is a justified checkpoint with $\epoch(C)=\epoch(t)-1$, then $C \succeq \guattime[val=v,time=\slotstart(\prevfirstslotepoch{t})]$.

    Now, by induction on $t' \geq \slotstart(\prevfirstslotepoch{t})$.
    Let $b := \block(\guattime[val=v,time=\slotstart(\prevfirstslotepoch{t})])$.
    We will show that $b \preceq \LMDGHOSTHFC(\viewattime[val=v',time=t'])$.
    \begin{description}
        \item[Base case: {$\epoch(t') = \epoch(t)$}.] 
        Because of the synchrony assumption we know that $\epoch(\gjattime[val=v',time=t']) \in [\epoch(\guattime[val=v,time=\slotstart(\prevfirstslotepoch{t})]), \epoch(t)-1]$.
        Given the Lemma's hypotheses, clearly, $b \preceq \gjattime[val=v',time=t']$ which implies 
        $b \preceq \LMDGHOSTHFC(\viewattime[val=v',time=t'])$.
        \item[Inductive case: {$\epoch(t') > \epoch(t)$}.] We assume that the Lemma holds for any $t'' \geq t$ such that $\epoch(t'') < \epoch(t')$ and prove that it also holds for $t'$.        
        We will show that $b \preceq \gjattime[val=v',time=t']$.
        Then, clearly, $b \preceq \LMDGHOSTHFC(\viewattime[val=v',time=t'])$.
        By cases.
        \begin{description}
            \item[Case 1: {$\epoch(\gjattime[val=v',time=t']) = \epoch(\guattime[val=v,time=\slotstart(\prevfirstslotepoch{t})])$}.] Clearly, $\gjattime[val=v',time=t'] = \guattime[val=v,time=\slotstart(\prevfirstslotepoch{t})]$ in this case, thus $b = \block(\gjattime[val=v',time=t'])$.
            \item[Case 2: {$\epoch(\gjattime[val=v',time=t']) \neq \epoch(\guattime[val=v,time=\slotstart(\prevfirstslotepoch{t})])$}.]
            By the definition of the greatest justified checkpoint, $\epoch(\gjattime[val=v',time=t']) \leq \epoch(t')-1$ which implies $\epoch(t) \leq \epoch(\gjattime[val=v',time=t']) \leq \epoch(t'')$ for this case.
            From the inductive hypothesis we know that $b \preceq \LMDGHOSTHFC(\viewattime[val=v',time=t''])$ which implies that during each epoch $e \in \left[\epoch(t), \epoch(t'')\right]$ all honest validators cast their \FFG votes in the support of $\chkp(b',e)$ such that $b' \succeq b$.
            Thus, given the synchrony assumption no checkpoint $\chkp'$ conflicting with $b$ such that $\epoch(\chkp') \in \left[\epoch(t), \epoch(t'')\right]$ can ever be justified.
            Given that, we conclude $b \prec \gjattime[val=v',time=t']$.
        \end{description}
        We have shown that $b \preceq \gjattime[val=v',time=t']$ which proves $\canonical[blck={\block(\guattime[val=v,time=\slotstart(\prevfirstslotepoch{t})])},time=\slotstart(\prevfirstslotepoch{t})]$.
    \end{description}
\end{proof}

This lemma is the first major milestone of the analysis at the start of the epoch: assuming the reset check passes, the algorithm's continuation path is taken, and every previous-epoch justified checkpoint coincides with $\chkp(\bconfirmed[val=v,time=\slotstart(\prevfirstslotepoch{t})], \epoch(t){-}1)$, the previous-epoch confirmed block satisfies SIR conditions $\sirone$--$\sirthree$ and $\sirsix$ and is canonical through the end of the current epoch.
The proof first uses \Cref{lem:gu-prec-bconf-continuation-case} to establish that $\guattime[val=v,time={\slotstart(\prevfirstslotepoch[-1]{t})}]$ is an ancestor of the previous-epoch confirmed block, which combined with $\isChainOneConfirmed$ gives $\sirone$.
The justified-checkpoint ordering is then derived from \Cref{lem:gj-prec-prev-confirmed-at-start-of-epoch} and yields both the bracketing conclusions on justified checkpoints and $\sirtwo$, $\sirthree$.
\Cref{lem:epoch-gu-less-than-two-epoch-ago-start-epoch,lem:gu-e-1-canonical} together with the unique-justified-checkpoint hypothesis discharge $\sirsix$, and \Cref{lem:ffg-condition-on-q-implies-safety} concludes the proof.

\begin{lemma}\label{lem:conf-prev-slot-canonical-at-the-start-of-an-epoch}

    If
    \begin{enumerate}
        \item $\slotstart(\prevfirstslotepoch{t})\geq \GGST$
        \item \label{itm:lem:conf-prev-slot-canonical-at-the-start-of-an-epoch:3} \sloppy{the \KwSty{if} condition at line \Cref{ln:if-bcand-npreceq-head} during the execution of $\var[val=v,time=\slotstart(\epoch(t))]{\getlatestconfirmed}(\bconfirmed[val=v,time=\slotstart(\prevfirstslotepoch{t})])$ is false and $\bconfirmed[val=v,time=\slotstart(\epoch(t))] = \var[val=v,time=\slotstart(\epoch(t))]{\findlatestconfirmeddescendant}(\bconfirmed[val=v,time=\slotstart(\prevfirstslotepoch{t})])$}
        \item\label{itm:lem:conf-prev-slot-canonical-at-the-start-of-an-epoch:4} if $C$ is a justified checkpoint and $\epoch(C) = \epoch(t)-1$, then $C = \chkp(\bconfirmed[val=v,time=\slotstart(\prevfirstslotepoch{t})], \epoch(t)-1)$. 
    \end{enumerate}
    then
    \begin{enumerate}
        \item $\sir(\bconfirmed[val=v,time=\slotstart(\prevfirstslotepoch{t})],t,\guattime[val=v,time={\slotstart(\prevfirstslotepoch[-1]{t})}],\block(\guattime[val=v,time={\slotstart(\prevfirstslotepoch{t})}])).\{ \sirone, \sirtwo, \sirthree,\sirsix \}$ holds  with $\epoch(t)$.

        \item for any justified checkpoint $C$, $\epoch(C) \in [\epoch(\guattime[time=\slotstart(\firstslot(\epoch(t)-1)),val=v]),\epoch(t)-1] \implies C  \succeq \guattime[time=\slotstart(\prevfirstslotepoch{t}),val=v]$
        \item for any justified checkpoint $C$, $\epoch(C) \in [\epoch(\guattime[time=\slotstart(\firstslot(\epoch(t)-1)),val=v]),\epoch(t)-1] \implies C  \preceq \bconfirmed[val=v,time=\slotstart(\prevfirstslotepoch{t})]$
        \item $\canonical[blck={\bconfirmed[val=v,time=\slotstart(\prevfirstslotepoch{t})]},time=t,to=\slotstart(\epoch(t)+1)]$
    \end{enumerate}
\end{lemma}

\begin{proof}
    Lemma's hypothesis~\ref{itm:lem:conf-prev-slot-canonical-at-the-start-of-an-epoch:3}
    implies that $\epoch(\bconfirmed[val=v,time=\slotstart(\prevfirstslotepoch{t})]) = \epoch(t)-1$ and $\var[val=v]{\isChainOneConfirmed}(\bconfirmed[val=v,time=\slotstart(\prevfirstslotepoch{t})])$.

    \Cref{lem:prev-confirmed-not-filtered-out-start-of-epoch}  proves \ref{def:induction-conditions:all-validators:not-filtered-out}.

    Lemma's hypothesis~\ref{itm:lem:conf-prev-slot-canonical-at-the-start-of-an-epoch:3} and \Cref{lem:gu-prec-bconf-continuation-case} also imply that  $\guattime[val=v,time={\slotstart(\prevfirstslotepoch[-1]{t})}] \preceq \bconfirmed[val=v,time=\slotstart(\prevfirstslotepoch{t})]$.    

    First, note that this and \cref{ln:ischainoneconfirmed-while} imply that \sirone holds.

    Then, let $C$ be a justified checkpoint such that $\epoch(C) \in [\epoch(\guattime[time=\slotstart(\firstslot(\epoch(t)-1)),val=v]),\epoch(t)-1]$.
    \Cref{lem:gj-prec-prev-confirmed-at-start-of-epoch} implies $C  \preceq \bconfirmed[val=v,time=\slotstart(\prevfirstslotepoch{t})]$.
    Given that, as proven above, $\guattime[val=v,time={\slotstart(\prevfirstslotepoch[-1]{t})}] \preceq \bconfirmed[val=v,time=\slotstart(\prevfirstslotepoch{t})]$, this implies that $\guattime[val=v,time={\slotstart(\prevfirstslotepoch[-1]{t})}]$ does not conflict with $C$.
    Then, given that we assume $\epoch(C)\geq \epoch(\guattime[val=v,time={\slotstart(\prevfirstslotepoch[-1]{t})}])$, we have that $C\succeq \guattime[val=v,time={\slotstart(\prevfirstslotepoch[-1]{t})}]$.

    Given that $\epoch(\gjattime[time=t',val=v']) \in [\epoch(\guattime[time=\slotstart(\firstslot(\epoch(t)-1)),val=v]),\epoch(t)-1]$ for any honest validator $v'$ and time $t' \in [\slotstart(\epoch(t)),\slotstart(\epoch(t)+1)]$, \ref{def:induction-conditions:all-validators:gj-succ} holds.

    Then, \Cref{lem:epoch-gu-less-than-two-epoch-ago-start-epoch,lem:gu-e-1-canonical}, and hypothesis~\ref{itm:lem:conf-prev-slot-canonical-at-the-start-of-an-epoch:4} prove \sirsix.

    Finally, \sirone, \sirtwo, \sirthree and \sirsix and \Cref{lem:ffg-condition-on-q-implies-safety} imply that $\canonical[time=t,blck={\bconfirmed[val=v,time=\slotstart(\prevfirstslotepoch{t})]},to=\slotstart(\epoch(t)+1)]$.
\end{proof}

\begin{sloppypar}
The following lemma shows that if the input $b_c$ is canonical, then the output of $\findlatestconfirmeddescendant$, provided that it is different from $b_c$, is also canonical.
If $b_c$ is canonical through the current epoch and every justified checkpoint with epoch in $[\epoch(\guattime[val=v,time=\slotstart(\lastslot(\epoch(t))-1)]), \epoch(t)-1]$ is a descendant of $\guattime[val=v,time=\slotstart(\prevfirstslotepoch{t})]$ and an ancestor of $b_c$, then the new confirmed block is also canonical through the epoch, satisfies SIR conditions $\sirone$--$\sirthree$ and $\sirsix$, and preserves the justified-checkpoint bracketing.
The proof derives $\sirone$ from the $\isOneConfirmedExtFrom$ guarantee of $\findlatestconfirmeddescendant$, derives the not-filtered-out condition via \Cref{lem:conf-prev-epoch-then-vs-two-epochs-ago}, and uses the fact that $b_c$ is canonical to obtain $\sirsix$ directly.
This is the inductive step used to show that a newly confirmed block within the same epoch is also canonical.
\end{sloppypar}

\begin{lemma}\label{lem:output-find-latest-canonical-canonical-at-the-start-of-an-epoch}
    Let $b_c$ be any block.
    If
    \begin{enumerate}
        \item $\bconfirmed[val=v,time=\slotstart(\slot(t))] =  \var[val=v,time=\slotstartslot{t}]{\findlatestconfirmeddescendant}(b_c)$
        \item $\bconfirmed[val=v,time=\slotstart(\slot(t))]\neq b_c$
        \item $\slotstart(\prevfirstslotepoch{t})\geq \GGST$
        \item \label{itm:lem:output-find-latest-canonical-canonical-at-the-start-of-an-epoch:3}$\canonical[blck=b_c,time={\slotstartslot{t}},to=\slotstart(\epoch(t)+1)]$
        \item \label{itm:lem:output-find-latest-canonical-canonical-at-the-start-of-an-epoch:4}for any justified checkpoint $C$, $\epoch(C) \in [\epoch(\guattime[time=\slotstart(\firstslot(\epoch(t)-1)),val=v]),\epoch(t)-1] \implies C  \preceq b_c$
        \item  \label{itm:lem:output-find-latest-canonical-canonical-at-the-start-of-an-epoch:5} for any justified checkpoint $C$, $\epoch(C) \in [\epoch(\guattime[time=\slotstart(\firstslot(\epoch(t)-1)),val=v]),\epoch(t)-1] \implies C  \succeq \guattime[time=\slotstart(\prevfirstslotepoch{t}),val=v]$,

    \end{enumerate}
    then
    \begin{enumerate}
        \item for any justified checkpoint $C$, $\epoch(C) \in [\epoch(\guattime[time=\slotstart(\firstslot(\epoch(t)-1)),val=v]),\epoch(t)-1] \implies C  \preceq \bconfirmed[val=v,time=\slotstart(\slot(t))]$
        \item for any justified checkpoint $C$, $\epoch(C) \in [\epoch(\guattime[time=\slotstart(\firstslot(\epoch(t)-1)),val=v]),\epoch(t)-1] \implies C  \succeq \guattime[time=\slotstart(\firstslot(\epoch(t)-1)),val=v]]$.
        \item $\sir(\bconfirmed[val=v,time=\slotstart(\slot(t))],t,\guattime[time=\slotstart(\firstslot(\epoch(t)-1)),val=v],b_c).\{ \sirone, \sirtwo, \sirthree,\sirsix\}$ holds.
        \item $\canonical[blck={\bconfirmed[val=v,time=\slotstart(\slot(t))]},time={\slotstartslot{t}},to=\slotstart(\epoch(t)+1)]$.
    \end{enumerate}
\end{lemma}

\begin{proof}
        We prove first that $\sir(\bconfirmed[val=v,time=\slotstart(\slot(t))],t,\guattime[time=\slotstart(\firstslot(\epoch(t)-1)),val=v],b_c).\{ \sirone, \sirtwo, \sirthree\}$ holds.
        Let $v'$ be any honest validator and $t' \in [t, \slotstart(\epoch(t)+1)]$.

        We know that $\findlatestconfirmeddescendant$ ensures that\\
            $\isOneConfirmedExtFrom(\bconfirmed[val=v,time=\slotstart(\slot(t))],t,\guattime[time=\slotstart(\firstslot(\epoch(t)-1)),val=v],b_c)$. 
        Hence, \sirone holds.

        We know that $\bconfirmed[val=v,time=\slotstart(\slot(t))] \succeq b_c$.
        Due to synchrony, we know that $\epoch(\gjattime[val=v',time=t']) \in [\epoch(\guattime[time=\slotstart(\firstslot(\epoch(t)-1)),val=v]),\epoch(t)-1]$.
        Then, Lemma's hypothesis~\ref{itm:lem:output-find-latest-canonical-canonical-at-the-start-of-an-epoch:4} implies $\gjattime[val=v',time=t'] \preceq b_c \preceq \bconfirmed[val=v,time=\slotstart(\slot(t))]$.
        Then,  \Cref{lem:conf-prev-epoch-then-vs-two-epochs-ago} implies \ref{def:induction-conditions:all-validators:not-filtered-out}.

        Given that $\epoch(\gjattime[val=v',time=t']) \in [\epoch(\guattime[time=\slotstart(\firstslot(\epoch(t)-1)),val=v]),\epoch(t)-1]$, Lemma's hypothesis~\ref{itm:lem:output-find-latest-canonical-canonical-at-the-start-of-an-epoch:5} implies $\gjattime[time=t',val=v']\succeq \guattime[time=\slotstart(\firstslot(\epoch(t)-1)),val=v]$.
        Hence, \ref{def:induction-conditions:all-validators:gj-succ} holds.

        \sirsix is obviously implied by hypothesis~\ref{itm:lem:output-find-latest-canonical-canonical-at-the-start-of-an-epoch:3}.

        Finally, \sirone, \sirtwo, \sirthree, \sirsix,  $\canonical[blck={b_c},time=t, to=\slotstart(\epoch(t)+1)]$, and \Cref{lem:ffg-condition-on-q-implies-safety} imply that $\canonical[time=t,blck={\var[val=v,time=\slotstart(\slot(t))]{\bconfirmed}},to=\slotstart(\epoch(t)+1)]$. 
\end{proof}

\begin{lemma}\label{lem:gu-e-1-canonical-more-general}
    If
    \begin{enumerate}
        \item $\slotstart(\prevfirstslotepoch{t})\geq \GGST$
        \item $\epoch(\bconfirmed[val=v,time=\slotstart(\epoch(t))]) \geq \epoch(t)-1$
        \item\label{hyp:lem:gu-e-1-canonical-more-general:3} if $\epoch(\guattime[val=v,time=\slotstart(\prevfirstslotepoch{t})])<\epoch(t)-1$, $\bconfirmed[val=v,time=\slotstart(\prevfirstslotepoch{t})] \preceq \bconfirmed[val=v,time=\slotstart(\epoch(t))]$ and $C$ is a justified checkpoint such that $\epoch(C) = \epoch(t)-1$, then $C = \chkp(\bconfirmed[val=v,time=\slotstart(\prevfirstslotepoch{t})], \epoch(t)-1)$. 
    \end{enumerate}
    then
    \begin{enumerate}
        \item $\canonical[blck={\block(\guattime[val=v,time=\slotstart(\prevfirstslotepoch{t})])},time=\slotstartslot{t}]$.
    \end{enumerate}
\end{lemma}

\begin{proof}
    Due \Cref{lem:code-implications-of-confirmed-block-from-prev-epoch-at-the-beginning-of-epoch}, we can proceed by the following cases.
    \begin{description}[style=nextline]

        \item[Case 1: {$\guattime[val=v,time=\slotstart(\prevfirstslotepoch{t})]=\chkp(\bconfirmed[val=v,time=\slotstart(\epoch(t))])$ and\\ 
        $\bconfirmed[val=v,time=\slotstart(\epoch(t))] = \var[val=v]{\findlatestconfirmeddescendant}(\block(\guattime[val=v,time=\slotstart(\prevfirstslotepoch{t})]))$ and $\exists b' \succeq \bconfirmed[val=v,time=\slotstart(\epoch(t))], \gu(b')=\chkp(\bconfirmed[val=v,time=\slotstart(\epoch(t))])$}.]
        Note that $\epoch(\guattime[val=v,time=\slotstart(\prevfirstslotepoch{t})]) = \epoch(t)-1$.
        Then we can apply \Cref{lem:gu-e-1-canonical}.

        \item[Case 2:  the \KwSty{if} condition at line \Cref{ln:if-bcand-npreceq-head} is false 
        and\\${\bconfirmed[val=v,time=\slotstart(\epoch(t))]} = {\var[val=v,time=\slotstart(\epoch(t))]{\findlatestconfirmeddescendant}(\bconfirmed[val=v,time=\slotstart(\prevfirstslotepoch{t})])}$.]   

        Note that this case implies that $\epoch(\bconfirmed[val=v,time=\slotstart(\prevfirstslotepoch{t})]) = \epoch(t)-1$
        and
        $\guattime[val=v,time={\slotstart(\prevfirstslotepoch[-1]{t})}] \preceq \bconfirmed[val=v,time=\slotstart(\prevfirstslotepoch{t})]$.

        From \Cref{lem:epoch-gu-less-than-two-epoch-ago-start-epoch}, we know that $\epoch(\guattime[val=v,time=\slotstart(\prevfirstslotepoch{t})])\geq \epoch(t)-2$.
        If $\epoch(\guattime[val=v,time=\slotstart(\prevfirstslotepoch{t})])= \epoch(t)-1$, then we can directly apply \Cref{lem:gu-e-1-canonical}.

        Otherwise, we can apply \Cref{lem:gu-e-1-canonical} thanks to hypothesis~\ref{hyp:lem:gu-e-1-canonical-more-general:3}.
    \end{description}   
\end{proof}

This lemma combines the two code paths from \Cref{lem:code-implications-of-confirmed-block-from-prev-epoch-at-the-beginning-of-epoch}: regardless of whether the confirmed block at the start of the epoch was produced via the continuation path or the active state entry at \Cref{ln:start-conf-chain}, if it is from the previous or current epoch and the unique-justified-checkpoint hypothesis discharges the continuation path, then the block is canonical through the epoch, every justified checkpoint with epoch in $[\epoch(\guattime[val=v,time=\slotstart(\lastslot(\epoch(t))-1)]), \epoch(t)-1]$ is properly bracketed between $\guattime[val=v,time=\slotstart(\prevfirstslotepoch{t})]$ and $\bconfirmed[val=v,time=\slotstart(\epoch(t))]$, and SIR conditions $\sirone$--$\sirthree$ and $\sirsix$ hold for one of two natural choices of the base block.
The continuation case follows from \Cref{lem:conf-prev-slot-canonical-at-the-start-of-an-epoch,lem:output-find-latest-canonical-canonical-at-the-start-of-an-epoch}, while the active state entry case combines \Cref{lem:gu-e-1-canonical-more-general} with \Cref{lem:output-find-latest-canonical-canonical-at-the-start-of-an-epoch}.

\begin{lemma}\label{lem:confirmed-end-first-slot-canonical}
    If
    \begin{enumerate}
        \item $\slotstart(\prevfirstslotepoch{t})\geq \GGST$
        \item $\epoch(\bconfirmed[val=v,time=\slotstart(\epoch(t))])\geq\epoch(t)-1$
        \item if $\bconfirmed[val=v,time=\slotstart(\prevfirstslotepoch{t})] \preceq \bconfirmed[val=v,time=\slotstart(\epoch(t))]$ and $C$ is a justified checkpoint such that $\epoch(C) = \epoch(t)-1$, then $C = \chkp(\bconfirmed[val=v,time=\slotstart(\prevfirstslotepoch{t})], \epoch(t)-1)$. 
    \end{enumerate}
    then
    \begin{enumerate}
        \item $\canonical[blck={\bconfirmed[val=v,time=\slotstart(\epoch(t))]},time=t,to=\slotstart(\epoch(t)+1)]$
        \item\label{conc:lem:confirmed-end-first-slot-canonical:2} for any justified checkpoint $C$, $\epoch(C) \in [\epoch(\guattime[time=\slotstart(\firstslot(\epoch(t)-1)),val=v]),\epoch(t)-1] \implies C  \preceq \bconfirmed[val=v,time=\slotstart(\epoch(t))]$
        \item\label{conc:lem:confirmed-end-first-slot-canonical:3}   for any justified checkpoint $C$, $\epoch(C) \in [\epoch(\guattime[time=\slotstart(\firstslot(\epoch(t)-1)),val=v]),\epoch(t)-1] \implies C  \succeq \guattime[time=\slotstart(\prevfirstslotepoch{t}),val=v]$        
        \item  If $\bconfirmed[val=v,time=\slotstart(\epoch(t))] \neq \block(\guattime[time=\slotstart(\firstslot(\epoch(t)-1)),val=v])$, then
        \begin{enumerate}[label*=\arabic*.]
            \item $\begin{aligned}[t]
                &\exists C^\mathcal{B} \in \{\guattime[time={\slotstart({\prevfirstslotepoch[-1]{t}})},val=v],\guattime[time=\slotstart(\prevfirstslotepoch{t}),val=v]\},\\
                &\quad \sir(\bconfirmed[val=v,time=\slotstart(\epoch(t))],t,C^\mathcal{B},\block(\guattime[time=\slotstart(\firstslot(\epoch(t)-1)),val=v])).\{\sirone,\sirtwo,\sirthree,\sirsix\}\text{ holds.}
            \end{aligned}$
        \end{enumerate}
    \end{enumerate}
\end{lemma}

\begin{proof}
    Due \Cref{lem:code-implications-of-confirmed-block-from-prev-epoch-at-the-beginning-of-epoch}, we can proceed by the following cases.
    \begin{description}[style=nextline]

        \item[Case 1: {$\guattime[val=v,time=\slotstart(\prevfirstslotepoch{t})]=\chkp(\bconfirmed[val=v,time=\slotstart(\epoch(t))])$ and\\ 
        $\bconfirmed[val=v,time=\slotstart(\epoch(t))] = \var[val=v]{\findlatestconfirmeddescendant}(\block(\guattime[val=v,time=\slotstart(\prevfirstslotepoch{t})]))$ and $\exists b' \succeq \bconfirmed[val=v,time=\slotstart(\epoch(t))], \gu(b')=\chkp(\bconfirmed[val=v,time=\slotstart(\epoch(t))])$}.]
        Note that $\epoch(\guattime[val=v,time=\slotstart(\prevfirstslotepoch{t})]) = \epoch(t)-1$.
        Then, clearly conclusions~\ref{conc:lem:confirmed-end-first-slot-canonical:2} and \ref{conc:lem:confirmed-end-first-slot-canonical:3} hold.
        Then, \Cref{lem:gu-e-1-canonical-more-general,lem:output-find-latest-canonical-canonical-at-the-start-of-an-epoch} conclude the proof.

        \item[Case 2:  the \KwSty{if} condition at line \Cref{ln:if-bcand-npreceq-head} is false 
        and\\${\bconfirmed[val=v,time=\slotstart(\epoch(t))]} = {\var[val=v,time=\slotstart(\epoch(t))]{\findlatestconfirmeddescendant}(\bconfirmed[val=v,time=\slotstart(\prevfirstslotepoch{t})])}$.] 
        \Cref{lem:conf-prev-slot-canonical-at-the-start-of-an-epoch,lem:output-find-latest-canonical-canonical-at-the-start-of-an-epoch} prove this case.  
    \end{description}
\end{proof}

The next lemma records a simple structural fact about mid-epoch slots: if a candidate differs from the previous confirmed block and is from the previous or current epoch, it must be the output of $\findlatestconfirmeddescendant$ applied to the previous confirmed block.
This holds because, at mid-epoch slots, neither the reset to the finalized block nor the active state entry condition at \Cref{ln:start-conf-chain} can fire.

\begin{lemma}\label{lem:if-new-bc-then-output-find-latest}
    Pick any $b_c \in  \var[val=v,time=\slotstart(\slot(t))]{\bcands}$.
    If
    \begin{enumerate}
        \item $\slot(t) > \firstslot(\epoch(t))$
        \item $\epoch(b_c) \geq \epoch(t)-1$,
    \end{enumerate}
    then
    \begin{enumerate}
        \item $b_c \neq \var[val=v,time=\slotstart(\slot(t)-1)]{\bconfirmed}  \implies b_c = \var[val=v]{\findlatestconfirmeddescendant}(\var[val=v,time=\slotstart(\slot(t)-1)]{\bconfirmed})$

    \end{enumerate}
\end{lemma}

\begin{proof}
    By \Cref{lem:prev-conf-at-least-e-1},
    $\epoch(\bconfirmed[val=v,time=\slotstart(\slot(t)-1)]) \geq \epoch(t){-}1$
    and
    $\bconfirmed[val=v,time=\slotstart(\slot(t)-1)] \preceq \LMDGHOSTHFC(\viewatstslottime[val=v,time=t])$.
    Combined with $\slot(t) > \firstslot(\epoch(t))$, all three disjuncts of the
    reset at Line~\ref{ln:if-bcand-npreceq-head} are false, and the restart at
    Line~\ref{ln:start-conf-chain} does not fire.
    Hence the only values $\bcand$ assumes are
    $\bconfirmed[val=v,time=\slotstart(\slot(t)-1)]$ and
    $\findlatestconfirmeddescendant_v(\bconfirmed[val=v,time=\slotstart(\slot(t)-1)])$.
\end{proof}

The following lemma shows that when the confirmed block advances to a previous-epoch block at a mid-epoch slot, the validator's view contains a block $b'$ that is a descendant of the confirmed block for which the algorithm's $\willNoConflictingChkpBeJustified$ predicate holds at the current epoch's checkpoint.
The proof derives the witness from Case~2 of $\findlatestconfirmeddescendant$ (which must have fired since the confirmed block changed and the slot is not the first in the epoch) and uses the descendant relation to place the witness as a descendant of the new confirmed block.
This guarantees that honest FFG votes will prevent conflicting current-epoch checkpoints from being justified.

\begin{lemma}\label{lem:willNoConflictingChkpBeJustified}
    If
    \begin{enumerate}
        \item $\slot(t)>\firstslot(\epoch(t))$
        \item $\epoch(\bconfirmed[val=v,time=\slotstartslot{t}]) = \epoch(t)-1$
        \item $\bconfirmed[val=v,time=\slotstartslot{t}]\neq\bconfirmed[val=v,time=\prevslotstartslot{t}]$,
    \end{enumerate}
    then
    \begin{enumerate}
        \item $\exists b', b' \succeq \bconfirmed[val=v,time=\slotstartslot{t}] \land \var[val=v,time=\slotstartslot{t}]{\willNoConflictingChkpBeJustified}(\chkp(b',\epoch(t)))$
    \end{enumerate}
\end{lemma}

\begin{proof}
    By \Cref{lem:if-new-bc-then-output-find-latest},
    $\bconfirmed[val=v,time=\slotstartslot{t}]
     = \findlatestconfirmeddescendant_v(\bconfirmed[val=v,time=\prevslotstartslot{t}])$.
    Since this differs from $\bconfirmed[val=v,time=\prevslotstartslot{t}]$,
    Case~2 of $\findlatestconfirmeddescendant$ did not fire.
    As $\slot(t) > \firstslot(\epoch(t))$, this means
    $\exists b' \succeq \bconfirmed[val=v,time=\prevslotstartslot{t}]$
    with
    $\var[val=v,time=\slotstartslot{t}]{\willNoConflictingChkpBeJustified}(\chkp(b',\epoch(t)))$.
    Since $\epoch(b') = \epoch(t) > \epoch(t){-}1
    = \epoch(\bconfirmed[val=v,time=\slotstartslot{t}])$
    and both are descendants of
    $\bconfirmed[val=v,time=\prevslotstartslot{t}]$
    on the canonical chain,
    $b' \succeq \bconfirmed[val=v,time=\slotstartslot{t}]$.
\end{proof}

Using the previous lemma together with the fact that the confirmed block is canonical through the end of the current epoch, the next result shows that no current-epoch justified checkpoint can conflict with the confirmed block: any justified checkpoint from the current epoch must be a descendant of the confirmed block.
The proof first promotes the witness $b'$ from the previous lemma to a current-epoch block that is a descendant of the confirmed block, using the fact that both lie on the canonical chain rooted at the previous slot's confirmed block.
It then uses the $\willNoConflictingChkpBeJustified$ bound to show that strictly more than $1/3$ of the validator weight has already committed FFG votes in support of a non-conflicting current-epoch checkpoint, so any conflicting checkpoint would lack the $2/3$ supermajority needed for justification.

\begin{lemma}\label{lem:conf-prev-epoch-in-current-epoch-no-nonsucc-chkp-can-be-justified}
    If
    \begin{enumerate}
        \item $\slot(t)>\firstslot(\epoch(t))$
        \item $\epoch(\bconfirmed[val=v,time=\slotstartslot{t}]) = \epoch(t)-1$
        \item $\bconfirmed[val=v,time=\slotstartslot{t}]\neq\bconfirmed[val=v,time=\prevslotstartslot{t}]$
        \item $\canonical[blck={\bconfirmed[val=v,time=\slotstartslot{t}]},time=\slotstartslot{t},to=\slotstart(\epoch(t)+1)]$,
    \end{enumerate}
    then
    \begin{enumerate}
        \item for any jusitified checkpoint $C$, if $\epoch(C) = \epoch(t)$, then $C\succeq \bconfirmed[val=v,time=\slotstartslot{t}]$. 
    \end{enumerate}
\end{lemma}

\begin{proof}
    From \Cref{lem:willNoConflictingChkpBeJustified} we know that $\exists b', b' \succeq \bconfirmed[val=v,time=\slotstartslot{t}] \land \var[val=v,time=\slotstartslot{t}]{\willNoConflictingChkpBeJustified}(\chkp(b',\epoch(t)))$.
    The above and the Lemma's hypotheses imply that $\chkp(b',\epoch(t)) \succeq \bconfirmed[val=v,time=\slotstartslot{t}]$.

    Then, give that $\canonical[blck={\bconfirmed[val=v,time=\slotstartslot{t}]},time=\slotstartslot{t}]$, from $\var[val=v,time=\slotstartslot{t}]{\willNoConflictingChkpBeJustified}(\chkp(b',\epoch(t)))$, we know that, for any block $b'$,
   $\weightofset[chkp=b']{\bigcup_{C \succeq \bconfirmed[val=v,time=\slotstartslot{t}] \land epoch(C) = \epoch(t)} \ffgvalsetallsentraw[source=,target=C,time=\slotstart(\epoch(t)+1)] \cap \totvalset[]{\honvals}} > \frac{1}{3} \totvalsetweight[chkp=b']{\allvals}$

    This implies that, for any checkpoint $C'$ such that $\epoch(C) =\epoch(t) \land C \nsucceq \bconfirmed[val=v,time=\slotstartslot{t}]$, for any time $t'$,
    $\weightofset[chkp={b'}]{\ffgvalsetallsentraw[time=t',source=,target=C']} < \frac{2}{3} \totvalsetweight[chkp=b']{\allvals} $ which concludes the proof.
\end{proof}

The following lemma provides the block needed to establish $\sirfour$ in the mid-epoch case: if the confirmed block advances to a previous-epoch block at a non-first slot, the validator's view contains a block $b'$ that is a descendant of the confirmed block whose greatest unrealized justified checkpoint has epoch at least that of the previous epoch.
The proof eliminates Cases~1 and~2 of $\findlatestconfirmeddescendant$ (which respectively require the input to be in the current epoch and return the input unchanged) and observes that the surviving Cases~3 and~4, when $\slot(t) > \firstslot(\epoch(t))$, both produce such a witness.
This ensures the unrealized-justification condition of SIR is satisfied.

\begin{lemma}\label{lem:conf-prev-epoch-in-current-epoch-exists-succ-with-gu-from-prev-epoch}
    If
    \begin{enumerate}
        \item $\slot(t)>\firstslot(\epoch(t))$
        \item $\epoch(\bconfirmed[val=v,time=\slotstartslot{t}]) = \epoch(t)-1$
        \item $\bconfirmed[val=v,time=\slotstartslot{t}]\neq\bconfirmed[val=v,time=\prevslotstartslot{t}]$,
    \end{enumerate}
    then
    \begin{enumerate}
        \item $\exists b' \in \viewattime[val=v,time=\slotstartslot{t}], b' \succeq \bconfirmed[val=v,time=\slotstartslot{t}], \epoch(\gu(b'))\geq \epoch(t)-1$. 
    \end{enumerate}
\end{lemma}

\begin{proof}
    By \Cref{lem:if-new-bc-then-output-find-latest}, $\bconfirmed[val=v,time=\slotstartslot{t}]
     = \findlatestconfirmeddescendant_v(\bconfirmed[val=v,time=\prevslotstartslot{t}])$.
    Since the output differs from the input and has epoch $\epoch(t){-}1$, it was not  produced by Case~1 (which requires epoch $\epoch(t)$) nor by Case~2 (which returns the input).
    Hence Case~3 or Case~4 produced it, and both require, when $\slot(t) > \firstslot(\epoch(t))$, $\exists b' \in \viewattime[val=v,time=\slotstartslot{t}]$ with $b' \succeq \bconfirmed[val=v,time=\slotstartslot{t}]$ and $\epoch(\gu(b')) \geq \epoch(t){-}1$.
\end{proof}

This lemma extends the previous result on the confirmed block being canonical at the start of the epoch to any slot within the epoch: if the confirmed block at the current slot is from the previous or current epoch and every previous-epoch justified checkpoint coincides with $\chkp(\bconfirmed[val=v,time=\slotstart(\prevfirstslotepoch{t})], \epoch(t){-}1)$ whenever the previous-epoch confirmed block precedes the current confirmed block at the start of the epoch, then the confirmed block is canonical through the end of the epoch.
The proof proceeds by induction on slots within the epoch, using \Cref{lem:confirmed-end-first-slot-canonical} as the base case and \Cref{lem:output-find-latest-canonical-canonical-at-the-start-of-an-epoch} for the inductive step, chaining the guarantee that the confirmed block is canonical from one slot to the next via the monotonicity of the confirmed block (\Cref{lem:conf-beginning-epoch-prec-bcand}).

\begin{lemma}\label{lem:canonical-for-current-epoch-with-extra-assum}
    If
    \begin{enumerate}
        \item $\slotstart(\prevfirstslotepoch{t})\geq \GGST$
        \item $\epoch(\bconfirmed[val=v,time=\slotstart(\slot(t))])\geq\epoch(t)-1$
        \item if $\bconfirmed[val=v,time=\slotstart(\prevfirstslotepoch{t})] \preceq \bconfirmed[val=v,time=\slotstart(\epoch(t))]$ and $C$ is a justified checkpoint such that $\epoch(C) = \epoch(t)-1$, then $C = \chkp(\bconfirmed[val=v,time=\slotstart(\prevfirstslotepoch{t})], \epoch(t)-1)$.
    \end{enumerate}
    then
    \begin{enumerate}
        \item $\canonical[blck={\bconfirmed[val=v,time=\slotstart(\slot(t))]},time=t, to=\slotstart(\epoch(t)+1)]$
    \end{enumerate}
\end{lemma}

\begin{proof}
    By induction on slot $\sind$.
    \begin{description}
        \item[Base Case: $\sind = \firstslot(\epoch(t))$.]  \Cref{lem:confirmed-end-first-slot-canonical} proves the base case.
        \item[Inductive Step: $\sind > \firstslot(\epoch(t))$.] 
        From \Cref{lem:conf-beginning-epoch-prec-bcand}, we know that $\bconfirmed[val=v,time=\slotstart(\sind)] \succeq \var[val=v,time=\slotstart(\sind-1)]{\bconfirmed}$.
        By cases.
        \begin{description}
            \item[Case 1: {$\var[val=v,time=\slotstart(\sind-1)]{\bconfirmed} = \bconfirmed[val=v,time=\slotstart(\sind)]$}.] 
            By the inductive hypothesis, $\canonical[blck={\var[val=v,time=\slotstart(\sind)]{\bconfirmed}},time=t, to=\slotstart(\epoch(t)+1)]$ follows.
            \item[Case 2: {$\var[val=v,time=\slotstart(\sind-1)]{\bconfirmed} \neq \bconfirmed[val=v,time=\slotstart(\sind)]$}.]  
            From \Cref{lem:if-new-bc-then-output-find-latest,lem:conf-beginning-epoch-prec-bcand}, we know that\\ $\bconfirmed[val=v,time=\slotstart(\sind)]=\findlatestconfirmeddescendant(\var[val=v,time=\slotstart(\sind-1)]{\bconfirmed})$ and $\bconfirmed[val=v,time=\slotstart(\sind)] \succeq \var[val=v,time=\slotstart(\firstslot(\epoch(\sind)))]{\bconfirmed}$.
            \Cref{lem:confirmed-end-first-slot-canonical} implies that
            \begin{enumerate}
                \item for any justified checkpoint $C$, $\epoch(C) \in [\epoch(\guattime[time=\slotstart(\firstslot(\epoch(t)-1)),val=v]),\epoch(t)-1] \implies C  \preceq \bconfirmed[val=v,time=\slotstart(\sind)]$
                \item   for any justified checkpoint $C$, $\epoch(C) \in [\epoch(\guattime[time=\slotstart(\firstslot(\epoch(t)-1)),val=v]),\epoch(t)-1] \implies C  \succeq \guattime[time=\slotstart(\prevfirstslotepoch{t}),val=v]$
            \end{enumerate}

            Given that, by the inductive hypothesis, $\canonical[blck={\var[val=v,time=\slotstart(\sind-1)]{\bconfirmed}},time=t, to=\slotstart(\epoch(t)+1)]$,
            we can apply \Cref{lem:output-find-latest-canonical-canonical-at-the-start-of-an-epoch} and conclude the proof.
        \end{description}
    \end{description}
\end{proof}

This is a key lemma that discharges the ``no conflicting checkpoint'' hypothesis used in the previous results.
It shows that if the reset check passes at the epoch boundary, $\GGST$ has passed two epochs ago, and no block in the view at the start of the previous epoch already unrealized-justifies the confirmed checkpoint, then the FFG support for $\chkp(\bconfirmed[val=v,time=\slotstart(\prevfirstslotepoch{t})])$ reaches the two-thirds threshold and any justified checkpoint of epoch $\epoch(t)-1$ must coincide with it.
The proof uses \Cref{lem:no-curr-epochconflict-chkp-is-justified} to extract an earlier slot at which $\willChkpBeJustified$ held, then splits on whether the confirmed block from two epochs ago is an ancestor of the witness block: in the non-ancestor case \Cref{lem:canonical-for-current-epoch-with-extra-assum} applies directly, while in the ancestor case hypothesis~\ref{itm:lem:no-conflicting-to-c-b-at-the-start-of-an-epoch:4} together with \Cref{lem:conf-current-epoch-then-gu-curr-epoch} pins down the unique justified checkpoint of epoch $\epoch(t)-2$.
The required FFG support bound then follows from \Cref{lem:sufficient-condition-for-justification}.

\begin{lemma}\label{lem:no-conflicting-to-c-b-at-the-start-of-an-epoch}
    If
    \begin{enumerate}
        \item $\slotstart(\prevfirstslotepoch[-1]{t})\geq \GGST$
        \item\label{itm:lem:no-conflicting-to-c-b-at-the-start-of-an-epoch:3} \sloppy{the \KwSty{if} condition at line \Cref{ln:if-bcand-npreceq-head} during the execution of $\var[val=v,time=\slotstart(\epoch(t))]{\getlatestconfirmed}(\bconfirmed[val=v,time=\slotstart(\prevfirstslotepoch{t})])$ is false}
        \item\label{itm:lem:no-conflicting-to-c-b-at-the-start-of-an-epoch:4} $\nexists b'  \in \viewattime[val=v,time=\slotstart(\prevfirstslotepoch{t})], b' \succeq \bconfirmed[val=v,time=\slotstart(\prevfirstslotepoch{t})] \land \gu(b') = \chkp(\bconfirmed[val=v,time=\slotstart(\prevfirstslotepoch{t})])$,
    \end{enumerate}
    then
    \begin{enumerate}
        \item\label{itm:lem:no-conflicting-to-c-b-at-the-start-of-an-epoch:conc1}  for any block $b' \succeq \chkp(\bconfirmed[val=v,time=\slotstart(\prevfirstslotepoch{t})])$ and time $t' \geq \slotstart(\epoch(t))$,$\weightofset[chkp=b']{\ffgvalsetallsentraw[source=,target={\chkp(\bconfirmed[val=v,time=\slotstart(\prevfirstslotepoch{t})])},time=t'] \setminus \slashedset[chkp=b']} \geq \frac{2}{3}\totvalsetweight[chkp=b'] {\allvals}$
        \item\label{itm:lem:no-conflicting-to-c-b-at-the-start-of-an-epoch:conc2}
        if $C$ is a justified checkpoint and $\epoch(C) = \epoch(t)-1$, then $C = \chkp(\bconfirmed[val=v,time=\slotstart(\prevfirstslotepoch{t})], \epoch(t)-1)$.
    \end{enumerate}
\end{lemma}

\begin{proof}
    Conclusion~\ref{itm:lem:no-conflicting-to-c-b-at-the-start-of-an-epoch:conc2} is implied by conclusion~\ref{itm:lem:no-conflicting-to-c-b-at-the-start-of-an-epoch:conc1}.
    So, we just prove the latter.
    Because of hypothesis~\ref{itm:lem:no-conflicting-to-c-b-at-the-start-of-an-epoch:3},
    we know that $\epoch(\bconfirmed[val=v,time=\slotstart(\prevfirstslotepoch{t})])\geq\epoch(t)-1$. 
    From \Cref{lem:no-curr-epochconflict-chkp-is-justified}, we know that there exists a time $t' \in \{\slotstart(s): s \in [\firstslot(\epoch(t)-1)+1,\lastslot(\epoch(t)-1)]\}$ such that $\var[val=v,time=t']{\willChkpBeJustified}(\bconfirmed[val=v,time=t'])$ and $\epoch(\bconfirmed[val=v,time=t'] ) = \epoch(t)-1$.
    \Cref{lem:conf-beginning-epoch-prec-bcand} then implies that $\bconfirmed[val=v,time=t'] \preceq \bconfirmed[val=v,time=\slotstart(\prevfirstslotepoch{t})]$ and $\epoch(\bconfirmed[val=v,time={\slotstart({\prevfirstslotepoch[-1]{t}})}] ) \geq \epoch(t)-2$.

    Then, given Lemma~\ref{lem:sufficient-condition-for-justification}, to conclude the proof it is sufficient to prove that  $\canonical[blck={\bconfirmed[val=v,time=t']},time=t',to=\slotstart(\epoch(t))]$.

    Now, by cases.
    
    \begin{description}
        \item[Case 1: {$\var[val=v,time={\slotstart(\prevfirstslotepoch[-1]{t})}]{\bconfirmed} \npreceq \bconfirmed[val=v,time=t']$}.]
        In this case, we can apply \Cref{lem:canonical-for-current-epoch-with-extra-assum} to conclude that $\canonical[blck={\bconfirmed[val=v,time=t']},time=t',to=\slotstart(\epoch(t))]$.
        \item[Case 2: {$\var[val=v,time={\slotstart(\prevfirstslotepoch[-1]{t})}]{\bconfirmed} \preceq \bconfirmed[val=v,time=t']$}.]
        In this case, to apply \Cref{lem:canonical-for-current-epoch-with-extra-assum} we need to show that if $C$ is a justified checkpoint such that $\epoch(C)=\epoch(t)-2$, then $C=\chkp(\bconfirmed[val=v,time=\slotstart(\prevfirstslotepoch{t})],\epoch(t)-2)$.
    \Cref{lem:conf-current-epoch-then-gu-curr-epoch} and hypothesis~\ref{itm:lem:no-conflicting-to-c-b-at-the-start-of-an-epoch:4} imply that $\exists b'  \in \viewattime[val=v,time=\slotstart(\prevfirstslotepoch{t})], b' \succeq \bconfirmed[val=v,time=\slotstart(\prevfirstslotepoch{t})] \land \epoch(\gu(b')) = \epoch(t)-2$.
    This implies that $\gu(b') = \chkp(\bconfirmed[val=v,time=\slotstart(\prevfirstslotepoch{t})],\epoch(t)-2)$.
    Hence, $C = \chkp(\bconfirmed[val=v,time=\slotstart(\prevfirstslotepoch{t})],\epoch(t)-2)$.  
    \end{description}
\end{proof}

The next lemma completes the SIR for a previous-epoch candidate $b_c$ at the epoch boundary: given that SIR conditions $\sirone$, $\sirtwo$, $\sirthree$ and $\sirsix$ already hold and $b_c$ is canonical throughout the current epoch, it establishes the remaining conditions ($\sirfour$ and $\sirfive$) and concludes that $b_c$ is canonical forever.
The proof splits into the two code paths (active state entry at \Cref{ln:start-conf-chain} where a block in the view already unrealized-justifies $\chkp(b_c)$, vs.\ continuation from the previous confirmed block).
In the active state entry case, $\sirfour$ follows directly from the witnessing block and no conflicting previous-epoch checkpoint can be justified.
In the continuation case, \Cref{lem:no-conflicting-to-c-b-at-the-start-of-an-epoch} forces $\chkp(b_c)$ to equal the confirmed checkpoint and \Cref{assum:ffg-assumptions:justified-checkpoint-next-epoch} delivers $\sirfour$.
Both paths combine with the hypothesis that $b_c$ is canonical to yield $\sirfive$, and \Cref{lem:ffg-safety-from-sir} then concludes that $b_c$ is canonical forever since $\epoch(b_c)+2 = \epoch(t)+1$.

\begin{lemma}\label{lem:helper-hard-to-name}
    Let $C^\mathcal{B}$ be any checkpoint and $b_c$ be any block.
    If
    \begin{enumerate}
        \item $\slotstart(\prevfirstslotepoch[-1]{t})\geq \GGST$
        \item $\epoch(b_c) = \epoch(t)-1$
        \item either
        \begin{enumerate}[label*=\arabic*.]
            \item\label{itm:lem:helper-hard-to-name:b}  $\exists b'  \in \viewattime[val=v,time=\slotstart(\prevfirstslotepoch{t})], b' \succeq b_c \land \gu(b') = \chkp(b_c)$
            \item\label{itm:lem:helper-hard-to-name:a}   \sloppy{the \KwSty{if} condition at line \Cref{ln:if-bcand-npreceq-head} during the execution of $\var[val=v,time=\slotstart(\epoch(t))]{\getlatestconfirmed}(\bconfirmed[val=v,time=\slotstart(\prevfirstslotepoch{t})])$ is false and $b_c \succeq \bconfirmed[val=v,time=\slotstart(\prevfirstslotepoch{t})]$.}

        \end{enumerate}
        \item \label{itm:lem:helper-hard-to-name:cond6} $\canonical[blck={b_c},time=\slotstart(\epoch(t)),to=\slotstart(\epoch(t)+1)]$
        \item $\sir(b_c,\slotstart(\epoch(t)),C^\mathcal{B},\block(\guattime[val=v,time={\slotstart(\prevfirstslotepoch{t})}])).\{\sirone,\sirtwo,\sirthree,\sirsix\}$ holds,
    \end{enumerate}
    then
    \begin{enumerate}
        \item $\sir(b_c,\slotstart(\epoch(t)),C^\mathcal{B},\block(\guattime[val=v,time={\slotstart(\prevfirstslotepoch{t})}]))$ holds
       
        \item $\canonical[blck={b_c},time=\slotstart(\epoch(t))]$
    \end{enumerate}
\end{lemma}

\begin{proof}
    By cases.
    \begin{description}[style=nextline]
        \item[Case 1: hypothesis~\ref{itm:lem:helper-hard-to-name:b} holds.]
        Given that $\epoch(b_c) = \epoch(t)-1$, \sirfour clearly holds in this case.
        Also, clearly no justified checkpoint $C$ with $\epoch(C)=\epoch(t)-1$ and $\block(C)$ conflicting with $b_c$ can exists.

        \item[Case 2: hypothesis~\ref{itm:lem:helper-hard-to-name:b} does not hold and hypothesis~\ref{itm:lem:helper-hard-to-name:a} holds.]
        This case implies that $\chkp(b_c) = \chkp(\bconfirmed[val=v,time=\slotstart(\prevfirstslotepoch{t})])$.
        This, Lemma's hypothesis~\ref{itm:lem:helper-hard-to-name:cond6} and \Cref{lem:no-conflicting-to-c-b-at-the-start-of-an-epoch} allows us to apply \Cref{assum:ffg-assumptions:justified-checkpoint-next-epoch} to conclude \ref{def:induction-conditions:ub}.

        Given that $\epoch(b_c) = \epoch(\bconfirmed[val=v,time=\slotstart(\prevfirstslotepoch{t})]) = \epoch(t)-1$,
        \Cref{lem:no-conflicting-to-c-b-at-the-start-of-an-epoch} also implies that no justified checkpoint $C$ with $\epoch(C)=\epoch(t)-1$ and $\block(C)$ conflicting with $b_c$ can exists
    \end{description}

    Given the above, \Cref{lem:no-conflicting-to-c-b-at-the-start-of-an-epoch} and $\canonical[blck={b_c},time=\slotstart(\epoch(t)),to=\slotstart(\epoch(t)+1)]$ also imply \sirfive.

    Now, $\epoch(b_c)+2=\epoch(t)+1$ and \Cref{lem:ffg-safety-from-sir} imply  $\canonical[blck={b_c},time=\slotstart(\epoch(t))]$.
\end{proof}

This lemma is the ``no extra assumption'' version of \Cref{lem:conf-prev-slot-canonical-at-the-start-of-an-epoch}: it removes the external no-conflicting-checkpoint hypothesis by appealing to \Cref{lem:no-conflicting-to-c-b-at-the-start-of-an-epoch} to discharge it internally.
The result is that if the reset check passes at the epoch boundary, the current confirmed block is the output of $\findlatestconfirmeddescendant$ on the previous confirmed block, and $\GGST$ has passed two epochs ago, then the only justified checkpoint of epoch $\epoch(t)-1$ is $\chkp(\bconfirmed[val=v,time=\slotstart(\epoch(t))], \epoch(t)-1)$ and the previous-epoch confirmed block is canonical forever.
The proof splits on whether some block in the view at the start of the previous epoch already unrealized-justifies $\chkp(\bconfirmed[val=v,time=\slotstart(\prevfirstslotepoch{t})])$: in the affirmative case the uniqueness conclusion is immediate, in the negative case \Cref{lem:no-conflicting-to-c-b-at-the-start-of-an-epoch} supplies it.
2\Cref{lem:conf-prev-slot-canonical-at-the-start-of-an-epoch} and \Cref{lem:helper-hard-to-name} then establish that the block is canonical forever.

\begin{lemma}\label{lem:conf-prev-slot-canonical-at-the-start-of-an-epoch-no-extra-assum}
    If
    \begin{enumerate}
        \item $\slotstart(\prevfirstslotepoch[-1]{t})\geq \GGST$
        \item\label{itm:lem:conf-at-the-start-of-an-epoch-no-extra-assum:cond3}  \sloppy{the \KwSty{if} condition at line \Cref{ln:if-bcand-npreceq-head} during the execution of $\var[val=v,time=\slotstart(\epoch(t))]{\getlatestconfirmed}(\bconfirmed[val=v,time=\slotstart(\prevfirstslotepoch{t})])$ is false and $\bconfirmed[val=v,time=\slotstart(\epoch(t))] = \var[val=v,time=\slotstart(\epoch(t))]{\findlatestconfirmeddescendant}(\bconfirmed[val=v,time=\slotstart(\prevfirstslotepoch{t})])$.},
    \end{enumerate}
    then
    \begin{enumerate}
        \item\label{conc:lem:conf-prev-slot-canonical-at-the-start-of-an-epoch-no-extra-assum:2} if $C$ is a justified checkpoint and $\epoch(C) = \epoch(t)-1$, then $C = \chkp(\bconfirmed[val=v,time=\slotstart(\epoch(t))], \epoch(t)-1)$.

        \item\label{conc:lem:conf-prev-slot-canonical-at-the-start-of-an-epoch-no-extra-assum:3} $\canonical[blck={\bconfirmed[val=v,time=\slotstart(\prevfirstslotepoch{t})]},time=\slotstart(\epoch(t))]$
    \end{enumerate}
\end{lemma}

\begin{proof}
    By cases.
    \begin{description}
        \item[{Case 1: $\exists b'  \in \viewattime[val=v,time=\slotstart(\prevfirstslotepoch{t})], b' \succeq \bconfirmed[val=v,time=\slotstart(\prevfirstslotepoch{t})] \land \gu(b') = \chkp(\bconfirmed[val=v,time=\slotstart(\prevfirstslotepoch{t})])$}]
        Conclusion~\ref{conc:lem:conf-prev-slot-canonical-at-the-start-of-an-epoch-no-extra-assum:2} clearly holds.
        \item[{Case 2: $\nexists b'  \in \viewattime[val=v,time=\slotstart(\prevfirstslotepoch{t})], b' \succeq \bconfirmed[val=v,time=\slotstart(\prevfirstslotepoch{t})] \land \gu(b') = \chkp(\bconfirmed[val=v,time=\slotstart(\prevfirstslotepoch{t})])$}]
        \Cref{lem:no-conflicting-to-c-b-at-the-start-of-an-epoch} allows us to prove conclusion~\ref{conc:lem:conf-prev-slot-canonical-at-the-start-of-an-epoch-no-extra-assum:2}, as $\bconfirmed[val=v,time=\slotstart(\epoch(t))]  \succeq \bconfirmed[val=v,time=\slotstart(\prevfirstslotepoch{t})]$.
    \end{description}
    Given that conclusion~\ref{conc:lem:conf-prev-slot-canonical-at-the-start-of-an-epoch-no-extra-assum:2} holds, we can apply
    \Cref{lem:conf-prev-slot-canonical-at-the-start-of-an-epoch}.
    In turn, this allows us to apply \Cref{lem:helper-hard-to-name} to prove conclusion~\ref{conc:lem:conf-prev-slot-canonical-at-the-start-of-an-epoch-no-extra-assum:3} and conclude the proof.
\end{proof}

\begin{lemma}\label{lem:gu-e-1-canonical-no-extra-assum}
    If
    \begin{enumerate}
        \item $\slotstart(\prevfirstslotepoch[-1]{t})\geq \GGST$
        \item $\epoch(\bconfirmed[val=v,time=\slotstart(\epoch(t))])=\epoch(t)-1$,
    \end{enumerate}
    then
    \begin{enumerate}
        \item $\canonical[blck={\block(\guattime[val=v,time=\slotstart(\prevfirstslotepoch{t})])},time=\slotstart(\epoch(t))]$.
    \end{enumerate}
\end{lemma}

\begin{proof}
        As per \Cref{lem:code-implications-of-confirmed-block-from-prev-epoch-at-the-beginning-of-epoch}, we can proceed by the following cases.
    \begin{description}[style=nextline]

        \item[Case 1: {$\guattime[val=v,time=\slotstart(\prevfirstslotepoch{t})]=\chkp(\bconfirmed[val=v,time=\slotstart(\epoch(t))])$ and\\ 
        $\bconfirmed[val=v,time=\slotstart(\epoch(t))] = \var[val=v]{\findlatestconfirmeddescendant}(\block(\guattime[val=v,time=\slotstart(\prevfirstslotepoch{t})]))$ and $\exists b' \succeq \bconfirmed[val=v,time=\slotstart(\epoch(t))], \gu(b')=\chkp(\bconfirmed[val=v,time=\slotstart(\epoch(t))])$}.]
        Note that $\epoch(\guattime[val=v,time=\slotstart(\prevfirstslotepoch{t})]) = \epoch(t)-1$.
        Then, we can apply \Cref{lem:gu-e-1-canonical-more-general} to prove that $\canonical[blck={\block(\guattime[val=v,time=\slotstart(\prevfirstslotepoch{t})])},time=\slotstart(\epoch(t))]$.

        \item[Case 2:  the \KwSty{if} condition at line \Cref{ln:if-bcand-npreceq-head} is false 
        and\\${\bconfirmed[val=v,time=\slotstart(\epoch(t))]} = {\var[val=v,time=\slotstart(\epoch(t))]{\findlatestconfirmeddescendant}(\bconfirmed[val=v,time=\slotstart(\prevfirstslotepoch{t})])}$.]

        Given that $\bconfirmed[time=\slotstart(\epoch(t)),val=v]\succeq \bconfirmed[val=v,time=\slotstart(\prevfirstslotepoch{t})]$, \Cref{lem:conf-prev-slot-canonical-at-the-start-of-an-epoch-no-extra-assum} implies that  if $C$ is a justified checkpoint and $\epoch(C) = \epoch(t)-1$, then $C = \chkp(\bconfirmed[val=v,time=\slotstart(\epoch(t))], \epoch(t)-1)$.
        This means that we can apply \Cref{lem:gu-e-1-canonical-more-general} in this case as well to prove that $\canonical[blck={\block(\guattime[val=v,time=\slotstart(\prevfirstslotepoch{t})])},time=\slotstart(\epoch(t))]$.
    \end{description}    
\end{proof}

This is the self-contained version of the result showing that the previous-epoch confirmed block is canonical forever at the start of the epoch: if the confirmed block at the first slot of the current epoch belongs to the previous epoch and $\slotstart(\prevfirstslotepoch[-1]{t}) \geq \GGST$, then the previous-epoch confirmed block is canonical forever, the only justified checkpoint of epoch $\epoch(t)-1$ is $\chkp(\bconfirmed[val=v,time=\slotstart(\epoch(t))], \epoch(t)-1)$, and the standard ordering of justified checkpoints between the previous-epoch GU and the confirmed block holds.
The proof proceeds via \Cref{lem:code-implications-of-confirmed-block-from-prev-epoch-at-the-beginning-of-epoch}: in the active state entry case the uniqueness conclusion is immediate; in the continuation case \Cref{lem:conf-prev-slot-canonical-at-the-start-of-an-epoch-no-extra-assum} supplies it.
In both cases \Cref{lem:gu-e-1-canonical-no-extra-assum} establishes $\sirsix$ for the previous-epoch GU block.
A final case split distinguishes whether the confirmed block coincides with the GU block: in the equality case the conclusion is immediate, otherwise \Cref{lem:confirmed-end-first-slot-canonical} and \Cref{lem:helper-hard-to-name} close the argument.

\begin{lemma}\label{lem:confirmed-end-first-slot-canonical-no-extra-assum}
    If
    \begin{enumerate}
        \item $\slotstart(\prevfirstslotepoch[-1]{t})\geq \GGST$
        \item $\epoch(\bconfirmed[val=v,time=\slotstart(\epoch(t))])=\epoch(t)-1$,
    \end{enumerate}
    then
    \begin{enumerate}
        \item\label{conc:lem:confirmed-end-first-slot-canonical-no-extra-assum:2} $\canonical[blck={\bconfirmed[val=v,time=\slotstart(\epoch(t))]},time=\slotstart(\epoch(t))]$
        \item\label{conc:lem:confirmed-end-first-slot-canonical-no-extra-assum:5} if $C$ is a justified checkpoint and $\epoch(C) = \epoch(t)-1$, then $C = \chkp(\bconfirmed[val=v,time=\slotstart(\epoch(t))], \epoch(t)-1)$.
        \item\label{conc:lem:confirmed-end-first-slot-canonical-no-extra-assum:3} for any justified checkpoint $C$, $\epoch(C) \in [\epoch(\guattime[time=\slotstart(\firstslot(\epoch(t)-1)),val=v]),\epoch(t)-1] \implies C  \succeq \guattime[time=\slotstart(\prevfirstslotepoch{t}),val=v]$,
        \item\label{conc:lem:confirmed-end-first-slot-canonical-no-extra-assum:4} for any justified checkpoint $C$, $\epoch(C) \in [\epoch(\guattime[time=\slotstart(\firstslot(\epoch(t)-1)),val=v]),\epoch(t)-1] \implies C  \preceq \bconfirmed[val=v,time=\slotstart(\epoch(t))]$        
    \end{enumerate}
\end{lemma}

\begin{proof}
    As per \Cref{lem:code-implications-of-confirmed-block-from-prev-epoch-at-the-beginning-of-epoch}, we can proceed by the following cases.
    \begin{description}[style=nextline]

        \item[Case 1: {$\guattime[val=v,time=\slotstart(\prevfirstslotepoch{t})]=\chkp(\bconfirmed[val=v,time=\slotstart(\epoch(t))])$ and\\ 
        $\bconfirmed[val=v,time=\slotstart(\epoch(t))] = \var[val=v]{\findlatestconfirmeddescendant}(\block(\guattime[val=v,time=\slotstart(\prevfirstslotepoch{t})]))$ and $\exists b' \succeq \bconfirmed[val=v,time=\slotstart(\epoch(t))], \gu(b')=\chkp(\bconfirmed[val=v,time=\slotstart(\epoch(t))])$}.]
        Note that $\epoch(\guattime[val=v,time=\slotstart(\prevfirstslotepoch{t})]) = \epoch(t)-1$.
        Then, this case clearly implies conclusion~\ref{conc:lem:confirmed-end-first-slot-canonical-no-extra-assum:5}.

        \item[Case 2:  the \KwSty{if} condition at line \Cref{ln:if-bcand-npreceq-head} is false 
        and\\${\bconfirmed[val=v,time=\slotstart(\epoch(t))]} = {\var[val=v,time=\slotstart(\epoch(t))]{\findlatestconfirmeddescendant}(\bconfirmed[val=v,time=\slotstart(\prevfirstslotepoch{t})])}$.]

        Given that $\bconfirmed[time=\slotstart(\epoch(t)),val=v]\succeq \bconfirmed[val=v,time=\slotstart(\prevfirstslotepoch{t})]$, \Cref{lem:conf-prev-slot-canonical-at-the-start-of-an-epoch-no-extra-assum} implies conclusion~\ref{conc:lem:confirmed-end-first-slot-canonical-no-extra-assum:5}.
    \end{description}    

    In both cases we prove conclusion~\ref{conc:lem:confirmed-end-first-slot-canonical-no-extra-assum:5}.    

    Also, \Cref{lem:gu-e-1-canonical-no-extra-assum} proves $\sirsix(\block(\guattime[val=v,time=\slotstart(\prevfirstslotepoch{t})]),\slotstart(\epoch(t)),\infty)$.

    Then, by cases again.
    \begin{description}
        \item[Case 1: {$\bconfirmed[val=v,time=\slotstart(\slot(t))]=\block(\guattime[val=v,time=\slotstart(\prevfirstslotepoch{t})])$}.] 
        Obvious given the above.
        \item[Case 2: {$\bconfirmed[val=v,time=\slotstart(\slot(t))]\neq\block(\guattime[val=v,time=\slotstart(\prevfirstslotepoch{t})])$}.]\leavevmode
        Given conclusion~\ref{conc:lem:confirmed-end-first-slot-canonical-no-extra-assum:5}, we can apply \Cref{lem:confirmed-end-first-slot-canonical}.
        Finally, this and \Cref{lem:code-implications-of-confirmed-block-from-prev-epoch-at-the-beginning-of-epoch} allow to apply \Cref{lem:helper-hard-to-name} which concludes the proof.
    \end{description}
\end{proof}

This is the key lemma showing that confirmed blocks with epoch $\geq \epoch(t)-1$ are canonical, with no extra hypotheses beyond $\GGST$ having passed two epochs ago.
The proof proceeds by induction on the slot within the current epoch.
The base case uses \Cref{lem:confirmed-end-first-slot-canonical-no-extra-assum}.
For the inductive step, when the confirmed block advances, the proof combines the inductive hypothesis with \Cref{lem:output-find-latest-canonical-canonical-at-the-start-of-an-epoch} to show that the block is canonical through $\slotstart(\epoch(t)+1)$ and to establish $\sirone$--$\sirthree$ and $\sirsix$.
For previous-epoch candidates, $\sirfour$ comes from \Cref{lem:conf-prev-epoch-in-current-epoch-exists-succ-with-gu-from-prev-epoch} and $\sirfive$ from \Cref{lem:conf-prev-epoch-in-current-epoch-no-nonsucc-chkp-can-be-justified}.
For current-epoch candidates, the $\willChkpBeJustified$ guarantee combined with \Cref{lem:ffg-condition-on-q-implies-safety} shows that the block is canonical through $\slotstart(\epoch(t)+2)$, and \Cref{assum:ffg-assumptions:justified-checkpoint-next-epoch} then delivers $\sirfour$.
In both cases, \Cref{lem:ffg-safety-from-sir} concludes the proof.

\begin{lemma}\label{lem:canonical-no-extra-assum}
    If
    \begin{enumerate}
        \item $\slotstart(\prevfirstslotepoch[-1]{t})\geq \GGST$
        \item $\epoch(\bconfirmed[val=v,time=\slotstartslot{t}]) \geq \epoch(t)-1$,
    \end{enumerate}
    then
    \begin{enumerate}
        \item $\canonical[blck={\bconfirmed[val=v,time=\slotstartslot{t}]},time=\slotstartslot{t}]$
    \end{enumerate}
\end{lemma}

\begin{proof}
    By induction on slot $\sind \in [\firstslot(\epoch(t)),\slot(t)]$ we show that\\ $\canonical[blck={\bconfirmed[val=v,time=\slotstart(\sind)]},time=\slotstart(\sind)]$.

    \begin{description}
        \item[Base Case: $\sind = \firstslot(\epoch(t))$.]  \Cref{lem:confirmed-end-first-slot-canonical-no-extra-assum} proves the base case.
    
        \item[Inductive Step: $\sind > \firstslot(\epoch(t))$.] 
        
        From \Cref{lem:conf-beginning-epoch-prec-bcand}, we know that $\bconfirmed[val=v,time=\slotstart(\sind)] \succeq \var[val=v,time=\slotstart(\sind-1)]{\bconfirmed}$.

        If $\var[val=v,time=\slotstart(\sind-1)]{\bconfirmed} = \bconfirmed[val=v,time=\slotstart(\sind)]$, then $\canonical[blck={\var[val=v,time=\slotstart(\sind)]{\bconfirmed}},time=\slotstart(\sind)]$ follows by the inductive hypothesis.

        Now, assume $\var[val=v,time=\slotstart(\sind-1)]{\bconfirmed} \neq \bconfirmed[val=v,time=\slotstart(\sind)]$.
        From \Cref{lem:if-new-bc-then-output-find-latest,lem:conf-beginning-epoch-prec-bcand}, we know that\\ $\bconfirmed[val=v,time=\slotstart(\sind)]=\findlatestconfirmeddescendant(\var[val=v,time=\slotstart(\sind-1)]{\bconfirmed})$ and $\bconfirmed[val=v,time=\slotstart(\sind)] \succ \var[val=v,time=\slotstart(\epoch(\sind))]{\bconfirmed}$.

        Then, given the above, \Cref{lem:confirmed-end-first-slot-canonical-no-extra-assum} implies that
        \begin{enumerate}
            \item \label{itm:lem:canonical-no-extra-assum:ind-1}  for any justified checkpoint $C$, $\epoch(C) \in [\epoch(\guattime[time=\slotstart(\firstslot(\epoch(t)-1)),val=v]),\epoch(t)-1] \implies C  \preceq \bconfirmed[val=v,time=\slotstart(\sind)]$
            \item   for any justified checkpoint $C$, $\epoch(C) \in [\epoch(\guattime[time=\slotstart(\firstslot(\epoch(t)-1)),val=v]),\epoch(t)-1] \implies C  \succeq \guattime[time=\slotstart(\prevfirstslotepoch{t}),val=v]$
            \item \label{itm:lem:canonical-no-extra-assum:ind-3}no checkpoint $C$ s.t.
            $\epoch(C) = \epoch(t)-1$ and $C$ conflicts with $\bconfirmed[val=v,time=\slotstart(\sind)]$ can ever be justified
        \end{enumerate}

        Given that $\bconfirmed[val=v,time=\slotstart(\sind)] \neq \var[val=v,time=\slotstart(\sind-1)]{\bconfirmed}$ and that, by the inductive hypothesis, $\canonical[blck={\var[val=v,time=\slotstart(\sind-1)]{\bconfirmed}},time=\slotstart(\sind),]$,
        we can apply \Cref{lem:output-find-latest-canonical-canonical-at-the-start-of-an-epoch} to conclude that
        \begin{enumerate}[resume]
            \item $\canonical[blck={ \bconfirmed[val=v,time=\slotstart(\sind)]},time=\slotstart(\sind),to=\slotstart(\epoch(t)+1)]$
            \item  $\sir({ \bconfirmed[val=v,time=\slotstart(\sind)]},\slotstart(\sind),\guattime[time=\slotstart(\prevfirstslotepoch{t}),val=v],{ \var[val=v,time=\slotstart(\sind-1)]{\bconfirmed}}).\{\sirone,\sirtwo,\sirthree,\sirsix\}$ holds.
            \item  \label{itm:lem:canonical-no-extra-assum:ind-2} for any justified checkpoint $C$, $\epoch(C) \in [\epoch(\guattime[time=\slotstart(\firstslot(\epoch(t)-1)),val=v]),\epoch(t)-1] \implies C  \succeq \guattime[time=\slotstart(\firstslot(\epoch(t)-1)),val=v]$.
        \end{enumerate}

        \begin{description}
            \item\item[Case 1: ${\epoch(\bconfirmed[val=v,time=\slotstart(\sind)])=\epoch(t)-1}$.]
            Given we assume $\bconfirmed[val=v,time=\slotstart(\sind)] \neq \var[val=v,time=\slotstart(\sind-1)]{\bconfirmed}$, we can apply \Cref{lem:conf-prev-epoch-in-current-epoch-exists-succ-with-gu-from-prev-epoch} to ensure $\sirfour(\bconfirmed[val=v,time=\slotstart(\sind)])$.
            
            Then,  $\canonical[blck={ \bconfirmed[val=v,time=\slotstart(\sind)]},time=\tind,to=\slotstart(\epoch(t)+1)]$, \Cref{lem:conf-prev-epoch-in-current-epoch-no-nonsucc-chkp-can-be-justified} and condition~\ref{itm:lem:canonical-no-extra-assum:ind-3} above imply $\sirfive(\bconfirmed[val=v,time=\slotstart(\sind)],\epoch(t)-1,\epoch(t)) = \sirfive(\bconfirmed[val=v,time=\slotstart(\sind)],\epoch(\bconfirmed[val=v,time=\slotstart(\sind)]),\epoch(\bconfirmed[val=v,time=\slotstart(\sind)])+1)$.
            Hence, $\sir({ \bconfirmed[val=v,time=\slotstart(\sind)]},\tind,\guattime[time=\slotstart(\prevfirstslotepoch{t}),val=v],{ \var[val=v,time=\slotstart(\sind-1)]{\bconfirmed}})$ holds.
            Now, $\epoch(\bconfirmed[val=v,time=\slotstart(\sind)])+2=\epoch(t)+1$ and \Cref{lem:ffg-safety-from-sir} imply  $\canonical[blck={\bconfirmed[val=v,time=\slotstart(\sind)]},time=t]$.

            \item [Case 2: ${\epoch(\bconfirmed[val=v,time=\slotstart(\sind)])=\epoch(t)}$.]
            Let $v'$ be any honest validator and $t' \in [\slotstart(\sind), \slotstart(\epoch(t)+2)]$.
            \Cref{lem:no-curr-epochconflict-chkp-is-justified,lem:conf-beginning-epoch-prec-bcand}, Lemma~\ref{lem:sufficient-condition-for-justification}, and $\canonical[blck={ \bconfirmed[val=v,time=\slotstart(\sind)]},time=\slotstart(\sind),to=\slotstart(\epoch(t)+1)]$ ensure that any checkpoint $C$ s.t.
            $C$ is justified and $\epoch(C)=\epoch(t)$, then $C = \chkp( \bconfirmed[val=v,time=\slotstart(\sind)])$.
            This and condition~\ref{itm:lem:canonical-no-extra-assum:ind-1} imply
            that, for any justified checkpoint $C$, $\epoch(C) \in [\epoch(\guattime[time=\slotstart(\firstslot(\epoch(t)-1)),val=v]),\epoch(t)] \implies C  \preceq \bconfirmed[val=v,time=\slotstartslot{t}]$.
            Then, this,
            $\epoch(\gjattime[time=t',val=v']) \in [\epoch(\guattime[time=\slotstart(\firstslot(\epoch(t)-1)),val=v]),\epoch(t)]$, the fact that \sirone implies $\guattime[time=\slotstart(\prevfirstslotepoch{t}),val=v]\preceq \bconfirmed[val=v,time=\slotstart(\sind)]$, 
            and condition~\ref{itm:lem:canonical-no-extra-assum:ind-2} above, imply that  $\gjattime[time=t',val=v']  \preceq \bconfirmed[val=v,time=\slotstart(\sind)]$ and $\gjattime[time=t',val=v']  \succeq \guattime[time=\slotstart(\prevfirstslotepoch{t}),val=v]$.
            Then, $\sirthree(\guattime[time=\slotstart(\prevfirstslotepoch{t}),val=v],\slotstart(\sind),\slotstart(\epoch(t)+2))$ holds.
            \Cref{lem:conf-prev-epoch-then-vs-two-epochs-ago,lem:conf-current-epoch-then-gu-curr-epoch} then imply that $\sirtwo(\bconfirmed[val=v,time=\slotstart(\sind)],\slotstart(\sind),\slotstart(\epoch(t)+2))$ holds as well.
            Then, \Cref{lem:ffg-condition-on-q-implies-safety} implies that $\canonical[blck={ \bconfirmed[val=v,time=\slotstart(\sind)]},time=\tind,to=\slotstart(\epoch(t)+2)]$.
            All the above then implies that $\sirfive(\bconfirmed[val=v,time=\slotstart(\sind)],\epoch(t),\epoch(t)+1) = \sirfive(\bconfirmed[val=v,time=\slotstart(\sind)],\epoch(\bconfirmed[val=v,time=\slotstart(\sind)]),\epoch(\bconfirmed[val=v,time=\slotstart(\sind)])+1)$ holds.
            Finally, \Cref{lem:no-curr-epochconflict-chkp-is-justified,lem:conf-beginning-epoch-prec-bcand}, $\canonical[blck={ \bconfirmed[val=v,time=\slotstart(\sind)]},time=\tind,to=\slotstart(\epoch(t)+2)]$, \Cref{assum:ffg-assumptions:justified-checkpoint-next-epoch} imply $\sirfour(\bconfirmed[val=v,time=\slotstart(\sind)])$.
            Now, $\epoch(\bconfirmed[val=v,time=\slotstart(\sind)])+2=\epoch(t)+2$ and \Cref{lem:ffg-safety-from-sir} imply  $\canonical[blck={\bconfirmed[val=v,time=\slotstart(\sind)]},time=\slotstart(\sind)]$.
        \end{description}
    \end{description}        
\end{proof}

The following lemma handles the case where the confirmed block has epoch $< \epoch(t)-1$: it must then equal the block of the finalized checkpoint at the start of the current epoch.
The proof inspects $\findlatestconfirmeddescendant$: if its output has epoch $< \epoch(t)-1$ then the guard at \Cref{ln:if-bcand-e-1} failed, ruling out both the no-reset path and the restart path that requires $\epoch(\bcand) = \epoch(t)-1$.
Hence the reset to the finalized block fired, so $\bcand$ equals the block of the finalized checkpoint, and this checkpoint coincides with the one at $\slotstart(\epoch(t))$ because finalizing epoch $\epoch(t)-1$ would require epoch-$\epoch(t)$ justification that cannot yet exist.

\begin{lemma}\label{lem:epoch-conf-less-prev-then-conf-finalized}
    If
    \begin{enumerate}
        \item $\epoch(\bconfirmed[val=v,time=\slotstart(\slot(t))]) < \epoch(t) - 1$,
    \end{enumerate}
    then
    \begin{enumerate}
        \item $\bconfirmed[val=v,time=\slotstart(\slot(t))] = \block(\gfattime[time=\slotstart(\epoch(t)),val=v])$
    \end{enumerate}
\end{lemma}

\begin{proof}
    If the output has epoch ${<}\; e{-}1$, then $\findlatestconfirmeddescendant$ was not invoked, so the guard at Line~\ref{ln:if-bcand-e-1} failed.
    This requires $\epoch(\bcand) < e{-}1$ at that point, which rules out both the no-reset path and the restart path (where $\epoch(\bcand) = e{-}1$).
    Hence only the reset fired: $\bcand = \block(\gfattime[val=v,time=\slotstart(\slot(t))])
            = \block(\gfattime[time=\slotstart(\epoch(t)),val=v])$, the last equality holding because finalizing epoch~$e{-}1$ requires epoch-$e$ justification, which cannot yet exist.
\end{proof}

The following theorem is the top-level safety result: sufficiently long after $\GGST$, the confirmed block is canonical at the next slot start.
It covers both cases: if the confirmed block is from the previous or current epoch, \Cref{lem:canonical-no-extra-assum} applies; if it is from earlier than the previous epoch, \Cref{lem:epoch-conf-less-prev-then-conf-finalized} shows it equals the finalized block, which is always canonical.

\begin{theorem}\label{thm:safety}
    If
    \begin{enumerate}
        \item $\slotstart(\prevfirstslotepoch[-1]{t})\geq \GGST$,

    \end{enumerate}
    then
    \begin{enumerate}
        \item $\canonical[blck={\bconfirmed[val=v,time=\slotstartslot{t}]},time=\slotstart(\slot(t)+1)]$
    \end{enumerate}
\end{theorem}
\begin{proof}
    Follows from \Cref{lem:epoch-conf-less-prev-then-conf-finalized,lem:canonical-no-extra-assum}.
\end{proof}

\subsection{Liveness of Reconfirmation}

The next lemma establishes an ordering between the greatest justified checkpoint and the greatest unrealized justified checkpoint when the confirmed block is from the current epoch: both are ancestors of the confirmed block, and $\gjattime[val=v,time=\slotstart(\slot(t)-1)]$ is an ancestor-or-equal of $\guattime[val=v,time=\slotstart(\prevfirstslot{\epoch(t)+1})]$.
This ordering is needed for the reconfirmation argument to show that the balance sources used at the beginning of each epoch are consistent.
The proof uses \Cref{lem:canonical-no-extra-assum} to place both checkpoints on the same chain as the confirmed block.

\begin{lemma}\label{lem:gu-prev-slot-descendant-of-gj-prev-slot}
    If
    \begin{enumerate}
        \item $\slotstart(\epoch(t)) \geq \GGST$
        \item $\epoch(\bconfirmed[val=v,time=\slotstart(\slot(t))]) = \epoch(t)$,
    \end{enumerate}
    then
    \begin{enumerate}
        \item $\gjattime[val=v,time=\slotstart(\slot(t)-1)] \preceq \guattime[val=v,time=\slotstart(\prevfirstslot{\epoch(t)+1})]$.
        \item $\gjattime[val=v,time=\slotstart(\slot(t)-1)] \preceq \bconfirmed[val=v,time=\slotstart(\slot(t))]$.
        \item $\guattime[val=v,time=\slotstart(\prevfirstslot{\epoch(t)+1})] \preceq \bconfirmed[val=v,time=\slotstart(\slot(t))]$.
    \end{enumerate}
\end{lemma}

\begin{proof}
    Condition $\epoch(\bconfirmed[val=v,time=\slotstart(\slot(t))]) = \epoch(t)$ implies that $\slot(t) > \firstslot(\epoch(t))$.
    This further implies that $\epoch(\gjattime[val=v,time=\slotstart(\slot(t)-1)]) \in [\epoch(\guattime[val=v,time=\slotstart(\prevfirstslot{\epoch(t)})]),\epoch(t)-1]$.
    We also know that $\epoch(\guattime[val=v,time=\slotstart(\prevfirstslot{\epoch(t)+1})]) \in [\epoch(\guattime[val=v,time=\slotstart(\prevfirstslot{\epoch(t)})]),\epoch(t)]$.
    \Cref{lem:canonical-no-extra-assum} then implies that $\gjattime[val=v,time=\slotstart(\slot(t)-1)] \preceq \bconfirmed[val=v,time=\slotstart(\slot(t))]$ and $\guattime[val=v,time=\slotstart(\prevfirstslot{\epoch(t)+1})]\preceq \bconfirmed[val=v,time=\slotstart(\slot(t))]$.
    Given that $\epoch(\guattime[val=v,time=\slotstart(\prevfirstslot{\epoch(t)+1})])\geq\epoch(\gjattime[val=v,time=\slotstart(\slot(t)-1)] )$, this
    concludes the proof.
\end{proof}

The following two lemmas show that every current-epoch ancestor of the confirmed block has passed the $\isOneConfirmed$ check at some prior slot.
This is because such ancestors are on the chain that Case~1 of $\findlatestconfirmeddescendant$ traversed, and that case requires $\isOneConfirmedExtFrom$ for every block it selects.
The result is used in the reconfirmation lemmas to ensure that the $Q$ indicator exceeds the safety threshold for all ancestors.
The first lemma handles the special case where $b$ is the first block of its epoch ($\epoch(\parent(b))<\epoch(b)$), using the committee weight from the first slot of the epoch in the coefficient of $\beta$.
The second lemma covers the general case, reducing to the first when applicable.

\begin{lemma}\label{lem:bconf-curr-epoch-ancestor-is-one-confirmed-fslot}
    Pick any $b \preceq \bconfirmed[val=v,time=\slotstart(\slot(t))]$.
    If
    \begin{enumerate}
        \item $\epoch(\bconfirmed[val=v,time=\slotstart(\slot(t))]) = \epoch(t)$
        \item $\epoch(b) = \epoch(t)$
        \item\label{hyp:lem:bconf-curr-epoch-ancestor-is-one-confirmed-fslot:3} $\epoch(\parent(b))<\epoch(b)$,
    \end{enumerate}
    then
    \begin{enumerate}
        \item There exists slot $s > \firstslot(\epoch(t))$ such that
        {$\indicatorfromblock[from=b,to=s-1,val=v,when=\slotstart(s),chkp={C}]{\indQ}
        > \frac{1}{2}\left(1 + \frac{\boostweight[chkp={C}] - \attsetweighttobediscsimplefromblock[from=b,to=\slot(b)-1,val=v,when=\slotstart(s),chkp={C}]{\honattsub}}{\commweightfromafterparentblock[from=b,to=s-1,chkp={C}]{\allvals}}\right)
        + \beta
        \frac
            {\commweightfromslot[from=\firstslot(\epoch(b)),to=s-1,chkp={C}]{\allvals}}
            {\commweightfromafterparentblock[from=b,to=s-1,chkp={C}]{\allvals}}
        - \frac
            {\attsetweightfromblock[from=b,to=s-1,val=v,when=\slotstart(s),chkp={C}]{\slashvals}}
            {\commweightfromafterparentblock[from=b,to=s-1,chkp={C}]{\allvals}}$},
        where $C = \guattime[val=v,time=\slotstart(\lastslot(\epoch(s)-1))]$.
    \end{enumerate}
\end{lemma}

\begin{proof}
    Since $\epoch(\bconfirmed[val=v,time=\slotstart(\slot(t))]) = \epoch(t)$, $\bconfirmed[val=v,time=\slotstart(\slot(t))]$ has been produced by Case~1 of $\findlatestconfirmeddescendant$
    (\Cref{ln:case1-pre,ln:case1}), the only case that can return a current-epoch block.
    Reaching \Cref{ln:case1-ret} requires the returned block to lie in $\btcands$, whose
    definition imposes
    $\isOneConfirmedExtFrom(b',\guattime[val=v,time=\prevfirstslotepoch{t}],\cdot)$
    for every $b'$ on the chain Case~1 considered.
    Since $b \preceq \bconfirmed[val=v,time=\slotstart(\slot(t))]$, $b$ lies on this
    chain, so
    $\isOneConfirmedExtFrom(b,\guattime[val=v,time=\prevfirstslotepoch{t}],\cdot)$ holds.

    By hypothesis~\ref{hyp:lem:bconf-curr-epoch-ancestor-is-one-confirmed-fslot:3}, $\epoch(\parent(b)) < \epoch(b)$, so $\isOneConfirmedExt$
    (Algorithm~\ref{alg:findlatestconf-helpers}) evaluates $\isOneConfirmedSpecial$ at $b$.
    Therefore, there exists a slot $s > \firstslot(\epoch(t))$ at which
    $\isOneConfirmedSpecial(b,\guattime[val=v,time=\prevfirstslotepoch{t}],\slotstart(s))$
    held.
    By Definition~\ref{def:isOneConfirmedSpecial}, this is the lemma's inequality with
    checkpoint $C = \guattime[val=v,time=\slotstart(\lastslot(\epoch(s)-1))]$.
\end{proof}

\begin{lemma}\label{lem:bconf-curr-epoch-ancestor-is-one-confirmed}
    Pick any $b \preceq \bconfirmed[val=v,time=\slotstart(\slot(t))]$.
    If
    \begin{enumerate}
        \item $\epoch(\bconfirmed[val=v,time=\slotstart(\slot(t))]) = \epoch(t)$
        \item $\epoch(b) = \epoch(t)$,
    \end{enumerate}
    then
    \begin{enumerate}
        \item There exists slot $s > \firstslot(\epoch(t))$ such that
        {$\indicatorfromblock[from=b,to=s-1,val=v,when=\slotstart(s),chkp={C}]{\indQ}
        > \frac{1}{2}\left(1 + \frac{\boostweight[chkp={C}]-\attsetweighttobediscsimplefromblock[from=b,to=\slot(b)-1,val=v,when=\slotstart(s),chkp={C}]{\honattsub}}{\commweightfromafterparentblock[from=b,to=s-1,chkp={C}]{\allvals}}\right)
        + \beta
        \frac
            {\commweightfromblock[from=b,to=s-1,chkp={C}]{\allvals}}
            {\commweightfromafterparentblock[from=b,to=s-1,chkp={C}]{\allvals}}
        - \frac
            {\attsetweightfromblock[from=b,to=s-1,val=v,when=\slotstart(s),chkp={C}]{\slashvals}}
            {\commweightfromafterparentblock[from=b,to=s-1,chkp={C}]{\allvals}}$},
        where $C = \guattime[val=v,time=\slotstart(\lastslot(\epoch(s)-1))]$.
    \end{enumerate}
\end{lemma}

\begin{proof}
    First, consider the case that $\epoch(\parent(b))<\epoch(b)$.
    Given that $\commweightfromslot[from=\firstslot(\epoch(b)),to=s-1,chkp={C}]{\allvals}\geq\commweightfromblock[from=b,to=s-1,chkp={C}]{\allvals}$, apply \Cref{lem:bconf-curr-epoch-ancestor-is-one-confirmed-fslot}.

    Then, consider the case that $\epoch(\parent(b))=\epoch(b)$.
    The proof proceeds as in \Cref{lem:bconf-curr-epoch-ancestor-is-one-confirmed-fslot}, except that $\isOneConfirmedExt$ (Algorithm~\ref{alg:findlatestconf-helpers}) now evaluates to $\isOneConfirmed$ at $b$ (since $\epoch(\parent(b)) = \epoch(b)$), so Definition~\ref{def:isOneConfirmed} yields the inequality directly.
\end{proof}

The following two lemmas are the core algebraic step of the reconfirmation argument under $\beta \leq 1/4$.
Both take a current-epoch ancestor $b$ of the confirmed block that already enjoys an $\indQ$ bound at some earlier slot $s$ (with the coefficient of $\beta$ scaled by a committee weight from a chosen slot $s'' \leq \slot(b)$) and propagate that bound to the epoch boundary $s' = \firstslot(\epoch(t)+1)$, paying for the additional new-epoch honest votes via $\beta \leq 1/4$ together with Assumptions~\ref{assum:no-change-to-the-validator-set} and~\ref{assum:beta}.
They differ only in the discount subtracted from the threshold's numerator: the first uses the (larger) discount of \Cref{def:attsetweighttobediscfromblock} introduced in Remark~\ref{app:disc-remark}, whereas the second uses the simple discount of \Cref{def:attsetweighttobediscsimplefromblock} adopted by the algorithm.
Only the second is invoked further in the appendix; the first is stated to make explicit that the same algebraic derivation goes through with either choice of discount.
The proof of each is a long algebraic chain that decomposes $\indQ$ at $s'-1$ into the votes already collected by $s-1$ and the votes from honest committee members in the interval $[s, s'-1]$, and applies hypothesis~\ref{itm:4:lem:beta-less-than-quarter-no-reconfirmation-required-ex} to the first piece.

\begin{lemma}\label{lem:beta-less-than-quarter-no-reconfirmation-required-ex}
    Let $s' := \firstslot(\epoch(t)+1)$.
    Pick any block $b$ such that $b \preceq \var[val=v,time=\slotstartslot{t})]{\bconfirmed}$.
    If
    \begin{enumerate}
        \item $\slotstart(\epoch(t)) \geq \GGST$
        \item $\epoch(b) = \epoch(t)$
        \item  $\beta  \leq \frac{1}{4}$
        \item\label{itm:4:lem:beta-less-than-quarter-no-reconfirmation-required-ex} there exists a slot $s > \firstslot(\epoch(t))$ and a slot $s''\leq\slot(b)$ such that
            {$\indicatorfromblock[from=b,to=s-1,val=v,when=\slotstart(s),chkp={C}]{\indQ}
        > \frac{1}{2}\left(1 + \frac{\boostweight[chkp={C}]-\attsetweighttobediscfromblock[from=b,to=\slot(b)-1,val=v,when=\slotstart(s),chkp={C}]{\honattsub}}{\commweightfromafterparentblock[from=b,to=s-1,chkp={C}]{\allvals}}\right)
        + \beta
        \frac
            {\commweightfromslot[from=s'',to=s-1,chkp=C]{\allvals}}
            {\commweightfromafterparentblock[from=b,to=s-1,chkp={C}]{\allvals}}
        - \frac
            {\attsetweightfromblock[from=b,to=s-1,val=v,when=\slotstart(s),chkp={C}]{\slashvals}}
            {\commweightfromafterparentblock[from=b,to=s-1,chkp={C}]{\allvals}}$},
        where $C = \guattime[val=v,time=\slotstart(\lastslot(\epoch(s)-1))]$.
    \end{enumerate}
    then
    \begin{enumerate}
        \item {$
            \indicatorfromblock[from=b,to=s'-1,val=v,when=\slotstart(s'),chkp={C}]{\indQ}
            > \frac{1}{2}\left(1 + \frac{\boostweight[chkp={C}]-\attsetweighttobediscfromblock[from=b,to=\slot(b)-1,val=v,when=\slotstart(s'),chkp=C]{\honattsub}}{\commweightfromafterparentblock[from=b,to=s'-1,chkp={C}]{\allvals}}\right)
            + \beta
                    \frac{\commweightfromslot[from=s'',to=s'-1,chkp=C]{\allvals}}
                        {\commweightfromafterparentblock[from=b,to=s'-1,chkp=C]{\allvals}}            
            - \frac
                {\attsetweightfromblock[from=b,to=s'-1,val=v,when=\slotstart(s'),chkp={C}]{\slashvals}}
                {\commweightfromafterparentblock[from=b,to=s'-1,chkp={C}]{\allvals}}$}
    \end{enumerate}
\end{lemma}

\begin{proof}
    Let $s:=\slot(t)$.
    \def\alignexplwidth{5cm}
    \allowdisplaybreaks
    \begin{align*}
        &\hspace{3ex} \indicatorfromblock[from=b,to=s'-1,val=v,when=\slotstart(s'),chkp={C}]{\indQ}\\  
        &=
        \frac
            {\attsetweightfromblock[from=b,to=s'-1,val=v,when=\slotstart(s'),chkp={C}]{\allatts}}
            {\commweightfromafterparentblock[from=b,to=s'-1,chkp={C}]{\allvals}}
        &&\alignexpl{By definition.}
        \\
        &\geq
        \frac
            {\commweightfromslot[from={s},to=s'-1,val=v,when=\slotstart(s'),chkp={C}]{\honvals}}
            {\commweightfromafterparentblock[from=b,to=s'-1,chkp={C}]{\allvals}}
        +
        \frac
            {\attsetweightfromblock[from=b,to=s-1,val=v,when=\slotstart(s'),chkp={C}]{\allatts}}
            {\commweightfromafterparentblock[from=b,to=s'-1,chkp={C}]{\allvals}}
        &&\alignexpl[\alignexplwidth]{First, $\attsetweightfromblock[from=b,to=s'-1,val=v,when=\slotstart(s'),chkp={C}]{\allatts} \geq \weightofset[chkp=C]{\attsetfromblock[from=b,to=s-1,val=v,when=\slotstart(s')]{\allatts}\cap \commfromblock[from=\slot(b),to=s-1]{\allvals}} + \commweightfromslot[from={s},to=s'-1,val=v,when=\slotstart(s'),chkp={C}]{\honvals}$ because $\epoch(b) = \epoch(t)$ and, due to Lemma's hypothesis~\ref{itm:4:lem:beta-less-than-quarter-no-reconfirmation-required-ex} and \Cref{lem:canonical-no-extra-assum}, $\canonical[blck=b,time=\slotstart(s)]$. 
        Then 
        $\weightofset[chkp=C]{\attsetfromblock[from=b,to=s-1,val=v,when=\slotstart(s')]{\allatts}\cap \commfromblock[from=\slot(b),to=s-1]{\allvals}} = \attsetweightfromblock[from=b,to=s-1,val=v,when=\slotstart(s'),chkp={C}]{\allatts}$ because no validators in $\commfromslot[from=\parentslotplusone(b),to=\slot(b)-1]{\allvals} \setminus \commfromslot[from=\slot(b),to=s-1]{\allvals}$
        can have cast a valid \LMDGHOST vote for $b$.
        }
        \\
        &=
        \begin{aligned}[t]
            &\frac
                {\commweightfromslot[from={s},to=s'-1,val=v,when=\slotstart(s'),chkp={C}]{\honvals}}
                {\commweightfromafterparentblock[from=b,to=s'-1,chkp={C}]{\allvals}}
            \\
            &+
            \frac
                {\attsetweightfromblock[from=b,to=s-1,val=v,when=\slotstart(s),chkp={C}]{\allatts} - \left(\attsetweightfromblock[from=b,to=s-1,val=v,when=\slotstart(s'),chkp={C}]{\slashvals}-\attsetweightfromblock[from=b,to=s-1,val=v,when=\slotstart(s),chkp={C}]{\slashvals}\right)}
                {\commweightfromafterparentblock[from=b,to=s'-1,chkp={C}]{\allvals}}
        \end{aligned}
        &&\alignexpl[\alignexplwidth]{As, $\attsetfromblockunfiltered[from=b,to=s-1,val=v,when=\slotstart(s')]{\allatts} = \attsetfromblockunfiltered[from=b,to=s-1,val=v,when=\slotstart(s)]{\allatts} \\ \setminus \left(\attsetfromblockunfiltered[from=b,to=s-1,val=v,when=\slotstart(s')]{\slashvals} \\ \setminus \attsetfromblockunfiltered[from=b,to=s-1,val=v,when=\slotstart(s)]{\slashvals}\right)$.}
        \\
        &=
        \begin{aligned}[t]
            &\frac
                {\commweightfromslot[from={s},to=s'-1,val=v,when=\slotstart(s'),chkp={C}]{\honvals} - \attsetweightfromblock[from=b,to=s-1,val=v,when=\slotstart(s'),chkp={C}]{\slashvals}}
                {\commweightfromafterparentblock[from=b,to=s'-1,chkp={C}]{\allvals}}
            \\
            &+
            \frac
                {\attsetweightfromblock[from=b,to=s-1,val=v,when=\slotstart(s),chkp={C}]{\allatts}}
                {\commweightfromafterparentblock[from=b,to=s'-1,chkp={C}]{\allvals}}
            +
            \frac
                {\attsetweightfromblock[from=b,to=s-1,val=v,when=\slotstart(s),chkp={C}]{\slashvals}}
                {\commweightfromafterparentblock[from=b,to=s'-1,chkp={C}]{\allvals}}
        \end{aligned}
        &&\alignexpl[\alignexplwidth]{By simplification.}
        \\
        &=
        \begin{aligned}[t]
            &\frac
                {\commweightfromslot[from={s},to=s'-1,val=v,when=\slotstart(s'),chkp={C}]{\honvals} - \attsetweightfromblock[from=b,to=s-1,val=v,when=\slotstart(s'),chkp={C}]{\slashvals}}
                {\commweightfromafterparentblock[from=b,to=s'-1,chkp={C}]{\allvals}}
            \\
            &+
            \frac
                {\indicatorfromblock[from=b,to=s-1,val=v,when=\slotstart(s),chkp={C}]{\indQ} \commweightfromafterparentblock[from=b,to=s-1,chkp={C}]{\allvals}
                }
                {\commweightfromafterparentblock[from=b,to=s'-1,chkp={C}]{\allvals}}
            +
            \frac
                {\attsetweightfromblock[from=b,to=s-1,val=v,when=\slotstart(s),chkp={C}]{\slashvals}}
                {\commweightfromafterparentblock[from=b,to=s'-1,chkp={C}]{\allvals}}
        \end{aligned}
        &&\alignexpl[\alignexplwidth]{By definition.}
        \\
        &>
        \begin{aligned}[t]
            &\frac
                {\commweightfromslot[from={s},to=s'-1,val=v,when=\slotstart(s'),chkp={C}]{\honvals} - \attsetweightfromblock[from=b,to=s-1,val=v,when=\slotstart(s'),chkp={C}]{\slashvals}}
                {\commweightfromafterparentblock[from=b,to=s'-1,chkp={C}]{\allvals}}
            \\
            &+
            \frac
                {\left(\frac{1}{2}\left(1 + \frac{\boostweight[chkp=C]-\attsetweighttobediscfromblock[from=b,to=\slot(b)-1,val=v,when=\slotstart(s),chkp=C]{\honattsub}}{\commweightfromafterparentblock[from=b,to=s-1,chkp={C}]{\allvals}}\right) 
                + \beta
                \frac
                    {\commweightfromslot[from=s'',to=s-1,chkp=C]{\allvals}}
                    {\commweightfromafterparentblock[from=b,to=s-1,chkp=C]{\allvals}}  
                - \frac{\attsetweightfromblock[from=b,to=s-1,val=v,when=\slotstart(s),chkp={C}]{\slashvals}}{\commweightfromafterparentblock[from=b,to=s-1,chkp={C}]{\allvals}}\right) \commweightfromafterparentblock[from=b,to=s-1,chkp={C}]{\allvals}}
                {\commweightfromafterparentblock[from=b,to=s'-1,chkp={C}]{\allvals}}
            \\
            &+
            \frac
                {\attsetweightfromblock[from=b,to=s-1,val=v,when=\slotstart(s),chkp={C}]{\slashvals}}
                {\commweightfromafterparentblock[from=b,to=s'-1,chkp={C}]{\allvals}}
        \end{aligned}
        &&\alignexpl[\alignexplwidth]{Due to Lemma's hypothesis~\ref{itm:4:lem:beta-less-than-quarter-no-reconfirmation-required-ex}.}
        \\
        &=
        \begin{aligned}[t]
            &\frac
                {\commweightfromslot[from={s},to=s'-1,val=v,when=\slotstart(s'),chkp={C}]{\honvals} - \attsetweightfromblock[from=b,to=s-1,val=v,when=\slotstart(s'),chkp={C}]{\slashvals}}
                {\commweightfromafterparentblock[from=b,to=s'-1,chkp={C}]{\allvals}}
            \\
            &+
            \frac
                {\frac{1}{2}\left(\commweightfromafterparentblock[from=b,to=s-1,chkp={C}]{\allvals} 
                + \boostweight[chkp=C]
                -\attsetweighttobediscfromblock[from=b,to=\slot(b)-1,val=v,when=\slotstart(s),chkp=C]{\honattsub}\right) 
                + \beta \commweightfromslot[from=s'',to=s-1,chkp=C]{\allvals}-\attsetweightfromblock[from=b,to=s-1,val=v,when=\slotstart(s),chkp={C}]{\slashvals}}
                {\commweightfromafterparentblock[from=b,to=s'-1,chkp={C}]{\allvals}}
            \\
            &+
            \frac
                {\attsetweightfromblock[from=b,to=s-1,val=v,when=\slotstart(s),chkp={C}]{\slashvals}}
                {\commweightfromafterparentblock[from=b,to=s'-1,chkp={C}]{\allvals}}
        \end{aligned}
        &&\alignexpl[\alignexplwidth]{By simplification.}
        \\       
        &\geq
        \begin{aligned}[t]
            &\frac
                {\commweightfromslot[from={s},to=s'-1,val=v,when=\slotstart(s'),chkp={C}]{\honvals} - \attsetweightfromblock[from=b,to=s-1,val=v,when=\slotstart(s'),chkp={C}]{\slashvals}}
                {\commweightfromafterparentblock[from=b,to=s'-1,chkp={C}]{\allvals}}
            \\
            &+
            \frac
                {\frac{1}{2}\left(\commweightfromafterparentblock[from=b,to=s-1,chkp={C}]{\allvals} 
                + \boostweight[chkp=C]
                -\attsetweighttobediscfromblock[from=b,to=\slot(b)-1,val=v,when=\slotstart(s),chkp=C]{\honattsub}\right) + \beta \commweightfromslot[from=s'',to=s-1,chkp=C]{\allvals}}
                {\commweightfromafterparentblock[from=b,to=s'-1,chkp={C}]{\allvals}}
        \end{aligned}
        &&\alignexpl[\alignexplwidth]{By simplification.}
        \\
        &\geq
        \begin{aligned}[t]
            &\frac
                {\commweightfromslot[from={s},to=s'-1,val=v,when=\slotstart(s'),chkp={C}]{\honvals} - \attsetweightfromblock[from=b,to=s-1,val=v,when=\slotstart(s'),chkp={C}]{\slashvals}}
                {\commweightfromafterparentblock[from=b,to=s'-1,chkp={C}]{\allvals}}
            \\
            &+
            \frac
                {\frac{1}{2}\left(\commweightfromafterparentblock[from=b,to=s-1,chkp={C}]{\allvals}
                -\attsetweighttobediscfromblock[from=b,to=\slot(b)-1,val=v,when=\slotstart(s'),chkp=C]{\honattsub}
                + \boostweight[chkp=C]
                \right) + \beta \commweightfromslot[from=s'',to=s-1,chkp=C]{\allvals}}
                {\commweightfromafterparentblock[from=b,to=s'-1,chkp={C}]{\allvals}}
        \end{aligned}
        &&\alignexpl[\alignexplwidth]{
            $
            \attsetweighttobediscfromblock[from=b',to=\slot(b')-1,val=v,when=\slotstart(s'),chkp=C]{\allatts}
            \geq 
            \attsetweighttobediscfromblock[from=b',to=\slot(b')-1,val=v,when=\slotstart(s),chkp=C]{\allatts}
            - \left(\attsetweightfromblock[from=b',to=\slot(b)-1,val=v,when=\slotstart(s'),chkp=C]{\slashvals}-
            \attsetweightfromblock[from=b',to=\slot(b)-1,val=v,when=\slotstart(s),chkp=C]{\slashvals}\right)
            $.
            So,
            $
            \attsetweighttobediscfromblock[from=b',to=\slot(b')-1,val=v,when=\slotstart(s),chkp=C]{\allatts} + \attsetweightfromblock[from=b',to=\slot(b)-1,val=v,when=\slotstart(s),chkp=C]{\slashvals}
            \leq
            \attsetweighttobediscfromblock[from=b',to=\slot(b')-1,val=v,when=\slotstart(s'),chkp=C]{\allatts}+ \attsetweightfromblock[from=b',to=\slot(b)-1,val=v,when=\slotstart(s'),chkp=C]{\slashvals}$.
        }           
        \\
        &\geq
        \begin{aligned}[t]
            &\frac
                {\left(1-\beta\right) \commweightfromblock[from=s,to=s'-1,chkp={C}]{\allvals} - \attsetweightfromblock[from=b,to=s-1,val=v,when=\slotstart(s'),chkp={C}]{\slashvals}}
                {\commweightfromafterparentblock[from=b,to=s'-1,chkp={C}]{\allvals}}
            \\
            &+
            \frac
                {\frac{1}{2}\left(\commweightfromafterparentblock[from=b,to=s-1,chkp={C}]{\allvals}
                -\attsetweighttobediscfromblock[from=b,to=\slot(b)-1,val=v,when=\slotstart(s'),chkp=C]{\honattsub} 
                + \boostweight[chkp=C]\right) + \beta \commweightfromslot[from=s'',to=s-1,chkp=C]{\allvals}}
                {\commweightfromafterparentblock[from=b,to=s'-1,chkp={C}]{\allvals}}
        \end{aligned}
        &&\alignexpl[\alignexplwidth]{By Assumption 2 of the paper.}
        \\
        &\geq
        \begin{aligned}[t]
            &\frac
                {\left(\frac{1}{2}+\beta \right) \commweightfromblock[from=s,to=s'-1,chkp={C}]{\allvals} - \attsetweightfromblock[from=b,to=s-1,val=v,when=\slotstart(s'),chkp={C}]{\slashvals}}
                {\commweightfromafterparentblock[from=b,to=s'-1,chkp={C}]{\allvals}}
            \\
            &+
            \frac
                {\frac{1}{2}\left(\commweightfromafterparentblock[from=b,to=s-1,chkp={C}]{\allvals}
                -\attsetweighttobediscfromblock[from=b,to=\slot(b)-1,val=v,when=\slotstart(s'),chkp=C]{\honattsub}  
                + \boostweight[chkp=C]\right) + \beta \commweightfromslot[from=s'',to=s-1,chkp=C]{\allvals}}
                {\commweightfromafterparentblock[from=b,to=s'-1,chkp={C}]{\allvals}}
        \end{aligned}
        &&\alignexpl[\alignexplwidth]{Given that $\beta \leq \frac{1}{4}$.}        
        \\
        &=
        \begin{aligned}[t]
            &\frac
                {\frac{1}{2}\left(\commweightfromafterparentblock[from=b,to=s-1,chkp={C}]{\allvals} + \commweightfromblock[from=s,to=s'-1,chkp={C}]{\allvals}
                -\attsetweighttobediscfromblock[from=b,to=\slot(b)-1,val=v,when=\slotstart(s'),chkp=C]{\honattsub}  
                + \boostweight[chkp=C]\right)}
                {\commweightfromafterparentblock[from=b,to=s'-1,chkp={C}]{\allvals}}
            \\
            &+
            \frac
                {\left(\commweightfromslot[from=s'',to=s-1,chkp=C]{\allvals}+ \commweightfromblock[from=s,to=s'-1,chkp={C}]{\allvals}\right) \beta - \attsetweightfromblock[from=b,to=s-1,val=v,when=\slotstart(s'),chkp={C}]{\slashvals}}
                {\commweightfromafterparentblock[from=b,to=s'-1,chkp={C}]{\allvals}}
        \end{aligned}
        &&\alignexpl[\alignexplwidth]{By simplification.}     
        \\        
        &\geq
        \begin{aligned}[t]
            &\frac
                {\frac{1}{2}\left(\commweightfromafterparentblock[from=b,to=s'-1,chkp={C}]{\allvals} 
                -\attsetweighttobediscfromblock[from=b,to=\slot(b)-1,val=v,when=\slotstart(s'),chkp=C]{\honattsub} 
                + \boostweight[chkp=C]\right)}
                {\commweightfromafterparentblock[from=b,to=s'-1,chkp={C}]{\allvals}}
            \\
            &+
            \frac
                {\commweightfromslot[from=s'',to=s'-1,chkp={C}]{\allvals} \beta - \attsetweightfromblock[from=b,to=s-1,val=v,when=\slotstart(s'),chkp={C}]{\slashvals}}
                {\commweightfromafterparentblock[from=b,to=s'-1,chkp={C}]{\allvals}}
        \end{aligned}
        &&\alignexpl[\alignexplwidth]{As, $\commfromafterparentblock[from=b,to=s'-1]{\allvals} = \commfromafterparentblock[from=b,to=s-1]{\allvals} \cup \commfromblock[from=s,to=s'-1]{\allvals}$ and 
        $\commfromblock[from=b,to=s'-1]{\allvals} = \commfromblock[from=b,to=s-1]{\allvals} \cup \commfromblock[from=s,to=s'-1]{\allvals}$.}         
        \\
        &=
            \frac{1}{2}\left(1 + \frac{\boostweight[chkp=C]-\attsetweighttobediscfromblock[from=b,to=\slot(b)-1,val=v,when=\slotstart(s'),chkp=C]{\honattsub} }
            {\commweightfromafterparentblock[from=b,to=s'-1,chkp={C}]{\allvals}
            }\right)
            + \beta
            \frac
                {\commweightfromslot[from=s'',to=s'-1,chkp={C}]{\allvals}}
                {\commweightfromafterparentblock[from=b,to=s'-1,chkp={C}]{\allvals}}
            - \frac
                {\attsetweightfromblock[from=b,to=s-1,val=v,when=\slotstart(s'),chkp={C}]{\slashvals}}
                {\commweightfromafterparentblock[from=b,to=s'-1,chkp={C}]{\allvals}}
        &&\alignexpl[\alignexplwidth]{By simplification.}
    \end{align*}
\end{proof}

\begin{lemma}\label{lem:beta-less-than-quarter-no-reconfirmation-required-ex-with-simple}
    Let $s' := \firstslot(\epoch(t)+1)$.
    Pick any block $b$ such that $b \preceq \var[val=v,time=\slotstartslot{t})]{\bconfirmed}$.
    If
    \begin{enumerate}
        \item $\slotstart(\epoch(t)) \geq \GGST$
        \item $\epoch(b) = \epoch(t)$
        \item  $\beta  \leq \frac{1}{4}$
        \item\label{itm:4:lem:beta-less-than-quarter-no-reconfirmation-required-ex-with-simple} there exists a slot $s > \firstslot(\epoch(t))$ and a slot $s''\leq\slot(b)$ such that
            {$\indicatorfromblock[from=b,to=s-1,val=v,when=\slotstart(s),chkp={C}]{\indQ}
        > \frac{1}{2}\left(1 + \frac{\boostweight[chkp={C}]-\attsetweighttobediscsimplefromblock[from=b,to=\slot(b)-1,val=v,when=\slotstart(s),chkp={C}]{\honattsub}}{\commweightfromafterparentblock[from=b,to=s-1,chkp={C}]{\allvals}}\right)
        + \beta
        \frac
            {\commweightfromslot[from=s'',to=s-1,chkp=C]{\allvals}}
            {\commweightfromafterparentblock[from=b,to=s-1,chkp={C}]{\allvals}}
        - \frac
            {\attsetweightfromblock[from=b,to=s-1,val=v,when=\slotstart(s),chkp={C}]{\slashvals}}
            {\commweightfromafterparentblock[from=b,to=s-1,chkp={C}]{\allvals}}$},
        where $C = \guattime[val=v,time=\slotstart(\lastslot(\epoch(s)-1))]$.
    \end{enumerate}
    then
    \begin{enumerate}
        \item {$
            \indicatorfromblock[from=b,to=s'-1,val=v,when=\slotstart(s'),chkp={C}]{\indQ}
            > \frac{1}{2}\left(1 + \frac{\boostweight[chkp={C}]-\attsetweighttobediscsimplefromblock[from=b,to=\slot(b)-1,val=v,when=\slotstart(s'),chkp=C]{\honattsub}}{\commweightfromafterparentblock[from=b,to=s'-1,chkp={C}]{\allvals}}\right)
            + \beta
                    \frac{\commweightfromslot[from=s'',to=s'-1,chkp=C]{\allvals}}
                        {\commweightfromafterparentblock[from=b,to=s'-1,chkp=C]{\allvals}}            
            - \frac
                {\attsetweightfromblock[from=b,to=s'-1,val=v,when=\slotstart(s'),chkp={C}]{\slashvals}}
                {\commweightfromafterparentblock[from=b,to=s'-1,chkp={C}]{\allvals}}$}
    \end{enumerate}
\end{lemma}

\begin{proof}
    Note that 
    $
            \attsetweighttobediscsimplefromblock[from=b',to=\slot(b')-1,val=v,when=\slotstart(s'),chkp=C]{\allatts}
            \geq 
            \attsetweighttobediscsimplefromblock[from=b',to=\slot(b')-1,val=v,when=\slotstart(s),chkp=C]{\allatts}
            - \left(\attsetweightfromblock[from=b',to=\slot(b)-1,val=v,when=\slotstart(s'),chkp=C]{\slashvals}-
            \attsetweightfromblock[from=b',to=\slot(b)-1,val=v,when=\slotstart(s),chkp=C]{\slashvals}\right)
            $.
            So,
            $
            \attsetweighttobediscsimplefromblock[from=b',to=\slot(b')-1,val=v,when=\slotstart(s),chkp=C]{\allatts} + \attsetweightfromblock[from=b',to=\slot(b)-1,val=v,when=\slotstart(s),chkp=C]{\slashvals}
            \leq
            \attsetweighttobediscsimplefromblock[from=b',to=\slot(b')-1,val=v,when=\slotstart(s'),chkp=C]{\allatts}+ \attsetweightfromblock[from=b',to=\slot(b)-1,val=v,when=\slotstart(s'),chkp=C]{\slashvals}$.
    Then, flollow the proof of \Cref{lem:beta-less-than-quarter-no-reconfirmation-required-ex} and use the condition above to conclude that $\attsetweighttobediscsimplefromblock[from=b,to=\slot(b)-1,val=v,when=\slotstart(s'),chkp=C]{\honattsub}\geq\attsetweighttobediscsimplefromblock[from=b,to=\slot(b)-1,val=v,when=\slotstart(s),chkp=C]{\honattsub}$ when required.
\end{proof}

The following reconfirmation lemma shows that every current-epoch ancestor of the confirmed block satisfies the $\isOneConfirmed$ threshold at the epoch boundary under $\beta \leq 1/4$.
The proof combines \Cref{lem:bconf-curr-epoch-ancestor-is-one-confirmed}, which provides the $Q$ bound at an earlier slot, with \Cref{lem:beta-less-than-quarter-no-reconfirmation-required-ex-with-simple}, which propagates that bound to the epoch boundary.

\begin{lemma}\label{lem:beta-less-than-quarter-no-reconfirmation-required-bconf-curr-epoch-ex}
    Let $s := \firstslot(\epoch(t)+1)$, $\chkp := \guattime[val=v,time=\slotstart(s-1)]$.
    If
    \begin{enumerate}
        \item $\slotstart(\epoch(t)) \geq \GGST$
        \item $\epoch(\var[val=v,time=\slotstartslot{t}]{\bconfirmed}) = \epoch(t)$
        \item $\beta \leq \frac{1}{4}$,
    \end{enumerate}
    then, $\forall b' \preceq \var[val=v,time=\slotstartslot{t}]{\bconfirmed}$ such that $\epoch(b') = \epoch(t)$,
    \begin{enumerate}
        \item {$\indicatorfromblock[from=b',to=s-1,val=v,when=\slotstart(s),chkp={C}]{\indQ}
        > \frac{1}{2}\left(1 + \frac{\boostweight[chkp={C}]-\attsetweighttobediscsimplefromblock[from=b,to=\slot(b)-1,val=v,when=\slotstart(s),chkp={C}]{\honattsub}}{\commweightfromafterparentblock[from=b',to=s-1,chkp={C}]{\allvals}}\right)
        + \beta
        \frac
            {\commweightfromblock[from=b',to=s-1,chkp={C}]{\allvals}}
            {\commweightfromafterparentblock[from=b',to=s-1,chkp={C}]{\allvals}}
        - \frac
            {\attsetweightfromblock[from=b',to=s-1,val=v,when=\slotstart(s),chkp={C}]{\slashvals}}
            {\commweightfromafterparentblock[from=b',to=s-1,chkp={C}]{\allvals}}$}
    \end{enumerate}
\end{lemma}

\begin{proof}
    Follows from \Cref{lem:bconf-curr-epoch-ancestor-is-one-confirmed,lem:beta-less-than-quarter-no-reconfirmation-required-ex-with-simple}.
\end{proof}

The following reconfirmation lemma extends the bound to previous-epoch ancestors of a current-epoch confirmed block.
When the confirmed block is from epoch~$\epoch(t)$ but some of its ancestors are from epoch~$\epoch(t)-1$, this lemma shows these ancestors also satisfy the $\isOneConfirmed$ threshold at the epoch boundary under $\beta \leq 1/4$, with after-parent committee weight in the denominator.
The key insight is that a previous-epoch ancestor $b'$ collects at least as many votes as the first current-epoch block~$b$ on the chain (since $b' \prec b$), and the after-parent committee weight for $b'$ coincides with the total committee weight from $b$ to the end of the current epoch because $\parentslotplusone(b') \leq \firstslot(\epoch(t))$.
The proof obtains the bound for $b$ from \Cref{lem:bconf-curr-epoch-ancestor-is-one-confirmed-fslot} and \Cref{lem:beta-less-than-quarter-no-reconfirmation-required-ex-with-simple}, transfers the extra votes that $b'$ inherits from $b$'s support, and finally collapses denominators using the equality $\commweightfromblock[from=b,to=s-1]{\allvals} = \commweightfromafterparentblock[from=b',to=s-1]{\allvals}$ to obtain the stated conclusion.

\begin{lemma}\label{lem:beta-less-than-quarter-no-reconfirmation-required-bconf-prev-epoch-ex}
    Let $s := \firstslot(\epoch(t)+1)$, $\chkp := \guattime[val=v,time=\slotstart(s-1)]$.
    If
    \begin{enumerate}
        \item $\slotstart(\epoch(t)) \geq \GGST$
        \item $\epoch(\var[val=v,time=\slotstartslot{t}]{\bconfirmed}) = \epoch(t)$
        \item $\beta \leq \frac{1}{4}$,
    \end{enumerate}
    then, $\forall b' \prec \var[val=v,time=\slotstartslot{t}]{\bconfirmed}$ such that $\epoch(b') = \epoch(t)-1$,
    \begin{enumerate}
        \item {
        $\begin{aligned}[t]
            \indicatorfromblock[from=b',to=s-1,val=v,when=\slotstart(s),chkp={C}]{\indQ}
            &> 
            \frac{1}{2}\left(1 + \frac{\boostweight[chkp={C}]-\attsetweighttobediscsimplefromblock[from=b,to=\slot(b)-1,val=v,when=\slotstart(s),chkp={C}]{\honattsub}}{\commweightfromafterparentblock[from=b',to=s-1,chkp={C}]{\allvals}}\right)
            + \beta
            - \frac
                {\attsetweightfromblock[from=b',to=s-1,val=v,when=\slotstart(s),chkp={C}]{\slashvals}}
                {\commweightfromafterparentblock[from=b',to=s-1,chkp={C}]{\allvals}}
                \\
            &\geq        
            \frac{1}{2}\left(1 + \frac{\boostweight[chkp={C}]-\attsetweighttobediscsimplefromblock[from=b,to=\slot(b)-1,val=v,when=\slotstart(s),chkp={C}]{\honattsub}}{\commweightfromafterparentblock[from=b',to=s-1,chkp={C}]{\allvals}}\right)
            + \beta
            \frac
                {\commweightfromblock[from=b',to=s-1,chkp={C}]{\allvals}}
                {\commweightfromafterparentblock[from=b',to=s-1,chkp={C}]{\allvals}}
            - \frac
                {\attsetweightfromblock[from=b',to=s-1,val=v,when=\slotstart(s),chkp={C}]{\slashvals}}
                {\commweightfromafterparentblock[from=b',to=s-1,chkp={C}]{\allvals}}
        \end{aligned}$}
    \end{enumerate}
\end{lemma}

\begin{proof}
    Let $b$ be the block in the chain of $\var[val=v,time=\slotstartslot{t}]{\bconfirmed}$ such that $\epoch(b)=\epoch(t) \land \epoch(\parent(b))<\epoch(t)$.
    With this in mind we proceed as the following:
    \def\alignexplwidth{5cm}
    \allowdisplaybreaks
    \begin{align*}
        &\hspace{3ex}\indicatorfromblock[from=b',to=s-1,val=v,when=\slotstart(s),chkp={C}]{\indQ}\\
        &=
        \frac
            {\attsetweightfromblock[from=b',to=s-1,val=v,when=\slotstart(s),chkp={C}]{\allatts}}
            {\commweightfromafterparentblock[from=b',to=s-1,chkp={C}]{\allvals}}
        &&\alignexpl{By definition.}
        \\
        &=
        \frac
            {\attsetweightfromblock[from=b',to=s-1,val=v,when=\slotstart(s),chkp={C}]{\allatts}}
            {\commweightfromafterparentblock[from=b,to=s-1,chkp={C}]{\allvals}}
        &&\alignexpl{As, $\commweightfromafterparentblock[from=b',to=s-1,chkp={C}]{\allvals} = \commweightfromafterparentblock[from=b,to=s-1,chkp={C}]{\allvals}$.}
        \\
        &\geq
        \frac
            {
                \attsetweightfromblock[from=b,to=s-1,val=v,when=\slotstart(s),chkp={C}]{\allatts}
                +\attsetweighttobediscsimplefromblock[from=b,to=\slot(b)-1,val=v,when=\slotstart(s),chkp=C]{\allatts}
            }
            {\commweightfromafterparentblock[from=b,to=s-1,chkp={C}]{\allvals}}
        &&\alignexpl[\alignexplwidth]{
        First, due to $\FILLMD$, 
        $\attsetfromblockunfiltered[from=b,to=s-1,val=v,when=\slotstart(s)]{\allatts} \cap \attsettobediscsimplefromblock[from=b,to=\slot(b)-1,val=v,when=\slotstart(s)]{\allatts}=\emptyset$.
        Then, given that $b' \preceq \parent(b)$, we have that $\attsetfromblockunfiltered[from=b',to=s-1,val=v,when=\slotstart(s)]{\allatts} \supseteq (\attsetfromblockunfiltered[from=b,to=s-1,val=v,when=\slotstart(s)]{\allatts} \cup \attsettobediscsimplefromblock[from=b,to=\slot(b)-1,val=v,when=\slotstart(s)]{\allatts})=(\attsetfromblockunfiltered[from=b,to=s-1,val=v,when=\slotstart(s)]{\allatts} \sqcup \attsettobediscsimplefromblock[from=b,to=\slot(b)-1,val=v,when=\slotstart(s)]{\allatts})$.
        }   
        \\
        &\geq
        \indicatorfromblock[from=b,to=s-1,val=v,when=\slotstart(s),chkp={C}]{\indQ}
        +
        \frac
            {\attsetweighttobediscsimplefromblock[from=b,to=\slot(b)-1,val=v,when=\slotstart(s),chkp=C]{\allatts}}
            {\commweightfromafterparentblock[from=b,to=s-1,chkp={C}]{\allvals}}
        &&\alignexpl{By definition.}
        \\
        &> 
        \begin{aligned}[t]
            &\frac{1}{2}\left(1 + 
            \frac
                {
                    \boostweight[chkp={C}]
                    -
                    \max
                    \left(
                        \attsetweighttobediscsimplefromblock[from=b,to=\slot(b)-1,val=v,when=\slotstart(s),chkp=C]{\allatts}
                        -\beta \commweightfromafterparentblock[from=b,to=\slot(b)-1,chkp=C]{\allvals}
                        +\attsetweightfromblock[from=b,to=\slot(b)-1,val=v,when=\slotstart(s),chkp=C]{\slashvals}
                        ,
                        0
                    \right)                
                }
                {\commweightfromafterparentblock[from=b,to=s-1,chkp={C}]{\allvals}}\right)
            \\
            &+ \beta
            - \frac
                {\attsetweightfromblock[from=b,to=s-1,val=v,when=\slotstart(s),chkp={C}]{\slashvals}}
                {\commweightfromafterparentblock[from=b,to=s-1,chkp={C}]{\allvals}}
            +
            \frac
                {\attsetweighttobediscsimplefromblock[from=b,to=\slot(b)-1,val=v,when=\slotstart(s),chkp=C]{\allatts}}
                {\commweightfromafterparentblock[from=b,to=s-1,chkp={C}]{\allvals}}                
        \end{aligned}        
        &&\alignexpl[\alignexplwidth]{
            By definition of $b$, $\epoch(\parent(b))<\epoch(b)$.
            Then, apply \Cref{lem:beta-less-than-quarter-no-reconfirmation-required-ex-with-simple,lem:bconf-curr-epoch-ancestor-is-one-confirmed-fslot}.
            Moreover, given that $s = \firstslot(\epoch(t)+1)$,  $\commweightfromslot[from=\firstslot(\epoch(t)),to=s-1,chkp={C}]{\allvals} =  \totvalsetweight[chkp=C]{\allvals}$.
            Finally, because $\parentslotplusone(b)\leq \firstslot(\epoch(t))$,  $\commweightfromafterparentblock[from=b,to=s-1,chkp={C}]{\allvals} =  \totvalsetweight[chkp=C]{\allvals}$.}
    \end{align*}

    \textbf{Case 1: $\attsetweighttobediscsimplefromblock[from=b,to=\slot(b)-1,val=v,when=\slotstart(s),chkp=C]{\allatts}
                        -\beta \commweightfromafterparentblock[from=b,to=\slot(b)-1,chkp=C]{\allvals}
                        +\attsetweightfromblock[from=b,to=\slot(b)-1,val=v,when=\slotstart(s),chkp=C]{\slashvals}
                        >
                        0$.}

    \begin{align*}
            &\hphantom{{}={}}
            \begin{aligned}[t]
            &\frac{1}{2}\left(1 + 
            \frac
                {
                    \boostweight[chkp={C}]
                    -
                    \max
                    \left(
                        \attsetweighttobediscsimplefromblock[from=b,to=\slot(b)-1,val=v,when=\slotstart(s),chkp=C]{\allatts}
                        -\beta \commweightfromafterparentblock[from=b,to=\slot(b)-1,chkp=C]{\allvals}
                        +\attsetweightfromblock[from=b,to=\slot(b)-1,val=v,when=\slotstart(s),chkp=C]{\slashvals}
                        ,
                        0
                    \right)                
                }
                {\commweightfromafterparentblock[from=b,to=s-1,chkp={C}]{\allvals}}\right)
            \\
            &+ \beta
            - \frac
                {\attsetweightfromblock[from=b,to=s-1,val=v,when=\slotstart(s),chkp={C}]{\slashvals}}
                {\commweightfromafterparentblock[from=b,to=s-1,chkp={C}]{\allvals}}
            +
            \frac
                {\attsetweighttobediscsimplefromblock[from=b,to=\slot(b)-1,val=v,when=\slotstart(s),chkp=C]{\allatts}}
                {\commweightfromafterparentblock[from=b,to=s-1,chkp={C}]{\allvals}}                
        \end{aligned}  
        \\              
        &= 
        \begin{aligned}[t]
            &\frac{1}{2}\left(1 + 
            \frac
                {
                    \boostweight[chkp={C}]
                    +
                    \attsetweighttobediscsimplefromblock[from=b,to=\slot(b)-1,val=v,when=\slotstart(s),chkp=C]{\allatts}
                    +
                    \left(
                        \beta \commweightfromafterparentblock[from=b,to=\slot(b)-1,chkp=C]{\allvals}
                        -\attsetweightfromblock[from=b,to=\slot(b)-1,val=v,when=\slotstart(s),chkp=C]{\slashvals}
                    \right)                
                }
                {\commweightfromafterparentblock[from=b,to=s-1,chkp={C}]{\allvals}}\right)
            \\
            &+ \beta
            - \frac
                {\attsetweightfromblock[from=b,to=s-1,val=v,when=\slotstart(s),chkp={C}]{\slashvals}}
                {\commweightfromafterparentblock[from=b,to=s-1,chkp={C}]{\allvals}}               
        \end{aligned}        
        &&\alignexpl{By simplification.}
        \\
        &\geq
        \begin{aligned}[t]
            &\frac{1}{2}\left(1 + 
            \frac
                {
                    \boostweight[chkp={C}]               
                }
                {\commweightfromafterparentblock[from=b,to=s-1,chkp={C}]{\allvals}}\right)
            + \beta
            - \frac
                {\attsetweightfromblock[from=b,to=s-1,val=v,when=\slotstart(s),chkp={C}]{\slashvals}}
                {\commweightfromafterparentblock[from=b,to=s-1,chkp={C}]{\allvals}}               
        \end{aligned}        
        &&\alignexpl[\alignexplwidth]{As $\attsetweighttobediscsimplefromblock[from=b,to=\slot(b)-1,val=v,when=\slotstart(s),chkp=C]{\allatts}\geq 0$ and $\beta \commweightfromafterparentblock[from=b,to=\slot(b)-1,chkp=C]{\allvals}
                        \geq\attsetweightfromblock[from=b,to=\slot(b)-1,val=v,when=\slotstart(s),chkp=C]{\slashvals}$.}        
    \end{align*}

    \textbf{Case 2: $\attsetweighttobediscsimplefromblock[from=b,to=\slot(b)-1,val=v,when=\slotstart(s),chkp=C]{\allatts}
                        -\beta \commweightfromafterparentblock[from=b,to=\slot(b)-1,chkp=C]{\allvals}
                        +\attsetweightfromblock[from=b,to=\slot(b)-1,val=v,when=\slotstart(s),chkp=C]{\slashvals}
                        \leq
                        0$.}

    \begin{align*}
        &\hphantom{{}={}}
            \begin{aligned}[t]
            &\frac{1}{2}\left(1 + 
            \frac
                {
                    \boostweight[chkp={C}]
                    -
                    \max
                    \left(
                        \attsetweighttobediscsimplefromblock[from=b,to=\slot(b)-1,val=v,when=\slotstart(s),chkp=C]{\allatts}
                        -\beta \commweightfromafterparentblock[from=b,to=\slot(b)-1,chkp=C]{\allvals}
                        +\attsetweightfromblock[from=b,to=\slot(b)-1,val=v,when=\slotstart(s),chkp=C]{\slashvals}
                        ,
                        0
                    \right)                
                }
                {\commweightfromafterparentblock[from=b,to=s-1,chkp={C}]{\allvals}}\right)
            \\
            &+ \beta
            - \frac
                {\attsetweightfromblock[from=b,to=s-1,val=v,when=\slotstart(s),chkp={C}]{\slashvals}}
                {\commweightfromafterparentblock[from=b,to=s-1,chkp={C}]{\allvals}}
            +
            \frac
                {\attsetweighttobediscsimplefromblock[from=b,to=\slot(b)-1,val=v,when=\slotstart(s),chkp=C]{\allatts}}
                {\commweightfromafterparentblock[from=b,to=s-1,chkp={C}]{\allvals}}                
        \end{aligned}  
        \\     
        &\geq\begin{aligned}[t]
            &\frac{1}{2}\left(1 + 
            \frac
                {
                    \boostweight[chkp={C}]               
                }
                {\commweightfromafterparentblock[from=b,to=s-1,chkp={C}]{\allvals}}\right)
            + \beta
            - \frac
                {\attsetweightfromblock[from=b,to=s-1,val=v,when=\slotstart(s),chkp={C}]{\slashvals}}
                {\commweightfromafterparentblock[from=b,to=s-1,chkp={C}]{\allvals}}
            +
            \frac
                {\attsetweighttobediscsimplefromblock[from=b,to=\slot(b)-1,val=v,when=\slotstart(s),chkp=C]{\allatts}}
                {\commweightfromafterparentblock[from=b,to=s-1,chkp={C}]{\allvals}}                
        \end{aligned}  
        \\                 
        &\geq
        \begin{aligned}[t]
            &\frac{1}{2}\left(1 + 
            \frac
                {
                    \boostweight[chkp={C}]               
                }
                {\commweightfromafterparentblock[from=b,to=s-1,chkp={C}]{\allvals}}\right)
            + \beta
            - \frac
                {\attsetweightfromblock[from=b,to=s-1,val=v,when=\slotstart(s),chkp={C}]{\slashvals}}
                {\commweightfromafterparentblock[from=b,to=s-1,chkp={C}]{\allvals}}               
        \end{aligned}        
        &&\alignexpl[\alignexplwidth]{As $\attsetweighttobediscsimplefromblock[from=b,to=\slot(b)-1,val=v,when=\slotstart(s),chkp=C]{\allatts}\geq 0$.}        
    \end{align*}                        
                        
    Then, continuing from what we concluded from the two cases above:

    \begin{align*}
        &\hphantom{{}={}}
        \frac{1}{2}\left(1 + 
            \frac
                {
                    \boostweight[chkp={C}]               
                }
                {\commweightfromafterparentblock[from=b,to=s-1,chkp={C}]{\allvals}}\right)
            + \beta
            - \frac
                {\attsetweightfromblock[from=b,to=s-1,val=v,when=\slotstart(s),chkp={C}]{\slashvals}}
                {\commweightfromafterparentblock[from=b,to=s-1,chkp={C}]{\allvals}}  
        \\
        &> \frac{1}{2}\left(1 + 
            \frac
                {\boostweight[chkp={C}]
                }
                {\commweightfromafterparentblock[from=b',to=s-1,chkp={C}]{\allvals}}
            \right)
        + \beta
        - \frac
            {\attsetweightfromblock[from=b,to=s-1,val=v,when=\slotstart(s),chkp={C}]{\slashvals}}
            {\commweightfromafterparentblock[from=b',to=s-1,chkp={C}]{\allvals}}
        &&\alignexpl{As, $\commweightfromafterparentblock[from=b,to=s-1,chkp={C}]{\allvals} = \commweightfromafterparentblock[from=b',to=s-1,chkp={C}]{\allvals}$.}
        \\
        &> \frac{1}{2}\left(1 + \frac{\boostweight[chkp={C}]}{\commweightfromafterparentblock[from=b',to=s-1,chkp={C}]{\allvals}}\right)
        + \beta
        - \frac
            {\attsetweightfromblock[from=b',to=s-1,val=v,when=\slotstart(s),chkp={C}]{\slashvals}}
            {\commweightfromafterparentblock[from=b',to=s-1,chkp={C}]{\allvals}}
        &&\alignexpl{As, $\attsetfromblockunfiltered[from=b,to=s-1,val=v,when=\slotstart(s)]{\slashvals} \subseteq \attsetfromblockunfiltered[from=b',to=s-1,val=v,when=\slotstart(s)]{\slashvals}$.}        
    \end{align*}

    Finally, 
    $        \frac
            {\commweightfromafterparentblock[from=b',to=s-1,chkp={C}]{\allvals}}
            {\commweightfromafterparentblock[from=b',to=s-1,chkp={C}]{\allvals}}\leq 1$ concludes the proof.
\end{proof}

\section{Iterative Version of $\findlatestconfirmeddescendant$}
\label{sec:iterative-version}

Algorithm~\ref{alg:findlatestconf-functional} presents
$\findlatestconfirmeddescendant$ in a declarative style: each case
selects the highest block satisfying a set of conditions from the
canonical chain.
This formulation is used in the main body because it makes the
structure of the five cases explicit and facilitates the safety and
monotonicity arguments.
Algorithm~\ref{alg:findlatestconf} presents an equivalent iterative
formulation that walks the canonical chain block by block, which is
closer to the actual reference implementation deployed in consensus
clients~\cite{fcr-impl}.

We now show that the two formulations produce the same output on every
input.
Both algorithms receive the same input $b_c$ and compute the same
$\head$ and $\phead$.
The iterative version consists of two phases: a
\emph{previous-epoch loop} (Lines~\ref{ln:if-prev-epoch}--\ref{ln:set-bcand-in-prev-epoch-loop})
and a \emph{current-epoch loop}
(Lines~\ref{ln:second-if}--\ref{ln:set-bcand-to-btcand}).
We describe how each phase corresponds to the cases of
Algorithm~\ref{alg:findlatestconf-functional}.

\begin{algorithm}
\caption{Find latest confirmed descendant (iterative version)}
\label{alg:findlatestconf}
\SetAlgoNoLine
\DontPrintSemicolon
\Fn{$\var[val=v]{\mathit{next\_child}}(b,\head)$}{
    $\mathit{extension} \gets \{ b' \in \viewattime[time={\now},val=v], b \prec b' \preceq \head \}$\\
    \uIf{$|\mathit{extension}| > 0$}
    {
        \Return{$\argmin_{b' \in \mathit{extension}} \slot(b')$}
    }
    \uElse{
        \Return{$\bot$}
    }
}

\Fn{$\var[val=v]{\findlatestconfirmeddescendant}(b_c)$}
{
    $\bcand \gets b_c$\nllabel{ln:bcand-set-beginning-of-find-latest}\\
    \Const $\head \gets \LMDGHOSTHFC(\viewattime[time={\now},val=v])$\\
    \Const $\phead \gets \LMDGHOSTHFC(\viewattime[time={\slotstart(\slot(\now)-1)},val=v])$\\

    \uIf{$\begin{aligned}[t]
        &\epoch(\bcand) = \epoch(\now) - 1\\
        &\text{\bf and}\ \epoch(\votsource[blck=\phead, time={\now}]) \geq \epoch(\now) - 2\\
        &\text{\bf and}\ (\slot(\now) = \firstslot(\epoch(\now))\\
        &\qquad \text{\bf or}\ \willNoConflictingChkpBeJustified_v(\chkp(\head,\epoch(\now)))\\
        &\qquad \text{\bf and}\ (\epoch(\gu(\phead)) \geq \epoch(\now) - 1\\
        &\qquad\qquad \text{\bf or}\ \epoch(\gu(\head)) \geq \epoch(\now) - 1)))
    \end{aligned}$\nllabel{ln:if-prev-epoch}\\}
    {
        \While{$\var[val=v]{\mathit{next\_child}}(\bcand,\head) \neq \bot$}
        {
            $\btemp \gets \var[val=v]{\mathit{next\_child}}(\bcand,\head)$\\
            \uIf{$\begin{aligned}[t]
                &\epoch(\btemp) < \epoch(\now)\\
                &\text{\bf and}\ \btemp \preceq \phead\\
                &\text{\bf and}\ \var[val=v]{\isOneConfirmedExt}(\btemp, \guattime[val=v,time={\prevfirstslotepoch{t}}])
            \end{aligned}$\\}
            {
                $\bcand \gets \btemp$\nllabel{ln:set-bcand-in-prev-epoch-loop}
            }
            \uElse { 
                \Break 
            }
        }
    }

    \uIf{$\begin{aligned}[t]
        &\slot(\now) = \firstslot(\epoch(\now))\\
        &\text{\bf or}\ \epoch(\gu(\head)) \geq \epoch(\now) - 1
    \end{aligned}$\nllabel{ln:second-if}\\}
    {
        $\btcand \gets \bcand$\;
        \While{$\var[val=v]{\mathit{next\_child}}(\btcand,\head) \neq \bot$\nllabel{ln:while-current}}
        {
            $\btemp \gets \var[val=v]{\mathit{next\_child}}(\btcand,\head)$\\
            \uIf{$\begin{aligned}[t]
                &\epoch(\btemp) > \epoch(\btcand)\\
                &\text{\bf and}\ \neg \willChkpBeJustified_v(\chkp(\btemp))
            \end{aligned}$\nllabel{ln:check-will-chkp-be-justified}\\}
            {
                \Break
            }
            \uIf{$\var[val=v]{\isOneConfirmedExt}(\btemp, \guattime[val=v,time={\prevfirstslotepoch{t}}])$\nllabel{ln:if-is-one-confirmed-second-loop}} {
                $\btcand \gets \btemp$\label{set-btcand-to-btemp}
            }
            \uElse { 
                \Break 
            }
        }

        \uIf{$\begin{aligned}[t]
            &\epoch(\btcand) = \epoch(\now)\\
            &\text{\bf or}\ (\epoch(\votsource[blck=\btcand, time={\now}]) \geq \epoch(\now) - 2\\
            &\qquad \text{\bf and}\ (\slot(\now) = \firstslot(\epoch(\now))\\
            &\qquad\qquad \text{\bf or}\ \willNoConflictingChkpBeJustified_v(\chkp(\head,\epoch(\now)))))
        \end{aligned}$\nllabel{ln:if-to-set-bcand-to-btcand}\\}
        {
            $\bcand \gets \btcand$\nllabel{ln:set-bcand-to-btcand}
        }
    }

    \Return{$\bcand$}
}
\end{algorithm}

\begin{algorithm}
    \caption{$\findlatestconfirmeddescendant$}
    \label{alg:findlatestconf-functional-with-head-and-phead}
    \SetAlgoNoLine
    \DontPrintSemicolon
    \continuefrom{lastline2}

    \Fn{$\var[val=v]{\findlatestconfirmeddescendant}(b_c)$}
    {
        \tcc{Compute the highest LMDGHOST-safe descendant of $b_c$ along the head chain}
        $\btcands \gets
            \{b' \in \var[val=v]{\canonChainFrom}(b_c),\, \var[val=v]{\isOneConfirmedExtFrom}(b',\guattime[val=v,time={\prevfirstslotepoch{t}}], b_c)\}$

        \BlankLine

        \tcc{Case 1: current-epoch confirmation}
        \uIf{$\btcands\neq\emptyset$}
        {
            $\btcand \gets \var[]{\getHighest}(\btcands)$\\
            \uIf{
                $\begin{aligned}[t]
                    &\epoch(\btcand) = \epoch(\now)\\
                    &\text{\bf and}\ \head \succeq \btcand\ \text{\bf and}\ \epoch(\gu(\head)) \geq \epoch(\now) - 1\\
                    &\text{\bf and}\ \epoch(\btcand)>\epoch(b_c) \implies \willChkpBeJustified_v(\chkp(\btcand))
                \end{aligned}$\nllabel{ln:case1-whp}\\
            }
            {
                \Return $\btcand$
            }
        }

        \BlankLine

        \tcc{Case 2: no progress -- cannot rule out conflicting checkpoint}
        \uIf{
            $\begin{aligned}[t]
            &\text{\bf not}\;\Bigl(
                \slot(\now) = \firstslot(\epoch(\now))\\
            &\phantom{\text{\bf not}\;\Bigl(}\text{\bf or}\;
                \willNoConflictingChkpBeJustified^v\!\bigl(\chkp(\head,\epoch(\now))\bigr)
            \Bigr),
            \end{aligned}$
            \\
        }
        {
                        \Return $b_c$\nllabel{ln:case2-whp}\\
        }
        \BlankLine

        \tcc{Recompute highest LMDGHOST-safe descendant, restricted to previous epoch}
        $\btcands \gets
            \begin{aligned}[t]
            &\{b' \in \var[val=v]{\canonChainFrom}(b_c),\,\\
            &\qquad\epoch(b')<\epoch(\now)\\
            &\qquad\text{\bf and}\ \var[val=v]{\isOneConfirmedExtFrom}(b',\guattime[val=v,time={\prevfirstslotepoch{t}}], b_c)\}
        \end{aligned}$\\

                \tcc{Case 3: previous-epoch confirmation with voting source from epoch $\epoch(\now) - 2$ or later}
        \uIf{$\btcands\neq\emptyset$}
        {
            $\btcand \gets \var[]{\getHighest}(\btcands)$\\
            \uIf{$\begin{aligned}[t]
                &\epoch(\votsource[blck=\btcand, time={\now}]) \geq \epoch(\now) - 2\\
                &\text{\bf and}\
                        (\slot(\now) \neq \firstslot(\epoch(\now)) \implies\\
                &\qquad \head \succeq \btcand\ \text{\bf and}\ \epoch(\gu(\head)) \geq \epoch(\now) - 1)
            \end{aligned}$\nllabel{ln:case3-whp}\\}
            {
                \Return $\btcand$
            }
        }

        \BlankLine

        \tcc{Case 4: previous-epoch confirmation with voting source verified from a block received by the previous slot}
        $\begin{aligned}[t]
            &\bcands \gets \{b' \in \var[val=v]{\canonChainFrom}(b_c),\\
            &\qquad\epoch(b') < \epoch(\now) \\
            &\qquad\text{\bf and}\
                (\phead \succeq b'\ \text{\bf and}\ \epoch(\votsource[blck=\phead, time={\now}]) \geq \epoch(\now) - 2)\\
            &\qquad\text{\bf and}\
                (\slot(\now) \neq \firstslot(\epoch(\now)) \implies\\
            &\qquad\qquad \exists b''\in \{\head,\phead\}, b''\succeq b'\ \text{\bf and}\ \epoch(\gu(b'')) \geq \epoch(\now) - 1)\\
            &\qquad\text{\bf and}\
                \var[val=v]{\isOneConfirmedExtFrom}(b',\guattime[val=v,time={\prevfirstslotepoch{t}}],b_c)\}\nllabel{ln:case4-whp}
        \end{aligned}$\\

        \uIf{$\bcands \neq \emptyset$}
        {
            \Return $\var{\getHighest}(\bcands)$
        }

        \tcc{Case 5: no progress}
        \Return $b_c$\\
    }
\end{algorithm}

\paragraph{Previous-epoch-only loop.}
The guard at \Cref{ln:if-prev-epoch} checks that $\bcand$ belongs to
epoch $\epoch(\now) - 1$, that the voting source epoch of $\phead$ is
at least $\epoch(\now) - 2$, and that either we are at the first slot
of the epoch or $\willNoConflictingChkpBeJustified$ holds and the
greatest unrealized justified checkpoint of $\phead$ or $\head$
belongs to epoch $\epoch(\now) - 1$ or later.
When these conditions are met, the loop walks forward from $b_c$
along the head chain, advancing $\bcand$ as long as the next block
belongs to the previous epoch, is an ancestor of $\phead$, and passes
$\isOneConfirmedExt$.
The block at which the loop stops is the highest previous-epoch block
satisfying all three conditions.

\paragraph{Current-and-previous-epoch loop.}
The guard at \Cref{ln:second-if} checks that either we are at the
first slot of the epoch or
$\epoch(\gu(\head)) \geq \epoch(\now) - 1$.
When this holds, the loop continues walking forward from $\bcand$
(which may have been advanced by the previous-epoch loop), now
allowing blocks from the current epoch.
At each step, if the block crosses an epoch boundary
($\epoch(\btemp) > \epoch(\btcand)$), the loop checks
$\willChkpBeJustified$ (\Cref{ln:check-will-chkp-be-justified})
and breaks if it fails.
Otherwise, the block must pass $\isOneConfirmedExt$
(\Cref{ln:if-is-one-confirmed-second-loop}).
The loop produces a tentative candidate $\btcand$.

\paragraph{Equivalence.}
In both formulations, the output is determined by the same conditions:
the $\isOneConfirmedExt$ predicate selects the highest block on the
head chain that passes the per-block confirmation check, and the
remaining conditions ($\willChkpBeJustified$,
$\willNoConflictingChkpBeJustified$, voting source epoch, and greatest
unrealized justified checkpoint checks) determine whether that block
is not removed by $\FILHFC$ and can therefore be returned.
The declarative version expresses these as set comprehensions with
$\getHighest$; the iterative version evaluates them incrementally
via two loops.

The two formulations are equivalent up to the following difference:
Algorithm~\ref{alg:findlatestconf-functional} uses existential
quantifiers to search for descendants of a block with the required
properties (e.g., a descendant with a sufficiently recent voting
source or greatest unrealized justified checkpoint).
Algorithm~\ref{alg:findlatestconf} replaces these existential
quantifiers with concrete blocks: $\phead$ (the fork-choice head at
the beginning of the previous slot) in place of arbitrary descendants
received by the previous slot, and $\head$ (the current fork-choice
head) in place of arbitrary descendants received by the current slot.
Since $\phead$ and $\head$ are themselves descendants of $b_c$ on the
head chain, any condition satisfied by $\phead$ or $\head$ is also
satisfied by the existential quantifier.
The converse does not hold in general: the declarative version may
confirm blocks in situations where the iterative version does not,
if a qualifying descendant exists that differs from $\phead$ or
$\head$.
In practice, $\head$ and $\phead$ are the most natural candidates and
the two formulations produce the same output in all realistic
scenarios.
The iterative formulation is preferred for implementation because it
avoids searching over all descendants and uses only values already
available in the fork-choice store.

Therefore, we argue the equivalence between \Cref{alg:findlatestconf} and \Cref{alg:findlatestconf-functional-with-head-and-phead}, with the latter corresponding to \Cref{alg:findlatestconf-functional}, but with the existential quantifiers replaced by $\phead$ and $\head$ as explained above.

To informally argue such an equivalence we proceed by exhaustive case analysis.
In \Cref{tab:equivalence}, for each Case combination of \Cref{alg:findlatestconf-functional-with-head-and-phead}, we identify conditions on \Cref{alg:findlatestconf}'s branch guards under which both algorithms output the same block, assuming the same input and the same \LMDGHOST weight for each block.
Note that in \Cref{tab:equivalence}, expressions of the form ``Condition~X \textbf{and} Condition~Y'' and ``Condition~X \textbf{or} Condition~Y'' use \textbf{and} and \textbf{or} as logical connectives (conjunction and disjunction), not as natural-language connectors. For example, ``Condition~X \textbf{and} Condition~Y: false'' means that the conjunction is false, i.e., at least one of the two conditions is false (and possibly both).
Then, in \Cref{tab:coverage}, we exhaustively enumerate all possible combinations of \Cref{alg:findlatestconf}'s conditions and show that each feasible combination maps to exactly one combination in \Cref{tab:equivalence} (infeasible combinations are marked accordingly).
Since every feasible combination of conditions is covered and each maps to a combination where both algorithms agree, the two algorithms produce the same output on all inputs.

\begin{table}[t]
  \centering
  \newlength{\rowgap}\setlength{\rowgap}{1em}
  \newcolumntype{L}[1]{>{\minipage[t]{#1}\RaggedRight\arraybackslash}l<{\endminipage}}
  \makeatletter\renewcommand{\Hy@raisedlink}[1]{}\makeatother
  \begin{threeparttable}
  \begin{tabular}{cL{3cm}L{8cm}}
    \toprule
    Comb. \# & \Cref{alg:findlatestconf-functional-with-head-and-phead}&\Cref{alg:findlatestconf} \\
    \midrule
    1 &%
    \begin{enumerate}[nosep, topsep=0pt, partopsep=0pt, leftmargin=*]
        \item Case~1
    \end{enumerate}
    &
    \begin{enumerate}[nosep, topsep=0pt, partopsep=0pt, leftmargin=*]
        \item $\isOneConfirmedExt(\mathit{child}(b_c), \guattime[val=v,time=\slotstart(\prevfirstslotepoch{t})])$\tnote{*}: true
        \item Condition at \Cref{ln:if-prev-epoch}: any
        \item Condition at \Cref{ln:second-if}: true
        \item At \Cref{ln:if-to-set-bcand-to-btcand}, $\epoch(\btcand)=\epoch(\now)$
    \end{enumerate}
    \\ \addlinespace[\rowgap]
    2 &
    \begin{enumerate}[nosep, topsep=0pt, partopsep=0pt, leftmargin=*]
        \item \textbf{not} Case~1
        \item \textbf{not} Case~2
        \item Case~3
    \end{enumerate}
    &
    \begin{enumerate}[nosep, topsep=0pt, partopsep=0pt, leftmargin=*]
        \item $\isOneConfirmedExt(\mathit{child}(b_c), \guattime[val=v,time=\slotstart(\prevfirstslotepoch{t})])$\tnote{*}: true
        \item Condition at \Cref{ln:if-prev-epoch}: any
        \item Condition at \Cref{ln:second-if}: true
        \item At \Cref{ln:if-to-set-bcand-to-btcand}, $\epoch(\btcand)<\epoch(\now)$
        \item Condition at \Cref{ln:if-to-set-bcand-to-btcand}: true
    \end{enumerate}
    \\ \addlinespace[\rowgap]
    3 &
    \begin{enumerate}[nosep, topsep=0pt, partopsep=0pt, leftmargin=*]
        \item \textbf{not} Case~1
        \item \textbf{not} Case~2
        \item \textbf{not} Case~3
        \item Case~4
    \end{enumerate}
    &
    \begin{enumerate}[nosep, topsep=0pt, partopsep=0pt, leftmargin=*]
        \item $\isOneConfirmedExt(\mathit{child}(b_c), \guattime[val=v,time=\slotstart(\prevfirstslotepoch{t})])$\tnote{*}: true
        \item Condition at \Cref{ln:if-prev-epoch}: true
        \item Condition at \Cref{ln:second-if} \textbf{and} Condition at \Cref{ln:if-to-set-bcand-to-btcand}: false
    \end{enumerate}
    \\ \addlinespace[\rowgap]
    4 &
    \begin{enumerate}[nosep, topsep=0pt, partopsep=0pt, leftmargin=*]
        \item \textbf{not} Case~1
        \item Case~2
    \end{enumerate}
    \textbf{or}
    \begin{enumerate}[nosep, topsep=0pt, partopsep=0pt, leftmargin=*]
        \item \textbf{not} Case~1
        \item \textbf{not} Case~2
        \item \textbf{not} Case~3
        \item \textbf{not} Case~4
        \item Case~5
    \end{enumerate}
    &
    \begin{enumerate}[nosep, topsep=0pt, partopsep=0pt, leftmargin=*]
        \item $\isOneConfirmedExt(\mathit{child}(b_c), \guattime[val=v,time=\slotstart(\prevfirstslotepoch{t})])$\tnote{*}: false
    \end{enumerate}
    \textbf{or}
    \begin{enumerate}[nosep, topsep=0pt, partopsep=0pt, leftmargin=*]
        \item Condition at \Cref{ln:if-prev-epoch}: false
        \item Condition at \Cref{ln:second-if} \textbf{and} Condition at \Cref{ln:if-to-set-bcand-to-btcand}: false
    \end{enumerate}
    \\
    \bottomrule
  \end{tabular}
  \begin{tablenotes}\RaggedRight
    \item[*] $\mathit{child}(b_c) \coloneqq \var[val=v]{\mathit{next\_child}}(b_c,\head)$ (the child of $b_c$ on the canonical chain toward $\head$); $\guattime[val=v,time=\slotstart(\prevfirstslotepoch{t})] \coloneqq \guattime[val=v,time={\prevfirstslotepoch{t}}]$.
  \end{tablenotes}
  \end{threeparttable}
  \caption{Equivalence between cases of \Cref{alg:findlatestconf-functional-with-head-and-phead} and condition combinations of \Cref{alg:findlatestconf}: each row pairs a case of \Cref{alg:findlatestconf-functional-with-head-and-phead} with the corresponding conditions of \Cref{alg:findlatestconf}, where each side holds if and only if the other does. In expressions of the form ``Condition~X \textbf{and}/\textbf{or} Condition~Y'', \textbf{and} and \textbf{or} denote logical conjunction and disjunction, respectively.}
  \label{tab:equivalence}
\end{table}

\begin{table}[t]
  \centering
  \begin{threeparttable}
  \begin{tabular}{ccccc@{\hspace{1em}}|@{\hspace{1em}}c}
    \toprule
    \parbox[t]{1.5cm}{\centering Cond.\ $(\S)$\tnote{$\S$}} & \parbox[t]{2cm}{\centering Cond.\ at \Cref{ln:if-prev-epoch}} & \parbox[t]{2cm}{\centering Cond.\ at \Cref{ln:second-if}} & \parbox[t]{2.5cm}{\centering $\epoch(\btcand)$ $\{=,<\}$ $\epoch(\now)$ at \Cref{ln:if-to-set-bcand-to-btcand}} & \parbox[t]{2cm}{\centering Cond.\ at \Cref{ln:if-to-set-bcand-to-btcand}} & \parbox[t]{1.5cm}{\centering Comb.\ \# in \Cref{tab:equivalence}} \\
    \midrule
    T & T & T & $=$ & T &  1\\
    T & T & T & $=$ & F &  not possible\tnote{$\ddagger$}\\
    T & T & T & $<$ & T &  2\\
    T & T & T & $<$ & F &  3\\
    T & T & F & any & any &  3\\
    T & F & T & $=$ & T &  1\\
    T & F & T & $=$ & F &  not possible\tnote{$\ddagger$}\\
    T & F & T & $<$ & T &  2\\
    T & F & any & any & F &  4\\
    T & F & F & any & any &  4\\
    F & any & any & any & any &  4\\
    \bottomrule
  \end{tabular}
  \begin{tablenotes}\RaggedRight
    \item[$\S$] $\isOneConfirmedExt(\mathit{child}(b_c), \guattime[val=v,time=\slotstart(\prevfirstslotepoch{t})])$, where $\mathit{child}(b_c) \coloneqq \var[val=v]{\mathit{next\_child}}(b_c,\head)$ (the child of $b_c$ on the canonical chain toward $\head$) and $\guattime[val=v,time=\slotstart(\prevfirstslotepoch{t})] \coloneqq \guattime[val=v,time={\prevfirstslotepoch{t}}]$ (see also \Cref{tab:equivalence}).
    \item[$\ddagger$] The combination $\epoch(\btcand)=\epoch(\now)$ with the condition at \Cref{ln:if-to-set-bcand-to-btcand} being false is not possible, because $\epoch(\btcand)=\epoch(\now)$ makes the first disjunct of that condition true.
  \end{tablenotes}
  \end{threeparttable}
  \caption{Exhaustive enumeration of condition combinations in \Cref{alg:findlatestconf}, each mapped to a combination in \Cref{tab:equivalence}.}
  \label{tab:coverage}
\end{table}

We now argue, for each row of \Cref{tab:equivalence}, that the case of \Cref{alg:findlatestconf-functional-with-head-and-phead} and the corresponding conditions of \Cref{alg:findlatestconf} imply one another.
Throughout the discussion, we use the fact that, by construction of $\getlatestconfirmed$ (the inactive-state guard at \Cref{ln:if-bcand-npreceq-head} and the condition at \Cref{ln:if-bcand-e-1}), the input $b_c$ passed to $\findlatestconfirmeddescendant$ satisfies $\epoch(b_c) \in \{\epoch(\now)-1, \epoch(\now)\}$.

\paragraph{Combination~1.}
We show that the conditions on \Cref{alg:findlatestconf} implied by this combination and Case~1 of \Cref{alg:findlatestconf-functional-with-head-and-phead} imply one another.

\emph{(\Cref{alg:findlatestconf}'s conditions $\Rightarrow$ Case~1 of \Cref{alg:findlatestconf-functional-with-head-and-phead}.)}
The guard at \Cref{ln:second-if} is true, so the second loop runs.
The loop starts from $\bcand$ (possibly advanced by the previous-epoch loop) and walks along the $\head$ chain via $\mathit{next\_child}(\cdot,\head)$, advancing $\btcand$ as long as $\isOneConfirmedExt(\btemp,\guattime[val=v,time=\slotstart(\prevfirstslotepoch{t})])$ holds (and \Cref{ln:check-will-chkp-be-justified} does not fire).
Since $\epoch(\btcand) = \epoch(\now)$, the first disjunct of \Cref{ln:if-to-set-bcand-to-btcand} is true, so $\bcand \gets \btcand$.

We now verify Case~1's guard.
The chain walked by the second loop, together with any blocks advanced through by the previous-epoch loop, covers every strict descendant of $b_c$ on the $\head$ chain up to $\btcand$.
For every such block the iterative algorithm checked $\isOneConfirmedExt(\cdot,\guattime[val=v,time=\slotstart(\prevfirstslotepoch{t})])$ and the check passed; combined with the hypothesis that $\isOneConfirmedExt(\mathit{child}(b_c), \guattime[val=v,time=\slotstart(\prevfirstslotepoch{t})])$ is true (covering the first step out of $b_c$), this gives $\isOneConfirmedExtFrom(\btcand,\guattime[val=v,time=\slotstart(\prevfirstslotepoch{t})],b_c)$.
Hence $\btcand \in \btcands$ in \Cref{alg:findlatestconf-functional-with-head-and-phead}, so $\btcands \neq \emptyset$.
For Case~1's third conjunct ($\epoch(\btcand) > \epoch(b_c) \implies \willChkpBeJustified_v(\chkp(\btcand))$), recall that $\epoch(b_c) \in \{\epoch(\now)-1, \epoch(\now)\}$.
If $\epoch(b_c) = \epoch(\now)$, then $\epoch(b_c) = \epoch(\btcand)$ and the implication is vacuously true.
If $\epoch(b_c) = \epoch(\now)-1$, the second loop's walk from $\bcand$ to $\btcand$ crossed exactly one epoch boundary (entering $\epoch(\now)$); the loop did not break at \Cref{ln:check-will-chkp-be-justified}, so $\willChkpBeJustified_v(\chkp(\btemp))$ held for the first block $\btemp$ in $\epoch(\now)$ that the loop visited.
Because $\btemp$ and $\btcand$ belong to the same epoch on the same chain, $\chkp(\btemp) = \chkp(\btcand)$, so $\willChkpBeJustified_v(\chkp(\btcand))$ holds.
$\head \succeq \btcand$ holds because $\btcand$ lies on the $\head$ chain by construction of the loop.
Because $\epoch(\btcand) = \epoch(\now)$, the current slot cannot be the first slot of the current epoch (honest validators only accept current-epoch blocks after the first slot, as already noted on page~\pageref{rem:current-epoch-block-not-at-first-slot}).
Hence the first disjunct of the guard at \Cref{ln:second-if} fails, and its second disjunct supplies $\epoch(\gu(\head)) \geq \epoch(\now)-1$.
Hence Case~1 of \Cref{alg:findlatestconf-functional-with-head-and-phead} applies.

\emph{(Case~1 of \Cref{alg:findlatestconf-functional-with-head-and-phead} $\Rightarrow$ \Cref{alg:findlatestconf}'s conditions.)}
Case~1 gives $\btcands \neq \emptyset$ with $\btcand = \getHighest(\btcands)$ satisfying $\isOneConfirmedExtFrom(\btcand,\guattime[val=v,time=\slotstart(\prevfirstslotepoch{t})],b_c)$, $\epoch(\btcand) = \epoch(\now)$, $\head \succeq \btcand$, $\epoch(\gu(\head)) \geq \epoch(\now)-1$, and $\willChkpBeJustified_v(\chkp(\btcand))$ (since $\epoch(\btcand)>\epoch(b_c)$).

The guard at \Cref{ln:second-if} is satisfied by $\epoch(\gu(\head)) \geq \epoch(\now)-1$, so the second loop fires (this is this combination's third condition; its second condition at \Cref{ln:if-prev-epoch} is unconstrained).
Since $\isOneConfirmedExtFrom(\btcand,\guattime[val=v,time=\slotstart(\prevfirstslotepoch{t})],b_c)$ holds, every strict descendant of $b_c$ on the $\head$ chain up to $\btcand$ passes $\isOneConfirmedExt(\cdot,\guattime[val=v,time=\slotstart(\prevfirstslotepoch{t})])$.
In particular, $\isOneConfirmedExt(\mathit{child}(b_c), \guattime[val=v,time=\slotstart(\prevfirstslotepoch{t})])$ is true --- this combination's first condition.
Recall that $\epoch(b_c) \in \{\epoch(\now)-1, \epoch(\now)\}$.
If $\epoch(b_c) = \epoch(\now)-1$, the second loop crosses one epoch boundary entering $\epoch(\now)$; at that boundary, $\chkp(\btemp) = \chkp(\btcand)$, and Case~1's third conjunct (which then has a true antecedent) gives $\willChkpBeJustified_v(\chkp(\btcand))$, so the iterative check at \Cref{ln:check-will-chkp-be-justified} passes.
If $\epoch(b_c) = \epoch(\now)$, the loop stays within the current epoch and \Cref{ln:check-will-chkp-be-justified} never fires.
Hence the second loop advances $\btcand$ all the way to the block from Case~1.
At \Cref{ln:if-to-set-bcand-to-btcand}, $\epoch(\btcand) = \epoch(\now)$ makes the first disjunct true --- this combination's fourth condition.

\paragraph{Combination~2.}
We show that the conditions on \Cref{alg:findlatestconf} implied by this combination and the conjunction ``\textbf{not} Case~1, \textbf{not} Case~2, Case~3'' of \Cref{alg:findlatestconf-functional-with-head-and-phead} imply one another.

\emph{(\Cref{alg:findlatestconf}'s conditions $\Rightarrow$ \textbf{not} Case~1, \textbf{not} Case~2, Case~3.)}
The guard at \Cref{ln:second-if} is true, so the second loop runs.
The loop starts from $\bcand$ (possibly advanced by the previous-epoch loop) and walks the $\head$ chain via $\mathit{next\_child}(\cdot,\head)$, advancing $\btcand$ as long as $\isOneConfirmedExt(\btemp,\guattime[val=v,time=\slotstart(\prevfirstslotepoch{t})])$ holds (and \Cref{ln:check-will-chkp-be-justified} does not fire).
The loop produces $\btcand$ with $\epoch(\btcand) < \epoch(\now)$, and the condition at \Cref{ln:if-to-set-bcand-to-btcand} is true.

Since $\epoch(\btcand) < \epoch(\now)$, the first disjunct of \Cref{ln:if-to-set-bcand-to-btcand} is false, so its second disjunct holds:
\begin{itemize}
  \item $\epoch(\votsource[blck=\btcand,time={\now}]) \geq \epoch(\now)-2$ --- this is Case~3's first conjunct.
  \item $\slot(\now) = \firstslot(\epoch(\now))$ or $\willNoConflictingChkpBeJustified_v(\chkp(\head,\epoch(\now)))$ --- exactly the negation of Case~2's guard, so Case~2 does not apply.
\end{itemize}

The chain walked by the loops covers every strict descendant of $b_c$ on the $\head$ chain up to $\btcand$, and each block passed $\isOneConfirmedExt(\cdot,\guattime[val=v,time=\slotstart(\prevfirstslotepoch{t})])$.
Combined with the hypothesis $\isOneConfirmedExt(\mathit{child}(b_c), \guattime[val=v,time=\slotstart(\prevfirstslotepoch{t})])$, this gives $\isOneConfirmedExtFrom(\btcand,\guattime[val=v,time=\slotstart(\prevfirstslotepoch{t})],b_c)$.
Since $\epoch(\btcand) < \epoch(\now)$, $\btcand$ belongs to the (previous-epoch-restricted) $\btcands$ of Case~3, so $\btcands \neq \emptyset$.

For Case~3's second conjunct ($\slot(\now) \neq \firstslot(\epoch(\now)) \implies \head \succeq \btcand \land \epoch(\gu(\head)) \geq \epoch(\now)-1$): if $\slot(\now) = \firstslot(\epoch(\now))$ the implication is vacuously true.
Otherwise the first disjunct of the guard at \Cref{ln:second-if} fails, so its second disjunct supplies $\epoch(\gu(\head)) \geq \epoch(\now)-1$; and $\head \succeq \btcand$ because the loop walked the $\head$ chain.
Hence Case~3 applies.

Case~1 does not apply.
Let $b' := \getHighest(\btcands)$ denote \Cref{alg:findlatestconf-functional-with-head-and-phead}'s (unrestricted) Case~1 candidate.
If $\epoch(b') < \epoch(\now)$, Case~1's first conjunct fails.
Otherwise $\epoch(b') = \epoch(\now)$; then $\isOneConfirmedExtFrom(b', \guattime[val=v,time=\slotstart(\prevfirstslotepoch{t})], b_c)$ ensures the per-block check passes for every block on the $\head$ chain from $b_c$ up to $b'$, including the first block in $\epoch(\now)$, so \Cref{alg:findlatestconf}'s second loop would not have stopped before that block due to $\isOneConfirmedExt$.
The only way for \Cref{alg:findlatestconf}'s loop to nonetheless end with $\epoch(\btcand) < \epoch(\now)$ is for the boundary check at \Cref{ln:check-will-chkp-be-justified} to have failed; since the first current-epoch block visited shares the checkpoint with $b'$, $\willChkpBeJustified_v(\chkp(b'))$ fails, so Case~1's third conjunct fails (its antecedent $\epoch(b') > \epoch(b_c)$ holds because $\epoch(b_c) \in \{\epoch(\now)-1, \epoch(\now)\}$ and the second loop crossed an epoch boundary).

\emph{(\textbf{not} Case~1, \textbf{not} Case~2, Case~3 $\Rightarrow$ \Cref{alg:findlatestconf}'s conditions.)}
Case~3 gives the (restricted) $\btcands \neq \emptyset$ with $\btcand = \getHighest(\btcands)$ satisfying $\isOneConfirmedExtFrom(\btcand,\guattime[val=v,time=\slotstart(\prevfirstslotepoch{t})],b_c)$, $\epoch(\btcand) < \epoch(\now)$, $\epoch(\votsource[blck=\btcand,time={\now}]) \geq \epoch(\now)-2$, and (if $\slot(\now) \neq \firstslot(\epoch(\now))$) $\head \succeq \btcand$ and $\epoch(\gu(\head)) \geq \epoch(\now)-1$.
``\textbf{not} Case~2'' gives $\slot(\now) = \firstslot(\epoch(\now))$ or $\willNoConflictingChkpBeJustified_v(\chkp(\head,\epoch(\now)))$.

From $\isOneConfirmedExtFrom(\btcand,\guattime[val=v,time=\slotstart(\prevfirstslotepoch{t})],b_c)$, every strict descendant of $b_c$ on the $\head$ chain up to $\btcand$ passes $\isOneConfirmedExt(\cdot,\guattime[val=v,time=\slotstart(\prevfirstslotepoch{t})])$.
In particular, $\isOneConfirmedExt(\mathit{child}(b_c), \guattime[val=v,time=\slotstart(\prevfirstslotepoch{t})])$ is true --- this combination's first condition.

The guard at \Cref{ln:second-if} is satisfied: if $\slot(\now) = \firstslot(\epoch(\now))$ the first disjunct holds; otherwise Case~3's second conjunct supplies $\epoch(\gu(\head)) \geq \epoch(\now)-1$.
This is this combination's third condition (its second condition at \Cref{ln:if-prev-epoch} is unconstrained).

\Cref{alg:findlatestconf}'s second loop walks the $\head$ chain advancing $\btcand$ as long as $\isOneConfirmedExt$ holds (and \Cref{ln:check-will-chkp-be-justified} does not break it).
By Combination~1's ($\Leftarrow$) argument, the loop advances to (at least) the block Case~3 returns.
``\textbf{not} Case~1'' forces the loop to stop with $\epoch(\btcand) < \epoch(\now)$: if instead the loop advanced into the current epoch, the highest $\isOneConfirmedExtFrom$ block of \Cref{alg:findlatestconf-functional-with-head-and-phead}'s (unrestricted) $\btcands$ would have $\epoch = \epoch(\now)$, $\head \succeq \btcand$ holds, $\epoch(\gu(\head)) \geq \epoch(\now)-1$ holds (as derived above), and $\willChkpBeJustified_v(\chkp(\btcand))$ holds (because the iterative loop's boundary check would have passed), making Case~1 applicable --- a contradiction.
So the loop produces $\btcand$ with $\epoch(\btcand) < \epoch(\now)$ --- this combination's fourth condition.

For this combination's fifth condition (\Cref{ln:if-to-set-bcand-to-btcand} true with $\epoch(\btcand) < \epoch(\now)$, i.e., its second disjunct): Case~3 gives $\epoch(\votsource[blck=\btcand,time={\now}]) \geq \epoch(\now)-2$, and ``\textbf{not} Case~2'' gives $\slot(\now) = \firstslot(\epoch(\now))$ or $\willNoConflictingChkpBeJustified_v(\chkp(\head,\epoch(\now)))$, jointly satisfying the second disjunct of \Cref{ln:if-to-set-bcand-to-btcand}.

\paragraph{Combination~3.}
We show that the conditions on \Cref{alg:findlatestconf} implied by this combination and the conjunction ``\textbf{not} Case~1, \textbf{not} Case~2, \textbf{not} Case~3, Case~4'' of \Cref{alg:findlatestconf-functional-with-head-and-phead} imply one another.

\emph{(\Cref{alg:findlatestconf}'s conditions $\Rightarrow$ \textbf{not} Case~1, \textbf{not} Case~2, \textbf{not} Case~3, Case~4.)}
The guard at \Cref{ln:if-prev-epoch} is true, so the previous-epoch loop runs.
It walks from $\bcand = b_c$ along the $\head$ chain via $\mathit{next\_child}(\cdot,\head)$, advancing $\bcand$ to $\btemp$ as long as $\epoch(\btemp) < \epoch(\now)$, $\btemp \preceq \phead$, and $\isOneConfirmedExt(\btemp,\guattime[val=v,time=\slotstart(\prevfirstslotepoch{t})])$ all hold.
Let $b'$ denote the final value of $\bcand$ after this loop.
Since the conjunction of \Cref{ln:second-if} and \Cref{ln:if-to-set-bcand-to-btcand} is false, $b'$ is not subsequently overwritten by the second loop's $\btcand$, and the algorithm returns $b'$.

By construction $b' \in \canonChainFrom(b_c)$, $\phead \succeq b'$, $\epoch(b') < \epoch(\now)$, and every step's $\isOneConfirmedExt(\cdot,\guattime[val=v,time=\slotstart(\prevfirstslotepoch{t})])$ check passed; together with the hypothesis $\isOneConfirmedExt(\mathit{child}(b_c), \guattime[val=v,time=\slotstart(\prevfirstslotepoch{t})])$ this gives $\isOneConfirmedExtFrom(b', \guattime[val=v,time=\slotstart(\prevfirstslotepoch{t})], b_c)$.

\Cref{ln:if-prev-epoch} also gives:
\begin{itemize}
  \item $\epoch(\votsource[blck=\phead,time={\now}]) \geq \epoch(\now)-2$ --- Case~4's voting-source conjunct.
  \item $\slot(\now) = \firstslot(\epoch(\now))$ or $\willNoConflictingChkpBeJustified_v(\chkp(\head,\epoch(\now)))$ --- this negates Case~2's guard, so Case~2 does not apply.
  \item When $\slot(\now) \neq \firstslot(\epoch(\now))$, in addition: $\willNoConflictingChkpBeJustified_v(\chkp(\head,\epoch(\now)))$ holds and $\epoch(\gu(\phead)) \geq \epoch(\now)-1$ or $\epoch(\gu(\head)) \geq \epoch(\now)-1$ holds.
  The latter supplies Case~4's existential ($\exists b'' \in \{\head, \phead\}, b'' \succeq b' \land \epoch(\gu(b'')) \geq \epoch(\now)-1$): pick $b''$ to be whichever of $\phead, \head$ has $\gu$ at the required epoch. Then $b'' \succeq b'$: if $b'' = \phead$ this is the loop's per-step check $\btemp \preceq \phead$; if $b'' = \head$ this follows from $b'$ lying on the $\head$ chain (the loop walks $\mathit{next\_child}(\cdot,\head)$).
\end{itemize}
Hence $b'$ belongs to Case~4's $\bcands$, so $\bcands \neq \emptyset$ and Case~4 applies.

\textbf{Not Case~1, not Case~3.}
Suppose Case~1 applied. Then \Cref{alg:findlatestconf-functional-with-head-and-phead}'s unrestricted $\btcands$ contains a current-epoch block $b''$ with Case~1's guard satisfied, in particular $\willChkpBeJustified_v(\chkp(b''))$ (assuming $\epoch(b'') > \epoch(b_c)$, which holds since $\epoch(b_c) \leq \epoch(\now)-1$) and $\epoch(\gu(\head)) \geq \epoch(\now)-1$.
In \Cref{alg:findlatestconf}, this means the second loop would run (\Cref{ln:second-if} holds via $\epoch(\gu(\head)) \geq \epoch(\now)-1$), advance $\btcand$ to $b''$ (per-block and boundary checks pass), and the first disjunct of \Cref{ln:if-to-set-bcand-to-btcand} would hold ($\epoch(\btcand) = \epoch(\now)$).
The conjunction \Cref{ln:second-if} $\land$ \Cref{ln:if-to-set-bcand-to-btcand} would then be true --- contradicting this combination's third condition.
Suppose now Case~3 applied. Then \Cref{alg:findlatestconf-functional-with-head-and-phead}'s previous-epoch-restricted $\btcands$ has a highest block $b''$ with $\epoch(\votsource[blck=b'',time={\now}]) \geq \epoch(\now)-2$ and (when $\slot \neq \firstslot$) $\head \succeq b''$, $\epoch(\gu(\head)) \geq \epoch(\now)-1$.
In \Cref{alg:findlatestconf}, the second loop would run (via $\slot(\now) = \firstslot(\epoch(\now))$ or $\epoch(\gu(\head)) \geq \epoch(\now)-1$), advance to $b''$, and the second disjunct of \Cref{ln:if-to-set-bcand-to-btcand} would hold (voting-source bound from Case~3, $\slot(\now) = \firstslot(\epoch(\now))$ or $\willNoConflictingChkpBeJustified_v(\chkp(\head,\epoch(\now)))$ from ``not Case~2'').
Again the conjunction would be true --- contradiction.

\emph{(\textbf{not} Case~1, \textbf{not} Case~2, \textbf{not} Case~3, Case~4 $\Rightarrow$ \Cref{alg:findlatestconf}'s conditions.)}
Case~4 gives a block $b' \in \canonChainFrom(b_c)$ with $\epoch(b') < \epoch(\now)$, $\phead \succeq b'$, $\epoch(\votsource[blck=\phead,time={\now}]) \geq \epoch(\now)-2$, the $\slot(\now) \neq \firstslot(\epoch(\now)) \implies \exists b'' \in \{\head,\phead\}, b'' \succeq b' \land \epoch(\gu(b'')) \geq \epoch(\now)-1$ implication, and $\isOneConfirmedExtFrom(b', \guattime[val=v,time=\slotstart(\prevfirstslotepoch{t})], b_c)$.
``\textbf{not} Case~2'' gives $\slot(\now) = \firstslot(\epoch(\now))$ or $\willNoConflictingChkpBeJustified_v(\chkp(\head,\epoch(\now)))$.

The existence of $b' \in \canonChainFrom(b_c)$ with $\epoch(b') < \epoch(\now)$ forces $\epoch(b_c) = \epoch(\now)-1$ (since otherwise $\epoch(b_c) = \epoch(\now)$ and no descendant of $b_c$ can have a smaller epoch).

From $\isOneConfirmedExtFrom(b', \cdot, b_c)$, we obtain $\isOneConfirmedExt(\mathit{child}(b_c), \guattime[val=v,time=\slotstart(\prevfirstslotepoch{t})])$ --- this combination's first condition.

For \Cref{ln:if-prev-epoch}: $\epoch(\bcand) = \epoch(b_c) = \epoch(\now)-1$ holds at entry; the voting-source bound comes from Case~4; the third disjunct splits as before --- $\slot(\now) = \firstslot(\epoch(\now))$ gives the first disjunct directly, otherwise Case~4's existential supplies $\epoch(\gu(\phead)) \geq \epoch(\now)-1$ or $\epoch(\gu(\head)) \geq \epoch(\now)-1$, and ``not Case~2'' supplies $\willNoConflictingChkpBeJustified_v(\chkp(\head,\epoch(\now)))$. Hence \Cref{ln:if-prev-epoch} is true --- this combination's second condition.

For the third condition (NOT (\Cref{ln:second-if} AND \Cref{ln:if-to-set-bcand-to-btcand})): if both held, then by the argument used in the ($\Rightarrow$) direction either Case~1 or Case~3 would apply in \Cref{alg:findlatestconf-functional-with-head-and-phead}, contradicting ``not Case~1, not Case~3''. Hence the conjunction is false.

\paragraph{Combination~4.}
We show that the conditions on \Cref{alg:findlatestconf} implied by this combination and the disjunction ``(\textbf{not} Case~1, Case~2) \textbf{or} (\textbf{not} Case~1, \textbf{not} Case~2, \textbf{not} Case~3, \textbf{not} Case~4, Case~5)'' of \Cref{alg:findlatestconf-functional-with-head-and-phead} imply one another.

This is the ``no progress'' combination: both algorithms return $b_c$.
\Cref{alg:findlatestconf-functional-with-head-and-phead} returns $b_c$ either via Case~2's early-return guard or by falling through to Case~5.
\Cref{alg:findlatestconf} returns $b_c$ either because the per-block check $\isOneConfirmedExt(\mathit{child}(b_c), \guattime[val=v,time=\slotstart(\prevfirstslotepoch{t})])$ fails at the first step or because the previous-epoch loop's guard at \Cref{ln:if-prev-epoch} is false and the second loop fails to promote $\btcand$ to $\bcand$.

\emph{(\Cref{alg:findlatestconf}'s conditions $\Rightarrow$ Case~2 or Case~5.)}
\textit{If \Cref{alg:findlatestconf}'s first disjunct holds (i.e., $\isOneConfirmedExt(\mathit{child}(b_c), \guattime[val=v,time=\slotstart(\prevfirstslotepoch{t})])$ is false):}
no strict descendant of $b_c$ on the $\head$ chain satisfies $\isOneConfirmedExtFrom(\cdot, \guattime[val=v,time=\slotstart(\prevfirstslotepoch{t})], b_c)$, so \Cref{alg:findlatestconf-functional-with-head-and-phead}'s Case~1, Case~3, and Case~4 sets are all empty and ``not Case~1, not Case~3, not Case~4'' all hold.
\Cref{alg:findlatestconf-functional-with-head-and-phead} then returns $b_c$ via Case~2 (if its guard fires) or Case~5.

\textit{If \Cref{alg:findlatestconf}'s second disjunct holds (i.e., \Cref{ln:if-prev-epoch} is false and the conjunction \Cref{ln:second-if} $\land$ \Cref{ln:if-to-set-bcand-to-btcand} is false):}
By the ($\Rightarrow$) arguments of Combinations~1, 2, and 3, neither Case~1, Case~3, nor Case~4 of \Cref{alg:findlatestconf-functional-with-head-and-phead} applies (each would require either the conjunction to hold or \Cref{ln:if-prev-epoch} to be true).
The algorithm therefore returns $b_c$ via Case~2 or Case~5.

\emph{(Case~2 or Case~5 $\Rightarrow$ \Cref{alg:findlatestconf}'s conditions.)}
\textit{If Case~2 applies:}
``not Case~1'' is given and Case~2's guard supplies $\neg(\slot(\now) = \firstslot(\epoch(\now)) \lor \willNoConflictingChkpBeJustified_v(\chkp(\head,\epoch(\now))))$.
This negation makes the third sub-condition of \Cref{ln:if-prev-epoch} false, so \Cref{ln:if-prev-epoch} is false.
It also makes the second disjunct of \Cref{ln:if-to-set-bcand-to-btcand} false; the first disjunct ($\epoch(\btcand) = \epoch(\now)$) would, by Combination~1's ($\Rightarrow$) argument, force Case~1 of \Cref{alg:findlatestconf-functional-with-head-and-phead} to apply, contradicting ``not Case~1''.
Hence \Cref{ln:if-to-set-bcand-to-btcand} is false, so the conjunction \Cref{ln:second-if} $\land$ \Cref{ln:if-to-set-bcand-to-btcand} is false.
This combination's second disjunct holds.

\textit{If Case~5 applies:}
``not Case~1, not Case~2, not Case~3, not Case~4'' all hold.
If $\isOneConfirmedExt(\mathit{child}(b_c), \guattime[val=v,time=\slotstart(\prevfirstslotepoch{t})])$ is false, this combination's first disjunct holds.
Otherwise, by Combination~3's ($\Leftarrow$) argument, the conditions implied by Combination~3 in \Cref{alg:findlatestconf} would force Case~4 of \Cref{alg:findlatestconf-functional-with-head-and-phead} to apply; ``not Case~4'' therefore rules them out, forcing \Cref{ln:if-prev-epoch} to be false.
Similarly, by the ($\Leftarrow$) arguments of Combinations~1 and 2, the conjunction \Cref{ln:second-if} $\land$ \Cref{ln:if-to-set-bcand-to-btcand} being true would force Case~1 or Case~3 to apply; ``not Case~1, not Case~3'' therefore rule out the conjunction.
This combination's second disjunct holds.

\section{Reference Implementation Difference}
\label{sec:implementation-note-chain}

\begin{algorithm}[t]
\caption{$\isChainOneConfirmedImpl$}
\label{alg:is-chain-one-confirmed-impl}
\SetAlgoNoLine
\Fn{$\var[val=v]{\isChainOneConfirmedImpl}(b_c)$}{
    $b \gets b_c$\\
    \While{$\begin{aligned}[t]
        &b \neq \block(\guattime[val=v,time=\slotstart(\prevfirstslotepoch{\now})]) \land 
        \epoch(b)=\epoch(\now)-1\\
        &\land \var[val=v]{\isOneConfirmedExt}(b, \guattime[val=v,time={\slotstart({\prevfirstslotepoch[-1]{\now}})}],\now)
    \end{aligned}$
    }
    {
        $b \gets \parent(b)$
    }

    \Return{$b = \block(\guattime[val=v,time=\slotstart(\prevfirstslotepoch{\now})]) \lor \epoch(b)<\epoch(\now)-1$}
}
\end{algorithm}

In the reference implementation~\cite{fcr-impl}, $\isChainOneConfirmed$ is replaced by $\isChainOneConfirmedImpl$ as per \Cref{alg:is-chain-one-confirmed-impl}.

The first difference, compared to $\isChainOneConfirmed$, is that $\isChainOneConfirmedImpl$ uses $\isOneConfirmedExt$ in place of $\isOneConfirmed$.
The second difference is that the while loop stops as soon as it reached a block belonging to two epochs ago (reconfirmation is executed at the beginning of the current epoch), or earlier.

To show that these changes do not affect the safety arguments of the protocol, it is sufficient to show that $\var[val=v,time=\slotstart(\epoch(t))]{\isChainOneConfirmedImpl}(b_c) \implies \var[val=v,time=\slotstart(\epoch(t))]{\isChainOneConfirmed}(b_c)$.
Then, assume that $\var[val=v,time=\slotstart(\epoch(t))]{\isChainOneConfirmedImpl}(b_c)$ holds and pick any $b\preceq b_c$.

If $\epoch(b) = \epoch(t)-1$, it must be that $\isOneConfirmedExt(b,\guattime[val=v,time={\slotstart({\prevfirstslotepoch[-1]{t}})}],t)$ holds.
This implies that $\isOneConfirmed(b,\guattime[val=v,time={\slotstart({\prevfirstslotepoch[-1]{t}})}],t)$ holds as well, as $\isOneConfirmedExt$ evaluates to either $\isOneConfirmed$ or $\isOneConfirmedSpecial$, and $\isOneConfirmedSpecial\implies \isOneConfirmed$.

If $\epoch(b) < \epoch(t)-1$, it must be that $\isOneConfirmedExt$ holds for a block $b_0$ such that $b_0\preceq b_c$ and $\epoch(\parent(b_0))<\epoch(b_0)=\epoch(t)-1$.
This implies that $\isOneConfirmedSpecial$ holds for such a block.
Note that the committees between $\parent(b)$ and $\slotstart(\epoch(t))-1$ span exactly one full epoch and, for every ancestor $b' \preceq b$, the committees between $\parent(b')$ and $\slotstart(\epoch(t))-1$ cover the same slots plus additional earlier ones.
Under \Cref{assum:no-change-to-the-validator-set}, all these committees have the same total weight, so the committee weight multiplied by $\beta$ in $\isOneConfirmedSpecial(b,C)$'s threshold and in $\isOneConfirmed(b',C)$'s threshold are equal.
Under \Cref{assum:no-change-to-the-validator-set}, 
Every validator supporting $b$ also supports $b'$; in addition, the validators subtracted in the discount at $b$, those whose \LMDGHOST vote is exactly $\parent(b)$, support $b'$ as well, since $b' \preceq \parent(b)$.
This additional support is what turns the hypothesis $\isOneConfirmedSpecial(b, C)$ into $\isOneConfirmed(b', C)$.
The intuition above is formalized in the following lemma.

\begin{lemma}\label{lem:isOneConfirmedExt-implies-isOneConfirmed-prev-epoch-fslot}
    If
    \begin{enumerate}
        \item $\epoch(\parent(b))<\epoch(b)=\epoch(t)-1$
        \item\label{hyp:lem:isOneConfirmedExt-implies-isOneConfirmed-prev-epoch-fslot:2} $\isOneConfirmedExt(b,C,\slotstart(\epoch(t)))$,
    \end{enumerate}
    then, for any $b' \preceq b$, $\isOneConfirmed(b', C,\slotstart(\epoch(t)))$.
\end{lemma}

\begin{proof}
    Let $b'$ be any block such that $b'\preceq b$ and $s := \firstslot(\epoch(t))$.

    Note that, due to  \Cref{assum:no-change-to-the-validator-set}, $\parentslotplusone(b')<\parentslotplusone(b)\leq \firstslot(\epoch(t)-1)$ and $s-1 = \lastslot(\epoch(t)-1)$.
    Hence, $\commweightfromslot[from=\firstslot(\epoch(b)),to=s-1,chkp={C}]{\allvals}=\commweightfromafterparentblock[from=b,to=s-1,chkp={C}]{\allvals}=\commweightfromafterparentblock[from=b',to=s-1,chkp={C}]{\allvals}=\totvalsetweight[chkp=C]{\allvals}$.

    Then, proceed as follows.
    \def\alignexplwidth{5cm}
    \allowdisplaybreaks
    \begin{align*}
        &\hspace{3ex} \indicatorfromblock[from=b',to=s-1,val=v,when=\slotstart(s),chkp={C}]{\indQ}
        &&\alignexpl{By definition.}
        \\  
        &=
        \frac
            {\attsetweightfromblock[from=b',to=s-1,val=v,when=\slotstart(s),chkp={C}]{\allatts}}
            {\commweightfromafterparentblock[from=b',to=s-1,chkp={C}]{\allvals}}
        &&\alignexpl{By definition.}
        \\
        &\geq
        \frac
            {\attsetweightfromblock[from=b,to=s-1,val=v,when=\slotstart(s),chkp={C}]{\allatts}
            +\attsetweighttobediscsimplefromblock[from=b,to=\slot(b)-1,val=v,when=\slotstart(s),chkp=C]{\allatts}
            }
            {\commweightfromafterparentblock[from=b,to=s-1,chkp={C}]{\allvals}}  
        &&\alignexpl[\alignexplwidth]{  
        First, due to $\FILLMD$, 
        $\attsetfromblockunfiltered[from=b,to=s-1,val=v,when=\slotstart(s)]{\allatts} \cap \attsettobediscsimplefromblock[from=b,to=\slot(b)-1,val=v,when=\slotstart(s)]{\allatts}=\emptyset$.
        Then, given that $b' \preceq \parent(b)$, we have that $\attsetfromblockunfiltered[from=b',to=s-1,val=v,when=\slotstart(s)]{\allatts} \supseteq (\attsetfromblockunfiltered[from=b,to=s-1,val=v,when=\slotstart(s)]{\allatts} \cup \attsettobediscsimplefromblock[from=b,to=\slot(b)-1,val=v,when=\slotstart(s)]{\allatts})=(\attsetfromblockunfiltered[from=b,to=s-1,val=v,when=\slotstart(s)]{\allatts} \sqcup \attsettobediscsimplefromblock[from=b,to=\slot(b)-1,val=v,when=\slotstart(s)]{\allatts})$.
        }            
        \\
        &=
        \indicatorfromblock[from=b,to=s-1,val=v,when=\slotstart(s),chkp={C}]{\indQ}
        +
        \frac
            {\attsetweighttobediscsimplefromblock[from=b,to=\slot(b)-1,val=v,when=\slotstart(s),chkp=C]{\allatts}
            }
            {\commweightfromafterparentblock[from=b,to=s-1,chkp={C}]{\allvals}}     
         &&\alignexpl[\alignexplwidth]{By definition.}     
        \\
        &> 
        \begin{aligned}[t]
            &\frac{1}{2}\left(1 + 
            \frac
                {
                    \boostweight[chkp={C}]
                    -
                    \max
                    \left(
                        \attsetweighttobediscsimplefromblock[from=b,to=\slot(b)-1,val=v,when=\slotstart(s),chkp=C]{\allatts}
                        -\beta \commweightfromafterparentblock[from=b,to=\slot(b)-1,chkp=C]{\allvals}
                        +\attsetweightfromblock[from=b,to=\slot(b)-1,val=v,when=\slotstart(s),chkp=C]{\slashvals}
                        ,
                        0
                    \right)                
                }
                {\commweightfromafterparentblock[from=b,to=s-1,chkp={C}]{\allvals}}\right)
            \\
            &+ \beta
            \frac
                {\commweightfromslot[from=\firstslot(\epoch(b)),to=s-1,chkp={C}]{\allvals}}
                {\commweightfromafterparentblock[from=b,to=s-1,chkp={C}]{\allvals}}
            - \frac
                {\attsetweightfromblock[from=b,to=s-1,val=v,when=\slotstart(s),chkp={C}]{\slashvals}}
                {\commweightfromafterparentblock[from=b,to=s-1,chkp={C}]{\allvals}}
            +
            \frac
                {\attsetweighttobediscsimplefromblock[from=b,to=\slot(b)-1,val=v,when=\slotstart(s),chkp=C]{\allatts}}
                {\commweightfromafterparentblock[from=b,to=s-1,chkp={C}]{\allvals}}                
        \end{aligned}
        &&\alignexpl[\alignexplwidth]{By hypothesis~\ref{hyp:lem:isOneConfirmedExt-implies-isOneConfirmed-prev-epoch-fslot:2} and the fact that, because $\epoch(\parent(b))<\epoch(b)$, $\isOneConfirmed$ evaluates to $\isOneConfirmedSpecial$.}   
        \\
        &=
        \begin{aligned}[t]
            &\frac{1}{2}\left(1 + 
            \frac
                {
                    \boostweight[chkp={C}]
                    -
                    \max
                    \left(
                        \attsetweighttobediscsimplefromblock[from=b,to=\slot(b)-1,val=v,when=\slotstart(s),chkp=C]{\allatts}
                        -\beta \commweightfromafterparentblock[from=b,to=\slot(b)-1,chkp=C]{\allvals}
                        +\attsetweightfromblock[from=b,to=\slot(b)-1,val=v,when=\slotstart(s),chkp=C]{\slashvals}
                        ,
                        0
                    \right)                
                }
                {\commweightfromafterparentblock[from=b,to=s-1,chkp={C}]{\allvals}}\right)
            \\
            &+ \beta
            - \frac
                {\attsetweightfromblock[from=b,to=s-1,val=v,when=\slotstart(s),chkp={C}]{\slashvals}}
                {\commweightfromafterparentblock[from=b,to=s-1,chkp={C}]{\allvals}}
            +
            \frac
                {\attsetweighttobediscsimplefromblock[from=b,to=\slot(b)-1,val=v,when=\slotstart(s),chkp=C]{\allatts}}
                {\commweightfromafterparentblock[from=b,to=s-1,chkp={C}]{\allvals}}                
        \end{aligned} 
        &&\alignexpl[\alignexplwidth]{Given that $\epoch(b) = \epoch(t)-1$, as argued at the beginning of this proof, we have that  $\totvalsetweight[chkp=C]{\allvals}=\commweightfromslot[from=\firstslot(\epoch(b)),to=s-1,chkp={C}]{\allvals}=\commweightfromafterparentblock[from=b,to=s-1,chkp={C}]{\allvals}$.}               
    \end{align*}

    \textbf{Case 1: $\attsetweighttobediscsimplefromblock[from=b,to=\slot(b)-1,val=v,when=\slotstart(s),chkp=C]{\allatts}
                    -\beta \commweightfromafterparentblock[from=b,to=\slot(b)-1,chkp=C]{\allvals}
                    +\attsetweightfromblock[from=b,to=\slot(b)-1,val=v,when=\slotstart(s),chkp=C]{\slashvals}
                    >
                    0$.}
    \def\alignexplwidth{5cm}
    \allowdisplaybreaks
    \begin{align*}
        &\hspace{3ex} 
        \begin{aligned}[t]
            &\frac{1}{2}\left(1 + 
            \frac
                {
                    \boostweight[chkp={C}]
                    -
                    \max
                    \left(
                        \attsetweighttobediscsimplefromblock[from=b,to=\slot(b)-1,val=v,when=\slotstart(s),chkp=C]{\allatts}
                        -\beta \commweightfromafterparentblock[from=b,to=\slot(b)-1,chkp=C]{\allvals}
                        +\attsetweightfromblock[from=b,to=\slot(b)-1,val=v,when=\slotstart(s),chkp=C]{\slashvals}
                        ,
                        0
                    \right)                
                }
                {\commweightfromafterparentblock[from=b,to=s-1,chkp={C}]{\allvals}}\right)
            \\
            &+ \beta
            - \frac
                {\attsetweightfromblock[from=b,to=s-1,val=v,when=\slotstart(s),chkp={C}]{\slashvals}}
                {\commweightfromafterparentblock[from=b,to=s-1,chkp={C}]{\allvals}}
            +
            \frac
                {\attsetweighttobediscsimplefromblock[from=b,to=\slot(b)-1,val=v,when=\slotstart(s),chkp=C]{\allatts}}
                {\commweightfromafterparentblock[from=b,to=s-1,chkp={C}]{\allvals}} 
        \end{aligned} 
        \\
        &=  
        \begin{aligned}[t]
        &\frac{1}{2}\left(1 + 
            \frac
                {
                    \boostweight[chkp={C}]
                    -
                        \attsetweighttobediscsimplefromblock[from=b,to=\slot(b)-1,val=v,when=\slotstart(s),chkp=C]{\allatts}
                        +
                        (
                            \beta \commweightfromafterparentblock[from=b,to=\slot(b)-1,chkp=C]{\allvals}
                            -\attsetweightfromblock[from=b,to=\slot(b)-1,val=v,when=\slotstart(s),chkp=C]{\slashvals}
                        )               
                }
                {\commweightfromafterparentblock[from=b,to=s-1,chkp={C}]{\allvals}}\right)
            \\
            &+ \beta
            - \frac
                {\attsetweightfromblock[from=b,to=s-1,val=v,when=\slotstart(s),chkp={C}]{\slashvals}}
                {\commweightfromafterparentblock[from=b,to=s-1,chkp={C}]{\allvals}}
            +
            \frac
                {\attsetweighttobediscsimplefromblock[from=b,to=\slot(b)-1,val=v,when=\slotstart(s),chkp=C]{\allatts}}
                {\commweightfromafterparentblock[from=b,to=s-1,chkp={C}]{\allvals}}              
        \end{aligned} 
        &&\alignexpl[\alignexplwidth]{By simplification.}   
        \\
        &=  
        \begin{aligned}[t]
        &\frac{1}{2}\left(1 + 
            \frac
                {
                    \boostweight[chkp={C}]
                    +
                        \attsetweighttobediscsimplefromblock[from=b,to=\slot(b)-1,val=v,when=\slotstart(s),chkp=C]{\allatts}
                        +
                        (
                            \beta \commweightfromafterparentblock[from=b,to=\slot(b)-1,chkp=C]{\allvals}
                            -\attsetweightfromblock[from=b,to=\slot(b)-1,val=v,when=\slotstart(s),chkp=C]{\slashvals}
                        )               
                }
                {\commweightfromafterparentblock[from=b,to=s-1,chkp={C}]{\allvals}}\right)
            \\
            &+ \beta
            - \frac
                {\attsetweightfromblock[from=b,to=s-1,val=v,when=\slotstart(s),chkp={C}]{\slashvals}}
                {\commweightfromafterparentblock[from=b,to=s-1,chkp={C}]{\allvals}}          
        \end{aligned}
        &&\alignexpl[\alignexplwidth]{By simplification.} 
        \\
        &\geq        
\frac{1}{2}\left(1 + 
            \frac
                {
                    \boostweight[chkp={C}]    
                }
                {\commweightfromafterparentblock[from=b,to=s-1,chkp={C}]{\allvals}}\right)
            + \beta
            - \frac
                {\attsetweightfromblock[from=b,to=s-1,val=v,when=\slotstart(s),chkp={C}]{\slashvals}}
                {\commweightfromafterparentblock[from=b,to=s-1,chkp={C}]{\allvals}} 
        &&\alignexpl[\alignexplwidth]{As $\attsetweighttobediscsimplefromblock[from=b,to=\slot(b)-1,val=v,when=\slotstart(s),chkp=C]{\allatts}\geq 0$ and 
                        $
                        (
                            \beta \commweightfromafterparentblock[from=b,to=\slot(b)-1,chkp=C]{\allvals}
                            -\attsetweightfromblock[from=b,to=\slot(b)-1,val=v,when=\slotstart(s),chkp=C]{\slashvals}
                        )\geq 0$.
                }          
    \end{align*}   
    
    \textbf{Case 2: $\attsetweighttobediscsimplefromblock[from=b,to=\slot(b)-1,val=v,when=\slotstart(s),chkp=C]{\allatts}
                    -\beta \commweightfromafterparentblock[from=b,to=\slot(b)-1,chkp=C]{\allvals}
                    +\attsetweightfromblock[from=b,to=\slot(b)-1,val=v,when=\slotstart(s),chkp=C]{\slashvals}
                    \leq
                    0$.}

    \def\alignexplwidth{5cm}
    \allowdisplaybreaks
    \begin{align*}
        &\hspace{3ex} 
        \begin{aligned}[t]
            &\frac{1}{2}\left(1 + 
            \frac
                {
                    \boostweight[chkp={C}]
                    -
                    \max
                    \left(
                        \attsetweighttobediscsimplefromblock[from=b,to=\slot(b)-1,val=v,when=\slotstart(s),chkp=C]{\allatts}
                        -\beta \commweightfromafterparentblock[from=b,to=\slot(b)-1,chkp=C]{\allvals}
                        +\attsetweightfromblock[from=b,to=\slot(b)-1,val=v,when=\slotstart(s),chkp=C]{\slashvals}
                        ,
                        0
                    \right)                
                }
                {\commweightfromafterparentblock[from=b,to=s-1,chkp={C}]{\allvals}}\right)
            \\
            &+ \beta
            - \frac
                {\attsetweightfromblock[from=b,to=s-1,val=v,when=\slotstart(s),chkp={C}]{\slashvals}}
                {\commweightfromafterparentblock[from=b,to=s-1,chkp={C}]{\allvals}}
            +
            \frac
                {\attsetweighttobediscsimplefromblock[from=b,to=\slot(b)-1,val=v,when=\slotstart(s),chkp=C]{\allatts}}
                {\commweightfromafterparentblock[from=b,to=s-1,chkp={C}]{\allvals}} 
        \end{aligned} 
        \\
        &\geq        
\frac{1}{2}\left(1 + 
            \frac
                {
                    \boostweight[chkp={C}]    
                }
                {\commweightfromafterparentblock[from=b,to=s-1,chkp={C}]{\allvals}}\right)
            + \beta
            - \frac
                {\attsetweightfromblock[from=b,to=s-1,val=v,when=\slotstart(s),chkp={C}]{\slashvals}}
                {\commweightfromafterparentblock[from=b,to=s-1,chkp={C}]{\allvals}}  
            +
            \frac
                {\attsetweighttobediscsimplefromblock[from=b,to=\slot(b)-1,val=v,when=\slotstart(s),chkp=C]{\allatts}}
                {\commweightfromafterparentblock[from=b,to=s-1,chkp={C}]{\allvals}} 
        &&\alignexpl[\alignexplwidth]{By simplification.}                                 
        \\
        &=        
\frac{1}{2}\left(1 + 
            \frac
                {
                    \boostweight[chkp={C}]    
                }
                {\commweightfromafterparentblock[from=b,to=s-1,chkp={C}]{\allvals}}\right)
            + \beta
            - \frac
                {\attsetweightfromblock[from=b,to=s-1,val=v,when=\slotstart(s),chkp={C}]{\slashvals}}
                {\commweightfromafterparentblock[from=b,to=s-1,chkp={C}]{\allvals}}    
        &&\alignexpl[\alignexplwidth]{As $\attsetweighttobediscsimplefromblock[from=b,to=\slot(b)-1,val=v,when=\slotstart(s),chkp=C]{\allatts}\geq 0$.
                }       
    \end{align*}   
    
    Then, continuing from what we concluded from the two cases above:

    \def\alignexplwidth{5cm}
    \allowdisplaybreaks
    \begin{align*}
        &\hspace{3ex}\frac{1}{2}\left(1 + 
            \frac
                {
                    \boostweight[chkp={C}]    
                }
                {\commweightfromafterparentblock[from=b,to=s-1,chkp={C}]{\allvals}}\right)
            + \beta
            - \frac
                {\attsetweightfromblock[from=b,to=s-1,val=v,when=\slotstart(s),chkp={C}]{\slashvals}}
                {\commweightfromafterparentblock[from=b,to=s-1,chkp={C}]{\allvals}}       
        \\        
        &=        
        \frac{1}{2}\left(1 + 
            \frac
                {
                    \boostweight[chkp={C}]    
                }
                {\commweightfromafterparentblock[from=b',to=s-1,chkp={C}]{\allvals}}\right)
            + \beta
            - \frac
                {\attsetweightfromblock[from=b',to=s-1,val=v,when=\slotstart(s),chkp={C}]{\slashvals}}
                {\commweightfromafterparentblock[from=b',to=s-1,chkp={C}]{\allvals}}
        &&\alignexpl[\alignexplwidth]{Because, as argued at the beginning of the proof, $\totvalsetweight[chkp=C]{\allvals}=\commweightfromafterparentblock[from=b,to=s-1,chkp={C}]{\allvals}=\commweightfromafterparentblock[from=b',to=s-1,chkp={C}]{\allvals}$.}
        \\        
        &\geq    
        \begin{aligned}[t]
            &\frac{1}{2}\left(1 + 
            \frac
                {
                    \boostweight[chkp={C}]
                    -
                    \max
                    \left(
                        \attsetweighttobediscsimplefromblock[from=b',to=\slot(b')-1,val=v,when=\slotstart(s),chkp=C]{\allatts}
                        -\beta \commweightfromafterparentblock[from=b',to=\slot(b')-1,chkp=C]{\allvals}
                        +\attsetweightfromblock[from=b',to=\slot(b')-1,val=v,when=\slotstart(s),chkp=C]{\slashvals}
                        ,
                        0
                    \right)                
                }
                {\commweightfromafterparentblock[from=b',to=s-1,chkp={C}]{\allvals}}\right)
            \\
            &+ \beta
            - \frac
                {\attsetweightfromblock[from=b,to=s-1,val=v,when=\slotstart(s),chkp={C}]{\slashvals}}
                {\commweightfromafterparentblock[from=b',to=s-1,chkp={C}]{\allvals}} 
        \end{aligned} 
        &&\alignexpl[\alignexplwidth]{As $\max
                    (
                        \attsetweighttobediscsimplefromblock[from=b',to=\slot(b')-1,val=v,when=\slotstart(s),chkp=C]{\allatts}
                        -\beta \commweightfromafterparentblock[from=b',to=\slot(b')-1,chkp=C]{\allvals}
                        +\attsetweightfromblock[from=b',to=\slot(b')-1,val=v,when=\slotstart(s),chkp=C]{\slashvals}
                        ,
                        0
                    )\geq 0$.}  
        \\        
        &\geq    
        \begin{aligned}[t]
            &\frac{1}{2}\left(1 + 
            \frac
                {
                    \boostweight[chkp={C}]
                    -
                    \max
                    \left(
                        \attsetweighttobediscsimplefromblock[from=b',to=\slot(b')-1,val=v,when=\slotstart(s),chkp=C]{\allatts}
                        -\beta \commweightfromafterparentblock[from=b',to=\slot(b')-1,chkp=C]{\allvals}
                        +\attsetweightfromblock[from=b',to=\slot(b')-1,val=v,when=\slotstart(s),chkp=C]{\slashvals}
                        ,
                        0
                    \right)                
                }
                {\commweightfromafterparentblock[from=b',to=s-1,chkp={C}]{\allvals}}\right)
            \\
            &+ \beta
            \frac
                {\commweightfromblock[from=b',to=s-1,chkp={C}]{\allvals}}
                {\commweightfromafterparentblock[from=b',to=s-1,chkp={C}]{\allvals}}
        \end{aligned}    
        &&\alignexpl[\alignexplwidth]{As $            \frac
                {\commweightfromblock[from=b',to=s-1,chkp={C}]{\allvals}}
                {\commweightfromafterparentblock[from=b',to=s-1,chkp={C}]{\allvals}}\leq 1$.}                       
    \end{align*}

    Hence, $\isOneConfirmed(b', C,\slotstart(\epoch(t)))$ holds.
\end{proof}

\medskip

Finally, liveness of reconfirmation when using $\isChainOneConfirmedImpl$ follows from \Cref{lem:bconf-curr-epoch-ancestor-is-one-confirmed-fslot,lem:bconf-curr-epoch-ancestor-is-one-confirmed,lem:beta-less-than-quarter-no-reconfirmation-required-ex-with-simple}.
\end{document}

%% file: preamble.tex
\usepackage{graphicx} 
\usepackage{amsmath}
\usepackage{amsthm}
\usepackage{mathtools}
\usepackage{amssymb}
\usepackage{amsfonts}
\usepackage{xcolor}
\usepackage{amsmath,amsfonts}
\usepackage{listings}
\usepackage{marginnote}
\usepackage[shortlabels,inline]{enumitem}
\usepackage{url}
\usepackage{xspace}
\usepackage[inkscapeformat=png]{svg}
\usepackage{keyval}
\usepackage{refcount}
\usepackage{xkeyval}
\usepackage{ifthen}
\usepackage{numbertabbing}
\usepackage{array}
\usepackage{booktabs}
\usepackage{ragged2e}
\usepackage{threeparttable}

\usepackage[linesnumbered,ruled]{algorithm2e}
\SetKwProg{Fn}{function}{}{end}
\SetKwProg{Proc}{procedure}{}{end}
\SetKwProg{Upon}{upon}{}{end}
\SetNlSty{textbf}{}{:}
\SetKw{KwStateK}{State}
\SetKwBlock{KwState}{State:}{}
\SetKwFor{While}{while}{}{}%
\SetKw{Break}{break}
\SetKw{Const}{const}
\SetKw{False}{False}
\SetKw{True}{True}

\newcommand{\slotattime}[1]{\ensuremath{\mathsf{slot}(#1)}}
\newcommand{\gujblock}{\ensuremath{\mathsf{GU}}}
\newcommand{\gjblock}{\ensuremath{\mathsf{GJ}}}
\newcommand{\gfblock}{\ensuremath{\mathsf{GF}}}
\makeatletter
\renewcommand{\nllabel}[1]
 {{\let\@currentlabel\algocf@currentlabel
  \let\@currentcounter\algocf@currentcounter
  \label{#1}}}%

\renewcommand{\algocf@nl@sethref}[1]{%
  \renewcommand{\theHAlgoLine}{\thealgocfproc.#1}%
  \hyper@refstepcounter{AlgoLine}%
  \gdef\algocf@currentlabel{#1}%
  \gdef\algocf@currentcounter{AlgoLine}%
 }%
\makeatother

\usepackage{hyperref}
\hypersetup{
    colorlinks=false, 
    linkbordercolor=green, 
    citebordercolor=green, 
    filebordercolor=green, 
    urlbordercolor=green 
}
\usepackage{cleveref}
\usepackage{ifthen}
\usepackage{varwidth}
\usepackage{witharrows}
\usepackage{afterpage}
\usepackage{xparse}

\usepackage{pgfplots}
\pgfplotsset{compat=newest}
\usepackage{tikzit}
\input{figures.tikzstyles}
\usetikzlibrary{arrows.meta, positioning}
\usepackage{etoolbox}

\newcommand{\continuefrom}[1]{%
  \ifcsname r@#1\endcsname
    \setcounterref{AlgoLine}{#1}%
    \stepcounter{AlgoLine}%
  \else
    \setcounter{AlgoLine}{10}%
  \fi
}

\newcommand{\marklastline}[1]{%
  \nllabel{#1}%
}

\makeatletter

\define@key{commfromslot}{from}{\def\commfromslot@From{#1}}
\define@key{commfromslot}{to}{\def\commfromslot@To{#1}}
\define@key{commfromslot}{val}{\def\commfromslot@Val{#1}}
\define@key{commfromslot}{when}{\def\commfromslot@When{#1}}
\newcommand{\commfromslot}[2][]{%
    \setkeys{commfromslot}{from=,to=,val=,when=}%
    \setkeys{commfromslot}{#1}%
    \ensuremath{\overline{#2}_{\commfromslot@From}^{\commfromslot@To%
    \ifthenelse{\equal{\commfromslot@Val}{}}{}{,\commfromslot@Val}%
    \ifthenelse{\equal{\commfromslot@When}{}}{}{,\commfromslot@When}%
    }}%
}%

\define@key{commfromblock}{from}{\def\commfromblock@From{#1}}
\define@key{commfromblock}{to}{\def\commfromblock@To{#1}}
\define@key{commfromblock}{val}{\def\commfromblock@Val{#1}}
\define@key{commfromblock}{when}{\def\commfromblock@When{#1}}
\newcommand{\commfromblock}[2][]{%
    \setkeys{commfromblock}{from=,to=,val=,when=}%
    \setkeys{commfromblock}{#1}%
    \ensuremath{{#2}_{\commfromblock@From}^{\commfromblock@To%
    \ifthenelse{\equal{\commfromblock@Val}{}}{}{,\commfromblock@Val}%
    \ifthenelse{\equal{\commfromblock@When}{}}{}{,\commfromblock@When}%
    }}%
}%

\define@key{commfromafterblock}{from}{\def\commfromafterblock@From{#1}}
\define@key{commfromafterblock}{to}{\def\commfromafterblock@To{#1}}
\define@key{commfromafterblock}{val}{\def\commfromafterblock@Val{#1}}
\define@key{commfromafterblock}{when}{\def\commfromafterblock@When{#1}}
\newcommand{\commfromafterblock}[2][]{%
    \setkeys{commfromafterblock}{from=,to=,val=,when=}%
    \setkeys{commfromafterblock}{#1}%
    \ensuremath{{#2}_{\commfromafterblock@From +}^{\commfromafterblock@To%
    \ifthenelse{\equal{\commfromafterblock@Val}{}}{}{,\commfromafterblock@Val}%
    \ifthenelse{\equal{\commfromafterblock@When}{}}{}{,\commfromafterblock@When}%
    }}%
}%

\define@key{commfromafterparentblock}{from}{\def\commfromafterparentblock@From{#1}}
\define@key{commfromafterparentblock}{to}{\def\commfromafterparentblock@To{#1}}
\define@key{commfromafterparentblock}{val}{\def\commfromafterparentblock@Val{#1}}
\define@key{commfromafterparentblock}{when}{\def\commfromafterparentblock@When{#1}}
\newcommand{\commfromafterparentblock}[2][]{%
    \setkeys{commfromafterparentblock}{from=,to=,when=,val=}%
    \setkeys{commfromafterparentblock}{#1}%
    \commfromafterblock[from=\parent(\commfromafterparentblock@From),to=\commfromafterparentblock@To,val=\commfromafterparentblock@Val,when=\commfromafterparentblock@When]{#2}%
}%

\define@key{commatepoch}{epoch}{\def\commatepoch@Epoch{#1}}
\define@key{commatepoch}{val}{\def\commatepoch@Val{#1}}
\define@key{commatepoch}{when}{\def\commatepoch@When{#1}}
\newcommand{\commatepoch}[2][]{%
    \setkeys{commatepoch}{epoch=,val=,when=}%
    \setkeys{commatepoch}{#1}%
    \ensuremath{{\hat{\overline{#2}}}{}^{\commatepoch@Epoch%
    \ifthenelse{\equal{\commatepoch@Val}{}}{}{,\commatepoch@Val}%
    \ifthenelse{\equal{\commatepoch@When}{}}{}{,\commatepoch@When}%
    }}
}%

\ExplSyntaxOn

\NewDocumentCommand{\addweight}{m}{
  \tl_set:Nx \l_tmpa_tl { \cs_to_str:N #1 }
  \tl_put_right:Nn \l_tmpa_tl { weight }
  \use:c { \l_tmpa_tl }
}

\ExplSyntaxOff

\define@key{chkpattime}{time}{\def\chkpattime@Time{#1}}
\define@key{chkpattime}{val}{\def\chkpattime@Val{#1}}
\setkeys{chkpattime}{time=,val=}
\newcommand{\chkpattime}[1][]{%
    \setkeys{chkpattime}{#1}%
    \ensuremath{\gjviewsym^{\chkpattime@Time\ifthenelse{\equal{\chkpattime@Val}{}}{}{,\chkpattime@Val}}}%
}%

\define@key{commweightatepoch}{epoch}{\def\commweightatepoch@Epoch{#1}}
\define@key{commweightatepoch}{val}{\def\commweightatepoch@Val{#1}}
\define@key{commweightatepoch}{when}{\def\commweightatepoch@When{#1}}
\define@key{commweightatepoch}{chkp}{\def\commweightatepoch@Chkp{#1}}
\newcommand{\commweightatepoch}[2][]{%
    \setkeys{commweightatepoch}{epoch=,val=,when=,chkp=}%
    \setkeys{commweightatepoch}{#1}%
    \ensuremath{\hat{\overline{\addweight{#2}}}{}^{\commweightatepoch@Epoch,%
    \ifthenelse{\equal{\commweightatepoch@Val}{}}{}{\commweightatepoch@Val,}%
    \ifthenelse{\equal{\commweightatepoch@When}{}}{}{\commweightatepoch@When,}%
    \commweightatepoch@Chkp%
    }}
}%

\define@key{commweightfromblock}{from}{\def\commweightfromblock@From{#1}}
\define@key{commweightfromblock}{to}{\def\commweightfromblock@To{#1}}
\define@key{commweightfromblock}{chkp}{\def\commweightfromblock@ValSetChkp{#1}}
\define@key{commweightfromblock}{val}{\def\commweightfromblock@Val{#1}}
\define@key{commweightfromblock}{when}{\def\commweightfromblock@When{#1}}
\newcommand{\commweightfromblock}[2][]{%
    \setkeys{commweightfromblock}{from=,to=,chkp=,when=,val=}
    \setkeys{commweightfromblock}{#1}%
    \ensuremath{\addweight{#2}_{\commweightfromblock@From}^{\commweightfromblock@To,%
    \ifthenelse{\equal{\commweightfromblock@Val}{}}{}{\commweightfromblock@Val,}%
    \ifthenelse{\equal{\commweightfromblock@When}{}}{}{\commweightfromblock@When,}%
    \commweightfromblock@ValSetChkp}}%
}%

\define@key{commweightfromafterblock}{from}{\def\commweightfromafterblock@From{#1}}
\define@key{commweightfromafterblock}{to}{\def\commweightfromafterblock@To{#1}}
\define@key{commweightfromafterblock}{chkp}{\def\commweightfromafterblock@ValSetChkp{#1}}
\define@key{commweightfromafterblock}{val}{\def\commweightfromafterblock@Val{#1}}
\define@key{commweightfromafterblock}{when}{\def\commweightfromafterblock@When{#1}}
\newcommand{\commweightfromafterblock}[2][]{%
    \setkeys{commweightfromafterblock}{from=,to=,chkp=,when=,val=}
    \setkeys{commweightfromafterblock}{#1}%
    \ensuremath{\addweight{#2}_{\commweightfromafterblock@From +}^{\commweightfromafterblock@To,%
    \ifthenelse{\equal{\commweightfromafterblock@Val}{}}{}{\commweightfromafterblock@Val,}%
    \ifthenelse{\equal{\commweightfromafterblock@When}{}}{}{\commweightfromafterblock@When,}%
    \commweightfromafterblock@ValSetChkp}}%
}%

\define@key{commweightfromafterparentblock}{from}{\def\commweightfromafterparentblock@From{#1}}
\define@key{commweightfromafterparentblock}{to}{\def\commweightfromafterparentblock@To{#1}}
\define@key{commweightfromafterparentblock}{chkp}{\def\commweightfromafterparentblock@ValSetChkp{#1}}
\define@key{commweightfromafterparentblock}{val}{\def\commweightfromafterparentblock@Val{#1}}
\define@key{commweightfromafterparentblock}{when}{\def\commweightfromafterparentblock@When{#1}}
\newcommand{\commweightfromafterparentblock}[2][]{%
    \setkeys{commweightfromafterparentblock}{from=,to=,chkp=,when=,val=}%
    \setkeys{commweightfromafterparentblock}{#1}%
    \commweightfromafterblock[from=\parent(\commweightfromafterparentblock@From),to=\commweightfromafterparentblock@To,val=\commweightfromafterparentblock@Val,chkp=\commweightfromafterparentblock@ValSetChkp,when=\commweightfromafterparentblock@When]{#2}%
}%

\define@key{commweightfromslot}{from}{\def\commweightfromslot@From{#1}}
\define@key{commweightfromslot}{to}{\def\commweightfromslot@To{#1}}
\define@key{commweightfromslot}{chkp}{\def\commweightfromslot@ValsetChkp{#1}}
\define@key{commweightfromslot}{val}{\def\commweightfromslot@Val{#1}}
\define@key{commweightfromslot}{when}{\def\commweightfromslot@When{#1}}
\newcommand{\commweightfromslot}[2][]{%
    \setkeys{commweightfromslot}{from=,to=,chkp=,when=,val=}%
    \setkeys{commweightfromslot}{#1}%
    \ensuremath{\overline{{\addweight{#2}}}_{\commweightfromslot@From}^{\commweightfromslot@To,%
    \ifthenelse{\equal{\commweightfromslot@Val}{}}{}{\commweightfromslot@Val,}%
    \ifthenelse{\equal{\commweightfromslot@When}{}}{}{\commweightfromslot@When,}%
    \commweightfromslot@ValsetChkp}}%
}%

\define@key{valsetfromslot}{from}{\def\valsetfromslot@From{#1}}
\define@key{valsetfromslot}{to}{\def\valsetfromslot@To{#1}}
\define@key{valsetfromslot}{chkp}{\def\valsetfromslot@Chkp{#1}}
\setkeys{valsetfromslot}{from=,to=,chkp=}
\newcommand{\valsetfromslot}[2][]{%
    \setkeys{valsetfromslot}{#1}%
    \ensuremath{\overline{{#2}}_{\valsetfromslot@From}^{\valsetfromslot@To,\valsetfromslot@Chkp}}%
}%

\define@key{valsetweightfromslot}{from}{\def\valsetweightfromslot@From{#1}}
\define@key{valsetweightfromslot}{to}{\def\valsetweightfromslot@To{#1}}
\define@key{valsetweightfromslot}{chkp}{\def\valsetweightfromslot@ValsetChkp{#1}}
\define@key{valsetweightfromslot}{weight chkp}{\def\valsetweightfromslot@WeightChkp{#1}}

\setkeys{valsetweightfromslot}{from=,to=,chkp=,weight chkp=}
\newcommand{\valsetweightfromslot}[2][]{%
    \setkeys{valsetweightfromslot}{#1}%
    \ensuremath{\overline{{\addweight{#2}}}_{\valsetweightfromslot@From}^{\valsetweightfromslot@To,\valsetweightfromslot@ValsetChkp\ifthenelse{\equal{\valsetweightfromslot@WeightChkp}{}}{}{,\valsetweightfromslot@WeightChkp}}}%
}%

\define@key{valsetfromblock}{from}{\def\valsetfromblock@From{#1}}
\define@key{valsetfromblock}{to}{\def\valsetfromblock@To{#1}}
\define@key{valsetfromblock}{chkp}{\def\valsetfromblock@Chkp{#1}}
\setkeys{valsetfromblock}{from=,to=,chkp=}
\newcommand{\valsetfromblock}[2][]{%
    \setkeys{valsetfromblock}{#1}%
    \ensuremath{{#2}_{\valsetfromblock@From}^{\valsetfromblock@To,\valsetfromblock@Chkp}}%
}%

\define@key{totvalset}{chkp}{\def\totvalset@Chkp{#1}}
\setkeys{totvalset}{chkp=}
\newcommand{\totvalset}[2][]{%
    \setkeys{totvalset}{#1}%
    \ensuremath{{#2}_{\mathsf{t}}^{\totvalset@Chkp}}%
}%

\define@key{weightofset}{chkp}{\def\weightofset@Chkp{#1}}
\setkeys{weightofset}{chkp=}
\newcommand{\weightofset}[2][]{%
    \setkeys{weightofset}{#1}%
    \ensuremath{\left|{#2}\right|^{\weightofset@Chkp}}%
}%

\define@key{totvalsetweight}{chkp}{\def\totvalsetweight@ValsetChkp{#1}}
\setkeys{totvalsetweight}{chkp=}
\newcommand{\totvalsetweight}[2][]{%
    \setkeys{totvalsetweight}{#1}%
    \ensuremath{{\addweight{#2}}_{\mathsf{t}}^{\totvalsetweight@ValsetChkp}}%
}%

\define@key{attsetfromblock}{from}{\def\attsetfromblock@From{#1}}
\define@key{attsetfromblock}{to}{\def\attsetfromblock@To{#1}}
\define@key{attsetfromblock}{val}{\def\attsetfromblock@Val{#1}}
\define@key{attsetfromblock}{when}{\def\attsetfromblock@When{#1}}
\setkeys{attsetfromblock}{from=,to=,val=,when=}
\newcommand{\attsetfromblock}[2][]{%
    \setkeys{attsetfromblock}{#1}%
    \ensuremath{{#2}_{\attsetfromblock@From}^{\attsetfromblock@To,\attsetfromblock@Val,\attsetfromblock@When}}
}%

\define@key{attsettobediscfromblock}{from}{\def\attsettobediscfromblock@From{#1}}
\define@key{attsettobediscfromblock}{to}{\def\attsettobediscfromblock@To{#1}}
\define@key{attsettobediscfromblock}{val}{\def\attsettobediscfromblock@Val{#1}}
\define@key{attsettobediscfromblock}{when}{\def\attsettobediscfromblock@When{#1}}
\setkeys{attsettobediscfromblock}{from=,to=,val=,when=}
\newcommand{\attsettobediscfromblock}[2][]{%
    \setkeys{attsettobediscfromblock}{#1}%
    \ensuremath{\overset{\nsucc}{#2}{}_{\parent(\attsettobediscfromblock@From)+}^{\attsettobediscfromblock@To,\attsettobediscfromblock@Val,\attsettobediscfromblock@When}}
}%

\define@key{attsettobediscsimplefromblock}{from}{\def\attsettobediscsimplefromblock@From{#1}}
\define@key{attsettobediscsimplefromblock}{to}{\def\attsettobediscsimplefromblock@To{#1}}
\define@key{attsettobediscsimplefromblock}{val}{\def\attsettobediscsimplefromblock@Val{#1}}
\define@key{attsettobediscsimplefromblock}{when}{\def\attsettobediscsimplefromblock@When{#1}}
\setkeys{attsettobediscsimplefromblock}{from=,to=,val=,when=}
\newcommand{\attsettobediscsimplefromblock}[2][]{%
    \setkeys{attsettobediscsimplefromblock}{#1}%
    \ensuremath{\overset{=}{#2}{}_{\parent(\attsettobediscsimplefromblock@From)+}^{\attsettobediscsimplefromblock@To,\attsettobediscsimplefromblock@Val,\attsettobediscsimplefromblock@When}}
}%

\define@key{attsettobediscforlatefromblock}{from}{\def\attsettobediscforlatefromblock@From{#1}}
\define@key{attsettobediscforlatefromblock}{to}{\def\attsettobediscforlatefromblock@To{#1}}
\define@key{attsettobediscforlatefromblock}{val}{\def\attsettobediscforlatefromblock@Val{#1}}
\define@key{attsettobediscforlatefromblock}{when}{\def\attsettobediscforlatefromblock@When{#1}}
\setkeys{attsettobediscforlatefromblock}{from=,to=,val=,when=}
\newcommand{\attsettobediscforlatefromblock}[2][]{%
    \setkeys{attsettobediscforlatefromblock}{#1}%
    \ensuremath{\overset{= \parent(\attsettobediscforlatefromblock@From)}{#2}{}_{\slot(\attsettobediscforlatefromblock@From)}^{\attsettobediscforlatefromblock@To,\attsettobediscforlatefromblock@Val,\attsettobediscforlatefromblock@When}}
}%

\newcommand{\ghostvotersfnname}{\ensuremath{GS}}

\define@key{ghostvoters}{block}{\def\ghostvoters@Block{#1}}
\define@key{ghostvoters}{view}{\def\ghostvoters@View{#1}}
\setkeys{ghostvoters}{block=,view=}
\newcommand{\ghostvoters}[1][]{%
    \setkeys{ghostvoters}{#1}%
    \ensuremath{{\ghostvotersfnname}(\ghostvoters@Block,\ghostvoters@View)}
}%

\define@key{attsetfromblockunfiltered}{from}{\def\attsetfromblockunfiltered@From{#1}}
\define@key{attsetfromblockunfiltered}{to}{\def\attsetfromblockunfiltered@To{#1}}
\define@key{attsetfromblockunfiltered}{val}{\def\attsetfromblockunfiltered@Val{#1}}
\define@key{attsetfromblockunfiltered}{when}{\def\attsetfromblockunfiltered@When{#1}}
\setkeys{attsetfromblockunfiltered}{from=,to=,val=,when=}
\newcommand{\attsetfromblockunfiltered}[2][]{%
    \setkeys{attsetfromblockunfiltered}{#1}%
    \ensuremath{{#2}_{\attsetfromblockunfiltered@From}^{\attsetfromblockunfiltered@To,\attsetfromblockunfiltered@Val,\attsetfromblockunfiltered@When}}%
    \setkeys{attsetfromblockunfiltered}{from=,to=,val=,when=}%
}%

\define@key{attsetweightfromblock}{from}{\def\attsetweightfromblock@From{#1}}
\define@key{attsetweightfromblock}{to}{\def\attsetweightfromblock@To{#1}}
\define@key{attsetweightfromblock}{val}{\def\attsetweightfromblock@Val{#1}}
\define@key{attsetweightfromblock}{when}{\def\attsetweightfromblock@When{#1}}
\define@key{attsetweightfromblock}{chkp}{\def\attsetweightfromblock@ValSetChkp{#1}}
\setkeys{attsetweightfromblock}{from=,to=,val=,when=}
\newcommand{\attsetweightfromblock}[2][]{%
    \setkeys{attsetweightfromblock}{#1}%
    \ensuremath{{\addweight{#2}}_{\attsetweightfromblock@From}^{\attsetweightfromblock@To,\attsetweightfromblock@Val,\attsetweightfromblock@When,\attsetweightfromblock@ValSetChkp}}%
    \setkeys{attsetweightfromblock}{from=,to=,val=,when=,chkp=}%
}%

\define@key{attsetweightfromafterblock}{from}{\def\attsetweightfromafterblock@From{#1}}
\define@key{attsetweightfromafterblock}{to}{\def\attsetweightfromafterblock@To{#1}}
\define@key{attsetweightfromafterblock}{val}{\def\attsetweightfromafterblock@Val{#1}}
\define@key{attsetweightfromafterblock}{when}{\def\attsetweightfromafterblock@When{#1}}
\define@key{attsetweightfromafterblock}{chkp}{\def\attsetweightfromafterblock@ValSetChkp{#1}}
\setkeys{attsetweightfromafterblock}{from=,to=,val=,when=}
\newcommand{\attsetweightfromafterblock}[2][]{%
    \setkeys{attsetweightfromafterblock}{#1}%
    \ensuremath{{\addweight{#2}}_{\attsetweightfromafterblock@From +}^{\attsetweightfromafterblock@To,\attsetweightfromafterblock@Val,\attsetweightfromafterblock@When,\attsetweightfromafterblock@ValSetChkp}}%
    \setkeys{attsetweightfromafterblock}{from=,to=,val=,when=,chkp=}%
}%

\define@key{attsetweighttobediscfromblock}{from}{\def\attsetweighttobediscfromblock@From{#1}}
\define@key{attsetweighttobediscfromblock}{to}{\def\attsetweighttobediscfromblock@To{#1}}
\define@key{attsetweighttobediscfromblock}{val}{\def\attsetweighttobediscfromblock@Val{#1}}
\define@key{attsetweighttobediscfromblock}{when}{\def\attsetweighttobediscfromblock@When{#1}}
\define@key{attsetweighttobediscfromblock}{chkp}{\def\attsetweighttobediscfromblock@ValSetChkp{#1}}
\setkeys{attsetweighttobediscfromblock}{from=,to=,val=,when=}
\newcommand{\attsetweighttobediscfromblock}[2][]{%
    \setkeys{attsetweighttobediscfromblock}{#1}%
    \ensuremath{\overset{\nsucc}{\addweight{#2}}{}_{\parent(\attsetweighttobediscfromblock@From)+}^{\attsetweighttobediscfromblock@To,\attsetweighttobediscfromblock@Val,\attsetweighttobediscfromblock@When,\attsetweighttobediscfromblock@ValSetChkp}}%
    \setkeys{attsetweighttobediscfromblock}{from=,to=,val=,when=,chkp=}%
}%

\define@key{attsetweighttobediscsimplefromblock}{from}{\def\attsetweighttobediscsimplefromblock@From{#1}}
\define@key{attsetweighttobediscsimplefromblock}{to}{\def\attsetweighttobediscsimplefromblock@To{#1}}
\define@key{attsetweighttobediscsimplefromblock}{val}{\def\attsetweighttobediscsimplefromblock@Val{#1}}
\define@key{attsetweighttobediscsimplefromblock}{when}{\def\attsetweighttobediscsimplefromblock@When{#1}}
\define@key{attsetweighttobediscsimplefromblock}{chkp}{\def\attsetweighttobediscsimplefromblock@ValSetChkp{#1}}
\setkeys{attsetweighttobediscsimplefromblock}{from=,to=,val=,when=}
\newcommand{\attsetweighttobediscsimplefromblock}[2][]{%
    \setkeys{attsetweighttobediscsimplefromblock}{#1}%
    \ensuremath{\overset{=}{\addweight{#2}}{}_{\parent(\attsetweighttobediscsimplefromblock@From)+}^{\attsetweighttobediscsimplefromblock@To,\attsetweighttobediscsimplefromblock@Val,\attsetweighttobediscsimplefromblock@When,\attsetweighttobediscsimplefromblock@ValSetChkp}}%
    \setkeys{attsetweighttobediscsimplefromblock}{from=,to=,val=,when=,chkp=}%
}%

\define@key{attsetweighttobediscforlatefromblock}{from}{\def\attsetweighttobediscforlatefromblock@From{#1}}
\define@key{attsetweighttobediscforlatefromblock}{to}{\def\attsetweighttobediscforlatefromblock@To{#1}}
\define@key{attsetweighttobediscforlatefromblock}{val}{\def\attsetweighttobediscforlatefromblock@Val{#1}}
\define@key{attsetweighttobediscforlatefromblock}{when}{\def\attsetweighttobediscforlatefromblock@When{#1}}
\define@key{attsetweighttobediscforlatefromblock}{chkp}{\def\attsetweighttobediscforlatefromblock@ValSetChkp{#1}}
\setkeys{attsetweighttobediscforlatefromblock}{from=,to=,val=,when=}
\newcommand{\attsetweighttobediscforlatefromblock}[2][]{%
    \setkeys{attsetweighttobediscforlatefromblock}{#1}%
    \ensuremath{\overset{=\parent(\attsetweighttobediscforlatefromblock@From)}{\addweight{#2}}{}_{\slot(\attsetweighttobediscforlatefromblock@From)}^{\attsetweighttobediscforlatefromblock@To,\attsetweighttobediscforlatefromblock@Val,\attsetweighttobediscforlatefromblock@When,\attsetweighttobediscforlatefromblock@ValSetChkp}}%
    \setkeys{attsetweighttobediscforlatefromblock}{from=,to=,val=,when=,chkp=}%
}%

\define@key{varforvalattime}{time}{\def\varforvalattime@Time{#1}}
\define@key{varforvalattime}{val}{\def\varforvalattime@Val{#1}}
\setkeys{varforvalattime}{time=,val=}
\newcommand{\varforvalattime}[2][]{%
    \setkeys{varforvalattime}{#1}%
    \ensuremath{%
        \ifthenelse%
        {\equal{\varforvalattime@Time}%
        {}}%
        {{#2}_{\varforvalattime@Val}}
        {{#2}_{\varforvalattime@Val}^{\varforvalattime@Time}}}%
        \setkeys{varforvalattime}{time=,val=}%
}%

\define@key{indicatorfromblock}{from}{\def\indicatorfromblock@From{#1}}
\define@key{indicatorfromblock}{to}{\def\indicatorfromblock@To{#1}}
\define@key{indicatorfromblock}{val}{\def\indicatorfromblock@Val{#1}}
\define@key{indicatorfromblock}{when}{\def\indicatorfromblock@When{#1}}
\define@key{indicatorfromblock}{chkp}{\def\indicatorfromblock@ValSetChkp{#1}}
\setkeys{indicatorfromblock}{from=,to=,val=,when=,chkp=}
\newcommand{\indicatorfromblock}[2][]{%
    \setkeys{indicatorfromblock}{#1}%
    \ensuremath{{{#2}}_{\indicatorfromblock@From}^{\indicatorfromblock@To,\indicatorfromblock@Val,\indicatorfromblock@When,\indicatorfromblock@ValSetChkp}}%
    \setkeys{indicatorfromblock}{from=,to=,val=,when=,chkp=}%
}%

\define@key{indicatordiscfromblock}{from}{\def\indicatordiscfromblock@From{#1}}
\define@key{indicatordiscfromblock}{to}{\def\indicatordiscfromblock@To{#1}}
\define@key{indicatordiscfromblock}{val}{\def\indicatordiscfromblock@Val{#1}}
\define@key{indicatordiscfromblock}{when}{\def\indicatordiscfromblock@When{#1}}
\define@key{indicatordiscfromblock}{chkp}{\def\indicatordiscfromblock@ValSetChkp{#1}}
\setkeys{indicatordiscfromblock}{from=,to=,val=,when=,chkp=}
\newcommand{\indicatordiscfromblock}[2][]{%
    \setkeys{indicatordiscfromblock}{#1}%
    \ensuremath{\overset{\mathsf{wdisc}}{#2}{}_{\indicatordiscfromblock@From}^{\indicatordiscfromblock@To,\indicatordiscfromblock@Val,\indicatordiscfromblock@When,\indicatordiscfromblock@ValSetChkp}}%
    \setkeys{indicatordiscfromblock}{from=,to=,val=,when=,chkp=}%
}%

\define@key{boostweight}{chkp}{\def\boostweight@Chkp{#1}}
\setkeys{boostweight}{chkp=}
\newcommand{\boostweight}[1][]{%
    \setkeys{boostweight}{#1}%
    \ensuremath{W_p^{\boostweight@Chkp}}%
}%

\define@key{ffgvote}{from}{\def\ffgvote@From{#1}}
\define@key{ffgvote}{to}{\def\ffgvote@To{#1}}
\setkeys{ffgvote}{from=,to=}
\newcommand{\ffgvote}[1][]{%
    \setkeys{ffgvote}{#1}%
    \ensuremath{{\ffgvote@From}\to{\ffgvote@To}}
}%

\define@key{ffgvalsetallsentraw}{source}{\def\ffgvalsetallsentraw@Source{#1}}
\define@key{ffgvalsetallsentraw}{target}{\def\ffgvalsetallsentraw@Target{#1}}
\define@key{ffgvalsetallsentraw}{time}{\def\ffgvalsetallsentraw@Time{#1}}
\setkeys{ffgvalsetallsentraw}{source=,target=,time=}
\newcommand{\ffgvalsetallsentraw}[1][]{%
    \setkeys{ffgvalsetallsentraw}{#1}%
    \ensuremath{\overset{T}{\mathcal{F}}{}_{\ffgvalsetallsentraw@Source \to \ffgvalsetallsentraw@Target}}^{\ffgvalsetallsentraw@Time}%
}%

\define@key{ffgvalsetraw}{source}{\def\ffgvalsetraw@Source{#1}}
\define@key{ffgvalsetraw}{target}{\def\ffgvalsetraw@Target{#1}}
\define@key{ffgvalsetraw}{time}{\def\ffgvalsetraw@Time{#1}}
\define@key{ffgvalsetraw}{val}{\def\ffgvalsetraw@Val{#1}}
\setkeys{ffgvalsetraw}{source=,target=,val=,time=}
\newcommand{\ffgvalsetraw}[1][]{%
    \setkeys{ffgvalsetraw}{#1}%
    \ensuremath{\mathcal{F}_{\ffgvalsetraw@Source \to \ffgvalsetraw@Target}^{\ffgvalsetraw@Val,\ffgvalsetraw@Time}}%
}%

\define@key{ffgvalsettoslot}{source}{\def\ffgvalsettoslot@Source{#1}}
\define@key{ffgvalsettoslot}{target}{\def\ffgvalsettoslot@Target{#1}}
\define@key{ffgvalsettoslot}{time}{\def\ffgvalsettoslot@Time{#1}}
\define@key{ffgvalsettoslot}{val}{\def\ffgvalsettoslot@Val{#1}}
\define@key{ffgvalsettoslot}{to}{\def\ffgvalsettoslot@To{#1}}
\setkeys{ffgvalsettoslot}{source=,target=,val=,time=,to=}
\newcommand{\ffgvalsettoslot}[1][]{%
    \setkeys{ffgvalsettoslot}{#1}%
    \ensuremath{\mathcal{F}_{\ffgvalsettoslot@Source \to \ffgvalsettoslot@Target}^{\ffgvalsettoslot@To,\ffgvalsettoslot@Val,\ffgvalsettoslot@Time}}%
}%

\define@key{ffgvalset}{target blck}{\def\ffgvalset@TargetBlck{#1}}
\define@key{ffgvalset}{time}{\def\ffgvalset@Time{#1}}
\define@key{ffgvalset}{val}{\def\ffgvalset@Val{#1}}
\setkeys{ffgvalset}{time=,val=,target blck=}
\newcommand{\ffgvalset}[1][]{%
    \setkeys{ffgvalset}{#1}%
    \ffgvalsetraw[source=\gjblock(\ffgvalset@TargetBlck),target=\chkp(\ffgvalset@TargetBlck),time=\ffgvalset@Time,val=\ffgvalset@Val]%
}%

\define@key{ffgvalsetallsent}{target blck}{\def\ffgvalsetallsent@TargetBlck{#1}}
\define@key{ffgvalsetallsent}{time}{\def\ffgvalsetallsent@Time{#1}}
\setkeys{ffgvalsetallsent}{time=,target blck=}
\newcommand{\ffgvalsetallsent}[1][]{%
    \setkeys{ffgvalsetallsent}{#1}%
    \ffgvalsetallsentraw[source={\votsource[blck=b,time=\epoch(b)]},target=\chkp(\ffgvalsetallsent@TargetBlck),time=\ffgvalsetallsent@Time]%
}%

\define@key{ffgvalsetweightraw}{source}{\def\ffgvalsetweightraw@Source{#1}}
\define@key{ffgvalsetweightraw}{target}{\def\ffgvalsetweightraw@Target{#1}}
\define@key{ffgvalsetweightraw}{time}{\def\ffgvalsetweightraw@Time{#1}}
\define@key{ffgvalsetweightraw}{val}{\def\ffgvalsetweightraw@Val{#1}}
\define@key{ffgvalsetweightraw}{weight chkp}{\def\ffgvalsetweightraw@WeightChkp{#1}}
\setkeys{ffgvalsetweightraw}{time=,val=,source=,target=,weight chkp=}
\newcommand{\ffgvalsetweightraw}[1][]{%
    \setkeys{ffgvalsetweightraw}{#1}%
    \ensuremath{F_{\ffgvalsetweightraw@Source \to \ffgvalsetweightraw@Target}^{\ffgvalsetweightraw@Val,\ffgvalsetweightraw@Time,\ffgvalsetweightraw@WeightChkp}}%
}%

\define@key{ffgvalsettoslotweight}{source}{\def\ffgvalsettoslotweight@Source{#1}}
\define@key{ffgvalsettoslotweight}{target}{\def\ffgvalsettoslotweight@Target{#1}}
\define@key{ffgvalsettoslotweight}{time}{\def\ffgvalsettoslotweight@Time{#1}}
\define@key{ffgvalsettoslotweight}{val}{\def\ffgvalsettoslotweight@Val{#1}}
\define@key{ffgvalsettoslotweight}{to}{\def\ffgvalsettoslotweight@To{#1}}
\define@key{ffgvalsettoslotweight}{weight chkp}{\def\ffgvalsettoslotweight@WeightChkp{#1}}
\setkeys{ffgvalsettoslotweight}{to=,time=,val=,source=,target=,weight chkp=}
\newcommand{\ffgvalsettoslotweight}[1][]{%
    \setkeys{ffgvalsettoslotweight}{#1}%
    \ensuremath{F_{\ffgvalsettoslotweight@Source \to \ffgvalsettoslotweight@Target}^{\ffgvalsettoslotweight@To, \ffgvalsettoslotweight@Val,\ffgvalsettoslotweight@Time,\ffgvalsettoslotweight@WeightChkp}}%
}%

\define@key{ffgvalsetweight}{target blck}{\def\ffgvalsetweight@TargetBlck{#1}}
\define@key{ffgvalsetweight}{time}{\def\ffgvalsetweight@Time{#1}}
\define@key{ffgvalsetweight}{val}{\def\ffgvalsetweight@Val{#1}}
\define@key{ffgvalsetweight}{weight chkp}{\def\ffgvalsetweight@WeightChkp{#1}}
\setkeys{ffgvalsetweight}{time=,val=,target blck=,weight chkp=}
\newcommand{\ffgvalsetweight}[1][]{%
    \setkeys{ffgvalsetweight}{#1}%
    \ffgvalsetweightraw[source=\gjblock(\ffgvalsetweight@TargetBlck),target=\chkp(\ffgvalsetweight@TargetBlck),time=\ffgvalsetweight@Time,val=\ffgvalsetweight@Val,weight chkp=\ffgvalsetweight@WeightChkp]%
}%

\define@key{slashedset}{chkp}{\def\slashedset@Chkp{#1}}
\setkeys{slashedset}{chkp=}
\newcommand{\slashedset}[1][]{%
    \setkeys{slashedset}{#1}%
    \ensuremath{\mathcal{D}^{\slashedset@Chkp}}%
}%

\define@key{slashedweight}{chkp}{\def\slashedweight@Chkp{#1}}
\setkeys{slashedweight}{chkp=}
\newcommand{\slashedweight}[1][]{%
    \setkeys{slashedweight}{#1}%
    \ensuremath{{D}^{\slashedweight@Chkp}}%
}%

\define@key{viewattime}{time}{\def\viewattime@Time{#1}}
\define@key{viewattime}{val}{\def\viewattime@Val{#1}}
\setkeys{viewattime}{time=,val=}
\newcommand{\viewattime}[1][]{%
    \setkeys{viewattime}{#1}%
    \ensuremath{\View^{\viewattime@Val,\viewattime@Time}}%
    \setkeys{viewattime}{time=,val=}%
}%

\define@key{viewatstslottime}{time}{\def\viewatstslottime@Time{#1}}
\define@key{viewatstslottime}{val}{\def\viewatstslottime@Val{#1}}
\setkeys{viewatstslottime}{time=,val=}
\newcommand{\viewatstslottime}[1][]{%
    \setkeys{viewatstslottime}{#1}%
    \ensuremath{\View^{\viewatstslottime@Val,\slotstart(\slot(\viewatstslottime@Time))}}%
    \setkeys{viewatstslottime}{time=,val=}%
}%

\define@key{gviewattime}{time}{\def\gviewattime@Time{#1}}
\setkeys{gviewattime}{time=}
\newcommand{\gviewattime}[1][]{%
    \setkeys{gviewattime}{#1}%
    \ensuremath{\View^{\mathsf{gbl},\gviewattime@Time}}%
}%

\newcommand{\gjviewsym}{\ensuremath{\mathsf{GJ}}}

\define@key{gjview}{time}{\def\gjview@Time{#1}}
\define@key{gjview}{view}{\def\gjview@View{#1}}
\setkeys{gjview}{time=,view=}
\newcommand{\gjview}[1][]{%
    \setkeys{gjview}{time=,view=}%
    \setkeys{gjview}{#1}%
    \ensuremath{\gjviewsym(\gjview@View, \gjview@Time)}%
}%

\define@key{gfview}{time}{\def\gfview@Time{#1}}
\define@key{gfview}{view}{\def\gfview@View{#1}}
\setkeys{gfview}{time=,view=}
\newcommand{\gfview}[1][]{%
    \setkeys{gfview}{#1}%
    \ensuremath{\mathit{GF}(\gfview@View, \gfview@Time)}%
}%

\define@key{gjattime}{time}{\def\gjattime@Time{#1}}
\define@key{gjattime}{val}{\def\gjattime@Val{#1}}
\setkeys{gjattime}{time=,val=}
\newcommand{\gjattime}[1][]{%
    \setkeys{gjattime}{time=,val=}%
    \setkeys{gjattime}{#1}%
    \ensuremath{\gjviewsym^{\gjattime@Time\ifthenelse{\equal{\gjattime@Val}{}}{}{,\gjattime@Val}}}%
}%

\define@key{gfattime}{time}{\def\gfattime@Time{#1}}
\define@key{gfattime}{val}{\def\gfattime@Val{#1}}
\setkeys{gfattime}{time=,val=}
\newcommand{\gfattime}[1][]{%
    \setkeys{gfattime}{time=,val=}%
    \setkeys{gfattime}{#1}%
    \ensuremath{\mathsf{GF}^{\gfattime@Time\ifthenelse{\equal{\gfattime@Val}{}}{}{,\gfattime@Val}}}%
}%

\newcommand{\guviewsym}{\ensuremath{\mathsf{GU}}}
\define@key{guattime}{time}{\def\guattime@Time{#1}}
\define@key{guattime}{val}{\def\guattime@Val{#1}}
\setkeys{guattime}{time=,val=}
\newcommand{\guattime}[1][]{%
    \setkeys{guattime}{time=,val=}%
    \setkeys{guattime}{#1}%
    \ensuremath{\guviewsym^{\guattime@Time\ifthenelse{\equal{\guattime@Val}{}}{}{,\guattime@Val}}}%
}%

\define@key{ceba}{time}{\def\ceba@Time{#1}}
\define@key{ceba}{val}{\def\ceba@Val{#1}}
\setkeys{ceba}{time=,val=}
\newcommand{\ceba}[1][]{%
    \setkeys{ceba}{time=,val=}%
    \setkeys{ceba}{#1}%
    \ensuremath{\chkp_\mathsf{EBA}^{\ceba@Time,\ceba@Val}}%
}%

\define@key{bconfirmed}{time}{\def\bconfirmed@Time{#1}}
\define@key{bconfirmed}{val}{\def\bconfirmed@Val{#1}}
\setkeys{bconfirmed}{time=,val=}
\newcommand{\bconfirmed}[1][]{%
    \setkeys{bconfirmed}{time=,val=}%
    \setkeys{bconfirmed}{#1}%
    \ifthenelse{\equal{\bconfirmed@Time}{}\and\equal{\bconfirmed@Val}{}}{\ensuremath{b_\mathsf{confirmed}}}%
    {{\ensuremath{b_\mathsf{bconfirmed}^{\bconfirmed@Time,\bconfirmed@Val}}}}%
}%

\define@key{finattime}{time}{\def\finattime@Time{#1}}
\define@key{finattime}{val}{\def\finattime@Val{#1}}
\setkeys{finattime}{time=,val=}
\newcommand{\finattime}[1][]{%
    \setkeys{finattime}{#1}%
    \ensuremath{\mathsf{GF}^{\finattime@Time\ifthenelse{\equal{\finattime@Val}{}}{}{,\finattime@Val}}}%
}%

\define@key{filtered}{time}{\def\filtered@Time{#1}}
\define@key{filtered}{val}{\def\filtered@Val{#1}}
\setkeys{filtered}{time=,val=}
\newcommand{\filtered}[1][]{%
    \setkeys{filtered}{time=,val=}%
    \setkeys{filtered}{#1}%
    \ensuremath{\mathsf{filt}_{\mathsf{hfc}}^{\filtered@Time\ifthenelse{\equal{\filtered@Val}{}}{}{,\filtered@Val}}}%
}%

\define@key{canonical}{time}{\def\canonical@Time{#1}}
\define@key{canonical}{blck}{\def\canonical@Blck{#1}}
\define@key{canonical}{to}{\def\canonical@To{#1}}
\setkeys{canonical}{time=,blck=}
\newcommand{\canonical}[1][]{%
    \setkeys{canonical}{time=,blck=,to=}%
    \setkeys{canonical}{#1}%
    \ensuremath{\square\mathsf{canonical}(\canonical@Blck,\canonical@Time\ifthenelse{\equal{\canonical@To}{}}{,\infty}{,\canonical@To})}%
}%

\define@key{votsource}{time}{\def\votsource@Time{#1}}
\define@key{votsource}{blck}{\def\votsource@Blk{#1}}
\define@key{votsource}{chkp}{\def\votsource@Chkp{#1}}
\newcommand{\votsource}[1][]{%
\setkeys{votsource}{time=,blck=,chkp=}%
    \setkeys{votsource}{#1}%
    \ensuremath{\mathsf{vs}(%
    \ifthenelse{\equal{\votsource@Chkp}{}}{%
    \votsource@Blk,\votsource@Time%
    }{%
    \votsource@Chkp}%
    )}%
}%

\define@key{blckandtime}{time}{\def\blckandtime@Time{#1}}
\define@key{blckandtime}{blck}{\def\blckandtime@Blk{#1}}
\newcommand{\blckandtime}[1][]{%
\setkeys{blckandtime}{time=,blck=}%
    \setkeys{blckandtime}{#1}%
    \ensuremath{(\textsf{blck: }\blckandtime@Blk,\textsf{ time: } \blckandtime@Time)}%
}%

\define@key{fcparam}{fc}{\def\fcparam@FC{#1}}
\define@key{fcparam}{balf}{\def\fcparam@BalF{#1}}
\define@key{fcparam}{val}{\def\fcparam@Val{#1}}
\setkeys{fcparam}{fc=,balf=,val=}
\newcommand{\fcparam}[1][]{%
    \setkeys{fcparam}{#1}%
    \ensuremath{\ifthenelse{\equal{\fcparam@Val}{}}{{\fcparam@FC}_{\fcparam@BalF}}{{\fcparam@FC}_{\fcparam@BalF}^{\fcparam@Val}}}%
}%

\newcommand{\customlabel}[3][]{%
   \protected@write \@auxout {}{\string \newlabel {#2}{{#3}{\thepage}{#3}{#2}{}} }%
   \protected@write \@auxout {}{\string \newlabel {#2@cref}{{[#1][#3][]#3}{[1][\thepage][]\thepage}}}%
   \hypertarget{#2}{}%
}

\define@key{var}{time}{\def\var@Time{#1}}
\define@key{var}{val}{\def\var@Val{#1}}
\setkeys{var}{time=,val=}
\newcommand{\var}[2][]{%
    \setkeys{var}{#1}%
    \ensuremath{%
        \ifthenelse%
        {\equal{\var@Time}%
        {}}%
        {#2^{\var@Val}}
        {#2^{\var@Val,\var@Time}}}%
        \setkeys{var}{time=,val=}%
}%

\define@key{children}{view}{\def\children@View{#1}}
\define@key{children}{blck}{\def\children@Blck{#1}}
\setkeys{children}{view=,blck=}
\newcommand{\children}[1][]{%
    \setkeys{children}{#1}%
    \ensuremath{\mathsf{children}(\children@Blck,\children@View)}%
}%

\makeatletter
\newcommand{\siritem}[1]{%
  \item[SIR.\number\numexpr\value{\@listctr}+1\relax#1]%
  \refstepcounter{\@listctr}%
  \edef\@currentlabel{SIR.\arabic{\@listctr}}%
}
\makeatother

\crefname{assumption}{assumption}{assumptions}
\Crefname{assumption}{Assumption}{Assumptions}
\crefname{condition}{condition}{conditions}
\crefname{algocfline}{line}{lines}
\Crefname{algocfline}{Line}{Lines}

\Crefname{AlgoLine}{Line}{Lines}

\newcounter{countersetupcrefforlines}
\loop
  \edef\temp{\noexpand\crefname{linecounter\arabic{countersetupcrefforlines}}{line}{lines}}
  \temp
  \stepcounter{countersetupcrefforlines}
  \ifnum\value{countersetupcrefforlines}<10
\repeat
\undef{\temp}

\makeatother

\newcommand{\chain}[0]{\ensuremath{\mathsf{chain}}}
\newcommand{\lmdghost}[0]{\ensuremath{\operatorname{\textsc{LMD-GHOST}}}}

\newcommand{\GHOST}[0]{\textsf{GHOST}\xspace}
\newcommand{\FFG}[0]{\textsf{FFG}\xspace}

\newcommand{\LMDGHOST}[0]{\textsf{LMD-GHOST}\xspace}
\newcommand{\FFGCASPER}[0]{\textsf{FFG-Casper}\xspace}
\newcommand{\LMDGHOSTHFC}[0]{\textsf{LMD-GHOST-HFC}\xspace}

\newcommand{\FIL}[0]{\textsf{FIL}\xspace}
\newcommand{\FILINV}[0]{\ensuremath{\FIL_{\lnot \mathsf{valid}}}\xspace}
\newcommand{\FILEQ}[0]{\ensuremath{\FIL_{\mathsf{eq}}}\xspace}
\newcommand{\FILLMD}[0]{\ensuremath{\FIL_{\mathsf{lmd}}}\xspace}
\newcommand{\FILCUR}[0]{\ensuremath{\FIL_{\mathsf{cur}}}\xspace}
\newcommand{\FILHFC}[0]{\ensuremath{\FIL_{\mathsf{hfc}}}\xspace}

\newcommand{\isOneConfirmed}{\ensuremath{\mathit{is\_one\_confirmed}}}
\newcommand{\isOneConfirmedWeaker}{\ensuremath{\mathit{is\_one\_confirmed\_weaker}}}
\newcommand{\isOneConfirmedSpecial}{\ensuremath{\mathit{is\_one\_confirmed\_special}}}
\newcommand{\isOneConfirmedExt}{\ensuremath{\mathit{is\_one\_confirmed\_ext}}}

\newcommand{\willChkpBeJustified}{\ensuremath{\mathit{will\_chkp\_be\_justified}}}
\newcommand{\willNoConflictingChkpBeJustified}{\ensuremath{\mathit{will\_no\_conflicting\_chkp\_be\_justified}}}

\newcommand{\getHighest}{\ensuremath{\mathit{get\_highest}}}
\newcommand{\canonChainFrom}{\ensuremath{\mathit{canon\_chain\_from}}}
\newcommand{\isOneConfirmedExtFrom}{\ensuremath{\mathit{is\_one\_confirmed\_ext\_from}}}
\newcommand{\isLMDGHOSTSafe}{\ensuremath{\mathit{is\_LMD\_GHOST\_safe}}}
\newcommand{\isChainOneConfirmed}{\ensuremath{\mathit{is\_chain\_one\_confirmed}}}
\newcommand{\isChainOneConfirmedImpl}{\ensuremath{\mathit{is\_chain\_one\_confirmed\_impl}}}
\newcommand{\Wslot}{\ensuremath{W_{\mathsf{slot}}}}

\newcommand{\FC}[0]{\ensuremath{\mathsf{FC}}}

\newcommand{\GST}[0]{\ensuremath{\mathsf{GST}}}
\newcommand{\GGST}[0]{\ensuremath{\mathbb{GST}}}

\newcommand{\View}{\ensuremath{\mathcal{V}}} 

\newcommand{\genesis}{b_{\texttt{gen}}}
\newcommand{\tuple}[1]{\langle #1\rangle}
\newcommand{\commentout}[1]{}
\DeclareMathOperator*{\argmin}{arg\,min}

\newcommand{\firstslot}[0]{\ensuremath{\mathsf{fslot}}}
\newcommand{\prevfirstslot}[1]{\ensuremath{\firstslot(#1)-1}}
\newcommand{\prevfirstslotepoch}[2][]{\ensuremath{\lastslot(\epoch(#2)\the\numexpr#1-1\relax)}}
\newcommand{\lastslot}[0]{\ensuremath{\mathsf{lslot}}}
\newcommand{\allU}[0]{\ensuremath{\mathsf{AU}}}
\newcommand{\allF}[0]{\ensuremath{\mathsf{AF}}}

\newcommand{\votingtime}[0]{{\mathsf{v}_{t}}}

\newcommand{\slotstart}[0]{\ensuremath{\mathsf{st}}}
\newcommand{\slot}[0]{\ensuremath{\mathsf{slot}}}
\newcommand{\slotstartslot}[1]{\slotstart(\slot(#1))}
\newcommand{\prevslotstartslot}[1]{\slotstart(\slot(#1)-1)}
\newcommand{\epoch}{\ensuremath{\mathsf{epoch}}}
\newcommand{\parent}{\ensuremath{\mathsf{p}}}
\newcommand{\block}{\ensuremath{\mathsf{block}}}
\newcommand{\chkp}{\ensuremath{\mathsf{C}}}
\newcommand{\latestchkp}{\ensuremath{\mathsf{C}}}
\newcommand{\effbalass}{\ensuremath{\mathsf{EBA}}}
\newcommand{\teffbalass}{\textit{effective-balance-assignment}\xspace}
\newcommand{\gu}{\ensuremath{\mathsf{GU}}}

\newcommand{\signer}{\ensuremath{\mathit{signer}}}
\newcommand{\blocksinview}{\ensuremath{\mathit{blocks}}}

\newcommand{\ghostsinview}{{\ensuremath{\mathit{GHOSTs}}}}
\newcommand{\parentslotplusone}{{\ensuremath{\mathsf{ps}^{+1}}}}

\newcommand{\bcand}{\ensuremath{b_\mathsf{cand}}}
\newcommand{\bcands}{\ensuremath{b_\mathsf{cands}}}
\newcommand{\btemp}{\ensuremath{b_\mathsf{temp}}}
\newcommand{\btcand}{\ensuremath{b_\mathsf{tentative\_cand}}}
\newcommand{\btcands}{\ensuremath{b_\mathsf{tentative\_cands}}}

\newcommand{\tinit}{\ensuremath{t_\mathsf{init}}}
\newcommand{\head}{\ensuremath{head}}
\newcommand{\phead}{\ensuremath{phead}}
\newcommand{\findlatestconfirmeddescendant}{\ensuremath{\mathit{find\_latest\_confirmed\_descendant}}}
\newcommand{\getlatestconfirmed}{\ensuremath{\mathit{get\_latest\_confirmed}}}
\newcommand{\sir}{\mathrm{SIR}}
\newcommand{\tind}{t_\mathsf{ind}}
\newcommand{\sind}{s_\mathsf{ind}}
\newcommand{\sirone}{\ref{def:induction-conditions:is-lmd-confirmed}\xspace}
\newcommand{\sirtwo}{\ref{def:induction-conditions:all-validators:not-filtered-out}\xspace}
\newcommand{\sirthree}{\ref{def:induction-conditions:all-validators:gj-succ}\xspace}
\newcommand{\sirfour}{\ref{def:induction-conditions:ub}\xspace}
\newcommand{\sirfive}{\ref{def:induction-conditions:no-conflicting}\xspace}
\newcommand{\sirsix}{\ref{def:induction-conditions:bs-canonical}\xspace}

\newcommand{\now}{\ensuremath{\mathsf{now}}}

\newcommand{\ujustify}[0]{u-justify\xspace}
\newcommand{\ujustifies}[0]{u-justifies\xspace}
\newcommand{\ujustified}[0]{u-justified\xspace}

\newcommand{\indP}{{\ensuremath{\mathit{P}}}}
\newcommand{\indQ}{{\ensuremath{\mathit{Q}}}}

\newcommand{\safetyinductionrequirement}[1]{Safety Induction Requirement{#1}\xspace}

\newcommand{\calX}{\mathcal{X}}

\newcommand{\calB}{\mathcal{B}}
\newcommand{\frakB}{\mathfrak{B}}

\newcommand{\slotsperepoch}[0]{{E}}

\newcommand{\honvalsweight}[0]{J}
\newcommand{\honvals}[0]{\mathcal{\honvalsweight}}

\newcommand{\allattsweight}[0]{S}
\newcommand{\allatts}[0]{\mathcal{\allattsweight}}

\newcommand{\advvalsweight}[0]{A}
\newcommand{\advvals}[0]{\mathcal{\advvalsweight}}

\newcommand{\slashvalsweight}[0]{I}
\newcommand{\slashvals}[0]{\mathcal{\slashvalsweight}}

\newcommand{\honattsweight}[0]{H}
\newcommand{\honatts}[0]{\mathcal{\honattsweight}}

\newcommand{\honattsub}[0]{\underset{\leq}{\mathcal{H}}}

\newcommand{\allvalsweight}[0]{W}
\newcommand{\allvals}[0]{\mathcal{\allvalsweight}}

\newcommand{\boostscore}[0]{p}

\newcommand{\safetydecay}[0]{\ensuremath{d}}

\newcommand{\alignexpl}[2][]{%
    \ifthenelse{\equal{#1}{}}{%
        \text{--- #2}%
        }{%
        \begin{minipage}[t]{#1}%
            \begin{itemize}[label=---,nosep,left=0pt]%
                \item #2%
            \end{itemize}%
        \end{minipage}%
    }
}

\newtheorem{theorem}{Theorem}

\newtheorem{remark}{Remark}
\newtheorem{lemma}{Lemma}

\newtheorem{definition}{Definition}

\newtheorem{assumption}{Assumption}
\newtheorem{property}{Property}

\newcommand{\ie}{\textit{i}.\textit{e}.}
\newcommand{\eg}{\textit{e}.\textit{g}.}

\renewlist{enumerate*}{enumerate*}{2}
\setlist[enumerate*,1,2]{label=(\roman*),itemjoin={{, }}, itemjoin*={{, and }}}

\DeclareMathOperator*{\argmax}{arg\,max}

%% file: figures.tikzstyles
\tikzstyle{block}=[draw, fill=lightgray, shape=rectangle, minimum size=0.8cm, thick]
\tikzstyle{confirmed}=[draw=blue, fill=lightgray, shape=rectangle, minimum size=0.8cm, thick]
\tikzstyle{candidate}=[draw=blue, fill=lightgray, dashed, shape=rectangle, minimum size=0.8cm, thick]
\tikzstyle{missing}=[draw, fill=white, dashed, shape=rectangle, minimum size=0.8cm, thick]
\tikzstyle{l}=[shape=rectangle]
\tikzstyle{lc}=[shape=rectangle, align=center]
\tikzstyle{ll}=[shape=rectangle, align=left]
\tikzstyle{llw}=[shape=rectangle, align=left, anchor=west]
\tikzstyle{rbrace}=[shape=rectangle, scale=5]

\tikzstyle{chain}=[->]
\tikzstyle{cchain}=[->, blue]
\tikzstyle{epochline}=[-, dashed]
\tikzstyle{slotarrow}=[dotted, ->]
\tikzstyle{nowarrow}=[->]
\tikzstyle{ujustifies}=[->, red]
\tikzstyle{ujustifies nohead}=[-, red]

%% file: predicates.tex
\begin{figure}[ht]
\centering
\begin{tikzpicture}[
    arr/.style={-{Stealth[length=5pt]}, thick},
    lbl/.style={font=\scriptsize},
    cond/.style={font=\small, text centered, inner sep=4pt,
                 rectangle, draw, rounded corners=2pt},
    slot/.style={font=\footnotesize, below=2pt}
]

\draw[thick] (0,0) -- (12,0);

\foreach \x/\lab in {0.5/{$s{-}1$}, 2.5/{$s$}, 4.5/{$s{+}1$}, 6.5/{\dots}, 8.5/{$s'{-}1$}, 10.5/{$s'$}} {
    \draw[thick] (\x, -0.15) -- (\x, 0.15);
    \node[slot] at (\x, -0.15) {\lab};
}

\node[cond] (Ps) at (2.5, 4.6)
    {$\indP > \frac{\Phi}{2J(1-\beta)}$};

\node[cond] (Ps') at (10.5, 4.6)
    {$\indP > \frac{\Phi}{2J(1-\beta)}$};

\node[cond] (Q) at (2.5, 1.0)
    {$\indQ > \frac{\Phi}{2W} + \beta$};

\node[cond] (H) at (10.5, 2.8)
    {$H > \frac{\Phi}{2}$};

\node[cond] (canon) at (10.5, 1.0)
    {$b$ canonical};

\draw[arr] (Q) -- (Ps)
    node[lbl, right, xshift=2pt, midway]
    {Lem.~\ref{lem:lmd-cond-on-q-implies-cond-on-p}};

\draw[arr] (Ps) -- (Ps')
    node[lbl, above, midway]
    {monotonicity (Lem.~\ref{lem:lmd-p-monotonic})};

\draw[arr] (Ps') -- (H)
    node[lbl, right, xshift=2pt, midway]
    {Lem.~\ref{lem:lmd-cond-on-p-implies-cond-on-h-ex}};

\draw[arr] (H) -- (canon)
    node[lbl, right, xshift=2pt, midway]
    {Lem.~\ref{lem:condition-on-h-for-canonical-ex}};

\end{tikzpicture}
\caption{Implication chain for the \LMDGHOST\ safety argument.
At slot~$s$, the observable condition on~$\indQ$
\eqref{eq:cond-on-q} implies the condition on the honest
indicator~$\indP$ \eqref{eq:cond-on-p}.
By monotonicity (Lemma~\ref{lem:lmd-p-monotonic}),
the condition on~$\indP$ persists at every later slot $s' > s$,
which implies the absolute honest support
inequality~\eqref{eq:abs-honest-support}
(Lemma~\ref{lem:lmd-cond-on-p-implies-cond-on-h-ex}) and hence
canonicity of~$b$ at slot~$s'$
(Lemma~\ref{lem:condition-on-h-for-canonical-ex}).
Here $W$ denotes the total committee weight, $J$ the honest committee
weight, and $H$ the honest attestation weight supporting~$b$;
all quantities are evaluated over the relevant slot range with
subscripts and superscripts omitted for readability.}
\label{fig:safety-chain}
\end{figure}

%% file: missingslots.tikz
\begin{tikzpicture}
	\begin{pgfonlayer}{nodelayer}
		\node [style=block] (0) at (0, 0) {$b'$};
		\node [style=missing] (1) at (2, 0) {};
		\node [style=missing] (2) at (4, 0) {};
		\node [style=block] (3) at (6, 0) {$b$};
		\node [style=l] (4) at (0, 1) {$7$};
		\node [style=l] (5) at (2, 1) {$8$};
		\node [style=l] (6) at (4, 1) {$9$};
		\node [style=l] (7) at (6, 1) {$10$};
		\node [style=none] (8) at (-1.5, 0) {};
	\end{pgfonlayer}
	\begin{pgfonlayer}{edgelayer}
		\draw [style=chain] (3) to (2);
		\draw [style=chain] (2) to (1);
		\draw [style=chain] (1) to (0);
		\draw [style=chain] (0) to (8.center);
		\node at (-2.2, 0) {\textbf{\ldots}};
	\end{pgfonlayer}
\end{tikzpicture}

%% file: case1a.tikz
\begin{tikzpicture}
	\begin{pgfonlayer}{nodelayer}
		\node [style=l] (0) at (-2.5, 1.8) {$\epoch(\now)-1$};
		\node [style=lc] (1) at (5.5, 1.8) {$\epoch(\now)$\\[2pt]$\epoch(b) > \epoch(b_c)$};
		\node [style=block] (2) at (5.25, 0) {$h$};
		\node [style=block] (3) at (3.75, -1.5) {$b'$};
		\node [style=candidate] (4) at (1.95, 0) {$b$};
		\node [style=confirmed] (5) at (0.1, 0) {$b_c$};
		\node [style=confirmed] (7) at (-1.6, 0) {};
		\node [style=confirmed] (8) at (-3.2, 0) {};
		\node [style=confirmed] (9) at (-4.8, 0) {};
		\node [style=none] (10) at (-5.6, 0) {};
		\node [style=none] (11) at (1.05, 2.4) {};
		\node [style=none] (12) at (1.05, -2) {};
		\node [style=lc] (14) at (4, -2.2) {$\epoch(\gu(b'))\geq\epoch(\now)-1$};
		\node [style=lc] (15) at (3.5, 0.8) {$\willChkpBeJustified_v(\chkp(b))$};
	\end{pgfonlayer}
	\begin{pgfonlayer}{edgelayer}
		\draw [style=epochline] (11.center) to (12.center);
		\draw [style=chain] (3) to (4);
		\draw [style=chain] (4) to (5);
		\draw [style=cchain] (7) to (8);
		\draw [style=cchain] (8) to (9);
		\draw [style=cchain] (9) to (10.center);
		\draw [style=chain] (2) to (4);
		\draw [style=chain] (5) to (7);
	\end{pgfonlayer}
\end{tikzpicture}

%% file: case1b.tikz
\begin{tikzpicture}
	\begin{pgfonlayer}{nodelayer}
		\node [style=l] (0) at (-2.5, 1.8) {$\epoch(\now)-1$};
		\node [style=lc] (1) at (5.5, 1.8) {$\epoch(\now)$\\[2pt]$\epoch(b) = \epoch(b_c) = \epoch(\now)$};
		\node [style=block] (2) at (8, 0) {$h$};
		\node [style=block] (3) at (6.5, -1.5) {$b'$};
		\node [style=candidate] (4) at (4.95, 0) {$b$};
		\node [style=confirmed] (5) at (3.35, 0) {$b_c$};
		\node [style=confirmed] (6) at (0, 0) {};
		\node [style=confirmed] (7) at (-1.6, 0) {};
		\node [style=confirmed] (8) at (-3.2, 0) {};
		\node [style=none] (10) at (-4.6, 0) {};
		\node [style=none] (11) at (0.8, 2.4) {};
		\node [style=none] (12) at (0.8, -2) {};
		\node [style=confirmed] (13) at (1.75, 0) {};
		\node [style=lc] (14) at (6, -2.2) {$\epoch(\gu(b'))\geq\epoch(\now)-1$};
	\end{pgfonlayer}
	\begin{pgfonlayer}{edgelayer}
		\draw [style=epochline] (11.center) to (12.center);
		\draw [style=chain] (3) to (4);
		\draw [style=chain] (4) to (5);
		\draw [style=cchain] (6) to (7);
		\draw [style=cchain] (7) to (8);
		\draw [style=chain] (2) to (4);
		\draw [style=chain] (5) to (13);
		\draw [style=chain] (13) to (6);
		\draw [style=chain] (8) to (10.center);
	\end{pgfonlayer}
\end{tikzpicture}

%% file: case3a.tikz
\begin{tikzpicture}
	\begin{pgfonlayer}{nodelayer}
		\node [style=l] (0) at (-0.8, 1.3) {$\epoch(\now)-1$};
		\node [style=lc] (1) at (5.5, 1.3) {$\epoch(\now)$};
		\node [style=block] (2) at (1.8, 0) {};
		\node [style=block] (3) at (0.4, 0) {};
		\node [style=candidate] (4) at (-1.2, 0) {$b$};
		\node [style=confirmed] (5) at (-2.8, 0) {};
		\node [style=confirmed] (6) at (-4.4, 0) {};
		\node [style=none] (7) at (-5.2, 0) {};
		\node [style=none] (10) at (2.8, 1.15) {};
		\node [style=none] (11) at (2.8, -1.45) {};
		\node [style=lc] (12) at (-1.25, -1.2) {$\epoch(\votsource[blck=b,time=\now])\ge\epoch(\now)-2$};
		\node [style=llw, inner xsep=-8pt] (now_label) at (2.8, 2.2) {$\now=\slotstart(\firstslot(\epoch(\now)))$};
		\node [style=none] (now_tip) at (2.8, 1.4) {};
	\end{pgfonlayer}
	\begin{pgfonlayer}{edgelayer}
		\draw [style=epochline] (10.center) to (11.center);
		\draw [style=chain] (2) to (3);
		\draw [style=chain] (3) to (4);
		\draw [style=chain] (4) to (5);
		\draw [style=cchain] (5) to (6);
		\draw [style=cchain] (6) to (7.center);
		\draw [style=nowarrow] (now_label.south west) to (now_tip.center);
	\end{pgfonlayer}
\end{tikzpicture}

%% file: case3b.tikz
\begin{tikzpicture}
	\begin{pgfonlayer}{nodelayer}
		\node [style=l] (0) at (-1.5, 2.3) {$\epoch(\now)-1$};
		\node [style=lc] (1) at (5.75, 2.3) {$\epoch(\now)$};
		\node [style=block] (2) at (6.25, 0) {$h$};
		\node [style=block] (3) at (3.25, 0) {};
		\node [style=block] (4) at (-0.55, 0) {};
		\node [style=candidate] (5) at (-2.15, 0) {$b$};
		\node [style=confirmed] (6) at (-3.75, 0) {};
		\node [style=none] (8) at (-4.9, 0) {};
		\node [style=none] (9) at (2.2, 2.6) {};
		\node [style=none] (10) at (2.2, -2) {};
		\node [style=l] (now_label) at (7, 1.7) {$\now$};
		\node [style=none] (now_tip) at (7, 0.65) {};
		\node [style=block] (11) at (4.75, -1.5) {$b'$};
		\node [style=l] (12) at (4.75, -2.3) {$\epoch(\gu(b'))\geq\epoch(\now)-1$};
		\node [style=llw] (14) at (7.75, 1.25) {$\slot(\now)>$\\[2pt]$\firstslot(\epoch(\now))$};
		\node [style=lc] (15) at (-2.25, -0.95) {$\epoch(\votsource[blck=b,time=\now])\ge\epoch(\now)-2$};
		\node [style=block] (17) at (1.25, 0) {};
	\end{pgfonlayer}
	\begin{pgfonlayer}{edgelayer}
		\draw [style=epochline] (9.center) to (10.center);
		\draw [style=chain] (2) to (3);
		\draw [style=chain] (4) to (5);
		\draw [style=chain] (5) to (6);
		\draw [style=nowarrow] (now_label) to (now_tip.center);
		\draw [style=chain] (3) to (17);
		\draw [style=chain] (17) to (4.east);
		\draw [style=chain] (11) to (4.east);
		\draw [style=cchain] (6) to (8.center);
	\end{pgfonlayer}
\end{tikzpicture}

%% file: case3b-noquorum.tikz
\begin{tikzpicture}
	\begin{pgfonlayer}{nodelayer}
		\node [style=l] (ep_prev) at (-4.25, 2.8) {$\epoch(\now)\!-\!2$};
		\node [style=lc] (ep_curr) at (0, 2.8) {$\epoch(\now)\!-\!1$};
		\node [style=lc] (ep_next) at (4.25, 2.8) {$\epoch(\now)$};
		\node [style=lc] (ep_next2) at (7.5, 2.8) {$\epoch(\now)\!+\!1$};
		\node [style=none] (ep0_top) at (-6.375, 2.6) {};
		\node [style=none] (ep0_bot) at (-6.375, -3.2) {};
		\node [style=none] (genesis) at (-6.8, 0) {};
		\node [style=confirmed] (pre1) at (-5.7, 0) {$C'$};
		\node [style=confirmed] (pre2) at (-3.7, 0) {};
		\node [style=none] (ep1_top) at (-2.125, 2.6) {};
		\node [style=none] (ep1_bot) at (-2.125, -3.2) {};
		\node [style=confirmed] (chkp) at (-1.4, 0) {$\chkp(b)$};
		\node [style=confirmed] (bpar) at (0.25, 0) {};
		\node [style=candidate] (b) at (1.45, 0) {$b$};
		\node [style=none] (ep2_top) at (2.125, 2.6) {};
		\node [style=none] (ep2_bot) at (2.125, -3.2) {};
		\node [style=block] (fk) at (3.2, 0) {};
		\node [style=block] (up1) at (4.4, 1.2) {};
		\node [style=block] (up1b) at (5.7, 1.2) {};
		\node [style=block] (up2) at (4.4, 0) {};
		\node [style=block] (up2b) at (5.7, 0) {};
		\node [style=block] (lo1) at (0.25, -2) {};
		\node [style=block] (bpp) at (1.45, -2) {};
		\node [style=block] (lo2) at (5.7, -2) {$b''$};
		\node [style=none] (ep3_top) at (6.375, 2.6) {};
		\node [style=none] (ep3_bot) at (6.375, -3.2) {};
		\node [style=l] (now_label) at (3.7, 2.25) {$\now$};
		\node [style=none] (now_tip) at (3.7, 0.65) {};
		\node [style=llw] (legend) at (-5.55, -2.75) {\small \textcolor{red}{$\rightarrow$}: \ujustifies};
		\node [style=llw] (cprime_label) at (-5.95, -0.9) {\small $C'\!=\!\chkp(b,\epoch(\now)\!-\!2)$};
	\end{pgfonlayer}
	\begin{pgfonlayer}{edgelayer}
		\draw [style=epochline] (ep0_top.center) to (ep0_bot.center);
		\draw [style=epochline] (ep1_top.center) to (ep1_bot.center);
		\draw [style=epochline] (ep2_top.center) to (ep2_bot.center);
		\draw [style=epochline] (ep3_top.center) to (ep3_bot.center);
		\draw [style=cchain] (pre1) to (genesis.center);
		\draw [style=cchain] (pre2) to (pre1);
		\draw [style=cchain] (chkp) to (pre2);
		\draw [style=cchain] (bpar) to (chkp);
		\draw [style=chain] (b) to (bpar);
		\draw [style=chain] (fk) to (b);
		\draw [style=chain] (up1) to (fk);
		\draw [style=chain] (up1b) to (up1);
		\draw [style=chain] (up2) to (fk);
		\draw [style=chain] (up2b) to (up2);
		\draw [style=chain] (lo1) to (bpar);
		\draw [style=chain] (bpp) to (lo1);
		\draw [style=chain] (lo2) to (bpp);
		\draw [style=ujustifies, bend left=345] (up1b) to (pre1);
		\draw [style=ujustifies, bend left=345] (up2b) to (pre1);
		\draw [style=ujustifies, bend left=15] (lo2) to (chkp);
		\draw [style=nowarrow] (now_label) to (now_tip.center);
		\draw [lightgray, decorate, decoration={brace, amplitude=5pt}] (6.5, 1.6) -- (6.5, -0.4) node [midway, right, xshift=4pt, font=\small, black] {filtered out by $\FILHFC$};
		\draw [lightgray] (6.5, 1.6) -- (6.2, 1.6);
		\draw [lightgray] (6.5, -0.4) -- (6.2, -0.4);
	\end{pgfonlayer}
\end{tikzpicture}

%% file: case4a.tikz
\begin{tikzpicture}
	\begin{pgfonlayer}{nodelayer}
		\node [style=l] (0) at (-0.8, 1.3) {$\epoch(\now)-1$};
		\node [style=lc] (1) at (5.5, 1.3) {$\epoch(\now)$};
		\node [style=block] (2) at (1.8, 0) {$b'$};
		\node [style=block] (3) at (0.4, 0) {};
		\node [style=candidate] (4) at (-1.2, 0) {$b$};
		\node [style=confirmed] (5) at (-2.8, 0) {};
		\node [style=confirmed] (6) at (-4.4, 0) {};
		\node [style=none] (7) at (-5.2, 0) {};
		\node [style=none] (10) at (2.8, 1.15) {};
		\node [style=none] (11) at (2.8, -2.2) {};
		\node [style=ll] (12) at (-1.25, -1.2) {$\epoch(\votsource[blck=b,time=\now])<\epoch(\now)-2$};
		\node [style=llw, inner xsep=-8pt] (now_label) at (2.8, 2.2) {$\now=\slotstart(\firstslot(\epoch(\now)))$};
		\node [style=none] (now_tip) at (2.8, 1.4) {};
		\node [style=llw] (13) at (3, 0.05) {$\epoch(\votsource[blck=b',time=\now])\geq\epoch(\now)-2$\\[2pt]$\lnot\isOneConfirmed(b',\guattime[val=v,time=\prevfirstslotepoch{\now}])$};
	\end{pgfonlayer}
	\begin{pgfonlayer}{edgelayer}
		\draw [style=epochline] (10.center) to (11.center);
		\draw [style=chain] (2) to (3);
		\draw [style=chain] (3) to (4);
		\draw [style=chain] (4) to (5);
		\draw [style=cchain] (5) to (6);
		\draw [style=cchain] (6) to (7.center);
		\draw [style=nowarrow] (now_label.south west) to (now_tip.center);
	\end{pgfonlayer}
\end{tikzpicture}

%% file: case4b.tikz
\begin{tikzpicture}
	\begin{pgfonlayer}{nodelayer}
		\node [style=l] (0) at (-1.5, 1.8) {$\epoch(\now)-1$};
		\node [style=lc] (1) at (5.5, 1.8) {$\epoch(\now)$};
		\node [style=block] (2) at (6.25, 0) {$h$};
		\node [style=block] (3) at (3.25, 0) {};
		\node [style=block] (4) at (-0.55, 0) {};
		\node [style=candidate] (5) at (-2.15, 0) {$b$};
		\node [style=confirmed] (6) at (-3.75, 0) {};
		\node [style=none] (8) at (-4.9, 0) {};
		\node [style=none] (9) at (2.2, 2.1) {};
		\node [style=none] (10) at (2.2, -2) {};
		\node [style=l] (now_label) at (7, 2.2) {$\now$};
		\node [style=none] (now_tip) at (7, 1.15) {};
		\node [style=block] (11) at (4.75, -1.5) {$b'$};
		\node [style=llw] (12) at (5.25, -1.55) {$\epoch(\gu(b'))\geq\epoch(\now)-1$\\[2pt]$\lnot \isOneConfirmed(b',C)$};
		\node [style=lc] (14) at (9.75, 1.75) {$\slot(\now) > \firstslot(\epoch(\now))$};
		\node [style=ll] (18) at (-2.25, -0.95) {$\epoch(\votsource[blck=b,time=\now])<\epoch(\now)-2$};
		\node [style=llw] (19) at (-1.5, -2.45) {$\epoch(\votsource[blck=b'',time=\now])\geq\epoch(\now)-2$\\[2pt]$\epoch(\gu(b''))<\epoch(\now)-1$};
		\node [style=block] (20) at (1.25, -1.5) {$b''$};
	\end{pgfonlayer}
	\begin{pgfonlayer}{edgelayer}
		\draw [style=epochline] (9.center) to (10.center);
		\draw [style=chain] (2) to (3);
		\draw [style=chain] (4) to (5);
		\draw [style=chain] (5) to (6);
		\draw [style=nowarrow] (now_label) to (now_tip.center);
		\draw [style=chain] (11) to (4.east);
		\draw [style=cchain] (6) to (8.center);
		\draw [style=chain] (20) to (4.east);
		\draw [style=chain] (3) to (4.east);
	\end{pgfonlayer}
\end{tikzpicture}

%% file: empty-slot-beginning.tikz
\begin{tikzpicture}
	\begin{pgfonlayer}{nodelayer}
		\node [style=l] (0) at (-2, 2) {$e-1$};
		\node [style=l] (1) at (7, 2) {$e$};
		\node [style=block] (2) at (-3, 0) {};
		\node [style=block] (3) at (-1.5, 0) {$\parent(b)$};
		\node [draw, dashed, shape=rectangle, minimum size=0.8cm, thick] (4) at (0, 0) {};
		\node [draw, dashed, shape=rectangle, minimum size=0.8cm, thick] (5) at (3.5, 0) {};
		\node [draw, dashed, shape=rectangle, minimum size=0.8cm, thick] (6) at (5, 0) {};
		\node [style=block] (7) at (6.5, 0) {$b$};
		\node [style=block] (8) at (8, 0) {};
		\node [style=block] (9) at (9.5, 0) {};
		\node [style=block] (10) at (11, 0) {};
		\node [style=l, font=\scriptsize, gray] (11) at (0, 1.2) {empty};
		\node [style=l, font=\scriptsize, gray] (12) at (3.5, 1.2) {empty};
		\node [style=l, font=\scriptsize, gray] (13) at (5, 1.2) {empty};
		\node [style=l, font=\footnotesize] (14) at (3.5, -1.0) {$\firstslot(e)$};
		\node [style=l, font=\footnotesize] (15) at (6.5, -1.0) {$\slot(b)$};
		\node [style=none] (16) at (1.75, 2.5) {};
		\node [style=none] (17) at (1.75, -1.5) {};
		\node [style=l, font=\footnotesize] (18) at (8.75, -2.8) {$\isOneConfirmed$: from $\slot(b)$};
		\node [style=l, font=\footnotesize] (19) at (7.25, -4.6) {$\isOneConfirmedSpecial$: from $\firstslot(e)$};
		\node [style=l, font=\scriptsize] (20) at (4.6, -3.2) {additional committees};
		\node [style=none] (21) at (-4.2, 0) {};
	\end{pgfonlayer}
	\begin{pgfonlayer}{edgelayer}
		\draw [style=epochline] (16.center) to (17.center);
		\draw [style=chain] (2) to (21.center);
		\draw [style=chain] (3) to (2);
		\draw [style=chain] (4) to (3);
		\draw [style=chain] (5) to (4);
		\draw [style=chain] (6) to (5);
		\draw [style=chain] (7) to (6);
		\draw [style=chain] (8) to (7);
		\draw [style=chain] (9) to (8);
		\draw [style=chain] (10) to (9);
		\draw [decorate, decoration={brace, amplitude=6pt, mirror, raise=4pt}]
			(6.1, -1.6) -- (11.4, -1.6);
		\draw [decorate, decoration={brace, amplitude=6pt, mirror, raise=4pt}]
			(3.1, -3.4) -- (11.4, -3.4);
		\draw [->, thick] (3.5, -3.0) -- (6.2, -3.0);
	\end{pgfonlayer}
\end{tikzpicture}